%% file: ms.tex
\documentclass[iop]{emulateapj}

\bibpunct{(}{)}{;}{a}{}{,}

\slugcomment{Submitted to the ApJ}

\shorttitle{ALMA Imaging of {\it Herschel} DSFGs}
\shortauthors{Bussmann et al.}

\begin{document}

\defcitealias{HB13}{HB13}
\defcitealias{Cowley:2015lr}{C15}

\title{HerMES: ALMA Imaging of {\it Herschel}$^{\dagger}$-selected Dusty
Star-forming Galaxies}


\author{R.~S.~Bussmann\altaffilmark{1},
D.~Riechers\altaffilmark{1},
A.~Fialkov\altaffilmark{2},
J.~Scudder\altaffilmark{3},
C.~C.~Hayward\altaffilmark{4,5},
W.~I.~Cowley\altaffilmark{6},
J.~Bock\altaffilmark{7,8},
J.~Calanog\altaffilmark{9},
S.~C.~Chapman\altaffilmark{10},
A.~Cooray\altaffilmark{9},
F.~De Bernardis\altaffilmark{9},
D.~Farrah\altaffilmark{11},
Hai~Fu\altaffilmark{12},
R.~Gavazzi\altaffilmark{13},
R.~Hopwood\altaffilmark{14},
R.~J.~Ivison\altaffilmark{15,16},
M.~Jarvis\altaffilmark{17,18},
C.~Lacey\altaffilmark{6},
A.~Loeb\altaffilmark{5},
S.~J.~Oliver\altaffilmark{3},
I.~P{\'e}rez-Fournon\altaffilmark{19,20},
D.~Rigopoulou\altaffilmark{21,17},
I.~G.~Roseboom\altaffilmark{3,16},
Douglas~Scott\altaffilmark{22},
A.~J.~Smith\altaffilmark{3},
J.~D.~Vieira\altaffilmark{23},
L.~Wang\altaffilmark{6,24},
J.~Wardlow\altaffilmark{25}}
\altaffiltext{$\dagger$}{Herschel is an ESA space observatory with science
instruments provided by European-led Principal Investigator consortia and with
important participation from NASA.}
\altaffiltext{1}{Department of Astronomy, Space Science Building, Cornell University, Ithaca, NY, 14853-6801}
\altaffiltext{2}{Departement de Physique, Ecole Normale Superieure, CNRS, 24 rue Lhomond, 75005 Paris, France}
\altaffiltext{3}{Astronomy Centre, Dept. of Physics \& Astronomy, University of Sussex, Brighton BN1 9QH, UK}
\altaffiltext{4}{TAPIR 350-17, California Institute of Technology, 1200 E. California Boulevard, Pasadena, CA 91125}
\altaffiltext{5}{Harvard-Smithsonian Center for Astrophysics, 60 Garden Street, Cambridge, MA 02138}
\altaffiltext{6}{Institute for Computational Cosmology, Department of Physics, University of Durham, South Road, Durham, DH1 3LE, UK}
\altaffiltext{7}{California Institute of Technology, 1200 E. California Blvd., Pasadena, CA 91125}
\altaffiltext{8}{Jet Propulsion Laboratory, 4800 Oak Grove Drive, Pasadena, CA 91109}
\altaffiltext{9}{Dept. of Physics \& Astronomy, University of California, Irvine, CA 92697}
\altaffiltext{10}{Institute of Astronomy, University of Cambridge, Madingley Road, Cambridge CB3 0HA, UK}
\altaffiltext{11}{Department of Physics, Virginia Tech, Blacksburg, VA 24061}
\altaffiltext{12}{Department of Physics and Astronomy, The University of Iowa, 203 Van Allen Hall, Iowa City, IA 52242}
\altaffiltext{13}{Institut d'Astrophysique de Paris, UMR 7095, CNRS, UPMC Univ. Paris 06, 98bis boulevard Arago, F-75014 Paris, France}
\altaffiltext{14}{Astrophysics Group, Imperial College London, Blackett Laboratory, Prince Consort Road, London SW7 2AZ, UK}
\altaffiltext{15}{UK Astronomy Technology Centre, Royal Observatory, Blackford Hill, Edinburgh EH9 3HJ, UK}
\altaffiltext{16}{Institute for Astronomy, University of Edinburgh, Royal Observatory, Blackford Hill, Edinburgh EH9 3HJ, UK}
\altaffiltext{17}{Department of Astrophysics, Denys Wilkinson Building, University of Oxford, Keble Road, Oxford OX1 3RH, UK}
\altaffiltext{18}{Astrophysics Group, Physics Department, University of the Western Cape, Private Bag X17, 7535, Bellville, Cape Town, South Africa}
\altaffiltext{19}{Instituto de Astrof{\'\i}sica de Canarias (IAC), E-38200 La Laguna, Tenerife, Spain}
\altaffiltext{20}{Departamento de Astrof{\'\i}sica, Universidad de La Laguna (ULL), E-38205 La Laguna, Tenerife, Spain}
\altaffiltext{21}{RAL Space, Rutherford Appleton Laboratory, Chilton, Didcot, Oxfordshire OX11 0QX, UK}
\altaffiltext{22}{Department of Physics \& Astronomy, University of British Columbia, 6224 Agricultural Road, Vancouver, BC V6T~1Z1, Canada}
\altaffiltext{23}{Department of Astronomy and Department of Physics, University of Illinois, 1002 West Green Street, Urbana, IL 61801}
\altaffiltext{24}{SRON Netherlands Institute for Space Research, Landleven 12, 9747 AD, Groningen, The Netherlands}
\altaffiltext{25}{Dark Cosmology Centre, Niels Bohr Institute, University of
Copenhagen, Juliane Maries Vej 30, 2100 Copenhagen, Denmark}


\begin{abstract}

The {\it Herschel} Multi-tiered Extragalactic Survey (HerMES) has identified large numbers of dusty star-forming galaxies (DSFGs) over a wide range in redshift.  A detailed understanding of these DSFGs is hampered by the limited spatial resolution of {\it Herschel}.  We present 870$\,\mu$m 0$\farcs$45 resolution imaging from the Atacama Large Millimeter/submillimeter Array (ALMA) of 29 HerMES DSFGs with far-infrared (FIR) flux densities in between the brightest of sources found by {\it Herschel} and fainter DSFGs found in ground-based sub-millimeter (sub-mm) surveys.  We identify 62 sources down to the $5\sigma$ point-source sensitivity limit in our ALMA sample ($\sigma\approx0.2\,$mJy), of which 6 are strongly lensed (showing multiple images) and 36 experience significant amplification ($\mu>1.1$).  To characterize the properties of the ALMA sources, we introduce and make use of {\sc uvmcmcfit}, a publicly available Markov chain Monte Carlo analysis tool for interferometric observations of lensed galaxies.  Our lens models tentatively favor intrinsic number counts for DSFGs with a steep fall off above 8$\,$mJy at 880$\,\mu$m.  Nearly 70\% of the {\it Herschel} sources comprise multiple ALMA counterparts, consistent with previous research indicating that the multiplicity rate is high in bright sub-mm sources.  Our ALMA sources are located significantly closer to each other than expected based on results from theoretical models as well as fainter DSFGs identified in the LABOCA ECDFS Submillimeter Survey.  The high multiplicity rate and low projected separations argue in favor of interactions and mergers driving the prodigious emission from the brightest DSFGs as well as the sharp downturn above $S_{880}=8\,$mJy.

\end{abstract}



\section{Introduction} \label{sec:intro} 

Galaxies selected in blind surveys at far-infrared (FIR) or sub-millimeter
(sub-mm) wavelengths are generally known as dusty star-forming galaxies
\citep[DSFGs; for a recent review, see][]{Casey:2014lr}.  They cover a wide
range in redshift from $z \sim 0.5$ to $z > 6$ \citep{2005ApJ...622..772C,
Casey:2012qy, Messias:2014fk, Riechers:2013lr}, with a significant component at
$z \sim 2$ \citep{Casey:2012uq, Bothwell:2013lr}, when they represent the most
FIR-luminous objects in existence during this epoch.   They are usually
signposts of significant over-densities \citep{Daddi:2009qy, Capak:2011qy}
\citep[c.f.][]{Robson:2014xy} and likely represent the formative stages of the
most massive elliptical galaxies found in the local Universe
\citep[e.g.,][]{Ivison:2013fk, Fu:2013lr}.  Moreover, they constitute an
important component of the overall galaxy population at $z \sim 2$
\citep[e.g.,][]{Magnelli:2011ul}, when the star-formation rate density in the
Universe peaked \citep[e.g.,][]{Lilly:1996uq, 1996MNRAS.283.1388M}.  

Our collective understanding of DSFGs is currently taking a dramatic leap
forward thanks in large part to the {\it Herschel Space Observatory}
\citep[{\it Herschel};][]{Pilbratt:2010fk}.  {\it Herschel} has revolutized the size and
depth of blind surveys at FIR wavelengths.  In particular, the {\it Herschel}
Multi-tiered Extragalactic Survey \citep[HerMES;][]{Oliver:2012lr} and the {\it
Herschel} Astrophysical Terahertz Large Area Survey
\citep[H-ATLAS;][]{2010PASP..122..499E} together have surveyed $\approx
650\,$deg$^2$ at 250$\,\mu$m, 350$\,\mu$m, and 500$\,\mu$m to the confusion
limit of {\it Herschel} \citep[$\sigma \approx 6-7\,$mJy in each
band][]{Nguyen:2010fk}, plus an additional $\approx 350\,$deg$^2$ to a
shallower level (approximately double the confusion limit).  A similar effort
to survey large areas of the sky has been undertaken at longer wavelengths by
the South Pole Telescope \citep[SPT;][]{Carlstrom:2011qy} and the Atacama
Cosmology Telescope \citep{Swetz:2011qy}.

Theoretical expectations based on the redshift distribution and luminosity
function of DSFGs suggested that HerMES and H-ATLAS would be efficient tools
for discovering strongly lensed DSFGs \citep[e.g.,][]{1996MNRAS.283.1340B,
2007MNRAS.377.1557N}.  Submillimeter Array \citep[SMA;][]{Ho:2004lr} imaging at
870$\,\mu$m with sub-arcsecond resolution has confirmed this, with $\geq 85\%$
of the brightest sources found by {\it Herschel} that satisfy $S_{500} >
100\,$mJy being gravitationally lensed by an intervening galaxy or group of
galaxies along the line of sight \citep{Negrello:2010fk, Conley:2011lr,
Riechers:2011uq, Bussmann:2012lr, Wardlow:2013lr, Bussmann:2013lr}.  Sources
discovered in SPT surveys have also been shown to have a high probability of
being strongly lensed \citep{Vieira:2013fk, Hezaveh:2013fk}.  However,
statistical models significantly over-predict the median magnification factor
experienced by a {\it Herschel} DSFG of a given $S_{500}$
\citep{Bussmann:2013lr}.  This could herald new insights in our understanding
of the bright end of the intrinsic DSFG number counts or in the nature of the
deflectors.

We here present Atacama Large Millimeter/submillimeter Array (ALMA) Cycle~0
imaging at 870$\,\mu$m of a sample of 29 HerMES DSFGs.
Three aspects of our dataset make it uniquely suited to improving our
understanding of the bright end of the intrinsic DSFG number counts.  First,
the sample occupies a distinct regime in flux density between the brightest
{\it Herschel} DSFGs (almost all of which are lensed) and much fainter DSFGs
found in ground-based surveys \citep[most of which are expected to be unlensed;
e.g.,][]{Hodge:2013qy}.  Second, the ALMA images are extremely sensitive (rms
point source sensitivity of $\sigma \approx 0.2\,$mJy) and all 29 HerMES DSFGs
are detected \citep[which was not the case in previous similar studies with
shallower imaging; e.g.,][]{Smolcic:2012zl, Barger:2012yg, Hodge:2013qy}.
Third, the typical angular resolution is $0\farcs45$ and nearly all sources
detected by ALMA are spatially resolved.

We also obtained Gemini-South optical imaging to complement our existing set
of ancillary multi-wavelength imaging.  We use those data in this paper to
identify lensing galaxies, which are typically early-types with little on-going
star-formation and therefore exhibit very weak sub-mm emission.

In Section~\ref{sec:obs}, we characterize our sample and present our ALMA and
Gemini-South imaging.  Section~\ref{sec:modelfits} presents our model fitting
methodology and model fits for all ALMA sources (lensed and unlensed) using
{\sc uvmcmcfit}, a publicly available \footnote{\url
https://github.com/sbussmann/uvmcmcfit} modified version of the visibility
plane lens modeling software used in \citet{Bussmann:2012lr, Bussmann:2013lr}.
Results on the effect of lensing for the observed properties of the {\it
Herschel} DSFGs in our sample, as well as the multiplicity rate and typical
angular separation between sources after delensing the ALMA sources, appear in
Section~\ref{sec:results}.  We scrutinize statistical predictions for the
magnification factor at 870$\,\mu$m ($\mu_{870}$) as a function of the flux
density at 870 $\,\mu$m ($S_{870}$) and discuss implications for the bright end
of the DSFG number counts in Section~\ref{sec:discuss}.  Finally, we present
our conclusions in Section~\ref{sec:conclusions}.

Throughout this paper, we assume a flat cosmology with
$H_0=$69~km~s$^{-1}$~Mpc$^{-1}$, $\Omega_{\rm m_0} = 0.29$
\citep{Hinshaw:2013ty}.

\section{Data}\label{sec:obs}

In this section, we describe the selection of our {\it Herschel} DSFG sample,
present our ALMA high-spatial resolution imaging of thermal dust emission, and
present Gemini-S optical imaging that we use to identify intervening galaxies
along the line of sight.  

\subsection{Selection of DSFG Sample}\label{sec:select}

The starting point for the sample selection is source extraction and
photometry.  For the objects in this paper, individual catalogs were generated
for each of the 250$\, \mu$m, 350$\,\mu$m, and 500$\,\mu$m {\it Herschel}
Spectral and Photometric Imaging REceiver \citep[SPIRE;][]{Griffin:2010lr}
channels using the SUSSEXtractor peak finder algorithm \citep{Savage:2007sf}.
Our sample includes 29 DSFGs drawn from five independent, confusion-limited
fields in HerMES with declinations below $+2 ^\circ$ and totaling $55\,$deg$^2$.

The sample was selected to be the 29 brightest DSFGs in the Southern sky that
are not known radio AGN, nearby late-type galaxies, or Galactic emission. The
selection was designed to assemble a large sample of lensed galaxies in the
ALMA-accessible HerMES fields, and was constructed from the SUSSEXtractor
catalogs \citep{Smith:2012uq}, which were available prior to the ALMA Cycle~0
deadline.  Subsequently, improved efforts to deblend SPIRE photometry at
500$\,\mu$m using StarFinder \citep{Wang:2014lr} were introduced that formed
the basis of the lens selection criteria used in \citet{Wardlow:2013lr}.  As a
result of the improved deblending algorithms in the StarFinder catalogues and
\citet{Wardlow:2013lr} a number of objects in our sample have significantly
lower $S_{500}$ values in the StarFinder catalog than in the original
SUSSEXtractor catalogs.  This and further investigation into the StarFinder
catalogs shows that their original flux was boosted by blending with nearby
sources rather than by gravitational lensing.  For this reason, the objects in
this sample comprise a combination of lenses and blends of multiple sources.

We used positional priors based on the ALMA data presented in this paper to
obtain the best possible estimates of the total SPIRE flux densities for each
{\it Herschel} source.  We also used {\it Spitzer}/MIPS
\citep{2004ApJS..154...25R} imaging to take into consideration the presence of
nearby 24$\,\mu$m sources that are not detected by ALMA but may still
contribute to the 250$\,\mu$m emission detected by {\it Herschel}.  Additional
details on our methodology are provided in Appendix~\ref{sec:photometry}.  The
SPIRE flux densities measured in this way represent our ``fiducial'' flux
densities and are presented in Table~\ref{tab:position}.  Interested readers
may refer to Table~\ref{tab:photometry} for a comparison of the fiducial,
StarFinder and SUSSEXtractor flux densities in tabular form. 


Figure~\ref{fig:sample} shows that the {\it Herschel}-ALMA sample is set
clearly apart from the very bright {\it Herschel} DSFGs that are selected to
have $S_{500} > 100 \, $mJy and have been shown to be almost entirely lensed
DSFGs \citep{Negrello:2010fk, Wardlow:2013lr, Bussmann:2013lr}.  In contrast,
the sample in this paper is expected to include a mix of lensed and unlensed
DSFGs.  On the other hand, the HerMES survey area is 200 times larger than that
of the Large Apex Bolometer Camera Extended Chandra Deep Field Survey
\citep[LESS][]{Weis:2009ly}.  This explains why the median $S_{500}$ in our
sample is $\sim4$ times brighter than the median $S_{500}$ in the sample of
ALMA-detected sources in LESS, known as ALESS \citep{Swinbank:2014lr}.  Our {\it
Herschel}-ALMA sample opens a new window of discovery space on the bright end
of the DSFG number counts.

\begin{figure}[!tbp] 
\includegraphics[width=\linewidth]{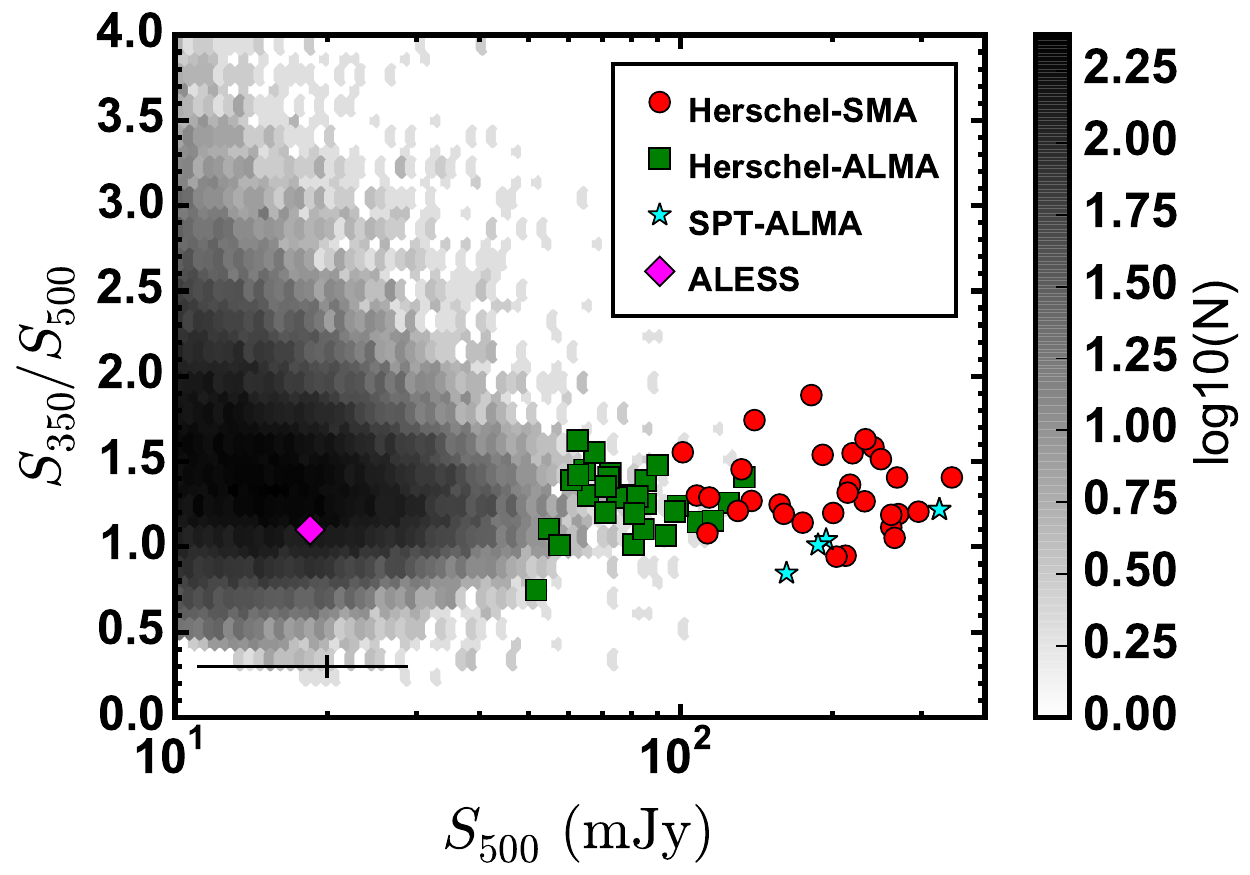}

\caption{ {\it Herschel}/SPIRE photometry of all galaxies in the HerMES phase~I
catalog with declination $< +2\,^\circ$ and signal to noise ratio greater than
5 at 350$\,\mu$m and 500$\,\mu$m (log of number of galaxies shown in
grayscale).  The sample of HerMES sources in this paper are shown with green
squares (``{\it Herschel}-ALMA'').  The very bright {\it Herschel} DSFGs from
\citet{Bussmann:2013lr} (``{\it Herschel}-SMA'') are shown by red circles, and
lensed SMGs discovered by the SPT that have published lens models
(``SPT-ALMA'') are represented by cyan stars \citep{Hezaveh:2013fk}.  A magenta
diamond shows the location in this diagram of the stacked signal from ALESS
DSFGs.  Representative error bars are shown in the lower left corner.  The {\it
Herschel}-ALMA sample fills the gap in 500$\,\mu$m flux density space between
50-100$\,$mJy.} \label{fig:sample}

\end{figure}

In detail, two of the sources in the {\it Herschel}-ALMA sample (HXMM01 and
HXMM02) overlap with the ``confirmed lensed'' sample in \citet{Wardlow:2013lr}
as well as with the {\it Herschel}-SMA sample in \citet{Bussmann:2013lr}.  A
further eight appear in the ``Supplementary sample'' of \citet{Wardlow:2013lr}.
The remainder have $S_{500} < 80\,$mJy and thus do not appear in
\citet{Wardlow:2013lr}.


Table~\ref{tab:position} provides reference data for the {\it Herschel}-ALMA
sample, including centroid positions measured from the ALMA 870$\, \mu$m
imaging (see Section~\ref{sec:almaobs}).  The centroid positions serve as the
reference point for subsequent offset positions of lenses and sources described
in later tables.  This is a useful choice (rather than the SPIRE centroid or
ALMA phase center) because it minimizes the number of pixels needed to generate
a simulated model of the source — and therefore minimized memory and cpu usage
when lens modeling.

\subsection{ALMA Observations}\label{sec:almaobs}

ALMA data were obtained during Cycle~0 over the from 2012 June to 2012 December
(Program 2011.0.00539.S; PI: D. Riechers).  The observations were carried out
in good 870$\,\mu$m weather conditions, which resulted in typical system
temperatures of $T_{\rm sys} \approx 130\,$K and phase fluctuations of $\sim
10\,^\circ$.  Each target was observed until an rms point-source noise level
near the phase center of $\sigma \approx 0.2\,$mJy per beam was achieved.  This
typically required 10 minutes of on-source integration time.  For the
observations targeting the CDFS, ELAISS, and COSMOS fields, the data reach
$\sigma \approx 0.14\,$mJy per beam.  The number of antennas used varied from
15 to 25.  The antennas were configured with baseline lengths of 20$\,$m to
400$\,$m, providing a synthesized beamsize of $\approx 0\farcs5 \times
0\farcs4$ FWHM while ensuring that no flux was resolved out by the
interferometer (since our targets all have size scales smaller than $1\arcsec-2
\arcsec$.  When possible, track-sharing of multiple targets in a single track
was used to optimize the {\it uv} coverage.

The quasars J0403$-$360, J2258$-$279, B0851$+$202, and J2258$-$279 were used
for bandpass and pointing calibration.  The quasars J0403$-$360, J0106$-$405,
J0519$-$454, J1008$+$063, and J0217$+017$ were used for amplitude and phase
gain calibration.  The following solar system objects were used for absolute
flux calibration: Callisto (CDFS targets); Neptune (XMM targets); Titan (COSMOS
targets); and Uranus (ADFS and XMM targets).  For HELAISS02, no solar system
object was observed.  Instead, J2258$-$279 was used for absolute flux
calibration, with the flux fixed according to a measurement made two days prior
to the observations of HELAISS02.

All observations were conducted with the correlator in ``Frequency Domain
Mode'', providing a total usable bandwidth of 7.5$\,$GHz with spectral windows
centered at 335.995$\,$GHz, 337.995$\,$GHz, 345.995$\,$GHz, 347.996$\,$GHz.  We
searched for evidence of serendipitous spectral lines but found none (typical
sensitivity is $\sigma \approx 8\,$mJy$\,$beam$^{-1}$ in 15$\,$km$\,$sec$^{-1}$
bins).  Given that our observations cover a total of 217.5$\,$GHz in bandwidth,
the lack of lines seems more likely to be due to limited sensitivity than
limited bandwidth.

We used the Common Astronomy Software Applications (CASA, version 4.2.1)
package to re-reduce the data provided by the North American ALMA Science
Center (NAASC).  We found that the quality of the processed data from the NAASC
was very high.  However, we achieved a significant improvement in the case of
the ADFS and XMM targets by excluding data sets with moderate $T_{\rm sys}$ and
poor phase fluctuations.  For a handful of targets with peak signal-to-noise
ratio (S/N) greater than 20, we obtained a $\approx 10\%$ improvement in S/N by
using the CASA {\sc selfcal} task with the clean component model as input to
improve the phase gain corrections.  Finally, we updated the absolute flux
calibration to use the Butler-JPL-Horizons 2012 solar system models
\footnote{\url
https://science.nrao.edu/facilities/alma/aboutALMA/Technology/ALMA\_Memo\_Series/alma594/abs594}.

For imaging, we used the CASA {\sc Clean} task with Briggs weighting and ``robust
= $+$0.5'' to achieve an optimal balance between sensitivity and spatial
resolution.  We selected the multi-frequency synthesis option to optimize {\it
uv} coverage.  We designed custom masks for each target in CASA to ensure that
only regions with high S/N were considered during the cleaning process.

Figure~\ref{fig:imaging} presents our ALMA images (color scale) in comparison to
the {\it Herschel} SPIRE images (black-white contours) originally used to
select the targets and noted in each panel as either 250$\,\mu$m, 350$\,\mu$m,
or 500$\,\mu$m.  Each panel is centered on the phase center of the ALMA
observations of that target and a white circle traces the FWHM of the primary
beam of an ALMA 12$\,$m antenna at 870$\,\mu$m.  All flux density measurements
given in this paper have been corrected for the primary beam by dividing the
total flux density by the primary beam correction factor at the center of the
source.  This is a valid approach because all sources have sizes $< 1\arcsec$,
such that the variation in the primary beam correction factor across the source
is insignificant.  A white dashed box represents the region of each image that
is shown in greater detail in Figure~\ref{fig:uvmodels}.

In most targets, the peak of the SPIRE map is spatially coincident with the
location of the ALMA sources.  In one case where two ALMA sources are separated
by $\approx 10\arcsec$ (HADFS08), the elongation in the SPIRE 250$\,\mu$m map
is consistent with the angular separation of the two ALMA counterparts.
Otherwise, the SPIRE imaging is consistent with a single component located at
the centroid of the ALMA sources.  This result is not a surprise, given the
typical angular separation of the ALMA sources ($\lesssim 5\arcsec$) and the
FWHM of the SPIRE beam at 250$\,\mu$m (18.1$\arcsec$).  We identify and catalog
by-eye all sources with peak flux density greater than 5$\sigma$.  

\subsection{Gemini-South Imaging}\label{sec:geminiobs}

Optical imaging observations using the Gemini Multi-Object Spectrograph-South
\citep[GMOS-S;][]{Hook:2004qy} were conducted in queue mode during the 2013B
semester as part of program GS-2013B-Q-77 (PI: R.~S.~Bussmann).  The goal of
the program is to use shallow $u$, $g$, $r$, $i$, and $z$ imaging to identify
structure at redshifts below unity and determine which of the ALMA sources are
affected by gravitational lensing.  Nearly half of the ALMA sources lie in
regions with existing deep optical imaging, thanks to the extensive
multi-wavelength dataset available in the HerMES fields --- these were excluded
from our Gemini-S program.  The remaining targets are: HADFS03, HADFS08,
HADFS09, HADFS10, HADFS02, HADFS04, HADFS01, HADFS11, HELAISS02, HXMM11,
HXMM12, HXMM22, HXMM07, HXMM30, and HXMM04.  Each of these targets were
observed for a total of 9$\,$minutes of on-source integration time in each of
$u$, $g$, $r$, $i$, and $z$.  The observations were obtained during dark time
in adequate seeing conditions (image quality in the 85th percentile,
corresponding to $\approx 1.1\arcsec$).

The data were reduced using the standard {\sc IRAF} Gemini GMOS reduction
routines, following the standard GMOS-S reduction steps in the example taken
from the Gemini observatory webpage
\footnote{\url http://www.gemini.edu/sciops/data-and-results/processing-software/getting-started\#gmos}.

We used the Sloan Digital Sky Survey (SDSS) or the 2 Micron All Sky Survey
(2MASS) to align the Gemini-S images to a common astrometric frame of
reference.  This imposes an rms uncertainty in the absolute astrometry of
$0\farcs2$ and $0\farcs4$ for SDSS and 2MASS, respectively.  The
astrometrically calibrated Gemini-S images served as the basis for aligning
higher resolution, smaller field-of-view imaging from {\it HST} or Keck (when
available), which were originally presented in \citet{Calanog:2014lr}.

\input{table_positions}

\begin{figure*}[!tbp] 
    \begin{centering}
\includegraphics[width=0.331\textwidth]{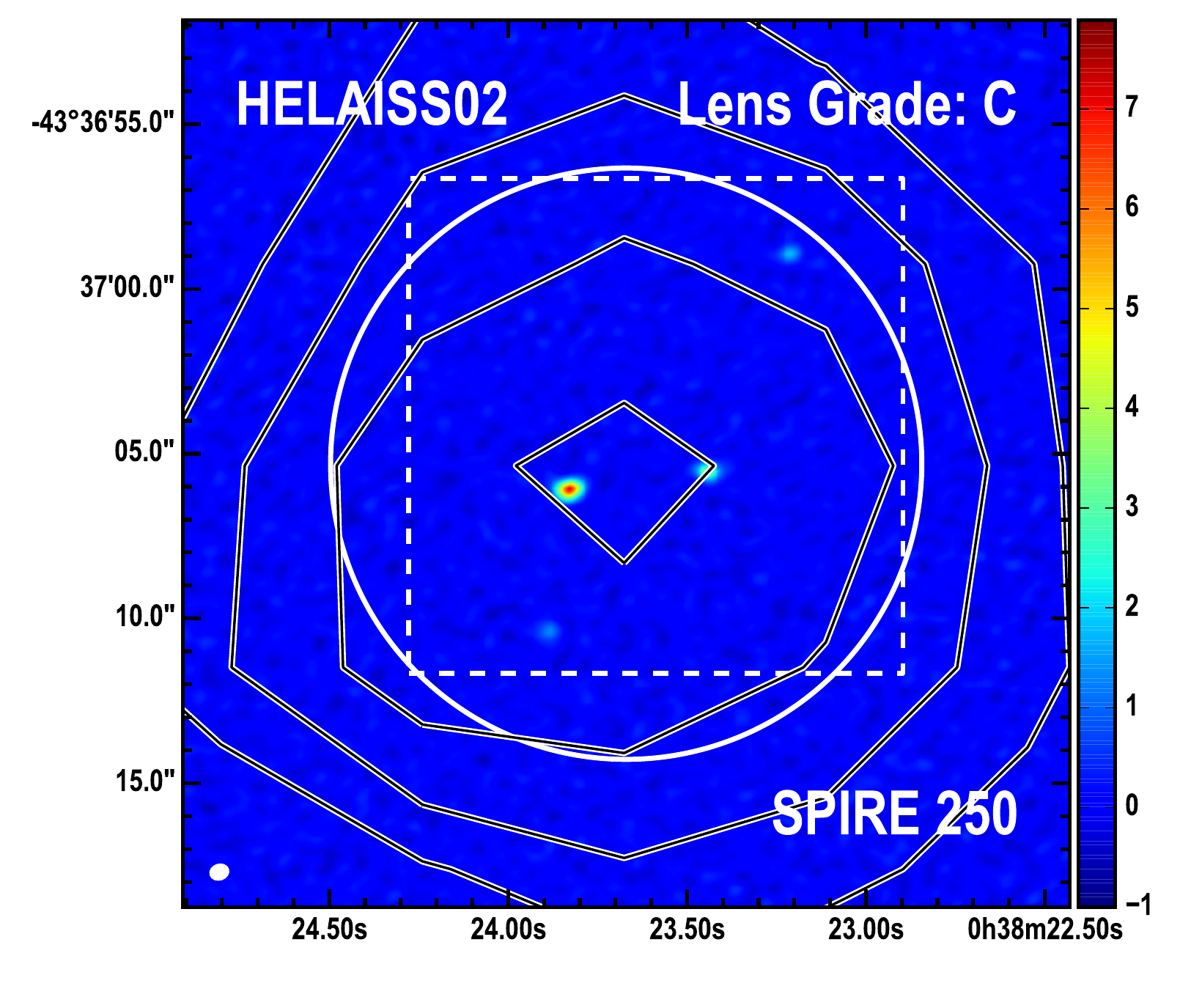}
\includegraphics[width=0.331\textwidth]{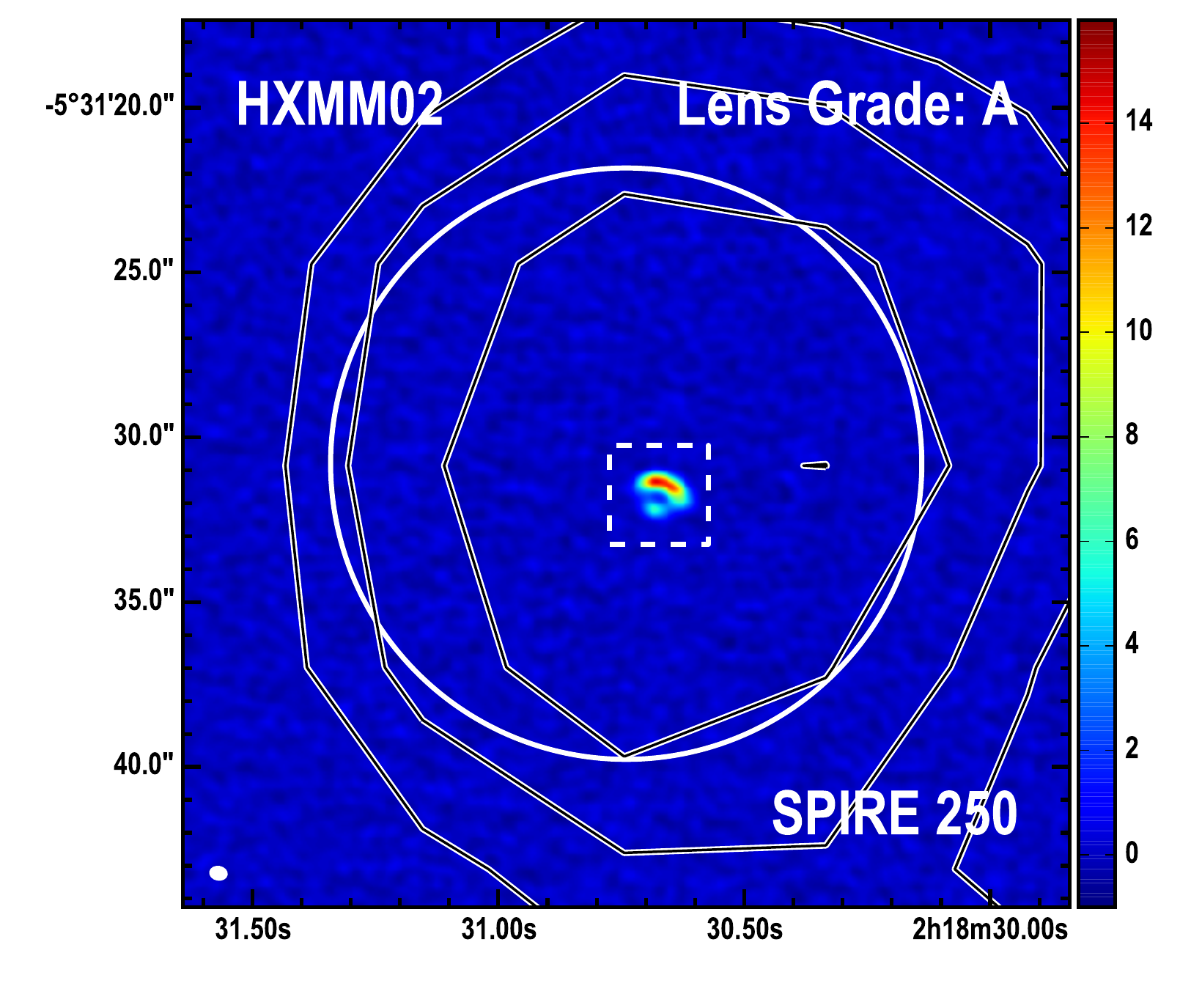}
\includegraphics[width=0.331\textwidth]{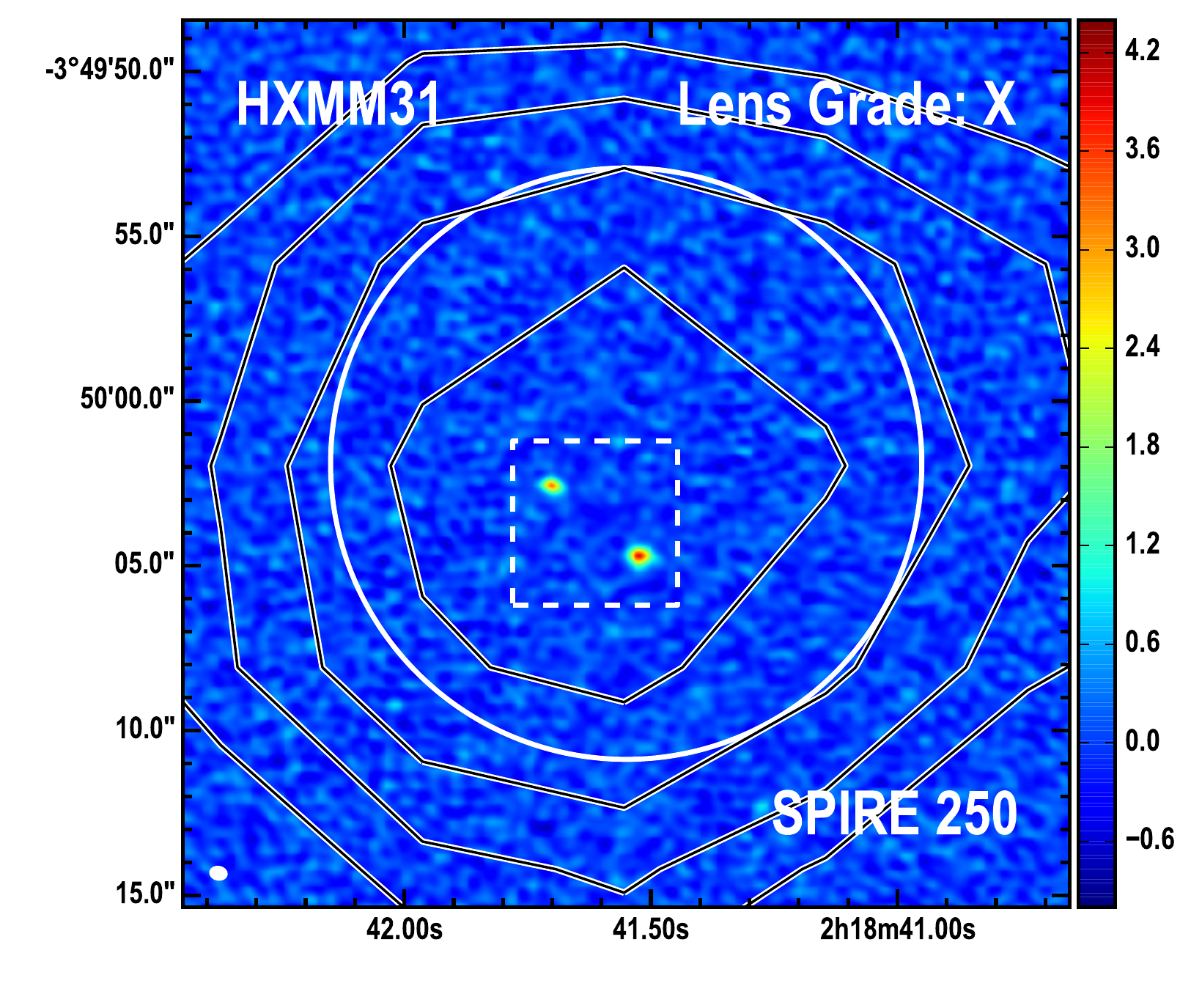}
\includegraphics[width=0.331\textwidth]{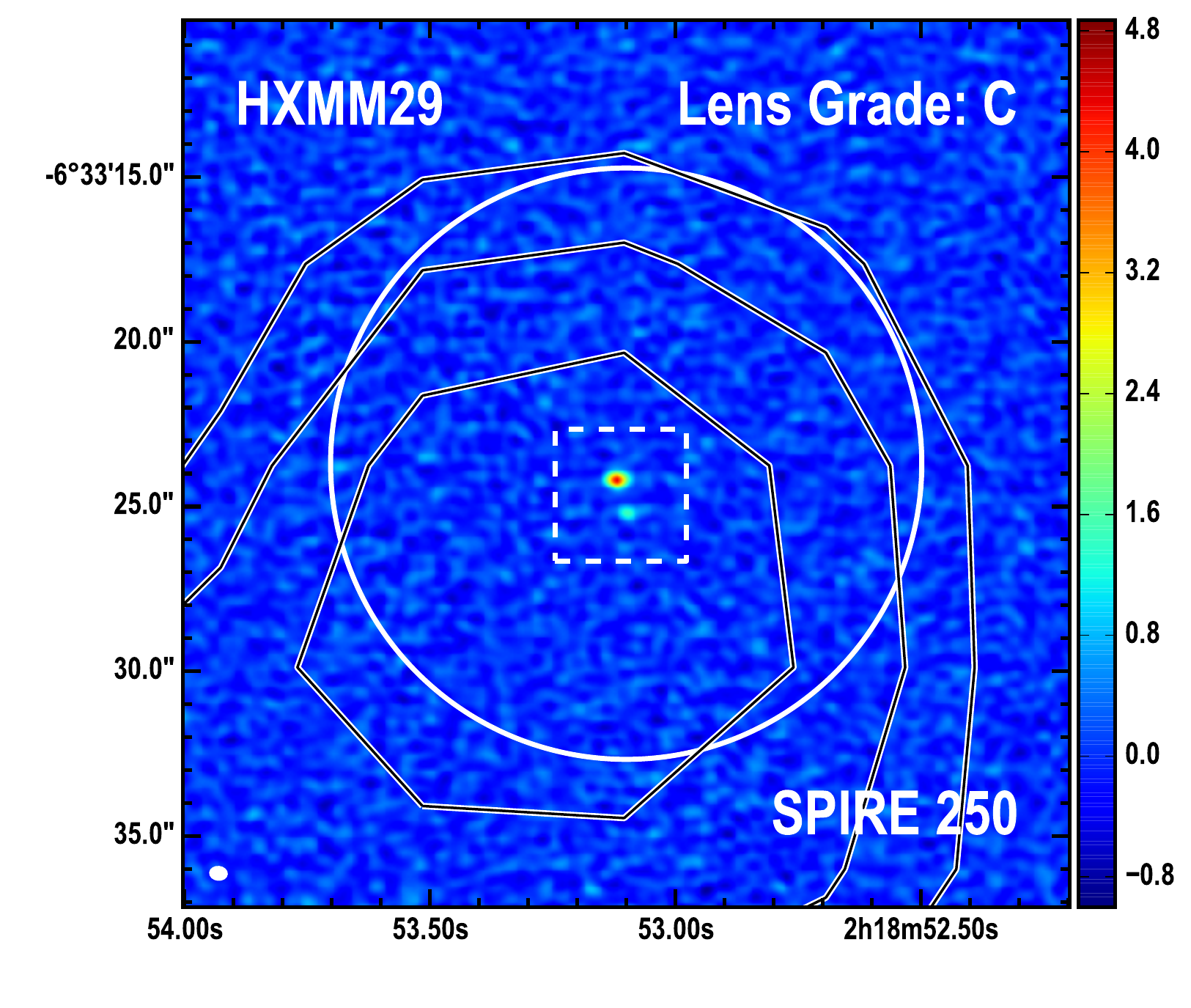}
\includegraphics[width=0.331\textwidth]{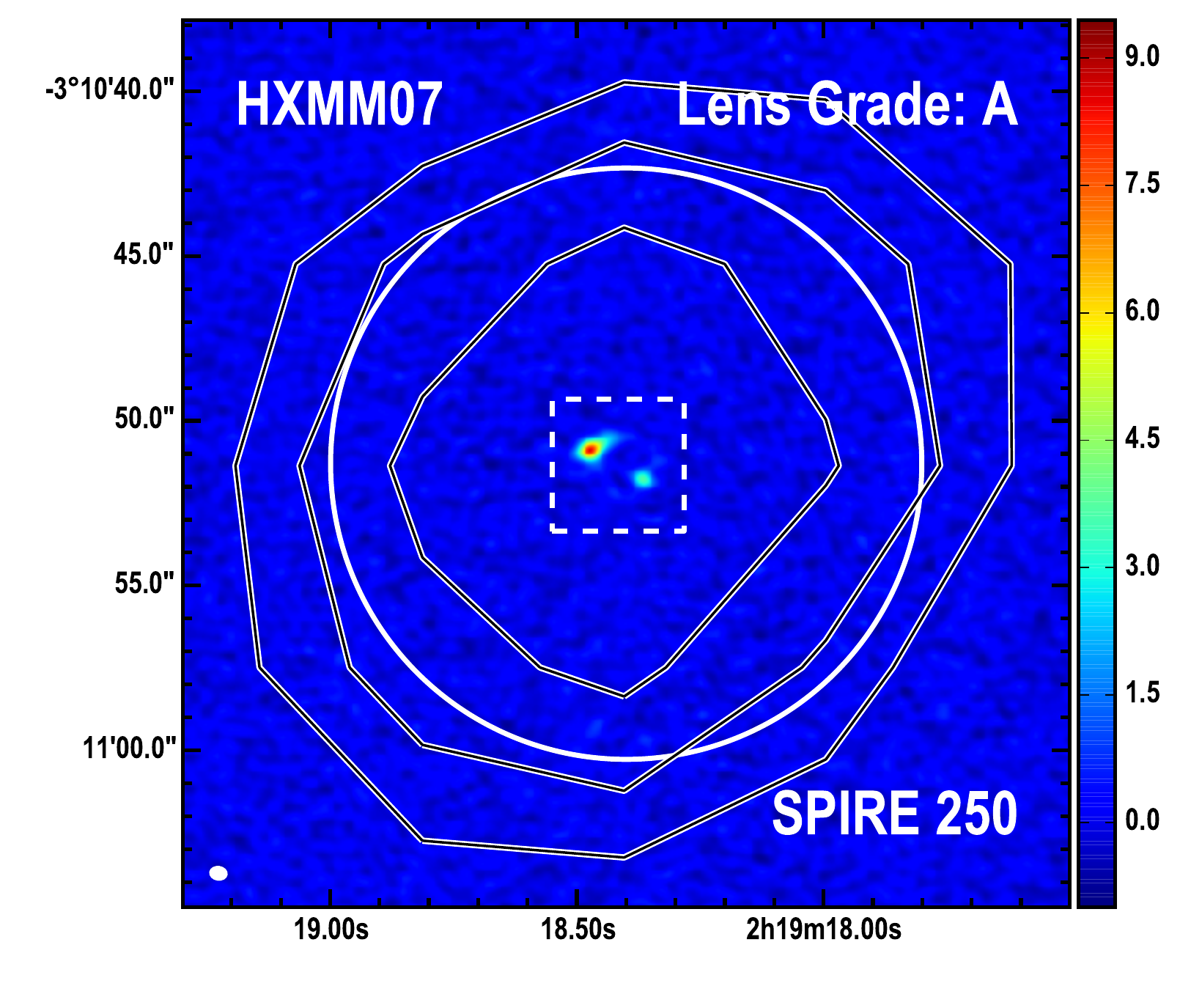}
\includegraphics[width=0.331\textwidth]{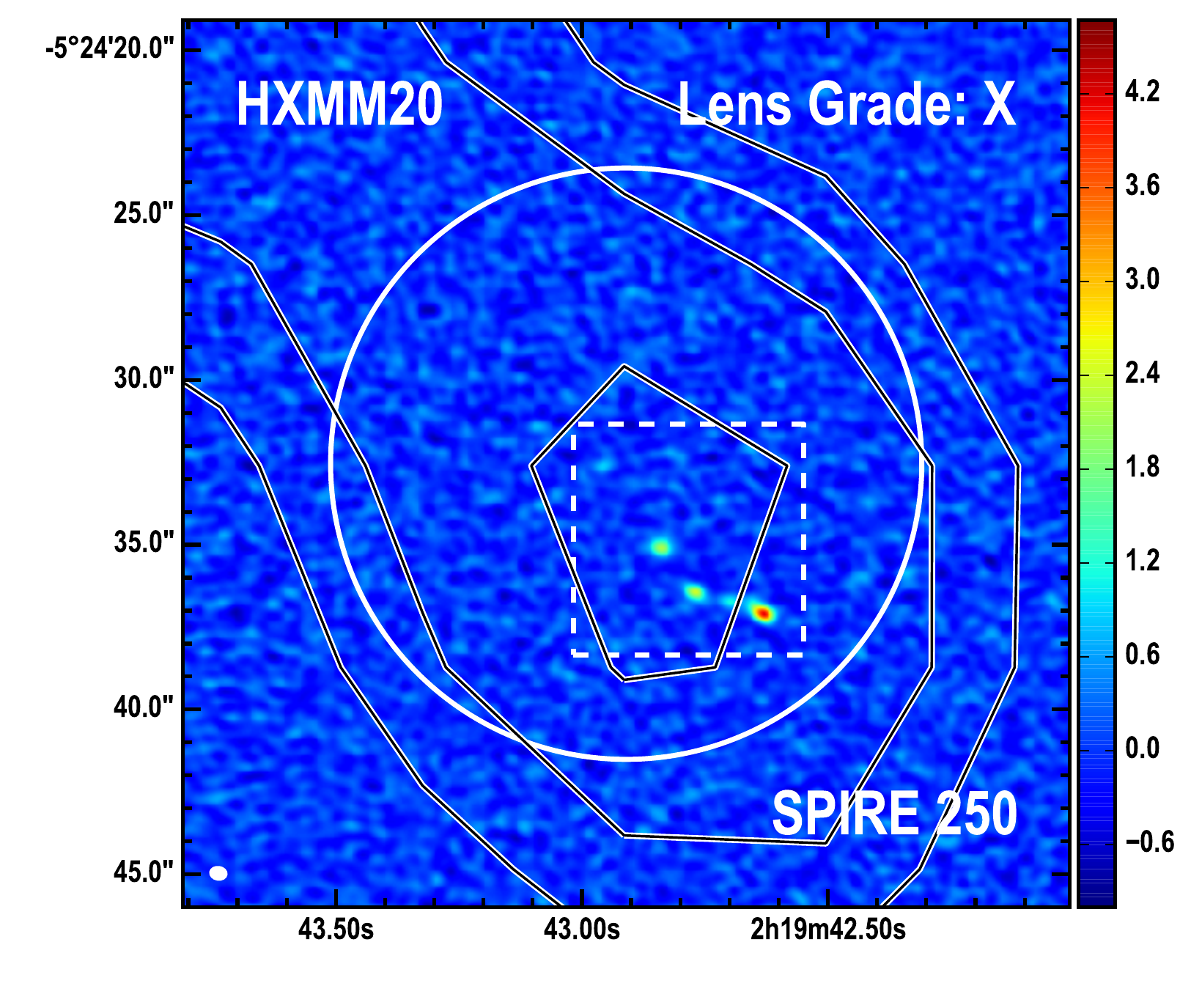}
\includegraphics[width=0.331\textwidth]{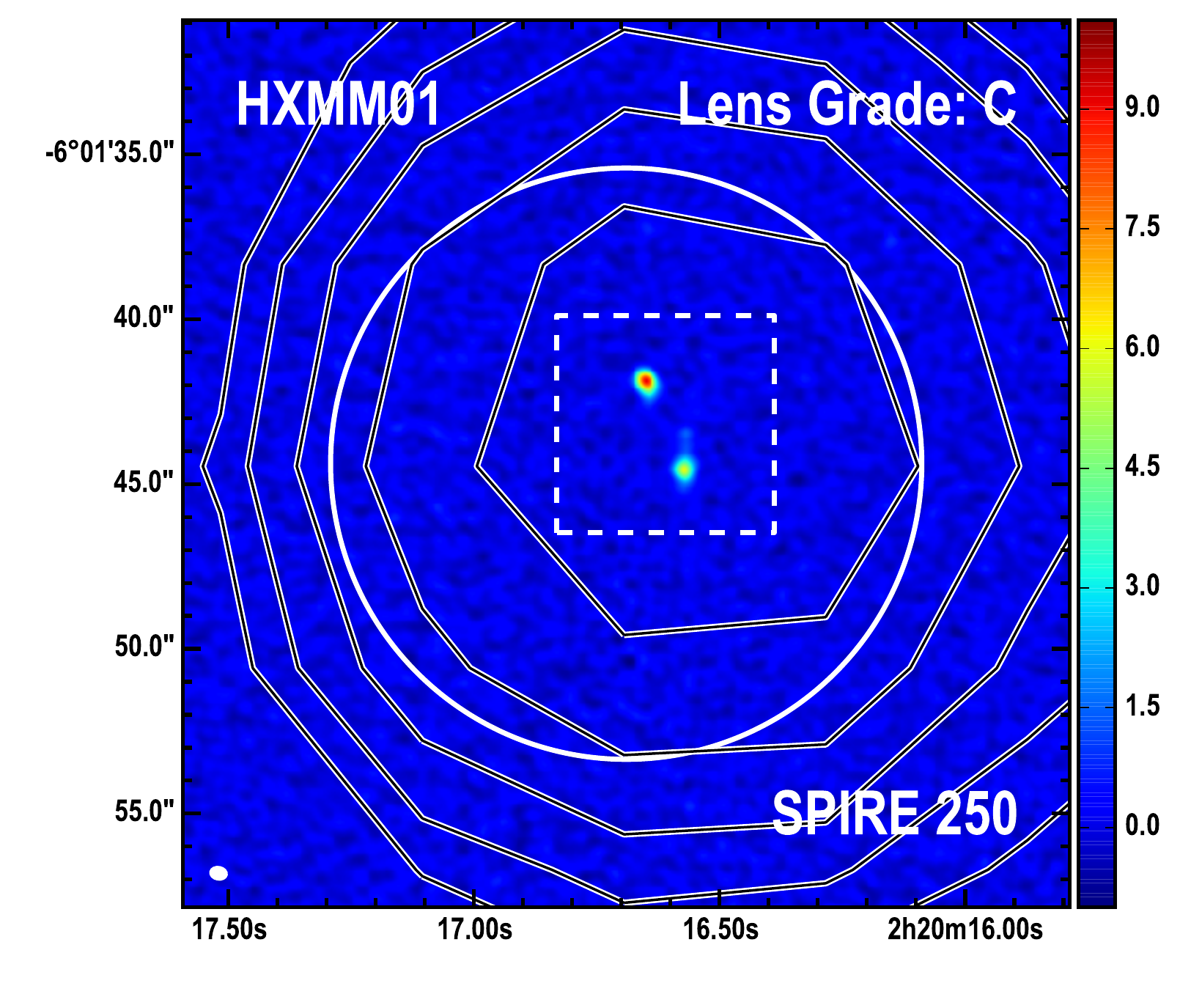}
\includegraphics[width=0.331\textwidth]{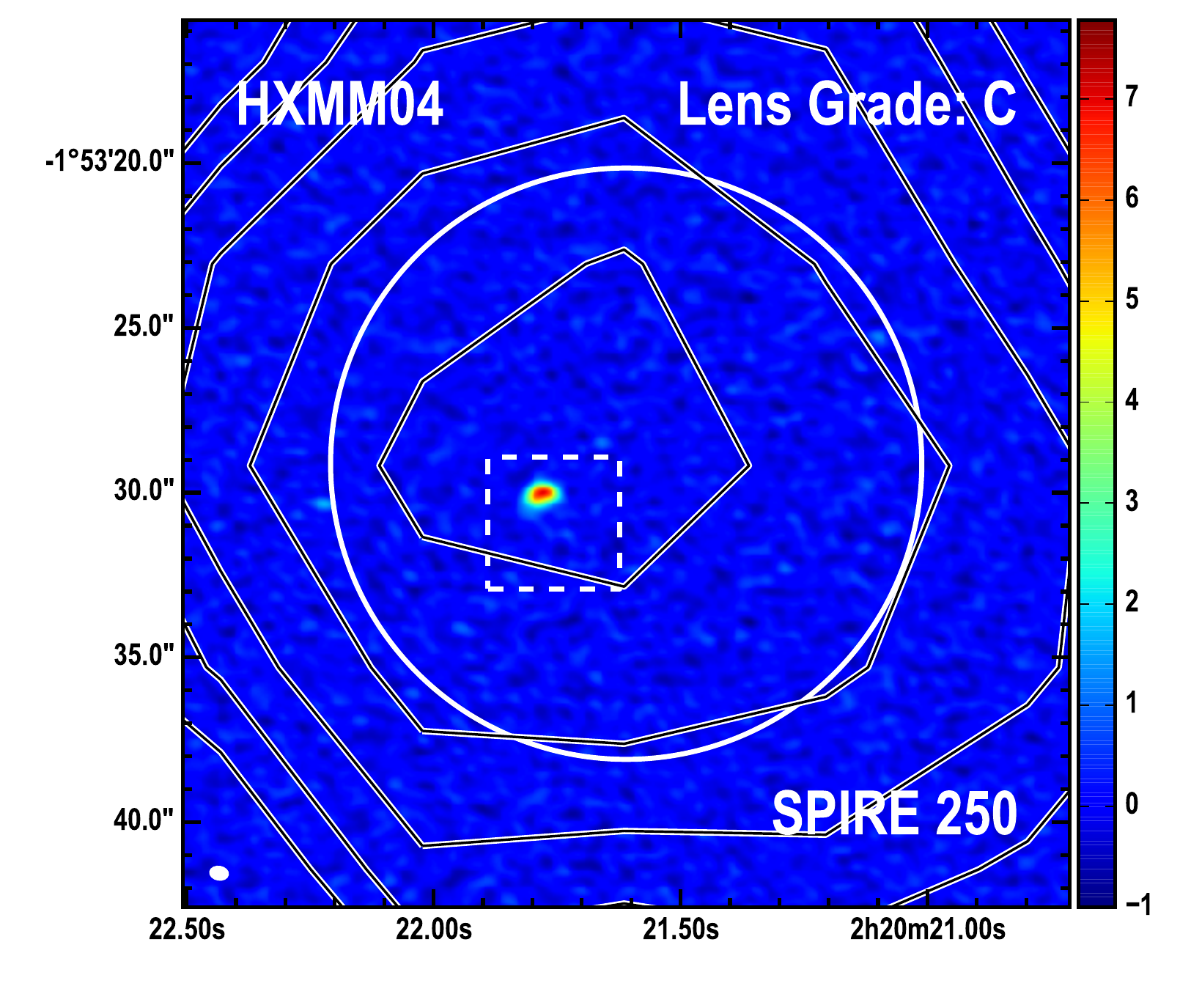}
\includegraphics[width=0.331\textwidth]{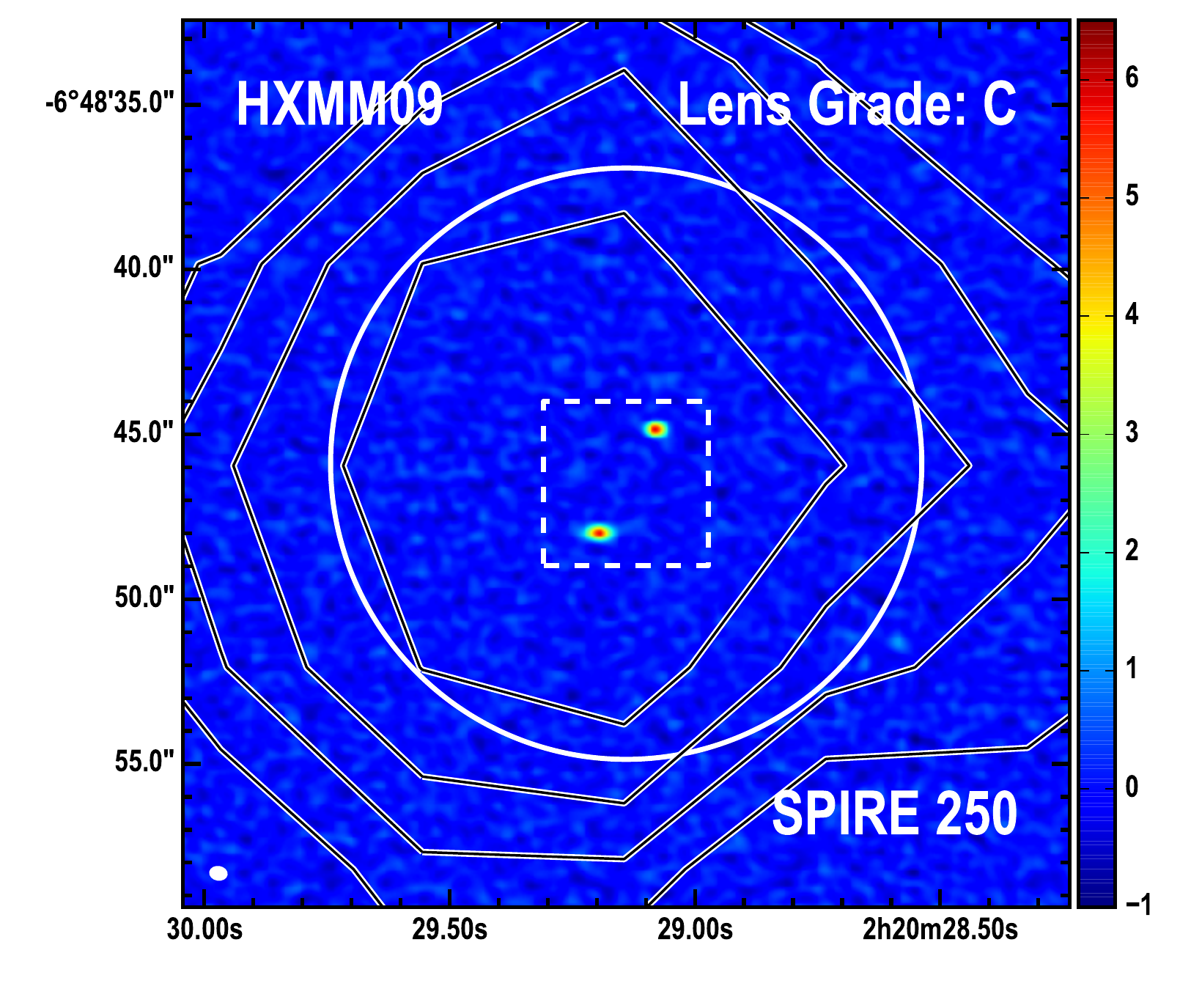}
\includegraphics[width=0.331\textwidth]{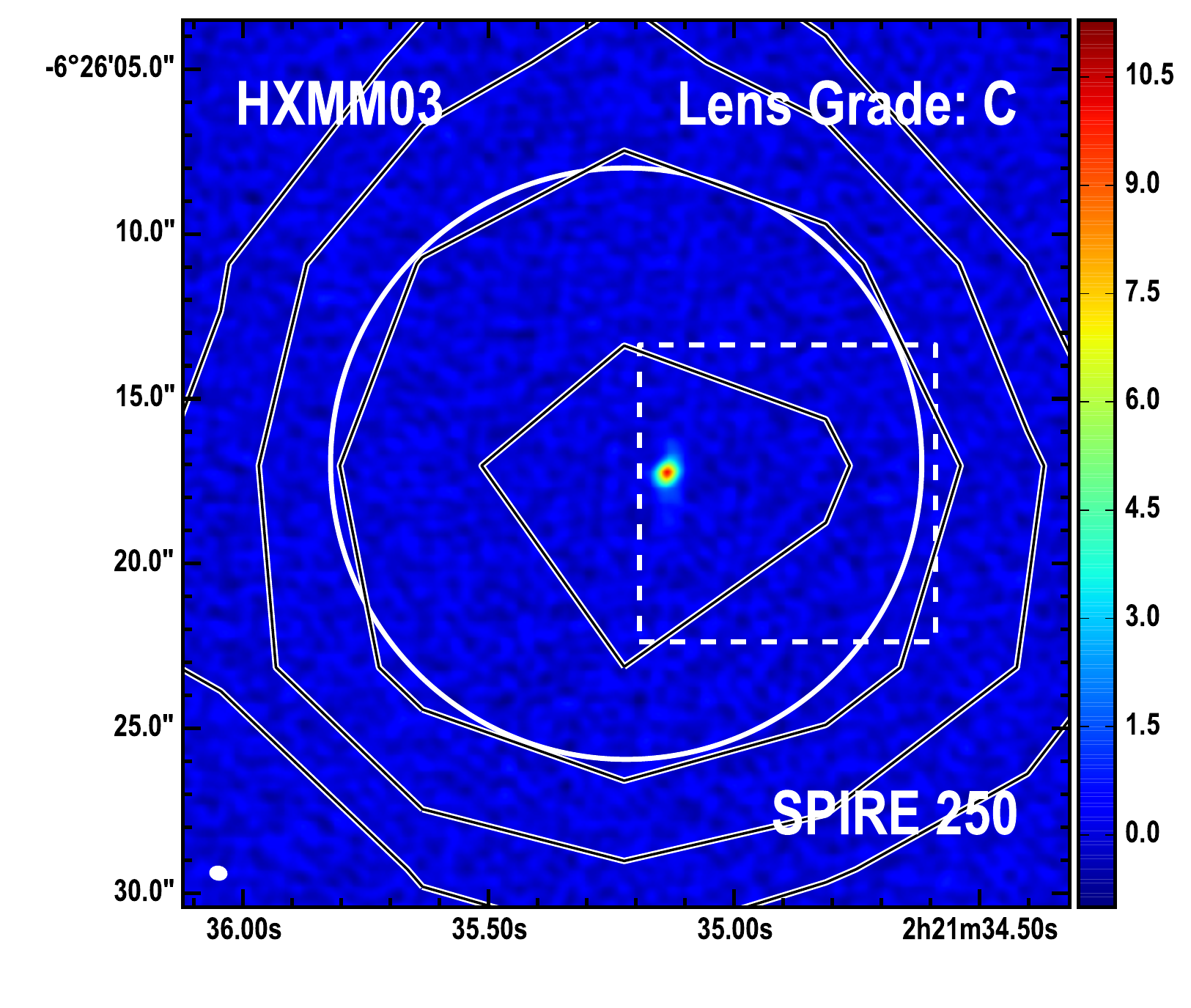}
\includegraphics[width=0.331\textwidth]{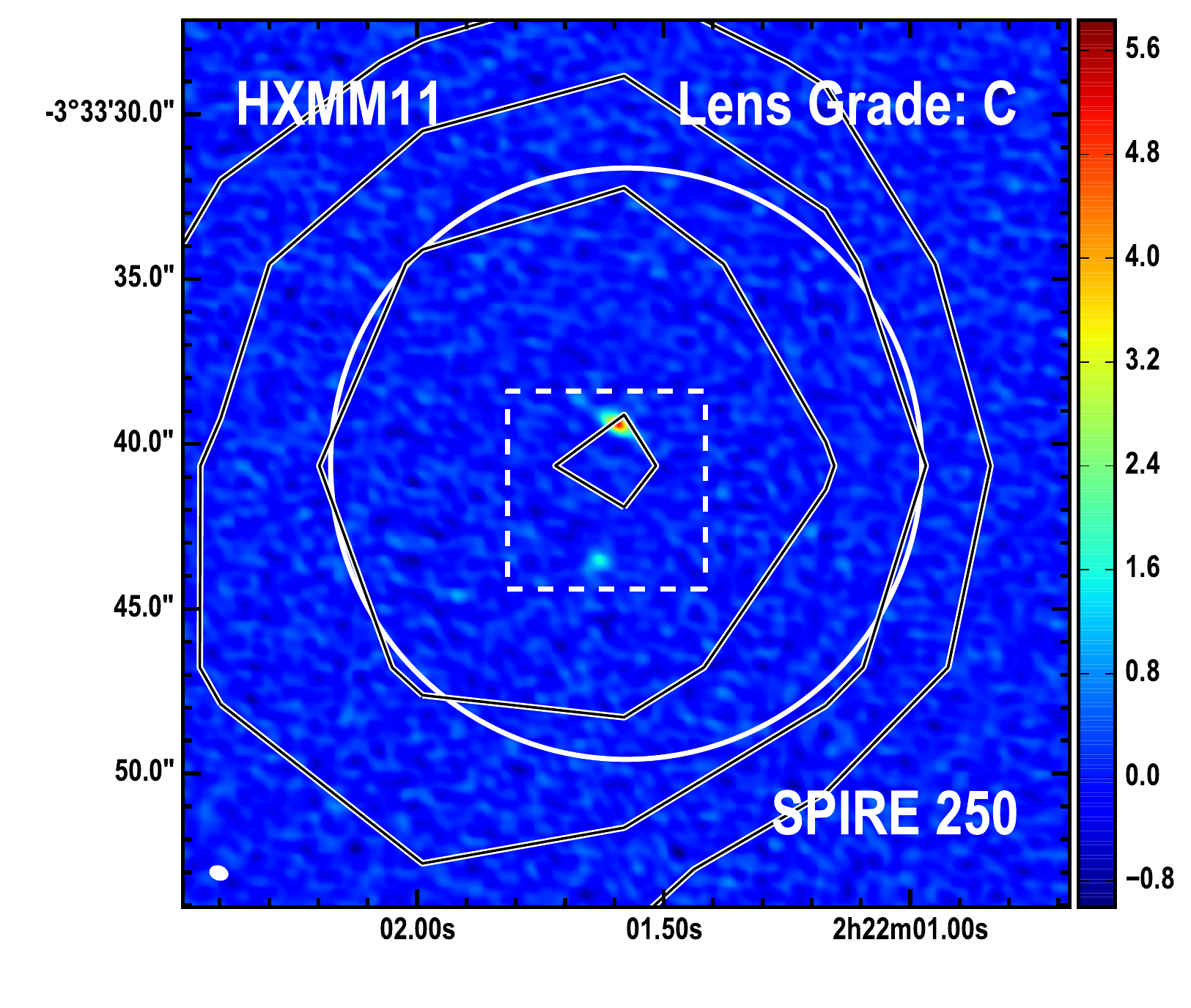}
\includegraphics[width=0.331\textwidth]{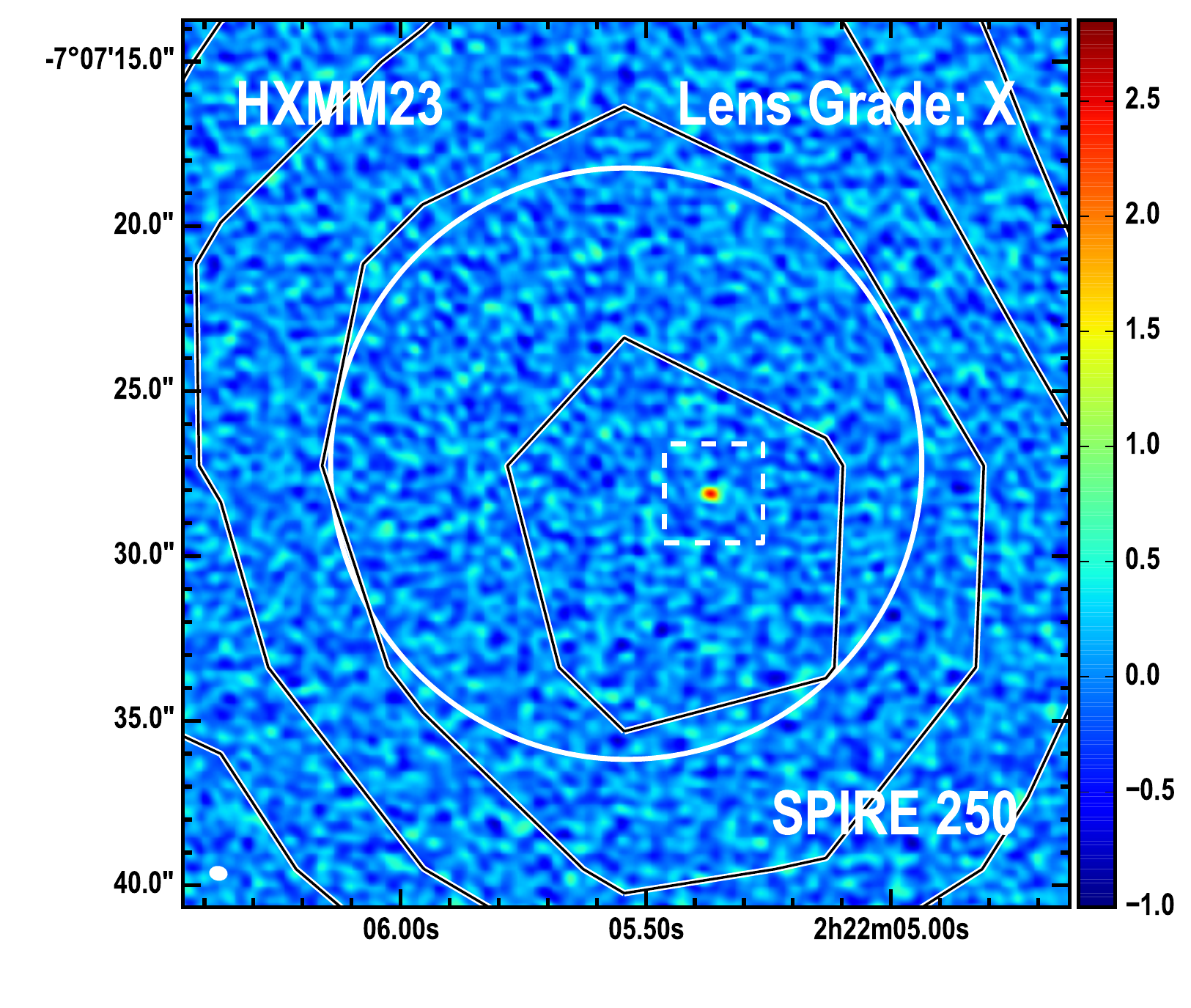}
\end{centering}

\caption{ ALMA 870$\,\mu$m images (color scale, units of mJy$\,$beam$^{-1}$) of
HerMES DSFGs (images have not been corrected for primary beam attenuation).
Contours (black and white) trace 250$\,\mu$m or 500$\,\mu$m emission from {\it
Herschel} (starting at 4$\sigma$ and increasing by factors of 2, where
$\sigma=7\,$mJy).  North is up, east is left.  The FWHM size of the ALMA
synthesized beam is shown in the lower left corner of each panel.  A solid
white circle shows the FWHM size of the primary beam.  Dashed squares identify
the regions of each image that are shown in greater detail in
Fig.~\ref{fig:uvmodels}.  \label{fig:imaging}} \addtocounter{figure}{-1}

\end{figure*}

\begin{figure*}[!tbp] 
    \begin{centering}
\includegraphics[width=0.331\textwidth]{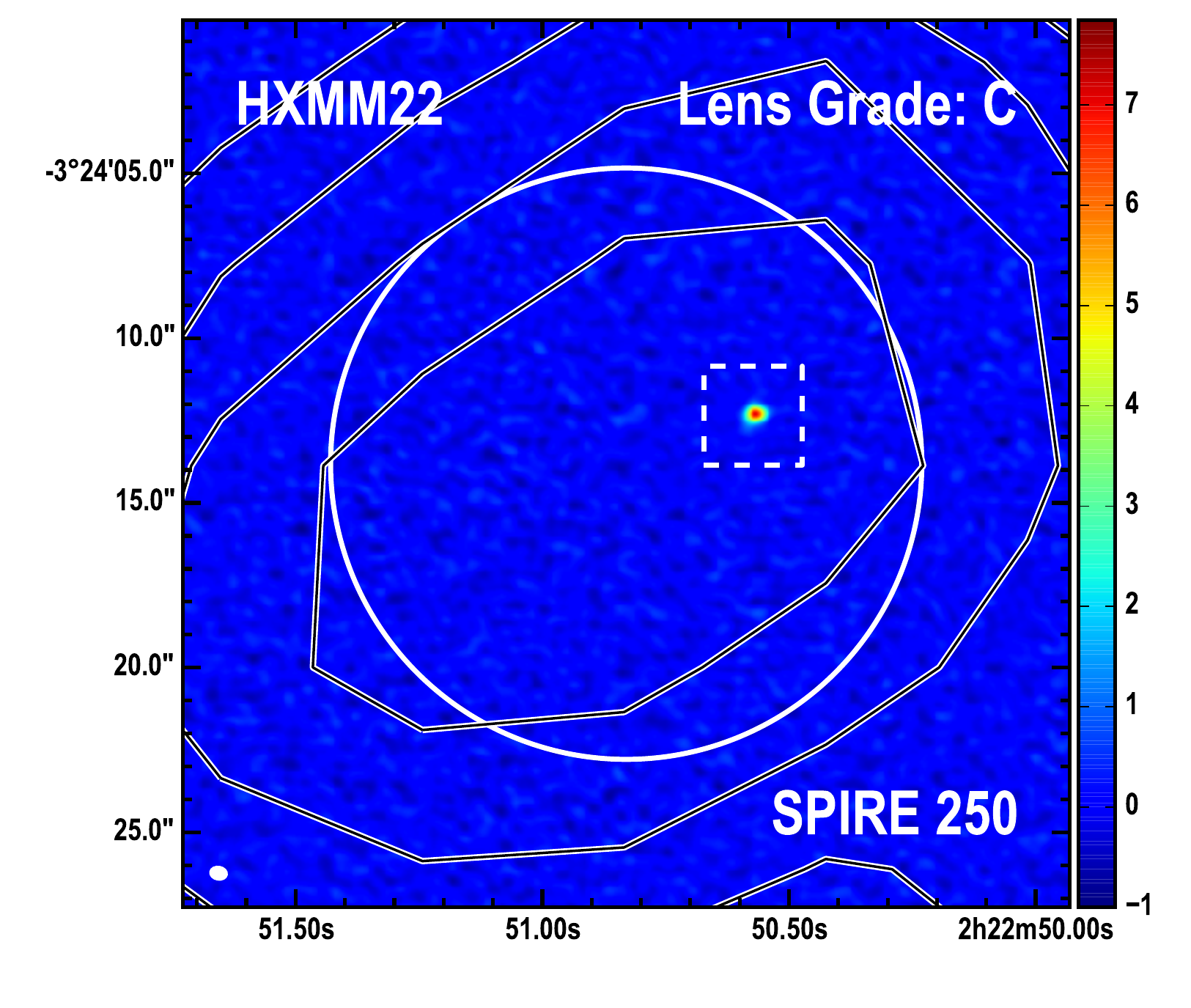}
\includegraphics[width=0.331\textwidth]{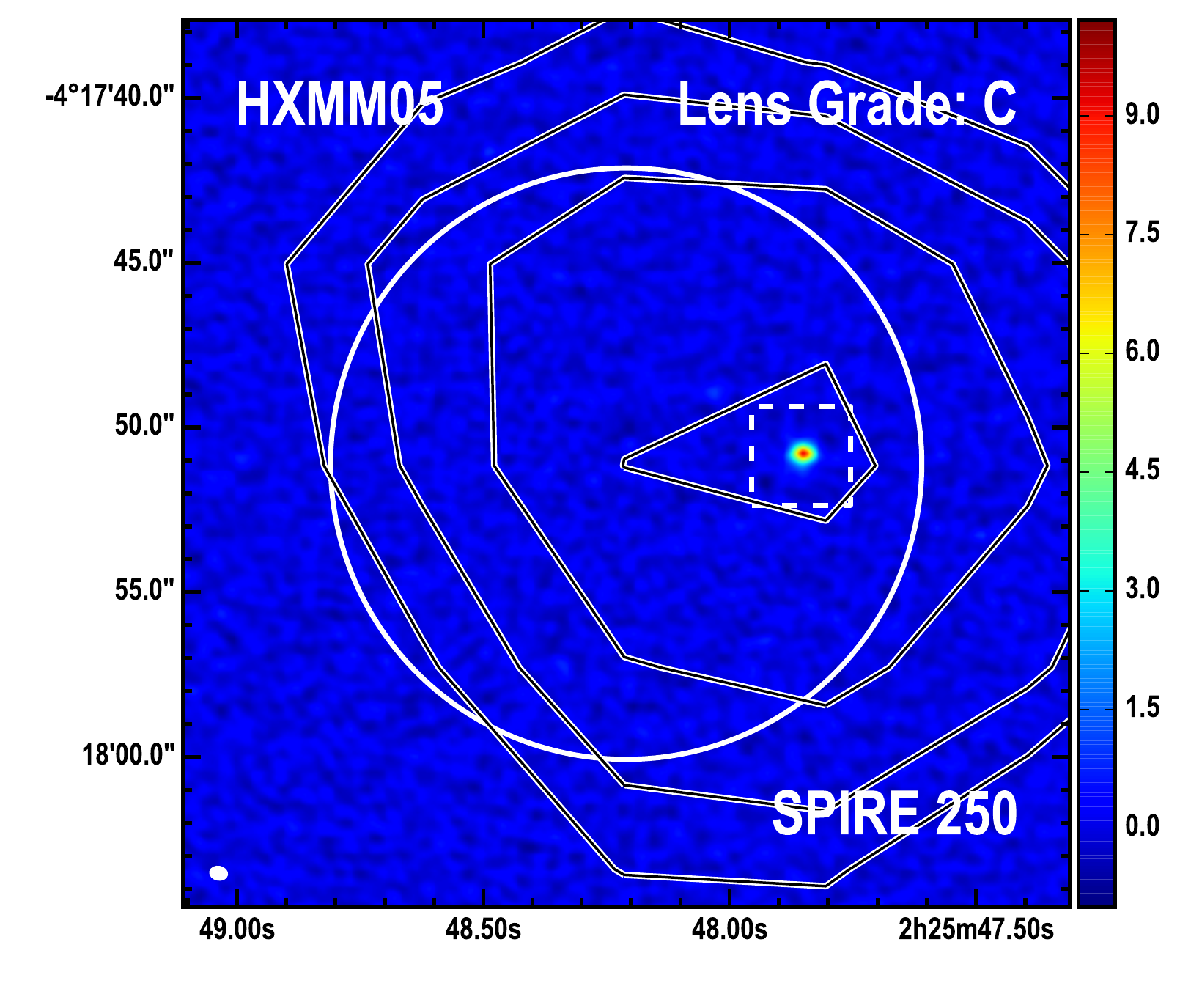}
\includegraphics[width=0.331\textwidth]{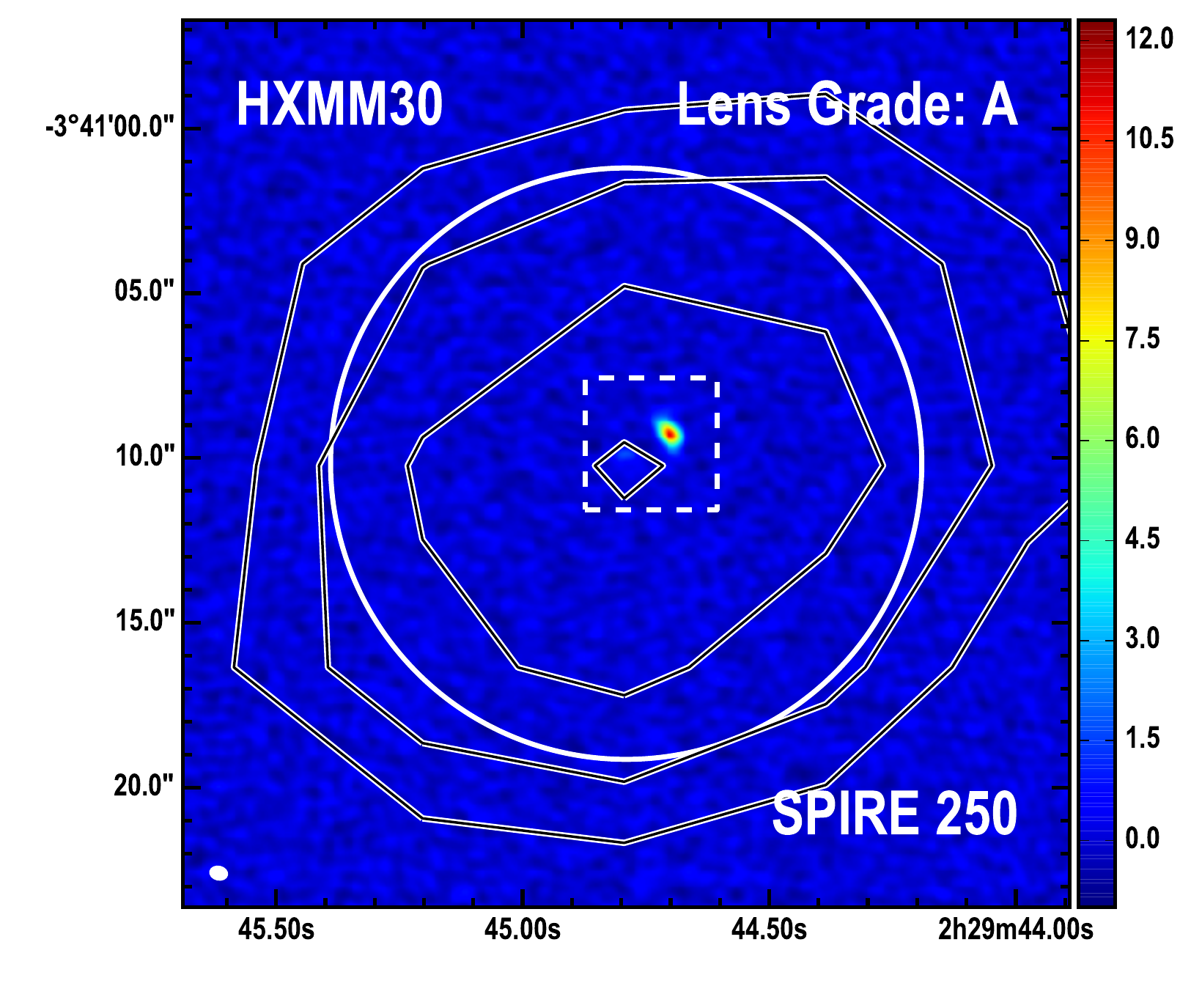}
\includegraphics[width=0.331\textwidth]{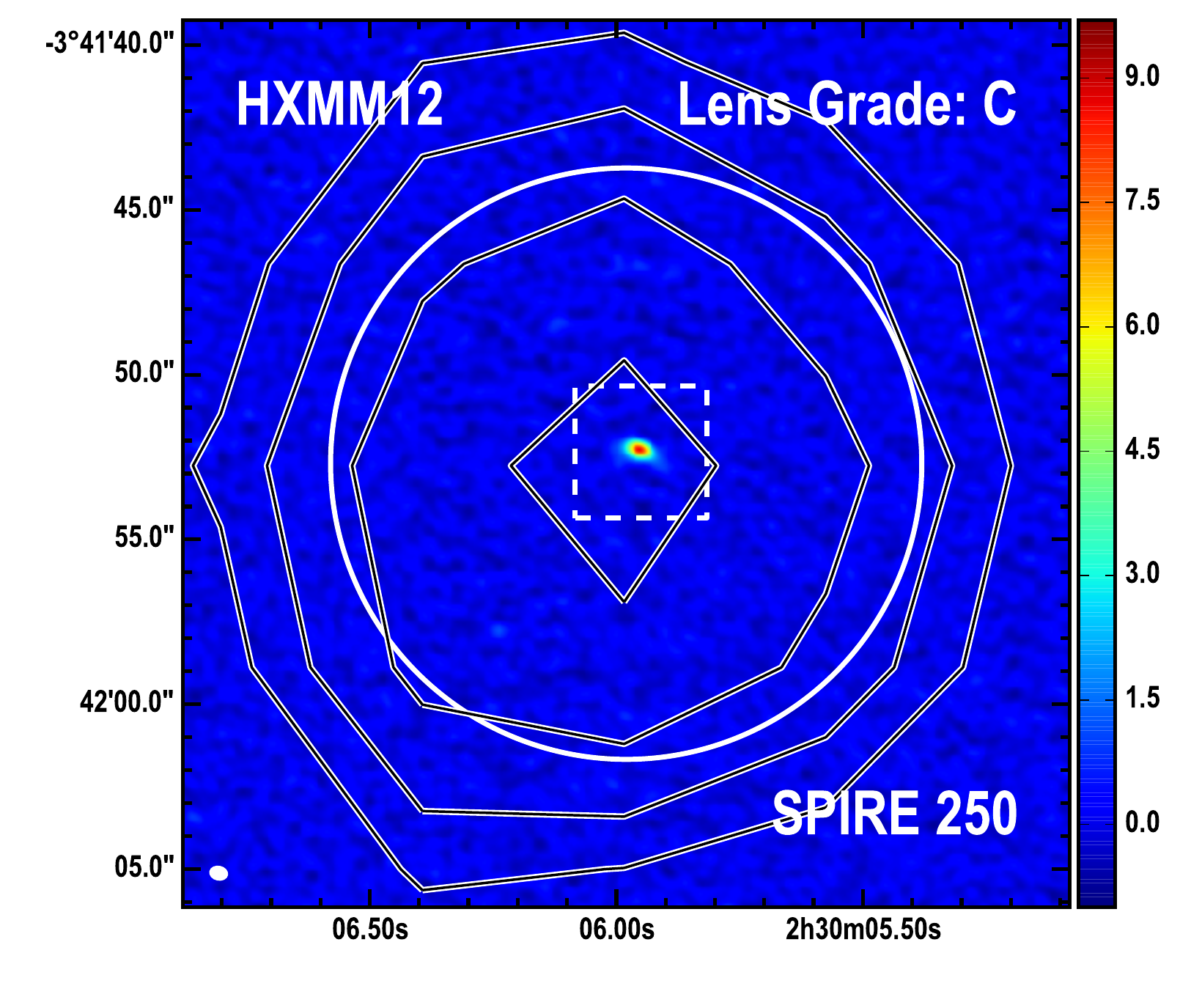}
\includegraphics[width=0.331\textwidth]{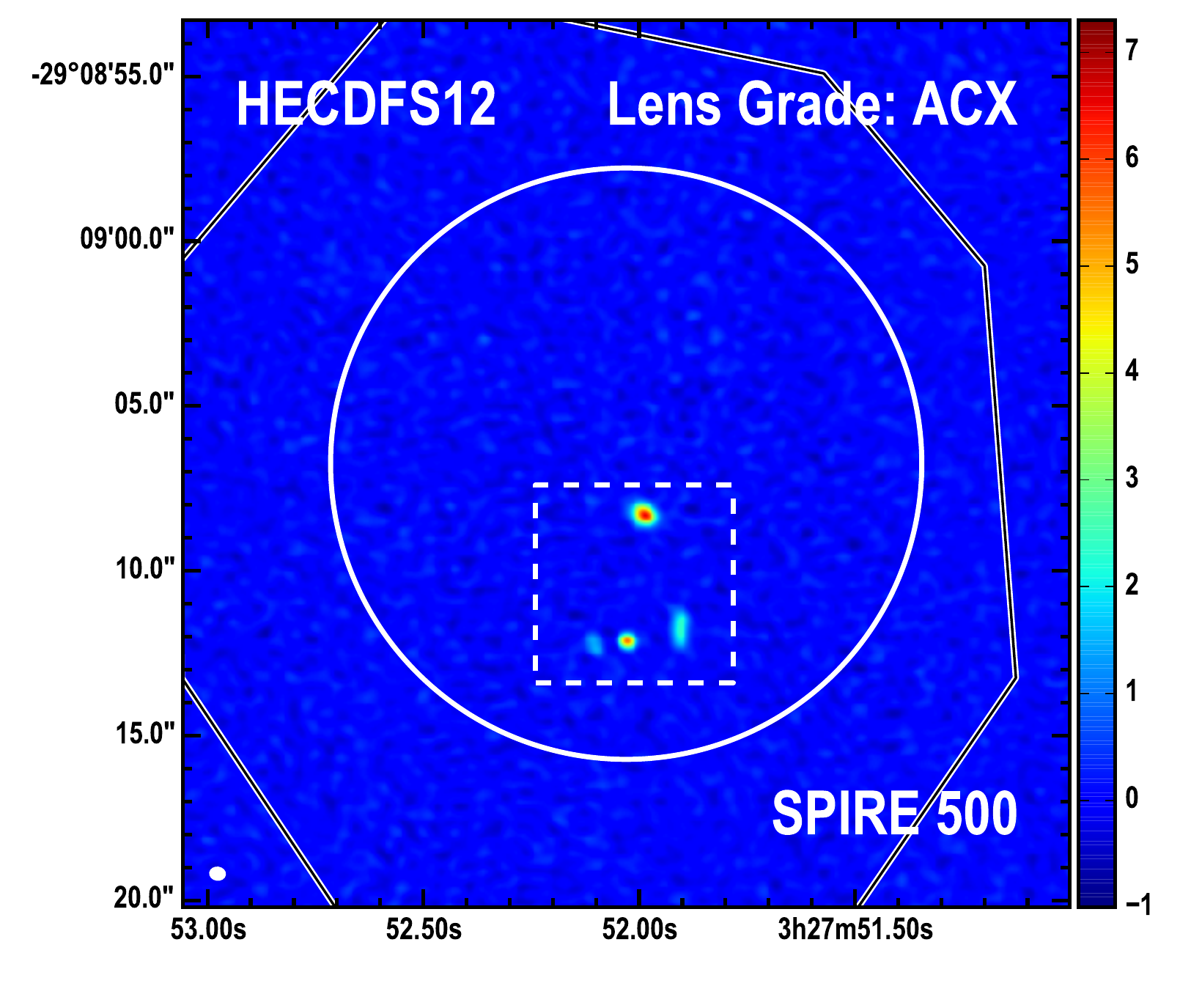}
\includegraphics[width=0.331\textwidth]{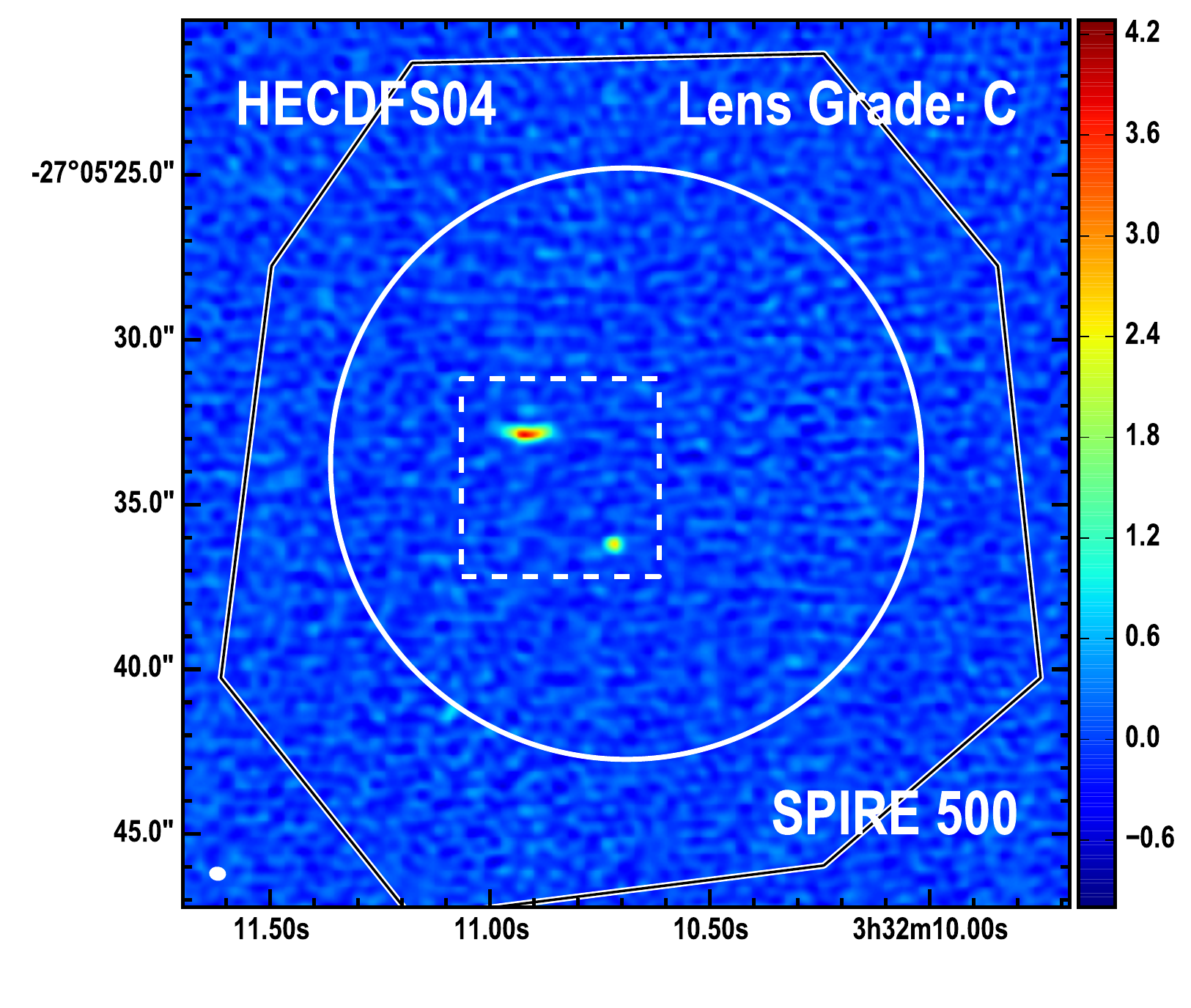}
\includegraphics[width=0.331\textwidth]{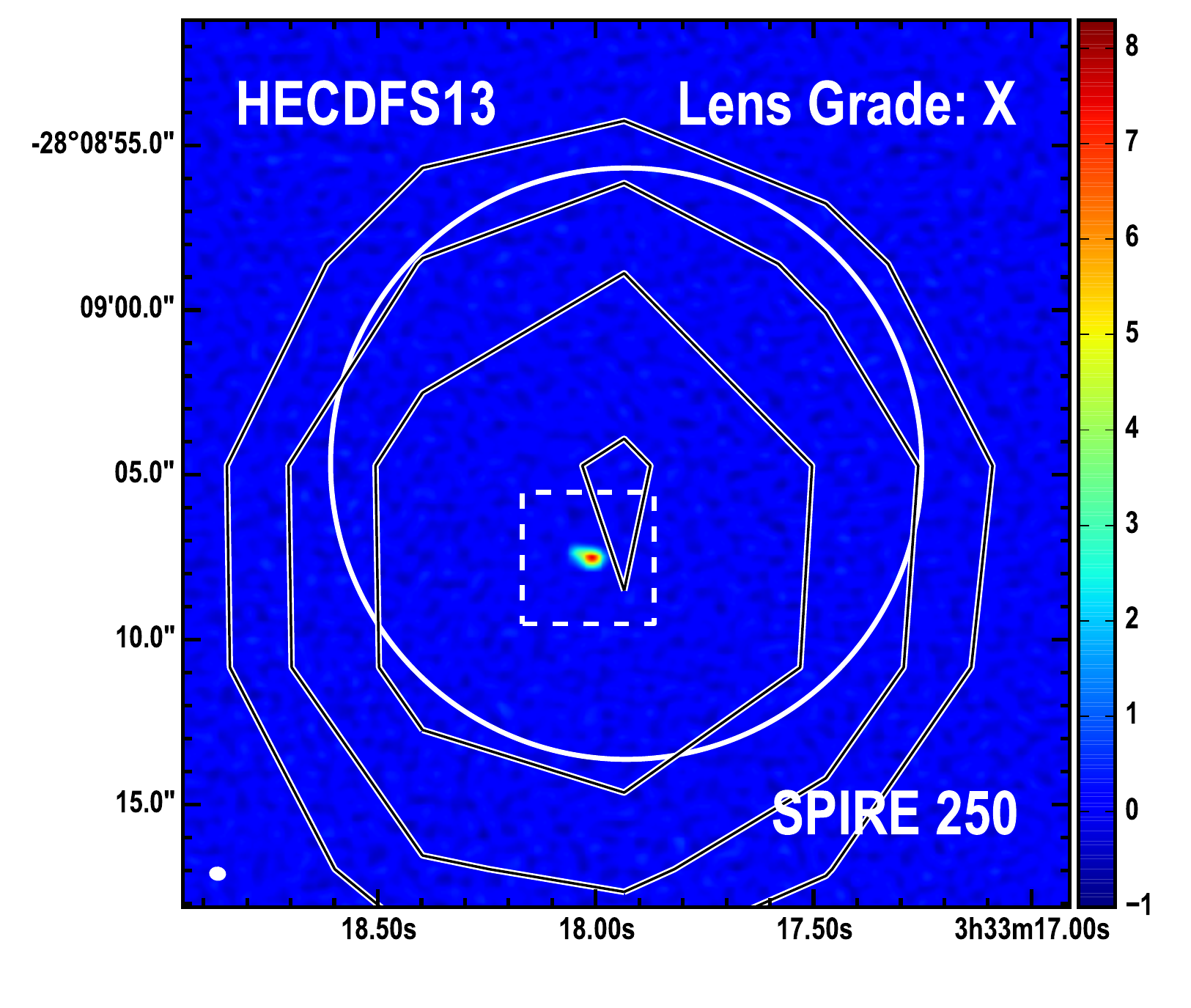}
\includegraphics[width=0.331\textwidth]{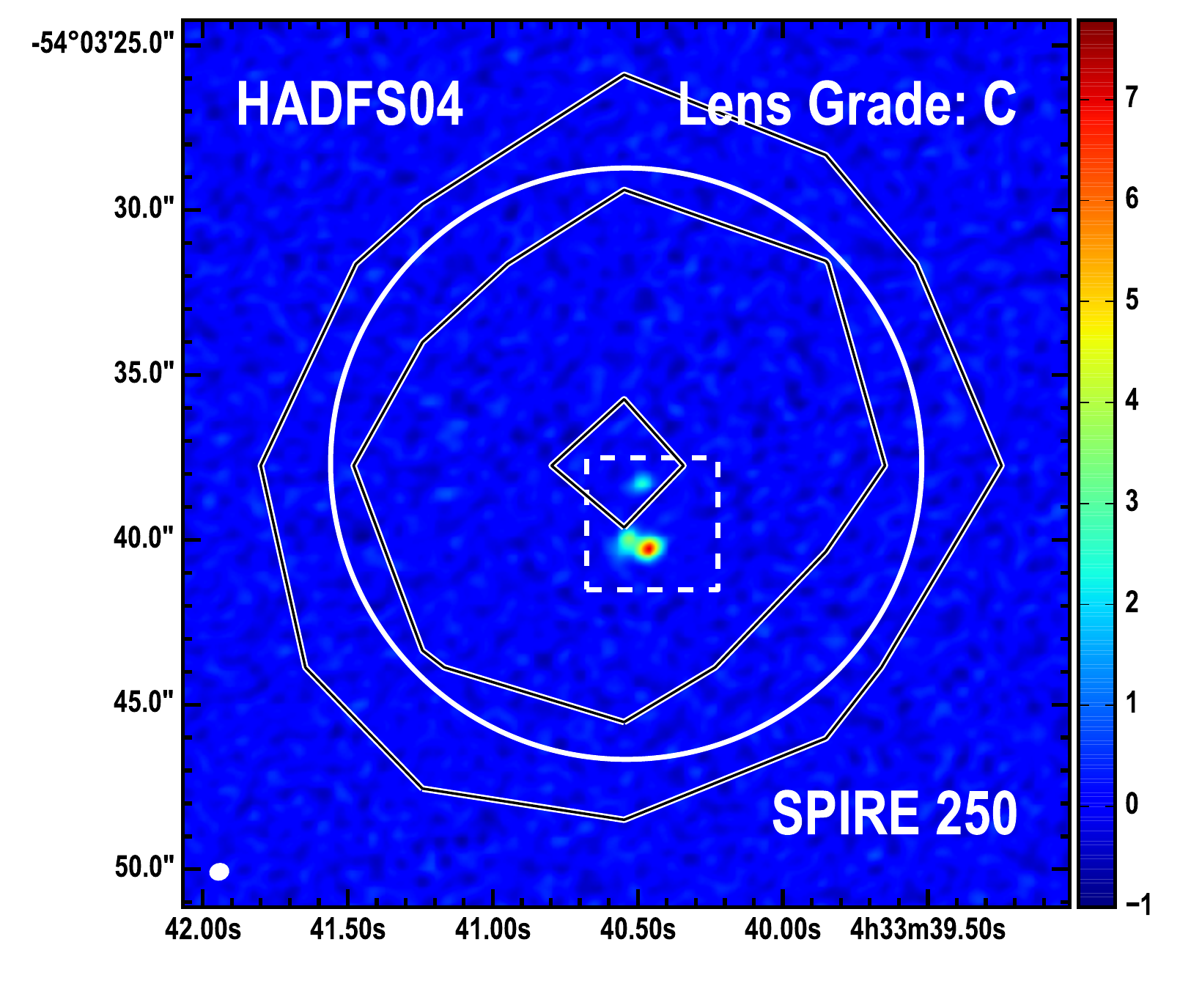}
\includegraphics[width=0.331\textwidth]{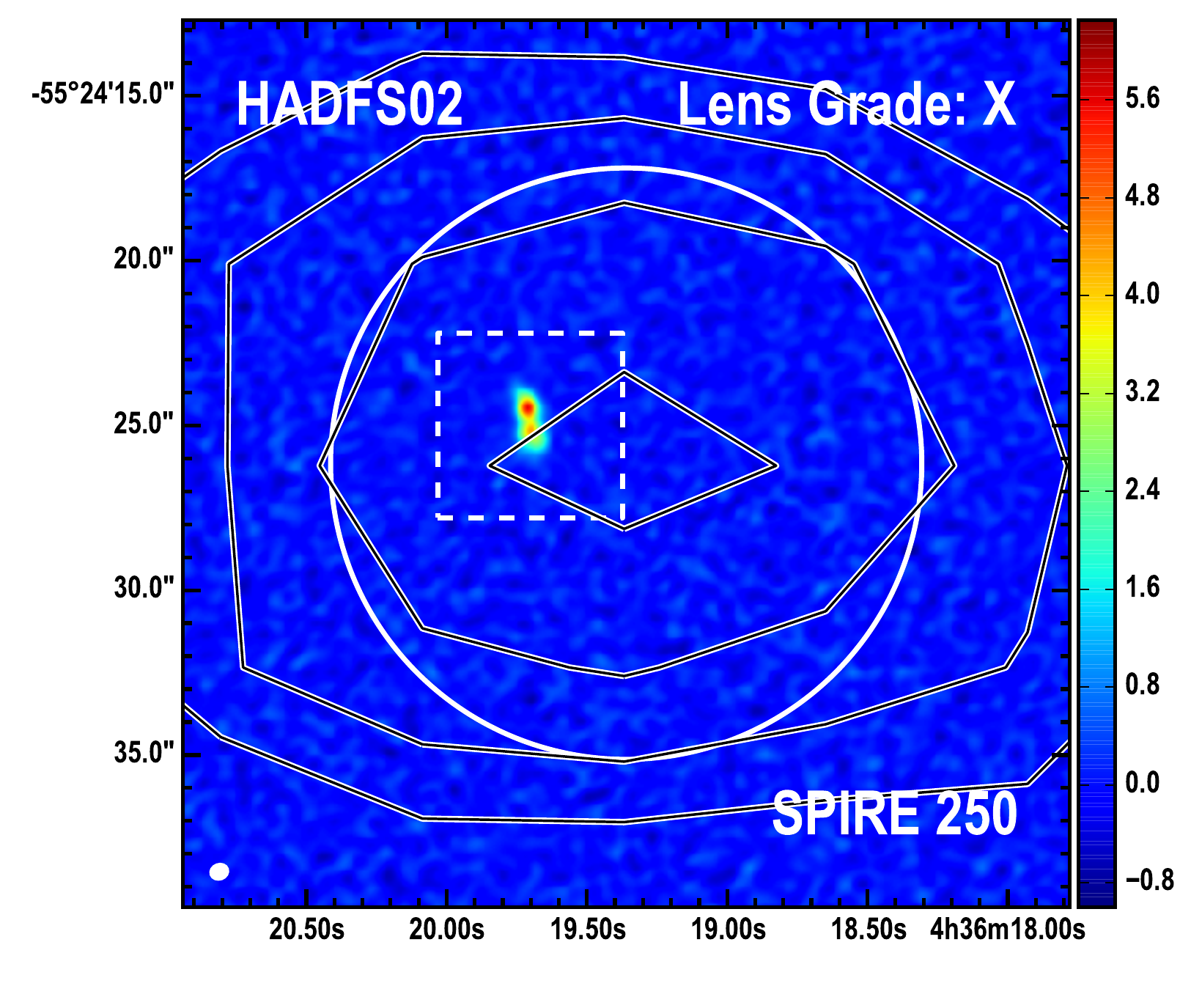}
\includegraphics[width=0.331\textwidth]{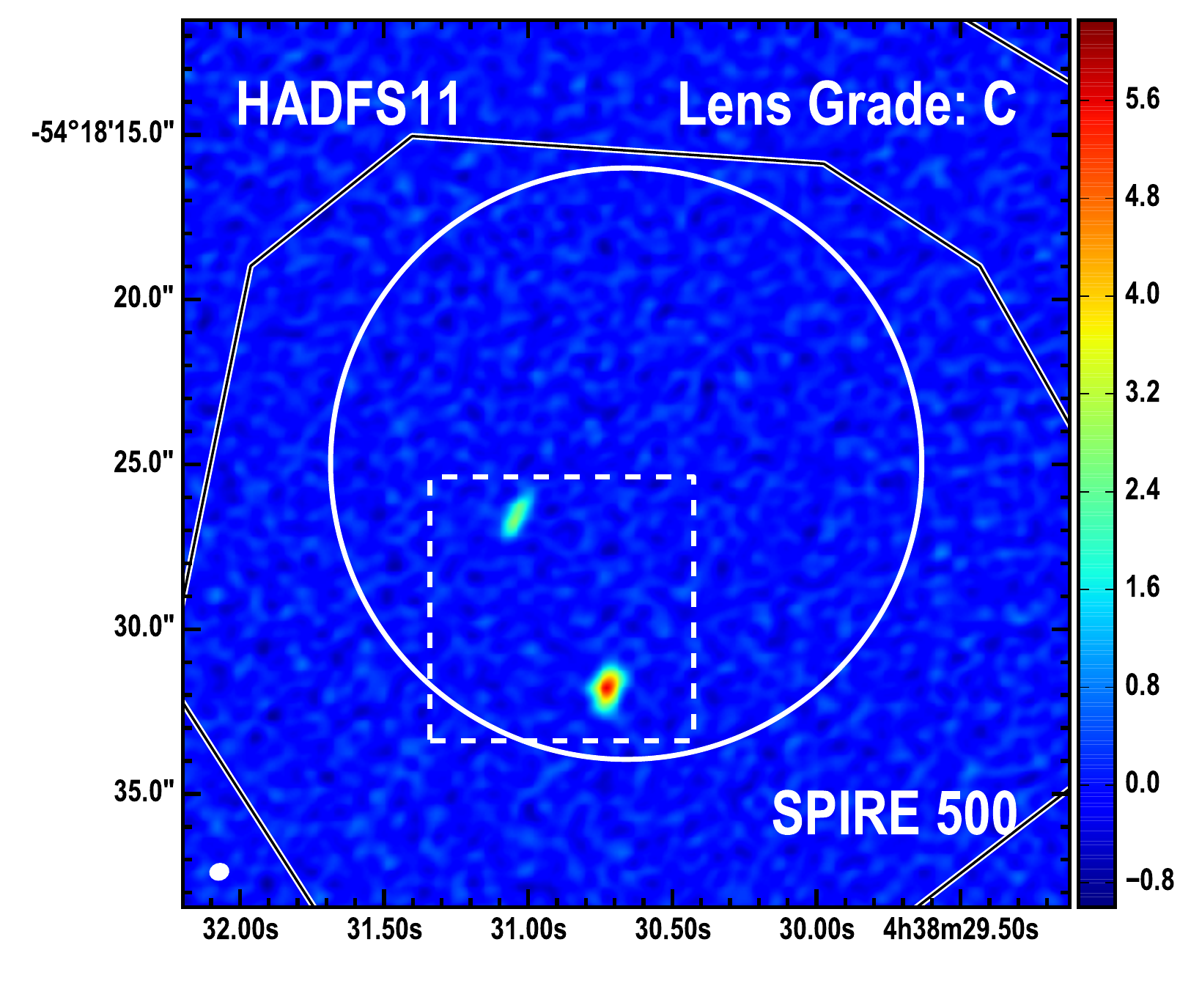}
\includegraphics[width=0.331\textwidth]{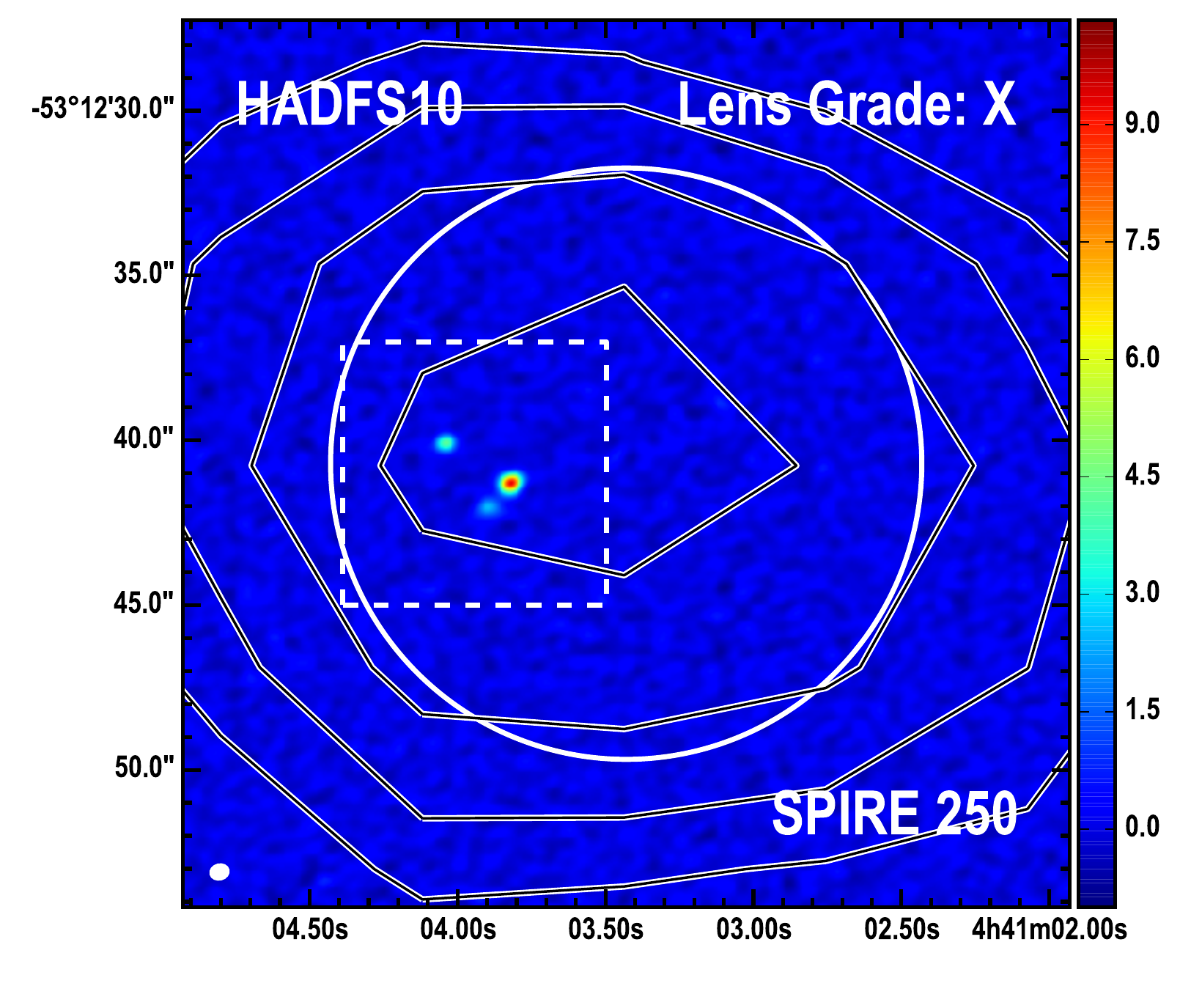}
\includegraphics[width=0.331\textwidth]{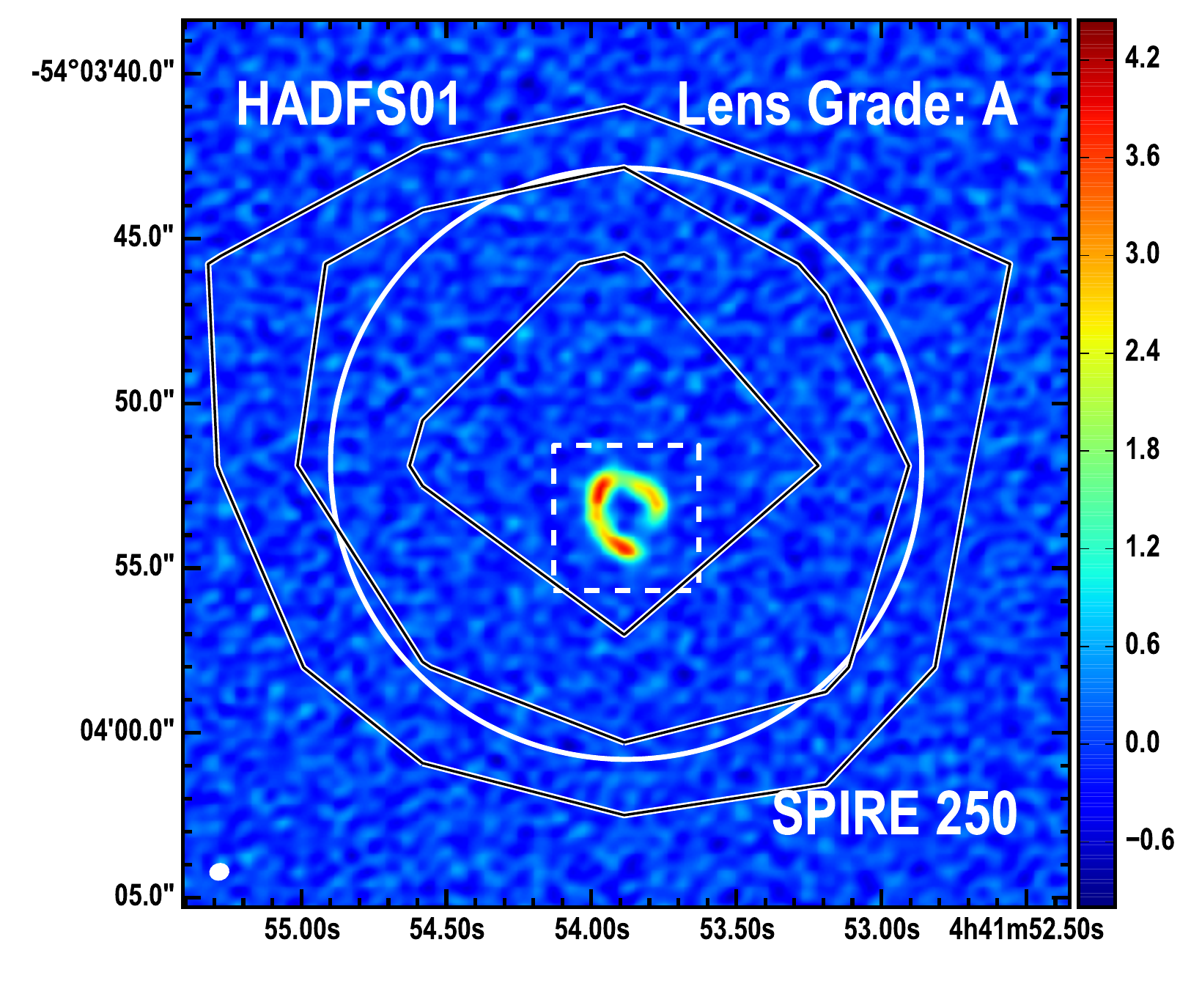}
\end{centering}

\caption{ Continued.}
\addtocounter{figure}{-1}

\end{figure*}

\begin{figure*}[!tbp] 
    \begin{centering}
\includegraphics[width=0.331\textwidth]{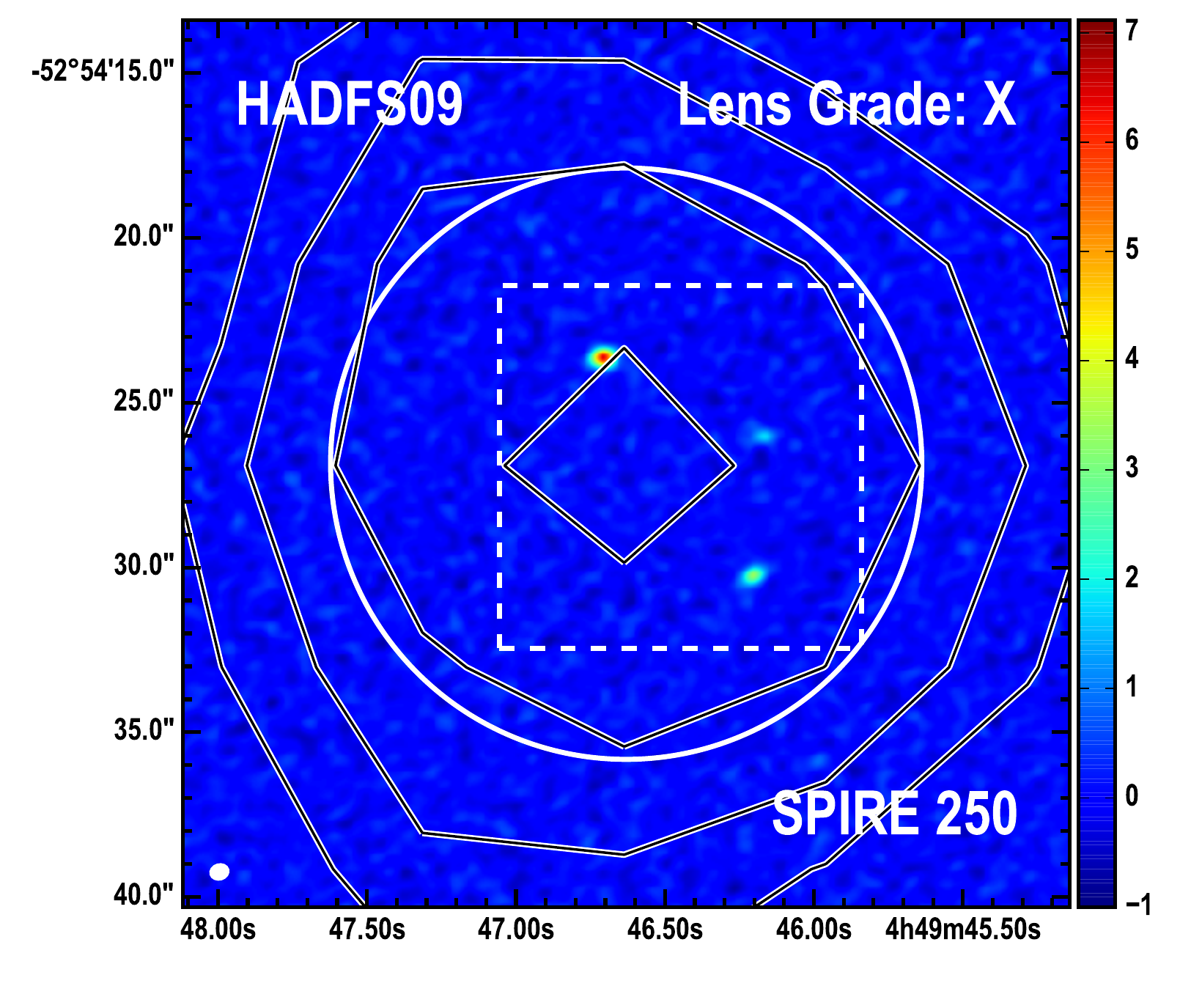}
\includegraphics[width=0.331\textwidth]{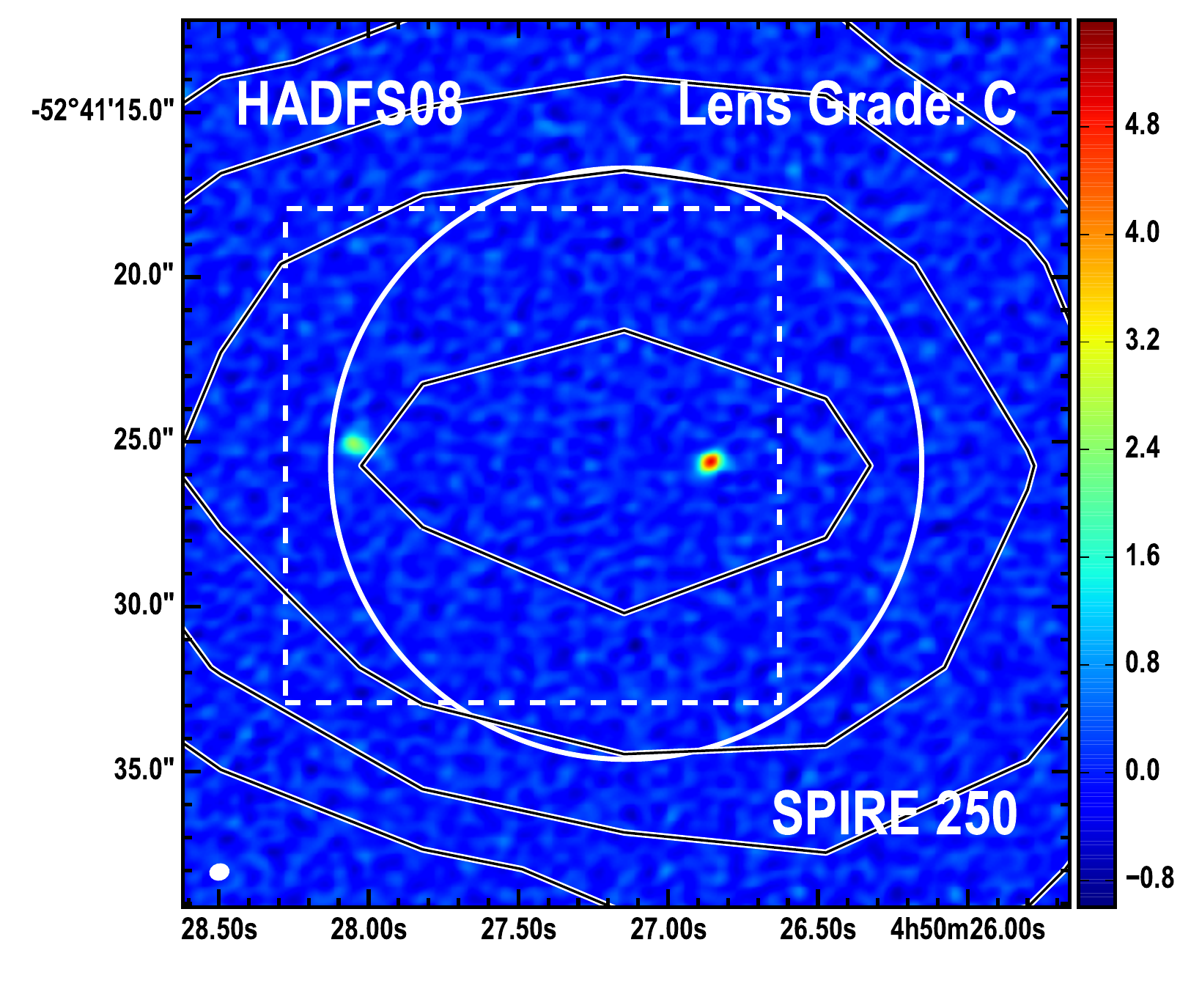}
\includegraphics[width=0.331\textwidth]{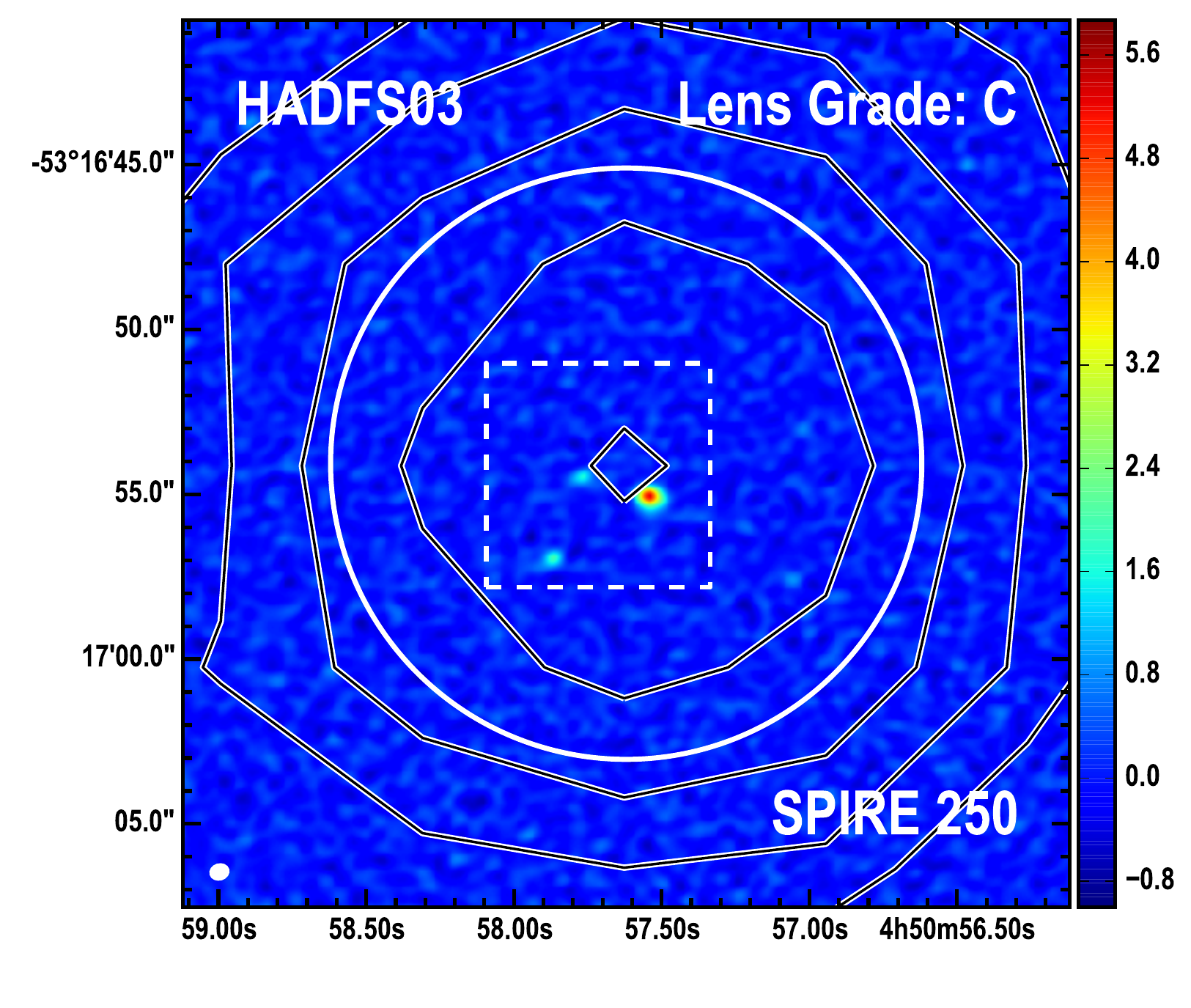}
\includegraphics[width=0.331\textwidth]{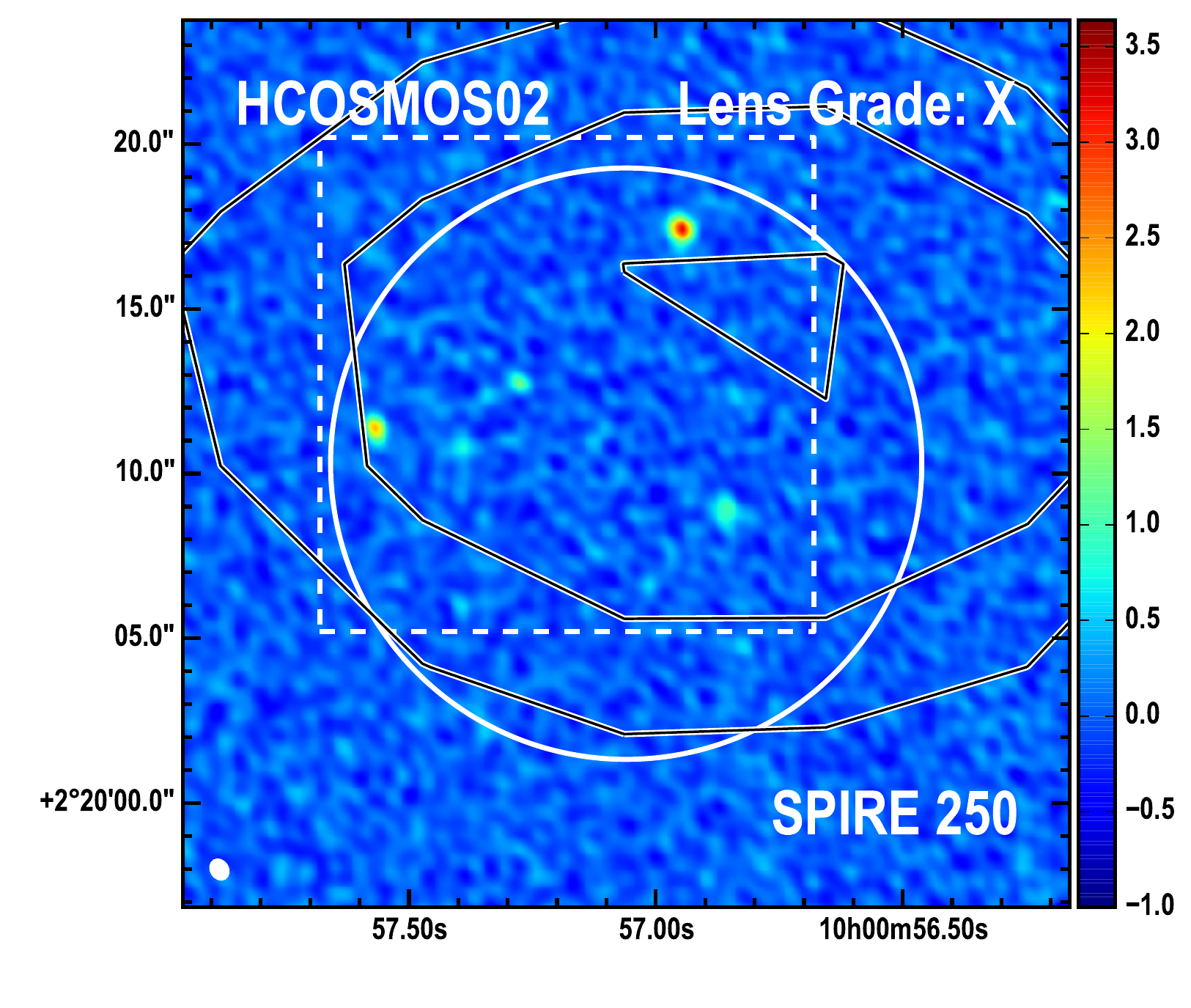}
\includegraphics[width=0.331\textwidth]{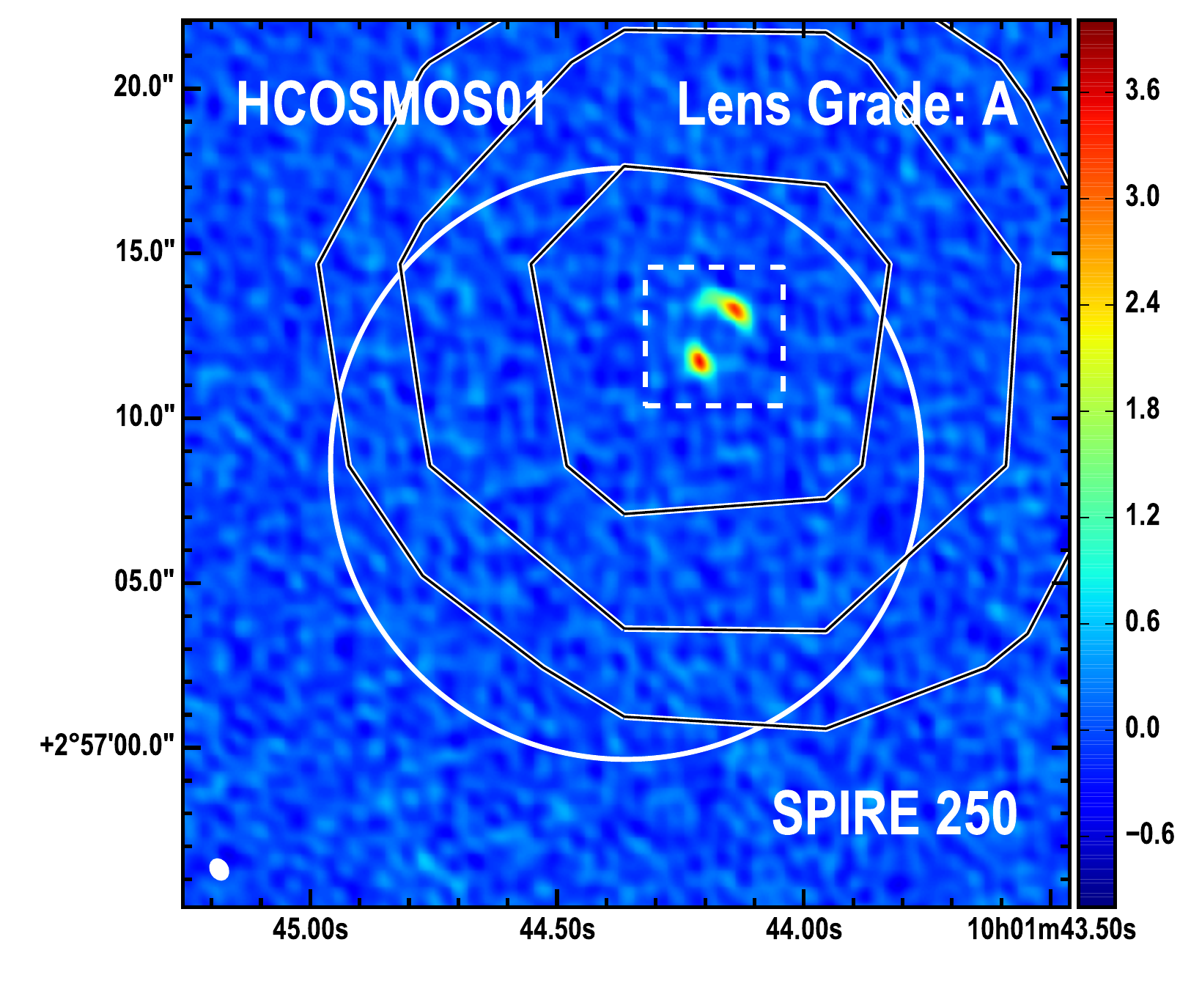}
\end{centering}

\caption{ Continued.}

\end{figure*}

\section{Model Fits}\label{sec:modelfits}

\subsection{Model Fitting Methodology}\label{sec:modelfitsmeth}

An interferometer measures visibilities at discrete points in the {\it uv}
plane.  This is why pixel-to-pixel errors in the inverted and deconvolved
surface brightness map of an astronomical source are correlated.  The best way
to deal with this situation is to compare model and data visibilities rather
than surface brightness maps.  The methodology used in this paper is similar in
many aspects to that used in \citet{Bussmann:2012lr}, who presented the first
lens model derived from a visibility-plane analysis of interferometric imaging
of a strongly lensed DSFG discovered in wide-field submm surveys as well as in
\citet{Bussmann:2013lr}, where this work was extended to a
statistically significant sample of 30 objects.  It also bears some resemblence
to the method used in \citet{Hezaveh:2013fk}, who undertake lens modeling of
interferometric data in the visibility plane.  We summarize important
information on the methodology here, taking care to highlight where any
differences occur between this work and that of our previous efforts.

We created and made publicly available custom software, called {\sc uvmcmcfit},
which is capable of modeling all of the ALMA sources in this paper efficiently.  

Sources are assumed to be elliptical Gaussians that are parameterized by the
following six free parameters: the position of the source (relative to the
primary lens if a lens is present, $\Delta \alpha_{\rm s}$ and $\Delta
\delta_{\rm s}$), the total intrinsic flux density ($S_{\rm in}$), the
effective radius length ($r_{\rm s} = \sqrt{a_{\rm s} b_{\rm s}}$), the axial
ratio ($q_{\rm s}$ =  $b_{\rm s}/a_{\rm s}$), and the position angle
($\phi_{\rm s}$, degrees east of north).  The use of an elliptical Gaussian
represents a simplification from the S\'ersic profile
\citep{1968adga.book.....S}  that is permitted based on the relatively weak
constraints on the S\'ersic index found in our previous work
\citep{Bussmann:2012lr, Bussmann:2013lr}.

When an intervening galaxy (or group of galaxies) is present along the line of
sight, {\sc uvmcmcfit} accounts for the deflection of light caused by this
structure using a simple ray-tracing routine that is adopted from a Python
routine written by A.~Bolton
\footnote{http://www.physics.utah.edu/$\sim$bolton/python\_lens\_demo/}.  This
represents a significant difference from \citet{Bussmann:2012lr} and
\citet{Bussmann:2013lr}, where we used the publicly available {\sc Gravlens}
software \citep{Keeton:2001lr} to map emission from the source plane to the
image plane for a given lensing mass distribution.  {\sc Gravlens} has a wide
range of lens mass profiles as well as a sophisticated algorithm for mapping
source plane emission to the image plane, but it also comes with a significant
input/output penalty that makes parallel computing prohibitively expensive.
For example, modeling a simple system comprising one lens and one source
typically required 24-48$\,$hours using the old software, whereas the same
system can be modeled in less than one hour with the pure-Python code (tests of
the Bolton ray-tracing routine indicate it produces results consistent with
{\sc Gravlens}).  The use of pure-Python code for tracing the deflection of
light rays is a critical component of making {\sc uvmcmcfit} computationally
feasible.

In {\sc uvmcmcfit}, lens mass profiles are represented by $N_{\rm lens}$
singular isothermal ellipsoid (SIE) profiles, where $N_{\rm lens}$ is the
number of lensing galaxies found from the best available optical or near-IR
imaging \citep[a multitude of evidence supports the SIE as a reasonable choice;
for a review, see][]{Treu:2010fk}.  Each SIE is fully described by the
following five free parameters: the position of the lens on the sky relative to
the arbitrarily chosen ``image center'' based on the ALMA 870$\,\mu$m emission
and any lensing galaxies seen in the optical or near-IR ($\Delta \alpha_{\rm
lens}$ and $\Delta \delta_{\rm lens}$; these can be compared with the position
of the optical or near-IR counterpart relative to the ``image center'': $\Delta
\alpha_{\rm NIR}$ and $\Delta \delta_{\rm NIR}$), the mass of the lens
(parameterized in terms of the angular Einstein radius, $\theta_{\rm E}$), the
axial ratio of the lens ($q_{\rm lens} = b_{\rm lens} / a_{\rm lens}$), and the
position angle of the lens ($\phi_{\rm lens}$; degrees east of north).  Unless
otherwise stated, when optical or near-IR imaging suggests the presence of
additional lenses (see Figure~\ref{fig:uvmodels}), we estimate centroids for
each lens by eye and fix the positions of the additional lenses with respect to
the primary lens.  Each additional lens thus has three parameters: $\theta_{\rm
E}$, $q_{\rm lens}$, and $\phi_{\rm lens}$.  

The total number of free parameters for any given system is $N_{\rm free} = 5 +
3 \times (N_{\rm lens} - 1) + 6 * N_{\rm source}$, where $N_{\rm source}$ is
the number of Gaussian profiles used.

We assume secondary, tertiary, etc., lenses are located at the same redshift as
the primary lens.  If this assumption were incorrect, to first order only the
conversion from an angular Einstein radius to a physical mass of the lensing
galaxy would be affected.  As the physical masses of the lensing galaxies are
not the focus of this work, this assumption is reasonable.

We use uniform priors for all model parameters.  The prior on the position of
the lenses covers $\pm0\farcs6$ ($1\farcs0$) in both RA and Dec, a value that
reflects the 1-$\sigma$ absolute astrometric solution between the ALMA and
optical/near-IR images of $0\farcs2$ ($0\farcs4$) for SDSS-based (2MASS-based)
astrometric calibration.  In Section~\ref{sec:objectbyobject}, we discuss the
level of agreement between the astrometry from the images and the astrometry
from the lens modeling on an object-by-object basis.  For $\theta_{\rm E}$, the
prior covers $0\farcs1 - 6\arcsec$.  The axial ratios of the lenses and sources
are restricted to be $q_{\rm lens} > 0.3$ and $q_{\rm s} > 0.2$. This
assumption is justified for the lenses because our optical observations reveal
lenses that are not highly elliptical and because we expect dark matter to be
more spherically distributed than the stars in lensing galaxies.  For the
sources, our ellipticity limit is primarily designed to aid numerical stability
in the lens modeling.  No prior is placed on the position angle of the lens or
source.  The intrinsic flux density for any source is allowed to vary from
0.1$\,$mJy to the total flux density observed by ALMA (we ensure that the
posterior PDF of the intrinsic flux density shows no signs of preferring a
value lower than 0.1$\,$mJy).  The source position is allowed to vary over any
reasonable range necessary to fit the data (typically, this is $\pm
$1$\arcsec$-2$\arcsec$).  The effective radius is allowed to vary from $0\farcs01 $-$
1\farcs5$.

The surface brightness map generated as part of {\sc uvmcmcfit} is then
converted to a ``simulated visibility'' dataset ($V_{\rm model}$) in much the
same way as MIRIAD's {\sc uvmodel} routine.  Indeed, the code used in {\sc
uvmcmcfit} is a direct Python port of {\sc uvmodel} (the use of {\sc uvmodel}
itself is not possible for the same reason as {\sc Gravlens}: constant
input/output makes parallel computing prohibitively expensive).  {\sc
uvmcmcfit} computes the Fourier transform of the surface brightness map and
samples the resulting visibilities in a way that closely matches the sampling
of the actual observed ALMA visibility dataset ($V_{\rm ALMA}$).

The goodness of fit for a given set of model parameters is determined from the
maximum likelihood estimate $L$ according to:

\begin{equation}
    L = \sum_{u, v}\left( \frac{|V_{\rm ALMA} - V_{\rm
    model}|^2}{\sigma^2} + {\rm log}(2 \pi \sigma^2) \right)
\end{equation}

\noindent where $\sigma$ is the 1$\sigma$ uncertainty level for each
visibility and is determined from the scatter in the visibilities within a
single spectral window (this is a natural weighting scheme).  

We use {\sc emcee} \citep{Foreman-Mackey:2013yq} to sample the posterior
probability density function (PDF) of our model parameters.  {\sc emcee} is a
Markov chain Monte Carlo (MCMC) code that uses an affine-invariant ensemble
sampler to obtain significant performance advantages over standard MCMC
sampling methods \citep{goodmanweare}.  

We employ a ``burn-in'' phase with 512 walkers and 500-1000 iterations (i.e.,
$\approx 250,000-500,000$ samplings of the posterior PDF) to identify the
best-fit model parameters.  This position then serves as the basis to
initialize the ``final'' phase with 512 walkers and 10 iterations (i.e., 5,120
samplings of the posterior PDF) to determine uncertainties on the best-fit
model parameters.  

During each MCMC iteration, we also measure the magnification factor at
870$\,\mu$m, $\mu_{870}$, for each source.  This is done simply by taking the
ratio of the total flux density in the lensed image of the model ($S_{\rm
out}$) to the total flux density in the unlensed, intrinsic source model
($S_{\rm in}$).  The use of an aperture when computing $\mu_{870}$ is important
when source profiles are used with significant flux at large radii (e.g., some
types of S\'ersic profiles).  For an elliptical Gaussian, such a step is
unneccessary (note that we did test this and found only $\approx 10\%$
difference between $\mu_{870}$ computed with and without an aperture).  The
best-fit value and 1$\sigma$ uncertainty on $\mu_{870}$ are drawn from the
posterior PDF, as with the other parameters of the model.  Exceptions are made
for cases of weakly lensed sources where we have only upper limits on the
Einstein radius (and hence upper limits on $\mu_{870}$).  In such instances, we
re-compute $\mu_{870}$ as the arithmetic mean of the limiting $\mu_{870}$ and
unity: $\mu_{870} = (\mu_{\rm 870-limit} + 1) / 2$.  The uncertainty in
$\mu_{870}$ is assumed to be equal to $\mu_{\rm 870-limit} - 1) / 2$.

Finally there are some important caveats to our approach. The spatial
resolution of the ALMA observations is $\approx 0\farcs45$, which is nearly
always sufficient to resolve the images of the lensed galaxy, but not always
sufficient to resolve the images themselves.  For this reason, in some cases
the lens models may moderately over-predict the intrinsic sizes of the lensed
galaxies and hence under-predict the magnification factors.  In addition, our
Gemini-S optical imaging may have missed optically faint lenses due to being at
high redshift or dust-obscured (but not sufficiently active to be detected
by ALMA).

\subsection{Individual Model Fits}\label{sec:objectbyobject}

In this section, we present our model fits (as shown in
Figure~\ref{fig:uvmodels}) and describe each source in detail.

\begin{figure*}[!tbp] 
    \begin{centering}
\includegraphics[width=0.195\textwidth]{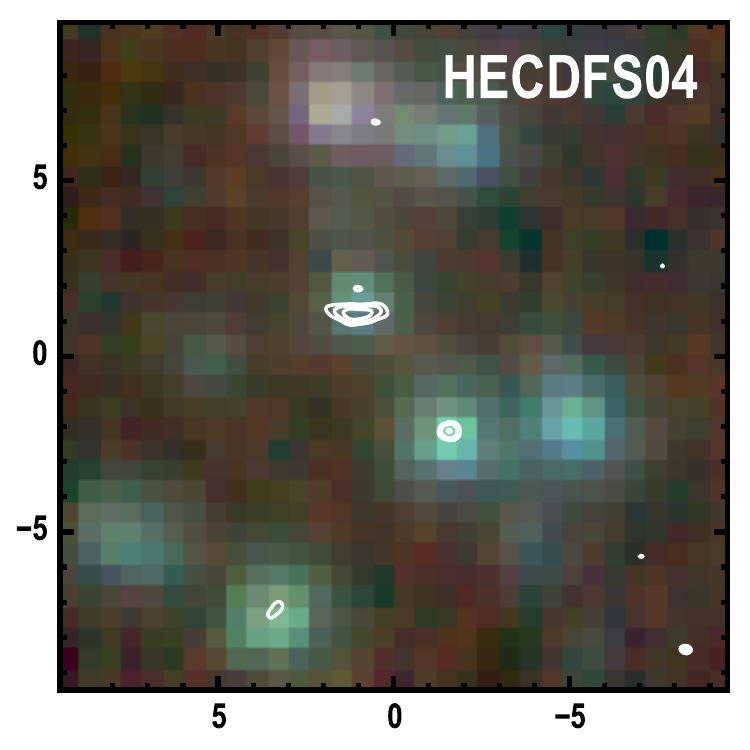}
\includegraphics[width=0.195\textwidth]{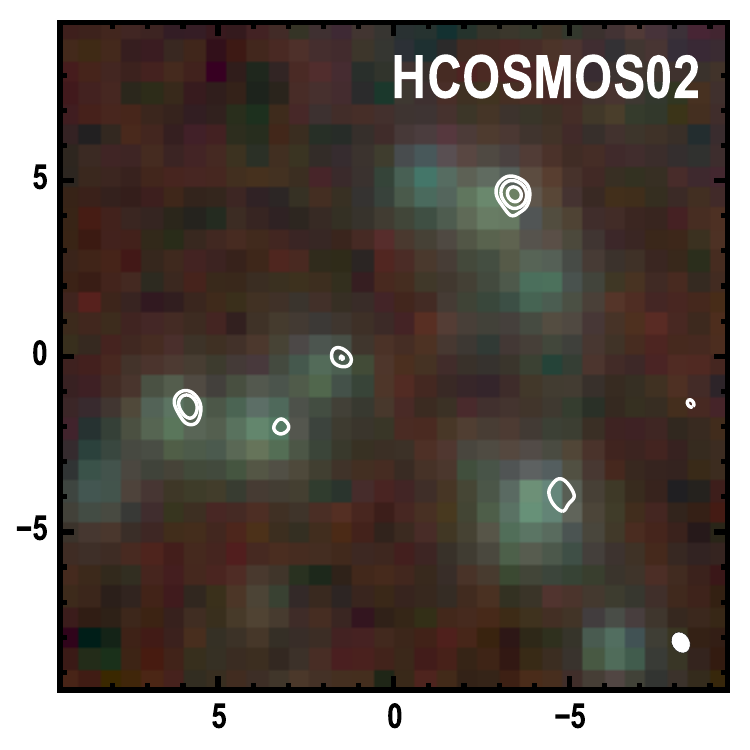}
\includegraphics[width=0.195\textwidth]{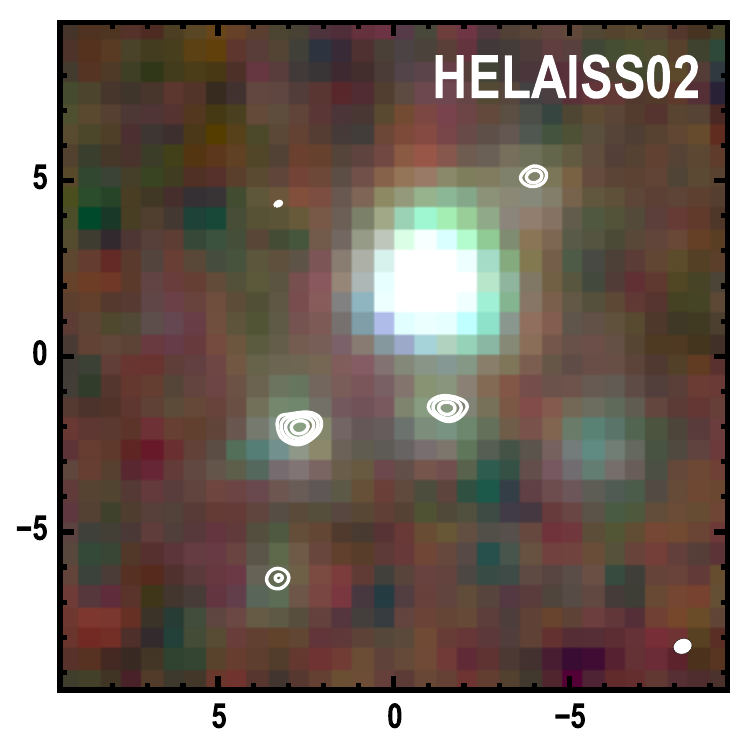}
\includegraphics[width=0.195\textwidth]{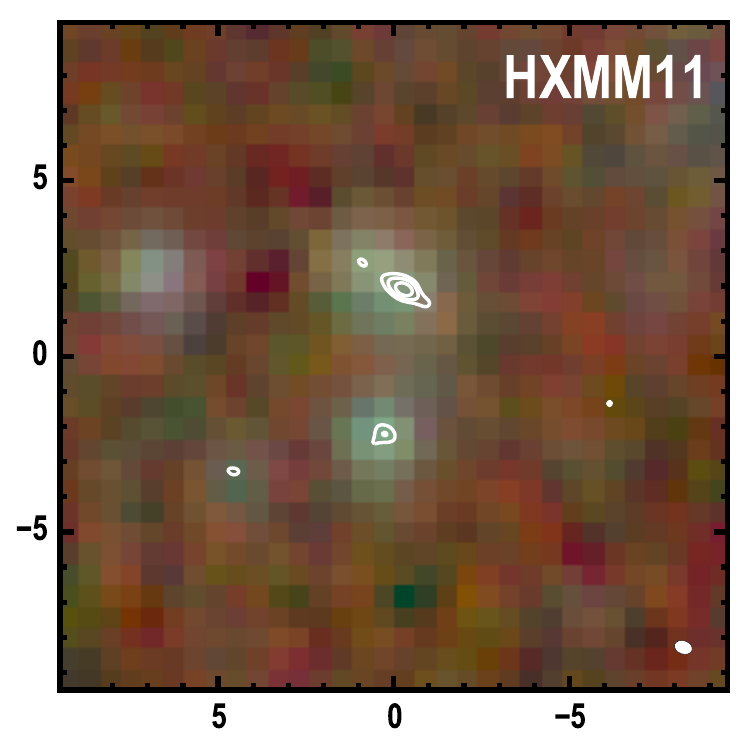}
\includegraphics[width=0.195\textwidth]{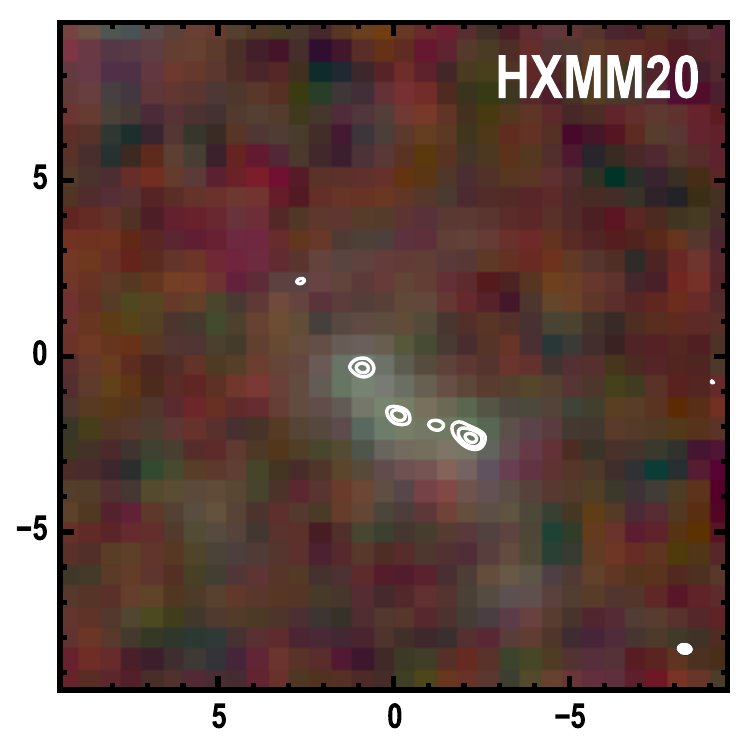}
\end{centering}

\caption{ ALMA 870$\,\mu$m imaging (white contours, starting at 4$\sigma$ and
increasing by factors of 2) overlaid on color composite IRAC imaging (blue =
3.6$\,\mu$m, green = 4.5$\,\mu$m, red = 8.0$\,\mu$m).  All panels are
$27\arcsec$ on a side.  North is up and east is left.  The synthesized beam is
represented in the lower right corner of each panel.  Each of the ALMA
counterparts are detected in the IRAC imaging.  In addition, the IRAC colors of
ALMA sources are broadly consistent, providing some evidence that they are at
the same redshift and not physically unassociated blends along the line of
sight.}\label{fig:iraccolor}

\end{figure*}

{\bf HELAISS02:} Four sources are detected by ALMA, all of which are weakly
lensed by a foreground galaxy seen in the {\it HST} image.  To estimate the
maximal magnification factors, we assume an Einstein radius of $1\farcs5$ for
the lens (larger values predict counter images that are not seen by ALMA).  The
ALMA sources are all detected by IRAC and their mid-IR colors are similar,
suggesting that they lie at the same redshift (see Figure~\ref{fig:iraccolor}).

{\bf HXMM02:} One source is detected by ALMA, and it is strongly lensed by one
foreground galaxy seen in the {\it HST} image.  The lensed source is not
detected in the {\it HST} image.  This object was first detected by
\citet{Ikarashi:2011qy} and also has high quality SMA imaging and an
accompanying lens model that produces consistent results with those given here
\citep{Bussmann:2013lr}.

{\bf HXMM31:} Two sources are detected by ALMA, neither of which are lensed.
The faint, diffuse emission seen in the CFHT $i$-band image is atypical of
lensing galaxies.  The nearest bright galaxy seen at $i$-band is located
$\approx 18\arcsec$ southeast of the ALMA sources.

{\bf HXMM29:} Two sources are detected by ALMA, none of which appear to be
lensed.  The brighter ALMA source is weakly detected in the CFHT $i$-band
image.  

{\bf HXMM07:} One source is detected by ALMA, and it is strongly lensed by one
foreground galaxy detected in the Gemini-S image.  There is a $\approx
0\farcs5$ offset in the position of the foreground galaxy between the lens
model and the Gemini-S image.  Given the absolute astrometric rms uncertainty
of $0\farcs2$ (based on SDSS), we do not consider this offset to be
significant.  The presence of a handful of $\pm3\sigma$ peaks in the residual
map is likely an indication that our assumption of a single Gaussian to
describe the source morphology is an oversimplification.

{\bf HXMM20:} Five sources are detected by ALMA, none of which appear to be
lensed.  There are a few faint smudges seen in the {\it HST} image which are
likely to be the rest-frame optical counterparts to the ALMA sources.  The ALMA
sources are all arranged in a chain like shape, possibly suggestive of a larger
filamentary overdensity in which they might reside.  IRAC imaging provides
support for this hypothesis (see Figure~\ref{fig:iraccolor}), as all of the ALMA
sources are detected and have similar mid-IR colors.

{\bf HXMM01:} Three sources are detected by ALMA, all of which are weakly
lensed by two foreground galaxies seen in the {\it HST} and Keck/NIRC2
imaging.  The ALMA imaging is broadly consistent with SMA data originally
presented in \citet{Fu:2013lr}, with two bright sources and a much fainter
third source very close to the more southern bright source.  We assume Einstein
radii of $0\farcs5$ for both lenses in order to reproduce the approach used in
\citet{Fu:2013lr}.  This results in magnification factors for the three sources
of $\mu_{870} \approx 1.6 - 1.7$, consistent with \citet{Fu:2013lr}.

{\bf HXMM04:} One source is detected by ALMA, and it is weakly lensed by a
foreground galaxy seen in the {\it HST} image.  We assume an Einstein radius of
$0\farcs5$ to represent the lensing scenario with maximum amplification.  Due
to the elliptical nature of the lens, this results in a maximum magnification
factor of $\mu_{\rm 870-limit} = 3.72 \pm 0.42$.  The {\it HST} morphology is
complex: diffuse emission to the north of the lens could be a detection of the
background source or could be a long spiral arm associated with the lensing
galaxy.

{\bf HXMM09:} Two sources are detected by ALMA, both of which are weakly lensed
by a single foreground galaxy detected in the {\it HST} image.  An Einstein
radius of $1\farcs5$ is used to represent the ``maximal lensing'' scenario and
results in maximal magnification factors of $\mu_{\rm 870-limit} = 2.25 \pm
0.17$ and $\mu_{\rm 870-limit} = 1.48 \pm 0.09$.

{\bf HXMM03:} Three sources are detected by ALMA, all of which are weakly
lensed by a foreground galaxy detected in the {\it HST} image and located
$\approx 6\arcsec$ from the ALMA sources.  The central source is much brighter
than the other two sources, which makes fitting a model challenging.  We forced
the positions of the second and third sources to be at least $0\farcs5$ and
$-0\farcs5$ away from the first source in declination, respectively.
Furthermore, we fixed the position of the lens to be located $2\farcs5$ west
and $0\farcs5$ south of the image centroid given in Table~\ref{tab:position}.
We also fixed the Einstein radius to be $1\farcs0$, a typical value for
isolated galaxies in this sample and in \citet{Bussmann:2013lr}.  Because the
source is so far from the lens, the maximal magnification factor is only
$\mu_{\rm 870-limit} = 2.0 \pm 0.1$.

{\bf HXMM11:} Two sources are detected by ALMA, both of which are weakly
lensed.  This system is similar to HADFS08, although the two ALMA sources are
much closer and the lens must be less massive in order to avoid producing
multiple images of the closest ALMA source.  The fainter ALMA source has a much
lower maximal magnification factor than the brighter source ($\mu_{\rm
870-limit} = 1.10 \pm 0.01$ vs.  $\mu_{\rm 870-limit} = 1.63 \pm 0.11$).  Both
ALMA sources are detected by IRAC and have similar mid-IR colors, suggesting
they lie at similar redshifts (see Figure~\ref{fig:iraccolor}).

{\bf HXMM23:} One source is detected by ALMA, and it is coincident (within the
astrometric uncertainty) with a late-type galaxy seen in the {\it HST} image.
Here, we assume that the {\it HST} source is the true counterpart to the ALMA
source, implying that no lensing is occuring.  Consistent with this hypothesis
is that the SPIRE photometry show blue colors that suggest this object is at low
redshift.  Note that models in which the late-type galaxy is lensing the ALMA
source by a modest amount ($\mu_{870} < 1.2$) cannot be ruled out with the
present data.

{\bf HXMM22:} One source is detected by ALMA, and it appears to be unlensed.  A
faint smudge seen in the {\it HST} image of this source is due to a star
located $3\farcs5$ northeast of the ALMA source.

{\bf HXMM05:} One source is detected by ALMA, and it is weakly lensed by two
foreground galaxies seen in the {\it HST} images.  To compute the maximum
magnification factor, we assume an Einstein radius of 1$\arcsec$ for the
foreground lenses and fix the positions of both lenses according to the
location of the foreground galaxies in the {\it HST} image.

{\bf HXMM30:} One source is detected by ALMA, and it is strongly lensed by one
foreground galaxy detected in the Gemini-S image.  As with HXMM07, there is a
$\approx 0\farcs5$ offset between the lens position according to the lens model
and the Gemini-S image.  We do not consider this offset significant.  An
alternative model in which the lens is sub-mm luminous cannot be ruled out, but
we consider this unlikely for a number of reasons.  First, it is a more complex
model (having two sources and one lens, rather than one source and one lens).
Second, lenses are very rarely detected in sub-mm imaging.  Third, the shape
and location of the ALMA sources relative to the Gemini-S source are typical of
strongly lensed objects (consistent with the very low residuals).  Fourth, the
alternative lens model predicts the lensed source to have an intrinsic flux
density of $\approx 13 \,$mJy, which would make it the brightest source in the
sample.

{\bf HXMM12:} One source is detected by ALMA, and it is weakly lensed by a
group of foreground galaxies seen in the {\it HST} image.  We assume an
Einstein radius of $0\farcs2$ for the nearest lensing galaxy and allow a
$\pm0\farcs4$ (i.e., 2$\sigma$) shift in its position relative to that
indicated by the {\it HST} image (which has its astrometry tied to SDSS).  We
represent the remaining members of the group as a single SIS (assumed to be
spherical to simplify the model) located $4\farcs5$ south and $4\farcs5$ east
of the image centroid and having an Einstein radius of $2\farcs0$.  This SIS is
justified by the presence of several sources in this region of the {\it HST}
image (not shown in Figure~\ref{fig:uvmodels}).  This is meant to represent the
``maximal lensing'' scenario.  The presence of two $3\sigma$ peaks located near
the center of the residual image indicates that the model does not fit the data
perfectly.  This could be an indication that either of our assumptions for the
lens potential or source structure are oversimplifications.  Higher resolution
imaging is needed to determine the most likely cause.

{\bf HECDFS12:} This is a complex, well-constrained system.  Two background
sources are detected by ALMA: one is multiply imaged and the other is singly
imaged.  In addition, the lens is detected by ALMA (this is one of two sources
in the entire {\it Herschel}-ALMA sample that is unresolved by ALMA).  These
facts work together to provide very tight constraints on the system.  Since the
lens is detected by ALMA, its position relative to the lensed images is
unambiguous.  Also, because there is a strongly lensed source with multiple
images, the Einstein radius of the lens is unambiguous.  Finally, this source
is detected (and unresolved) in the NRAO VLA Sky Survey \citep{Condon:1998uq},
having $S_{\rm 1.4\,GHz} = 21.8 \pm 0.8\,$mJy.  Assuming all of this radio
emission originates from the lens, this implies a spectral slope of $\alpha =
-0.24$ and is consistent with non-thermal emission from the lens.  For this
target, we show VIDEO $K_{\rm s}$ imaging \citep{Jarvis:2013lr}.

{\bf HECDFS04:} Two sources are detected by ALMA, both of which are weakly
lensed by a foreground galaxy seen in the {\it HST} image.  There is also a
3$\sigma$ peak coincident with an {\it HST} source that may be an indication
that the lens has been detected by ALMA.  We do not attempt to model this
3$\sigma$ peak.  We assume an Einstein radius of $0\farcs5$ for the lens, since
larger values predict the existence of counter images that are not seen by
ALMA.  The second ALMA source is located $\approx5\arcsec$ from the lens and
experiences a small but significant magnification of $\mu_{\rm 870-limit} =
1.12 \pm 0.02$.  Both ALMA sources appear to be detected by IRAC and have
similar mid-IR colors, suggesting they lie at the same redshift (see
Figure~\ref{fig:iraccolor}).  

{\bf HECDFS13:} This system is similar to HADFS02 (mentioned below), except
that here the two ALMA sources are separated by $\approx 0\farcs4$ rather than
$0\farcs8$ and one source is brighter than the other by a factor of 2.
Assuming the two sources have similar mass-to-light ratios, their brightness
ratios indicate major merger rather than minor merger activity.  The projected
physical distance is $\approx 2-3\,$kpc, assuming a redshift of $z=2$ for the
ALMA sources.  This could be an example of a major merger approaching final
coalescence and experiencing a significant boost in star-formation due to
enhancements in the local gas density brought about by tidal forces during the
merger.

{\bf HADFS04:} Three sources are detected by ALMA, all of which are weakly
lensed by a foreground galaxy seen in the {\it HST} image.  We assume an
Einstein radius of $0\farcs5$ for the lens, as values larger than this produce
multiple images of the ALMA sources.  Values for the Einstein radius that are
smaller than $0\farcs5$ are unlikely based on the brightness of the lens, so
the results we report for this object should be robust.

{\bf HADFS02:} Two sources are detected by ALMA.  The nearest possible lens is
located $\approx 8\arcsec$ from the ALMA sources, indicating that lensing is
likely to be irrelevant in this system.  The two ALMA sources are similarly
bright ($S_{870} = 8.27 \pm 0.53\,$mJy and $S_{870} = 9.07 \pm 0.27\,$mJy) and
separated by $\approx 0\farcs8$, corresponding to a projected physical distance
of $\approx 6\,$kpc. This distance is typical of the pericentric passage
distance in both hydrodynamical simulations of major mergers
\citep[e.g.,][]{Hayward:2012lr} and observations of major mergers
\citep[e.g.,][]{Ivison:2007qv, 2008ApJ...680..246T, 2010ApJ...724..233E,
2011ApJ...733L..11R, 2011MNRAS.412.1913I}.  Two plausible scenarios are that
HADFS02 represents a major merger that just experienced a first pass or is
approaching final coalescence, either of which would significantly enhance
star-formation in the system.

{\bf HADFS11:} Two sources are detected by ALMA, both of which are weakly
lensed by a group of small galaxies detected in the {\it HST} image.  To
estimate the maximum magnification factor, we represent the gravitational
potential of the group with a single SIE lens and an Einstein radius of
$1\farcs0$.  Values larger than this produce additional counter images that are
not seen in the ALMA imaging.

{\bf HADFS10:} Three sources are detected by ALMA.  We assume that all three
are unlensed.  There is a group of three sources detected in our Gemini-S
optical imaging located $\approx 7\arcsec$ east of the ALMA sources.  This
distance is so large that plausible mass ranges for the Gemini-S sources would
imply at most a factor of 1.1$-$1.2 boost in the apparent flux densities of the
ALMA sources.  We also tested a single-lens, single-source model in which the
source is triply-imaged in the manner that is observed.  The lens in this
hypothetical model has an Einstein radius of $\approx 1\farcs2$, requiring a
very high mass to light ratio or a very high lens redshift to be consistent
with the non-detection in the Gemini-S data.  Deep near-IR imaging is needed to
confirm that this target is unlensed.

{\bf HADFS01:} This is a single source that is strongly lensed by a foreground
galaxy seen in the {\it HST} image.  The lensed source is not detected by {\it
HST}.  The source is highly elongated ($q_{\rm s} = 0.31 \pm 0.01$), but fits the
data very well.  The position of the lens according to the lens model is
consistent with the position in the {\it HST} image, given the $0\farcs4$
fundamental uncertainty due to using the 2MASS system as the fundamental basis
for the astrometry.

{\bf HADFS09:} Three sources are detected by ALMA, none of which appear to be
lensed (the closest bright {\it HST} source is located $\approx 13\arcsec$ away
from the ALMA sources).

{\bf HADFS08:} Two sources are detected by ALMA, both of which are weakly
lensed by a foreground galaxy in the {\it HST} image.  The ALMA sources have
the largest separation of any in our sample overall, around $10\arcsec$.  We
assume an Einstein radius of $1\farcs5$ for the foreground lens as a ``maximal
lensing'' scenario.  This results in maximum magnification factors of $\mu_{\rm
870-limit} = 2.3 \pm 0.1$ and $\mu_{\rm 870-limit} = 1.2 \pm 0.1$ for the two
sources.

{\bf HADFS03:} Three sources are detected by ALMA, each of which is weakly
lensed by a single bright foreground galaxy seen in the {\it HST} image.
Alternative scenarios involving strong lensing can be ruled out by the location
of the lens: $\approx 2\arcsec-3\arcsec$ north of the centroid of the ALMA sources
(the rms error in the astrometry is set from 2MASS at a level of $\approx
0\farcs4$) as well as the atypical location and fluxes of the ALMA sources
relative to each other.  To obtain the maximum magnification factor, we assume
an Einstein radius of $0\farcs5$ and fix the position angle of the lens to be
between 40$\,^\circ$-50$\,^\circ$ to match the orientation seen in the {\it HST} image.
Larger Einstein radii can be ruled out by the absence of counter images north
of the lens.

\input{table_lenses}

{\bf HCOSMOS02:} Five sources are detected by ALMA \citep[the brightest of
which was already known;][]{Smolcic:2012zl}, none of which appear to be
lensed.  Previous research has shown this to be an overdense region \citep[this
object is called COSBO3 in][]{Aravena:2010fr} with an optical and near-IR
photometric redshift of $z=2.3-2.4$.  Our ALMA imaging offers the first
convincing evidence that the associated galaxies in the overdensity are sub-mm
bright and thus intensely star-forming.  There are a number of $2-3\sigma$
peaks in the map that could be real.  This would further increase the
multiplicity rate for this object, but we caution that there are also negative
peaks of similar amplitude (i.e., $2-3\sigma$) present in this map.  Some of
the ALMA sources have counterparts detected in the {\it HST} image, whereas all
of the ALMA sources are detected by IRAC (see Figure~\ref{fig:iraccolor}).
Their mid-IR colors are similar, providing further evidence that the ALMA
sources lie at the same redshift.  

{\bf HCOSMOS01:} This system is similar to HADFS01: a single source that is
strongly lensed by a foreground galaxy seen in the {\it HST} image.  In fact,
the background source is also detected by {\it HST} as well as Keck/NIRC2
adaptive optics imaging, and a lens model has been published based on these
data \citep{Calanog:2014lr}.  The morphology of the lensed emission is very
different between the Keck and ALMA imaging, suggesting differential
magnification is important in this object.  The very small sizes of the sources
are consistent with this as well ($r_{\rm s} = 0.023 \pm 0.003\arcsec$, Keck
and $r_{\rm s} = 0.055 \pm 0.007\arcsec$, ALMA).  Adopting a redshift of $z=2$
for the lensed source implies physical sizes of $\approx 150\,$pc and $\approx
300\,$pc for the rest-frame optical and rest-frame FIR emission, respectively.

\begin{figure*}[!tbp] 
    \begin{centering}
\includegraphics[width=0.162\textwidth]{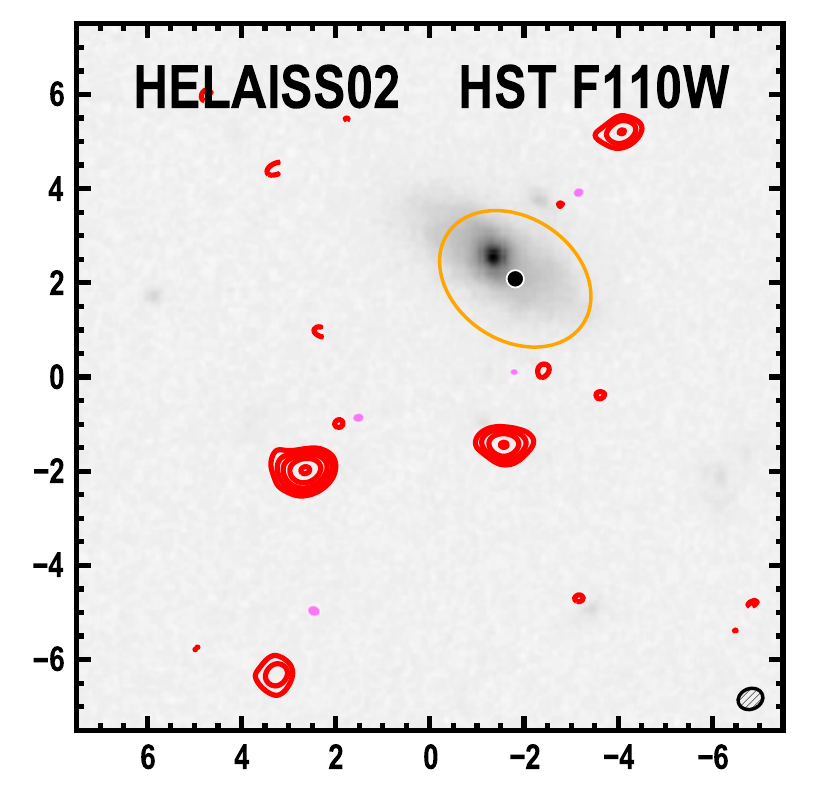}
\includegraphics[width=0.162\textwidth]{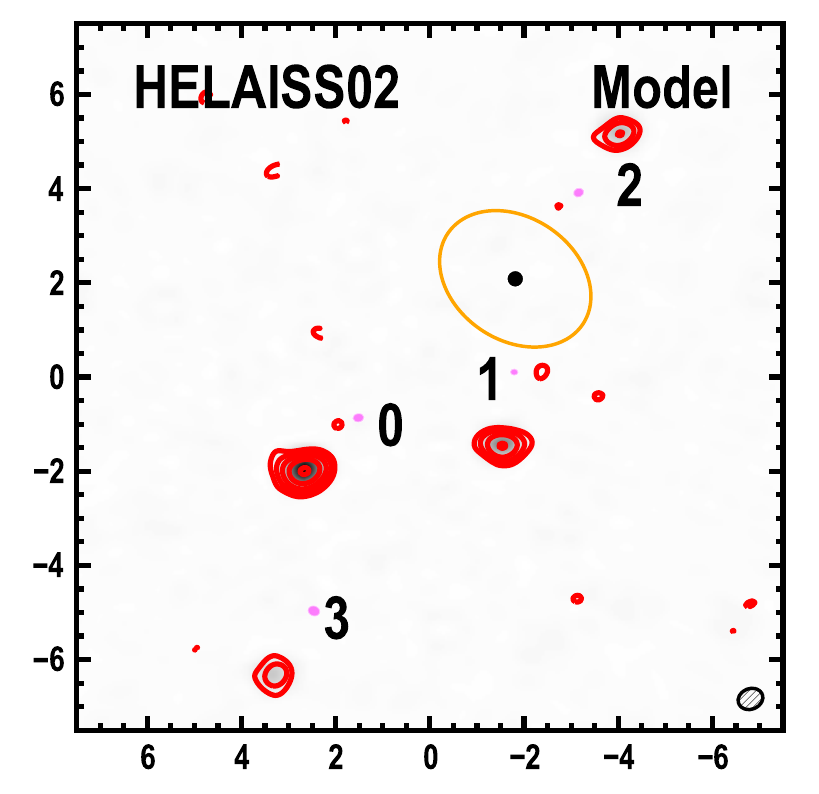}
\includegraphics[width=0.162\textwidth]{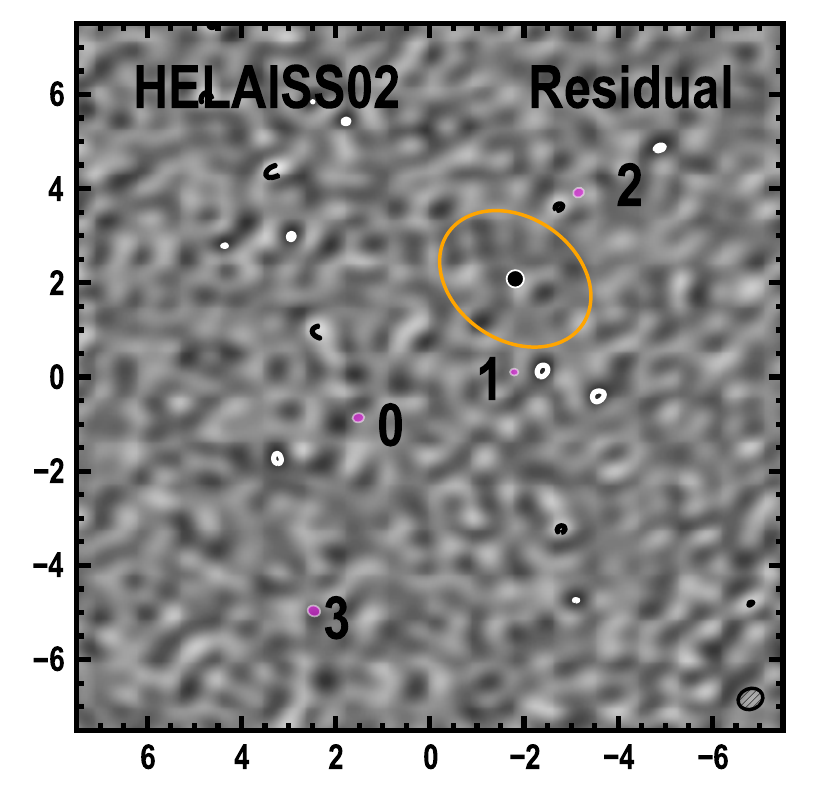}
\includegraphics[width=0.162\textwidth]{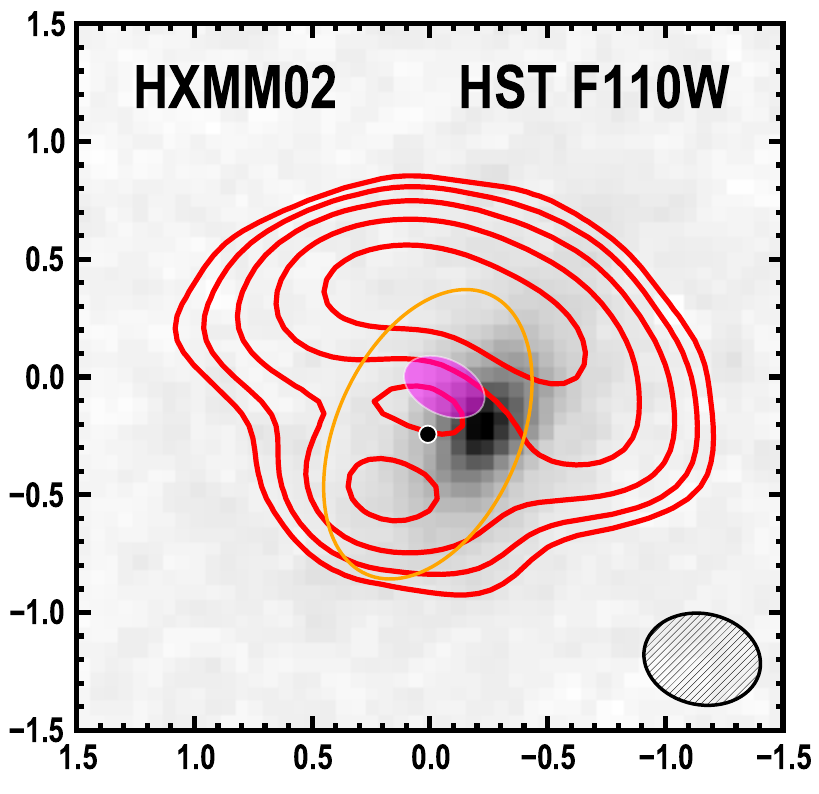}
\includegraphics[width=0.162\textwidth]{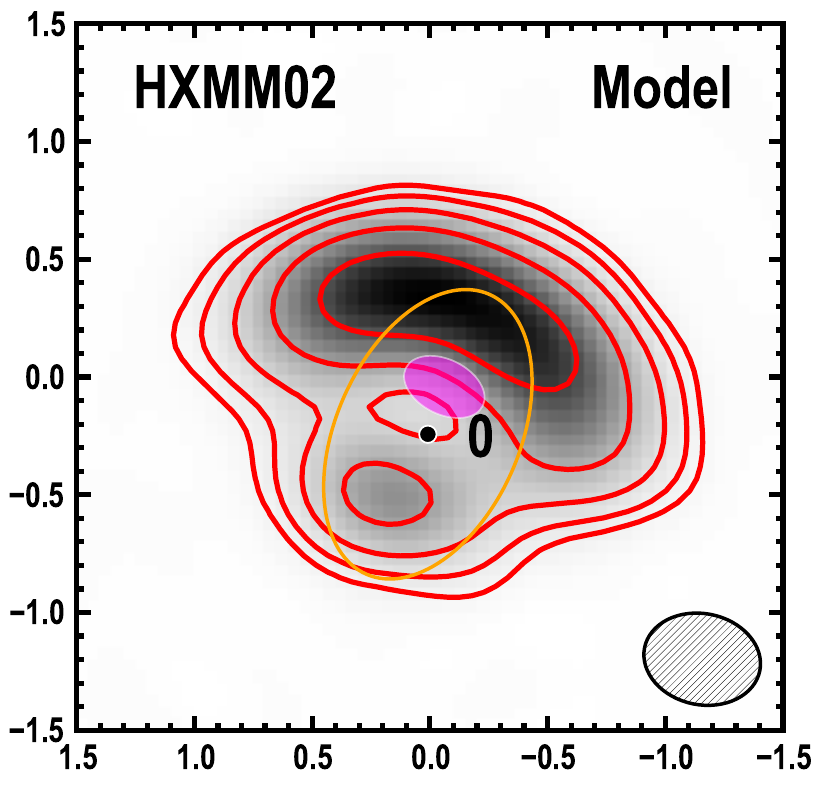}
\includegraphics[width=0.162\textwidth]{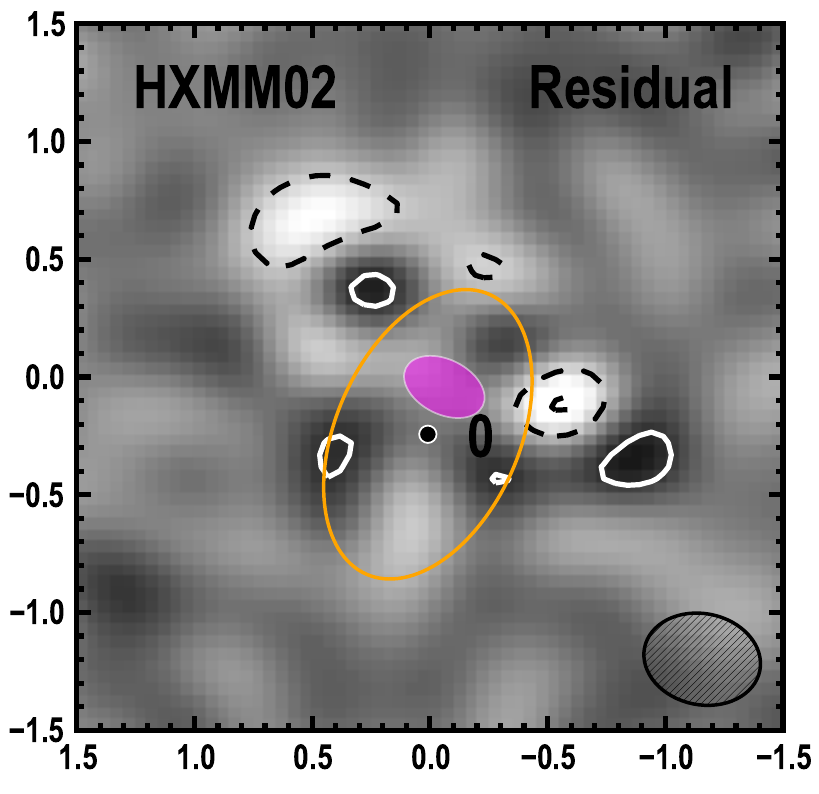}
\includegraphics[width=0.162\textwidth]{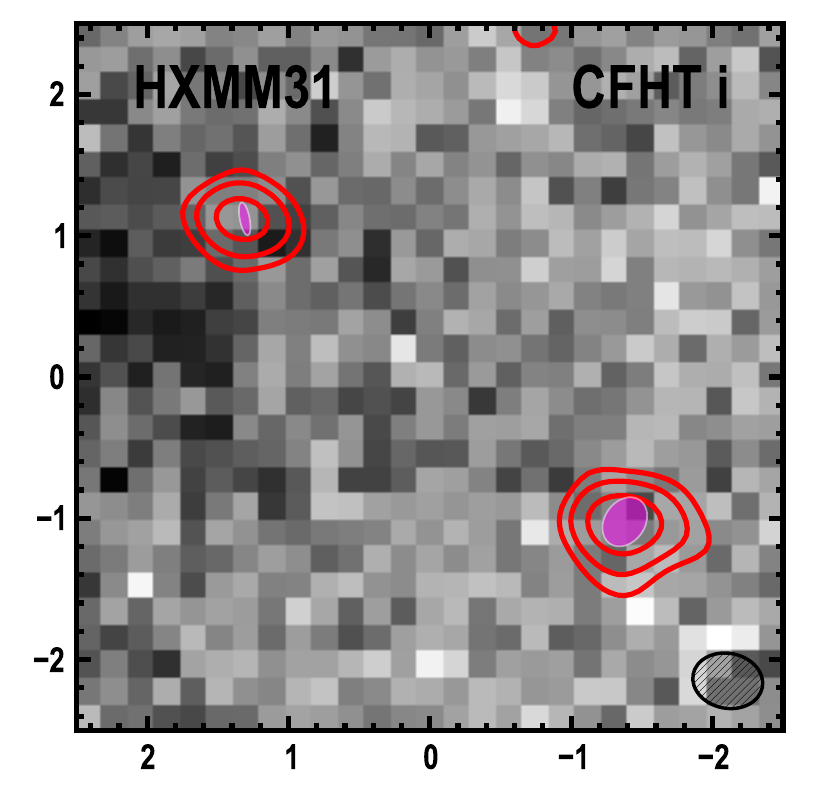}
\includegraphics[width=0.162\textwidth]{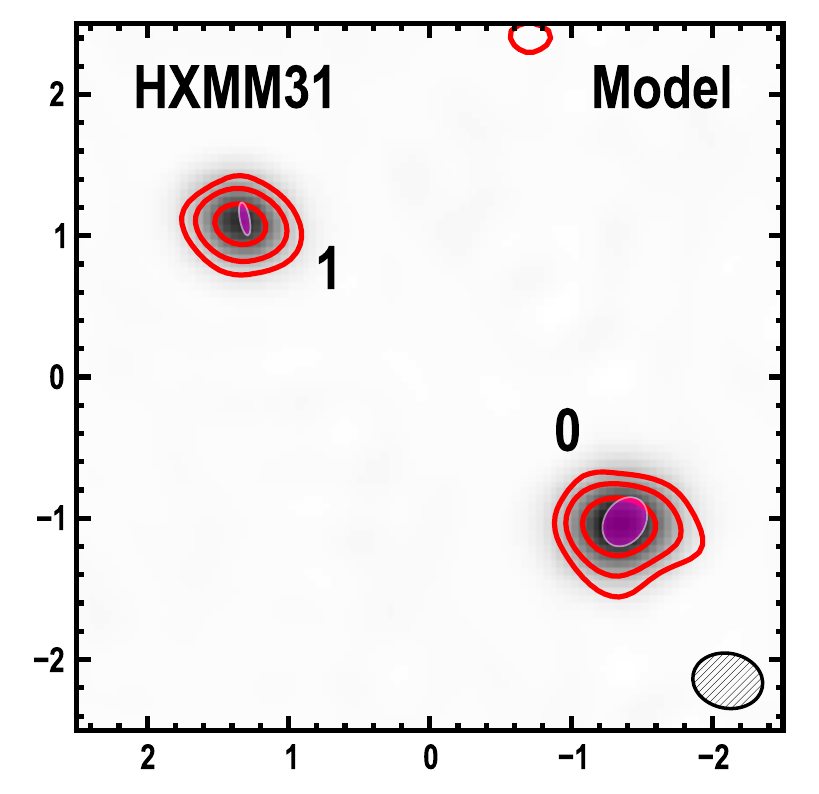}
\includegraphics[width=0.162\textwidth]{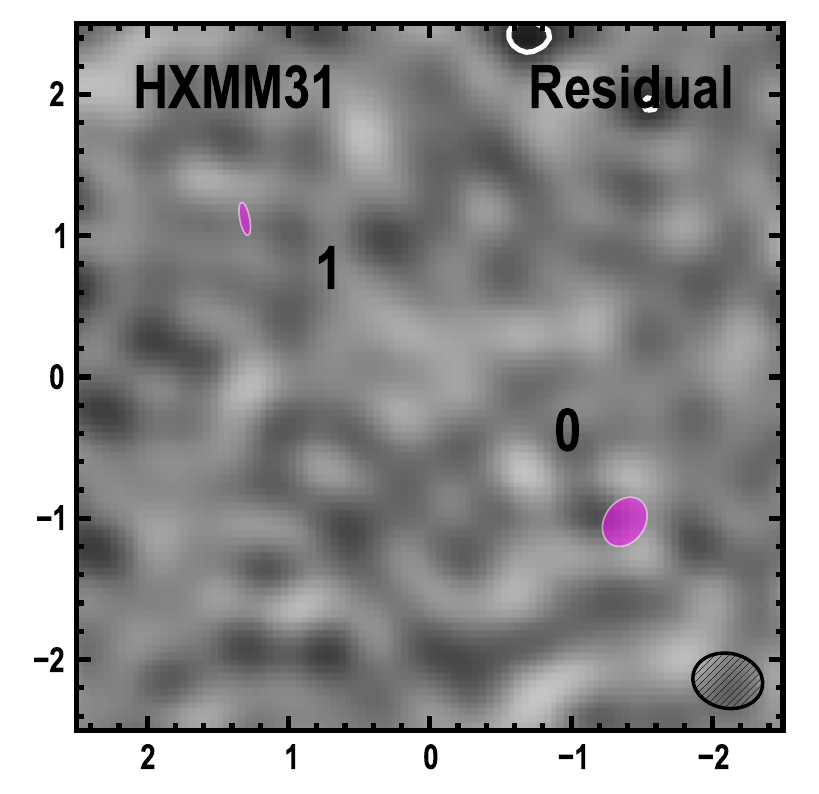}
\includegraphics[width=0.162\textwidth]{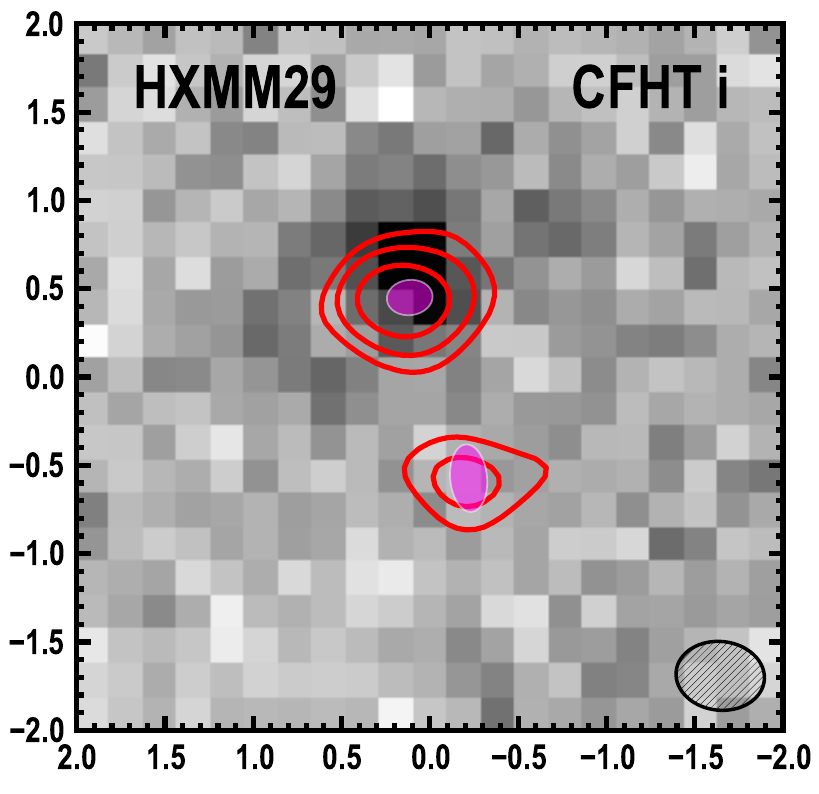}
\includegraphics[width=0.162\textwidth]{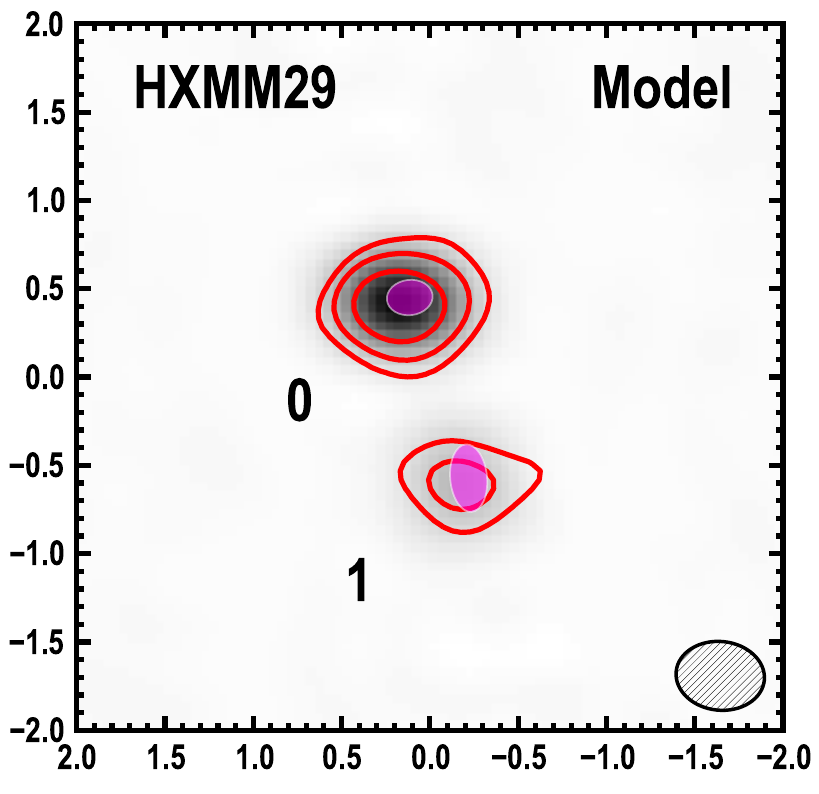}
\includegraphics[width=0.162\textwidth]{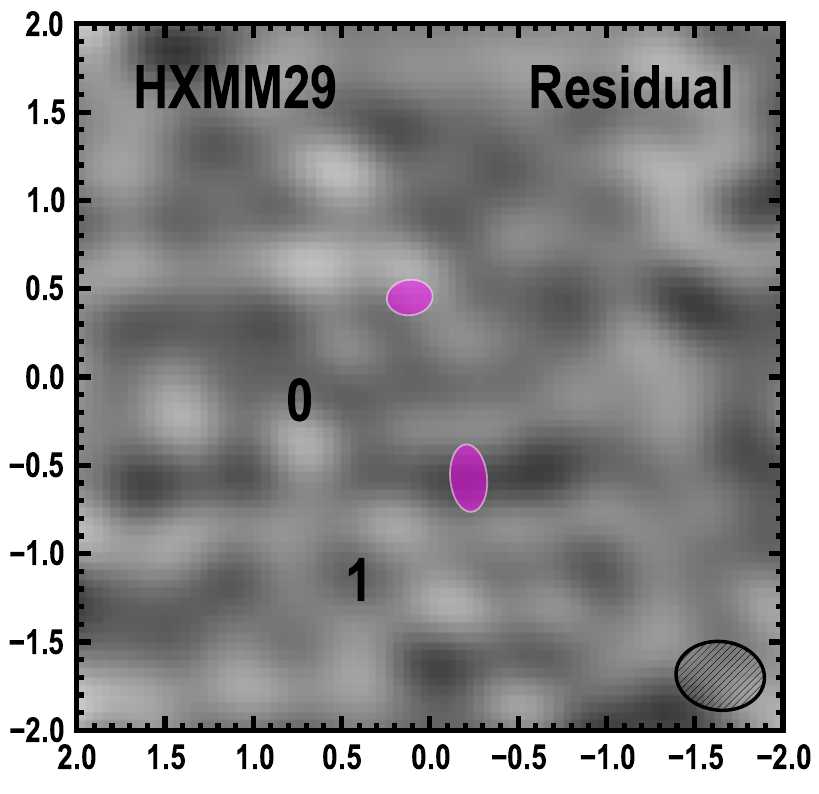}
\includegraphics[width=0.162\textwidth]{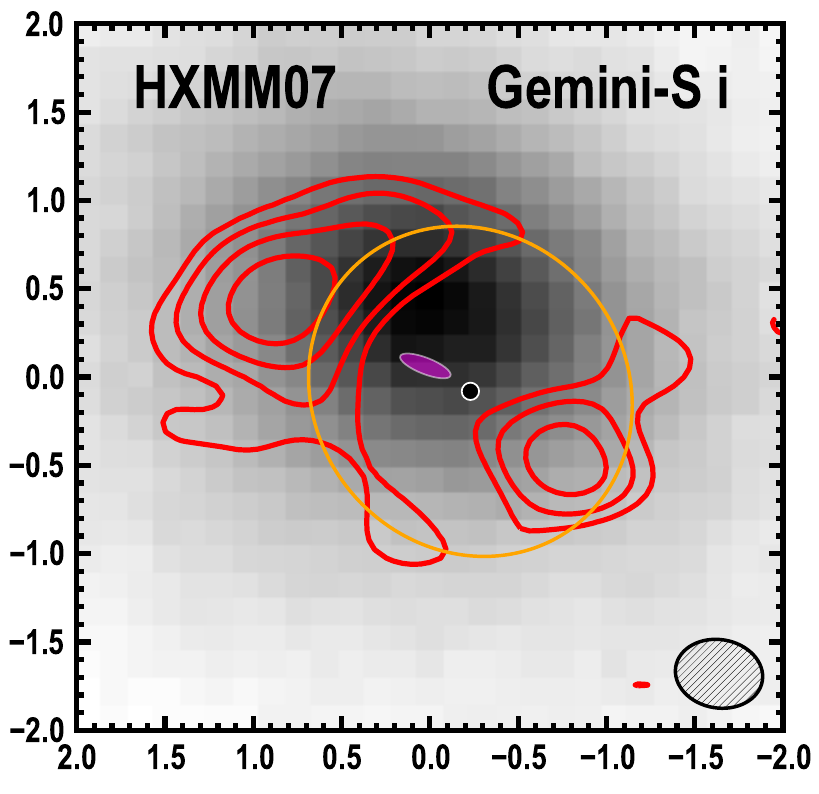}
\includegraphics[width=0.162\textwidth]{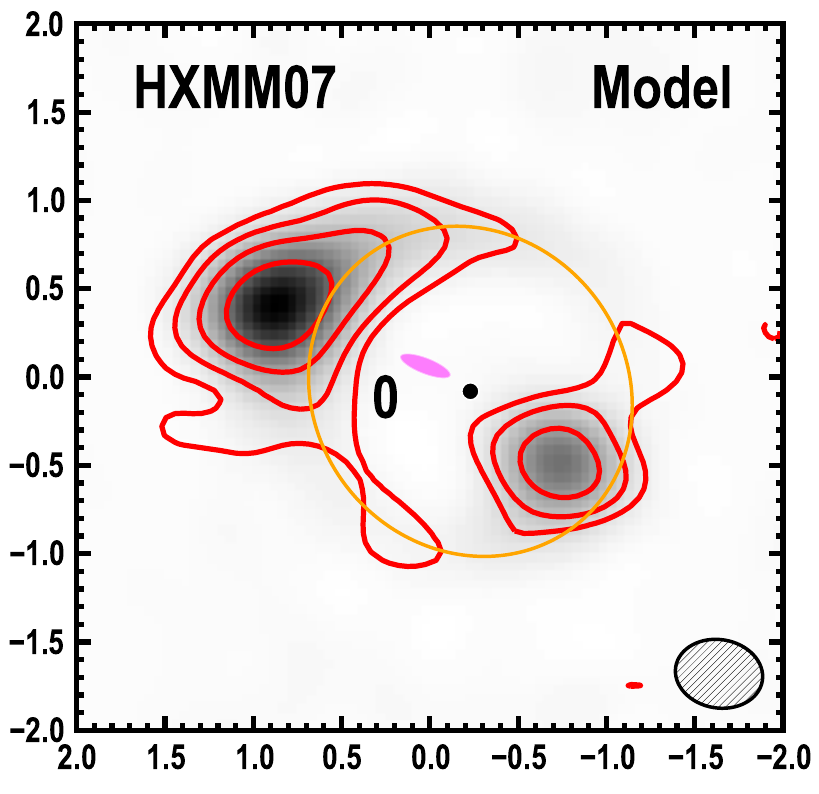}
\includegraphics[width=0.162\textwidth]{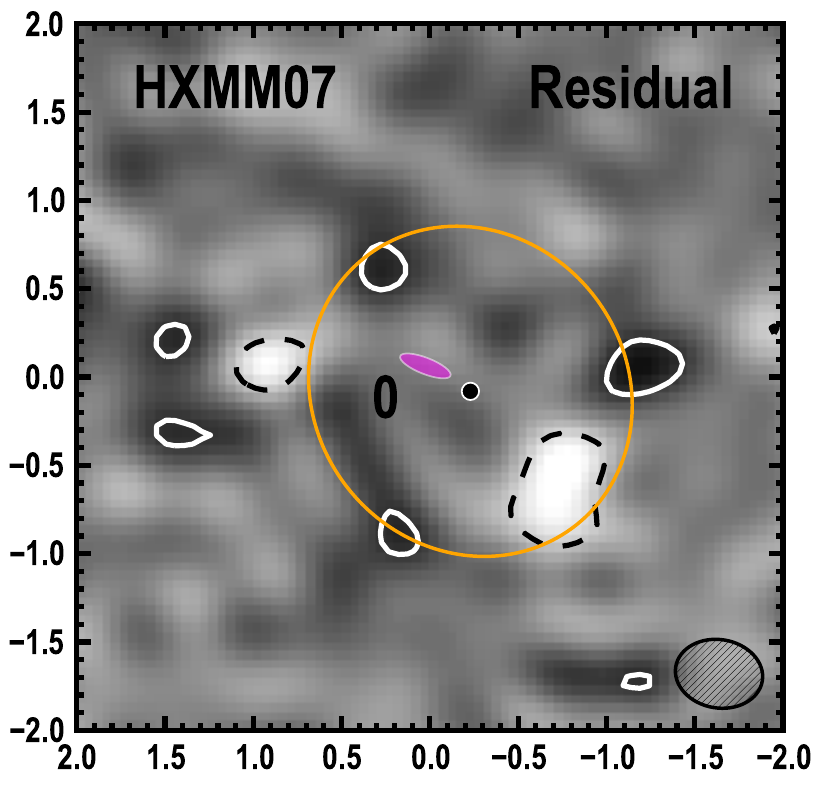}
\includegraphics[width=0.162\textwidth]{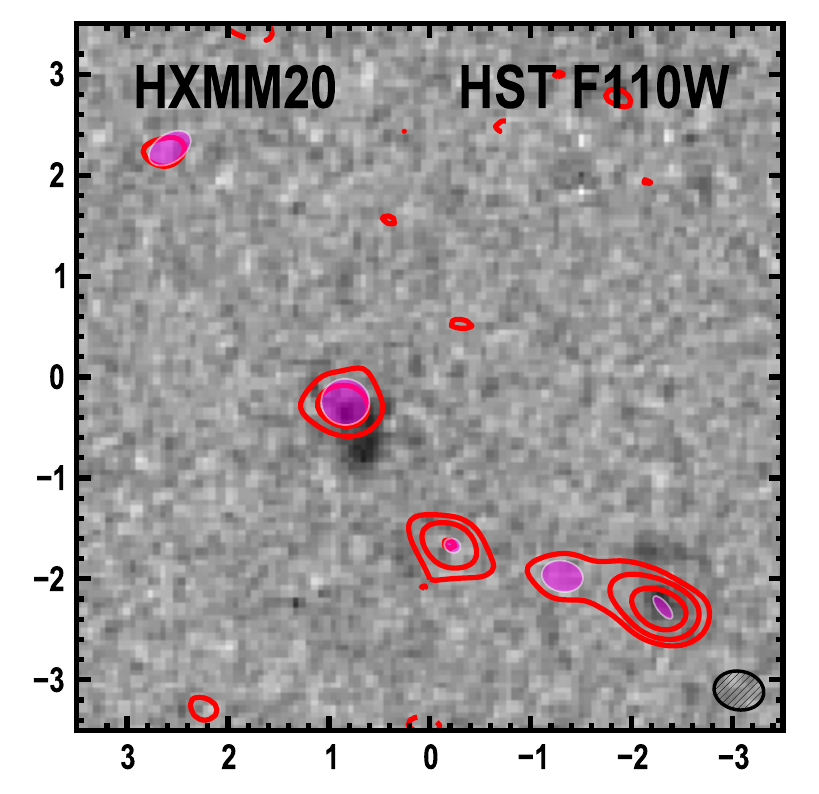}
\includegraphics[width=0.162\textwidth]{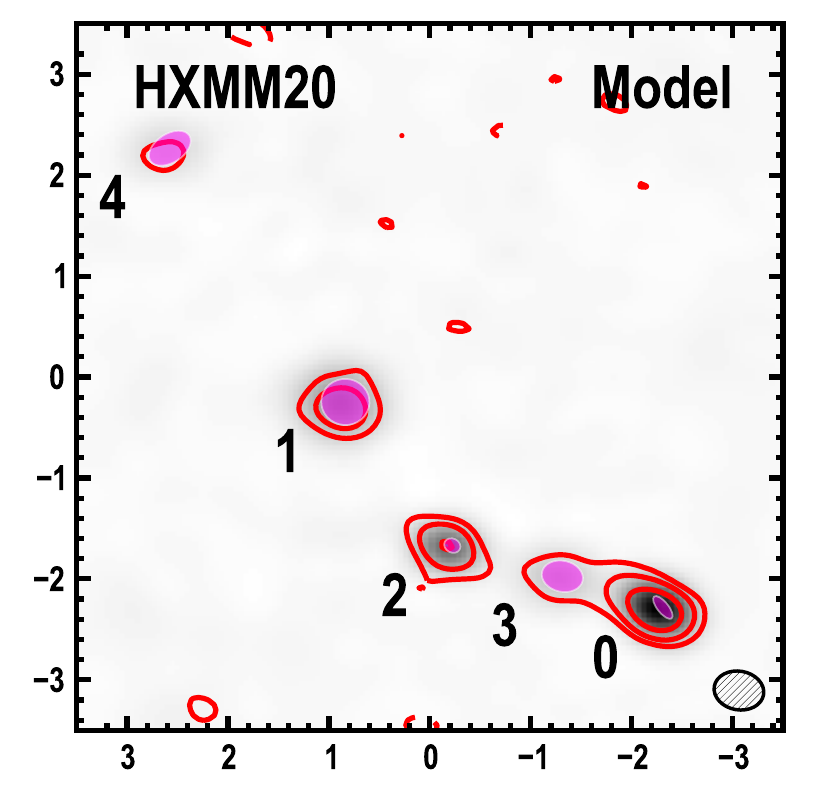}
\includegraphics[width=0.162\textwidth]{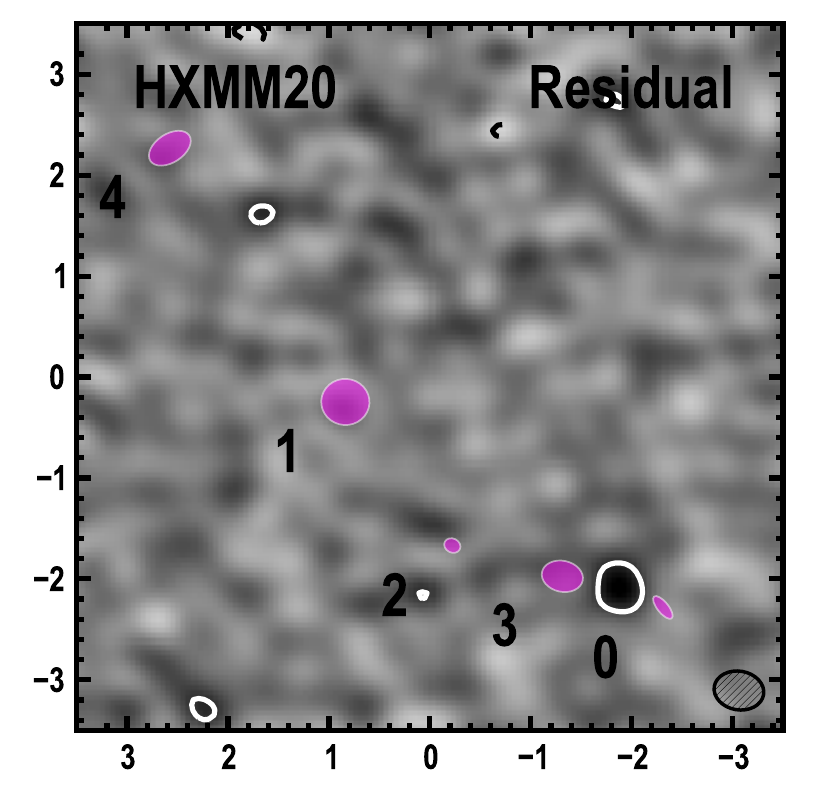}
\includegraphics[width=0.162\textwidth]{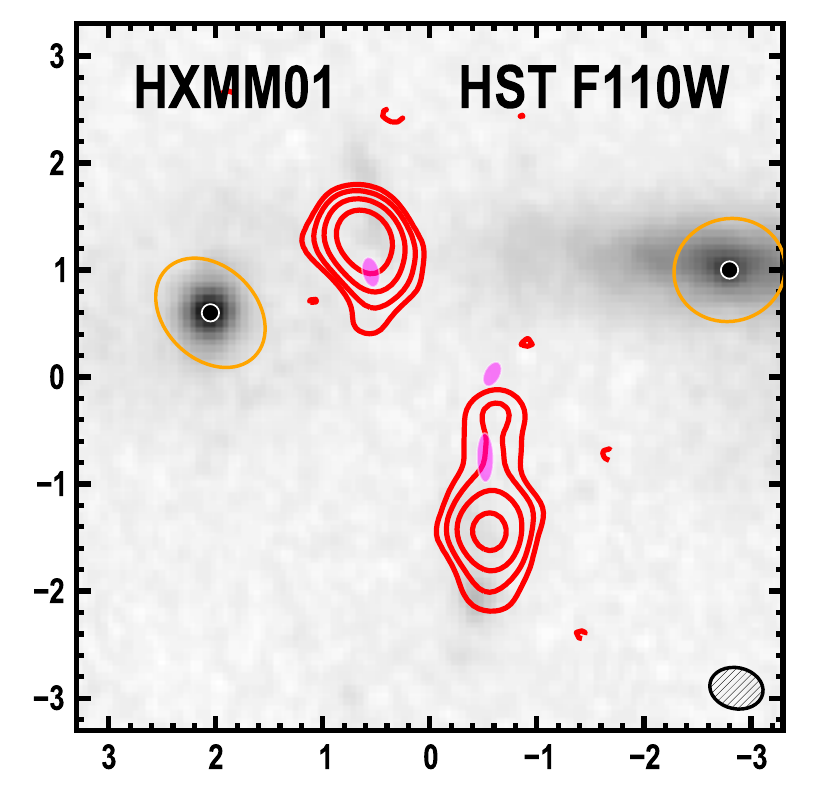}
\includegraphics[width=0.162\textwidth]{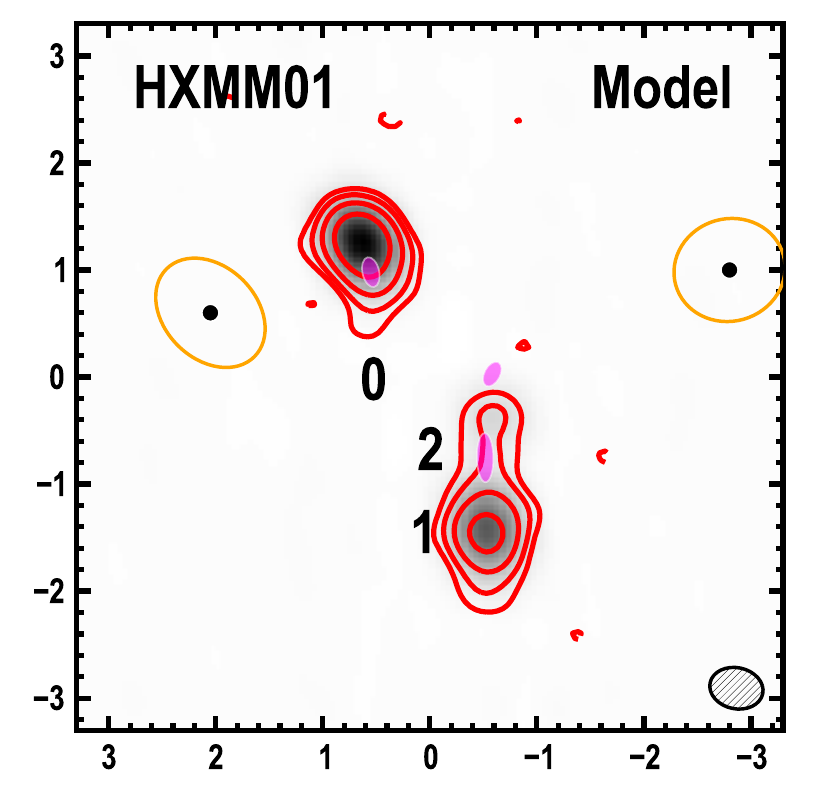}
\includegraphics[width=0.162\textwidth]{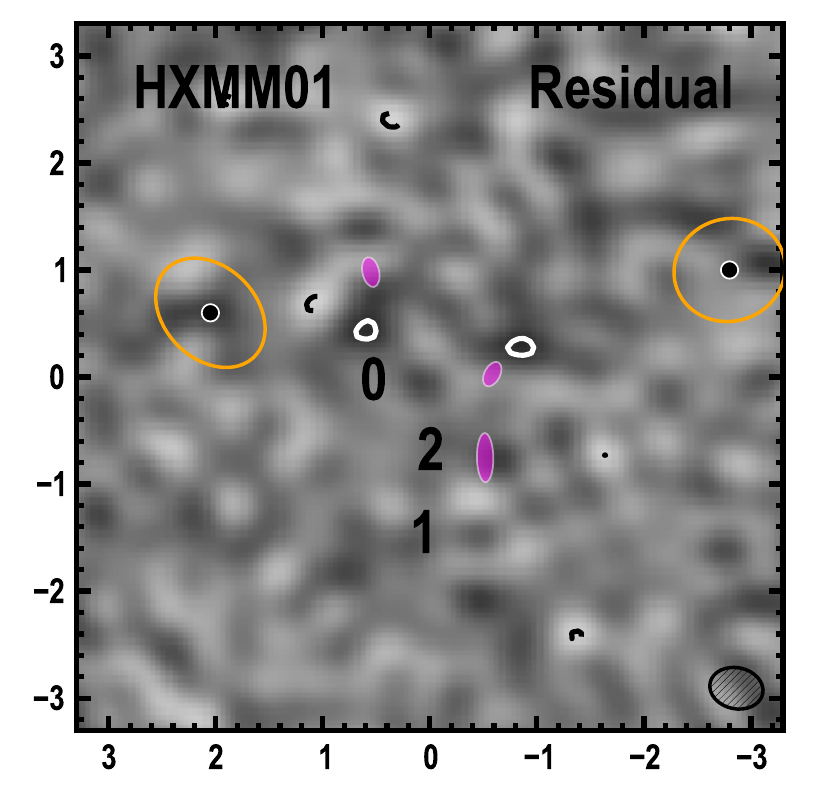}
\includegraphics[width=0.162\textwidth]{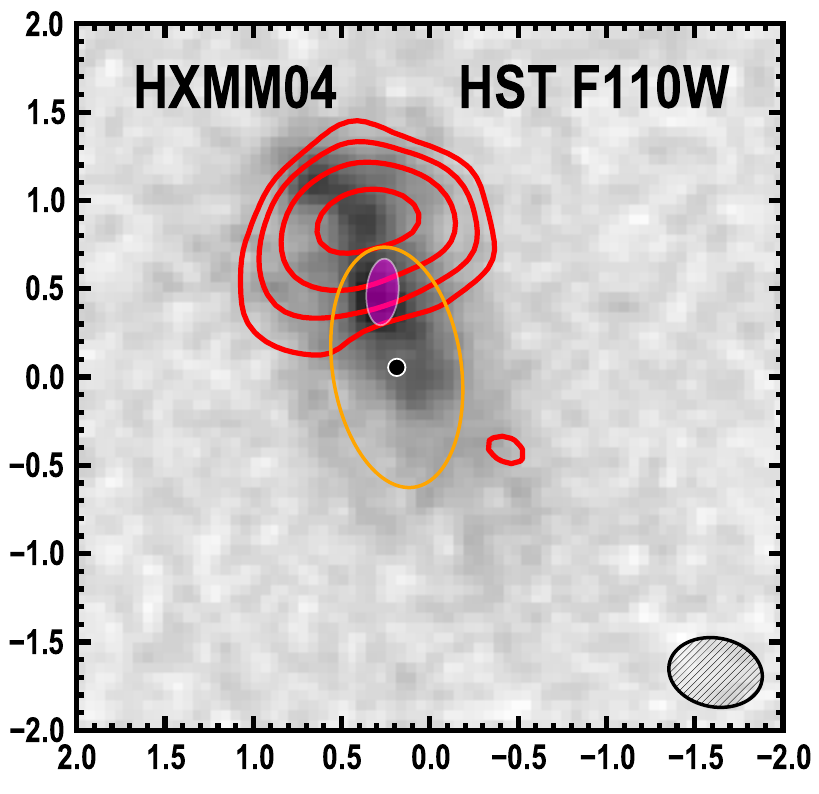}
\includegraphics[width=0.162\textwidth]{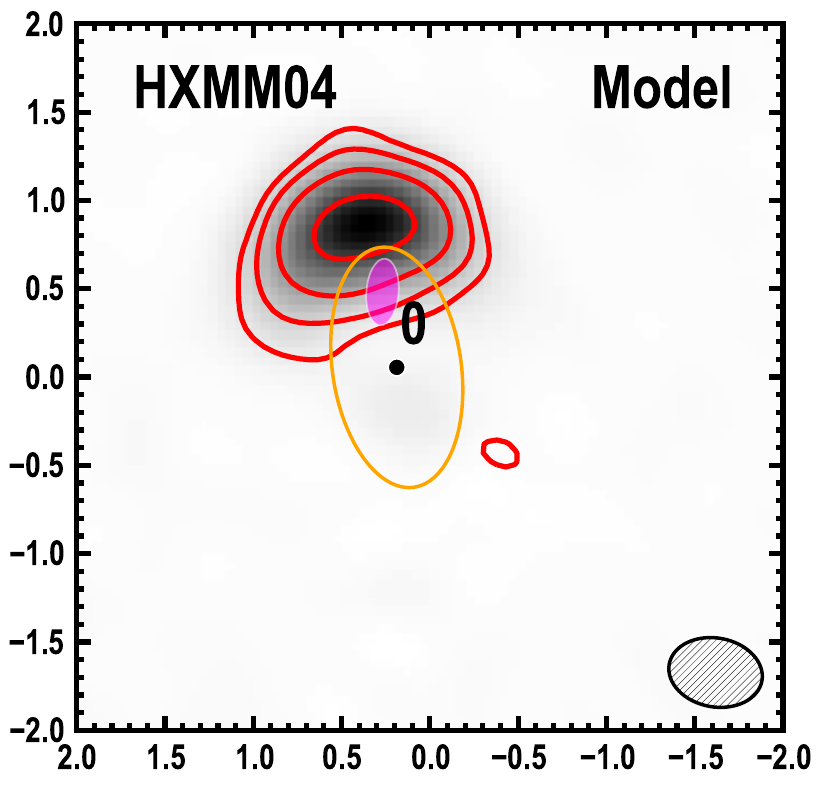}
\includegraphics[width=0.162\textwidth]{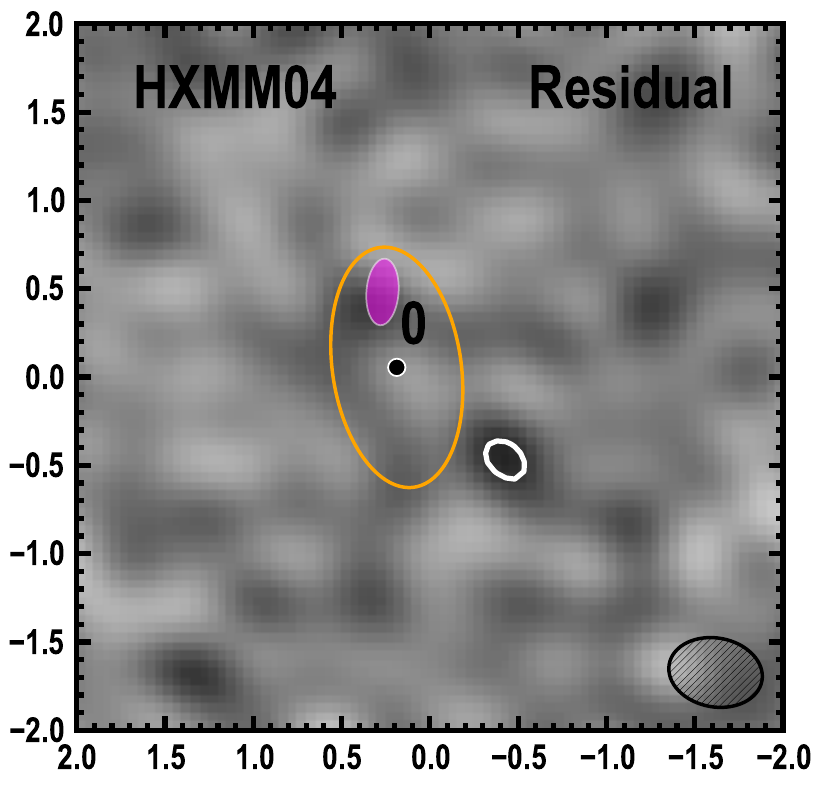}
\includegraphics[width=0.162\textwidth]{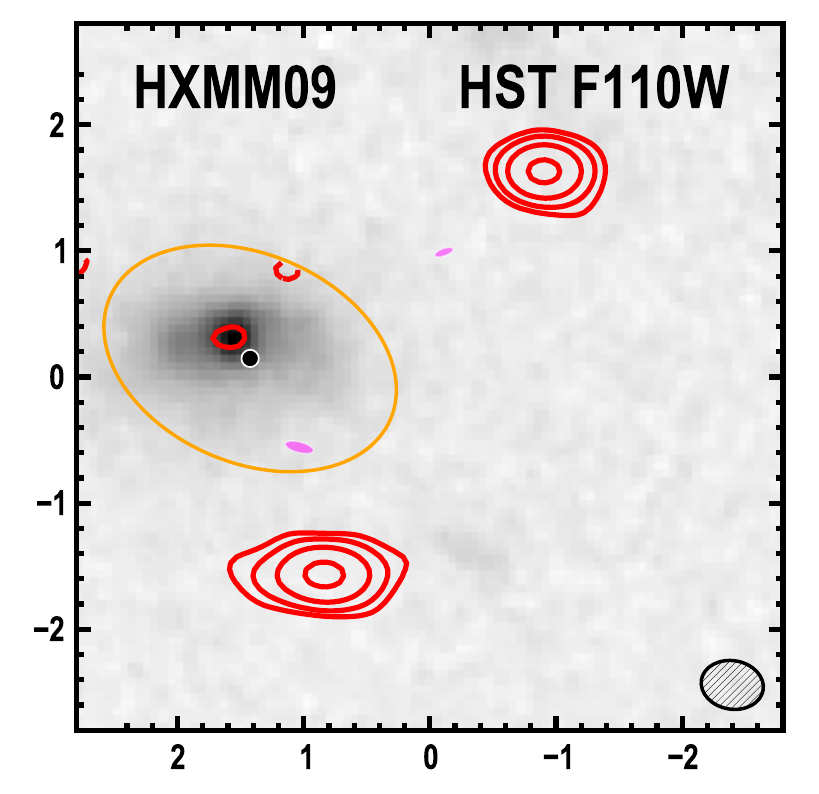}
\includegraphics[width=0.162\textwidth]{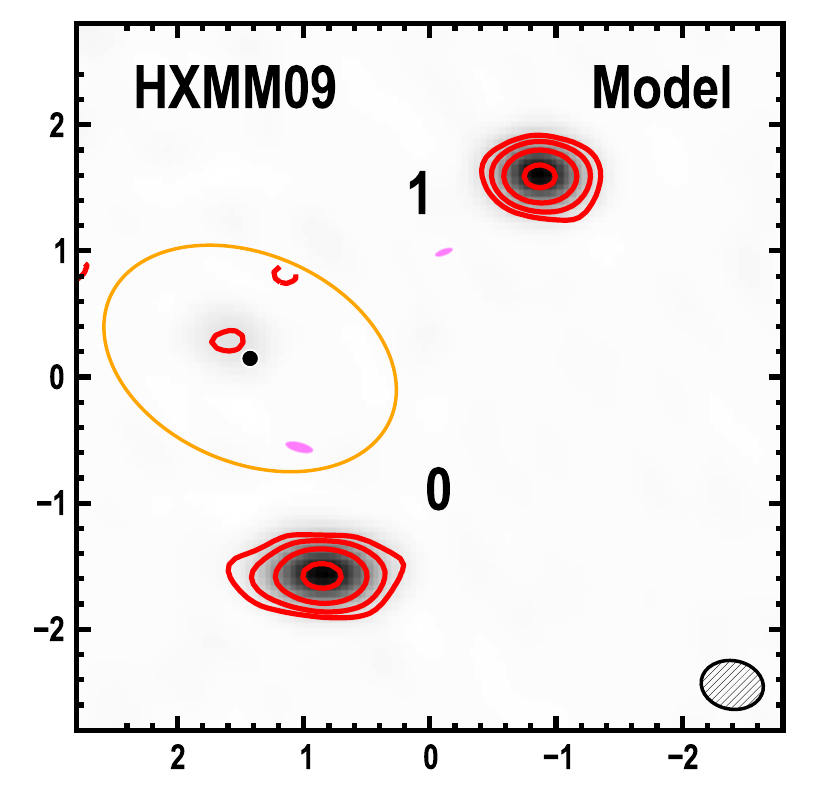}
\includegraphics[width=0.162\textwidth]{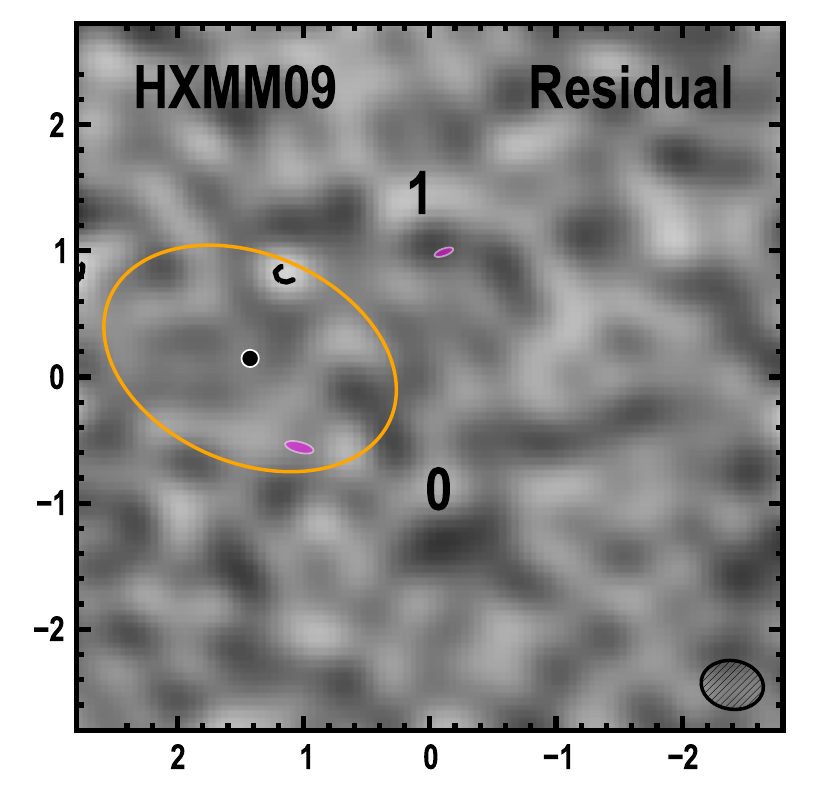}
\includegraphics[width=0.162\textwidth]{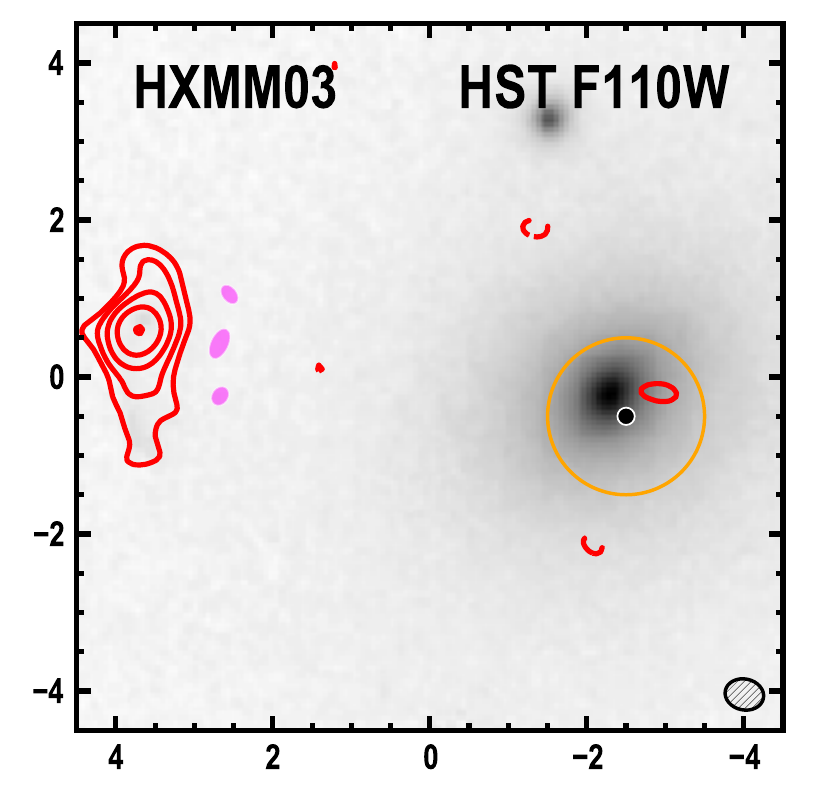}
\includegraphics[width=0.162\textwidth]{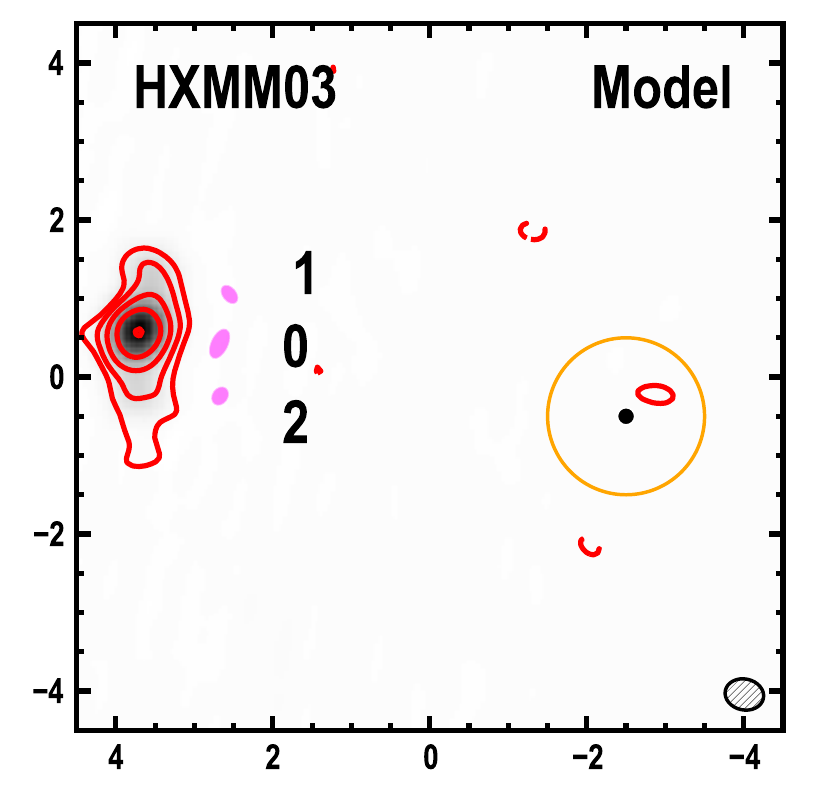}
\includegraphics[width=0.162\textwidth]{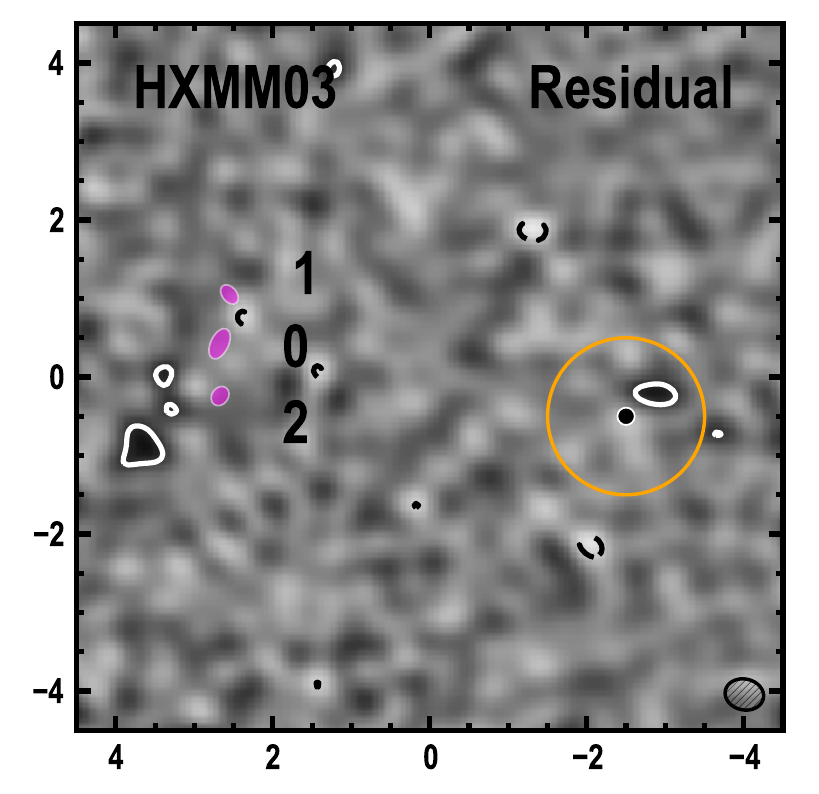}
\includegraphics[width=0.162\textwidth]{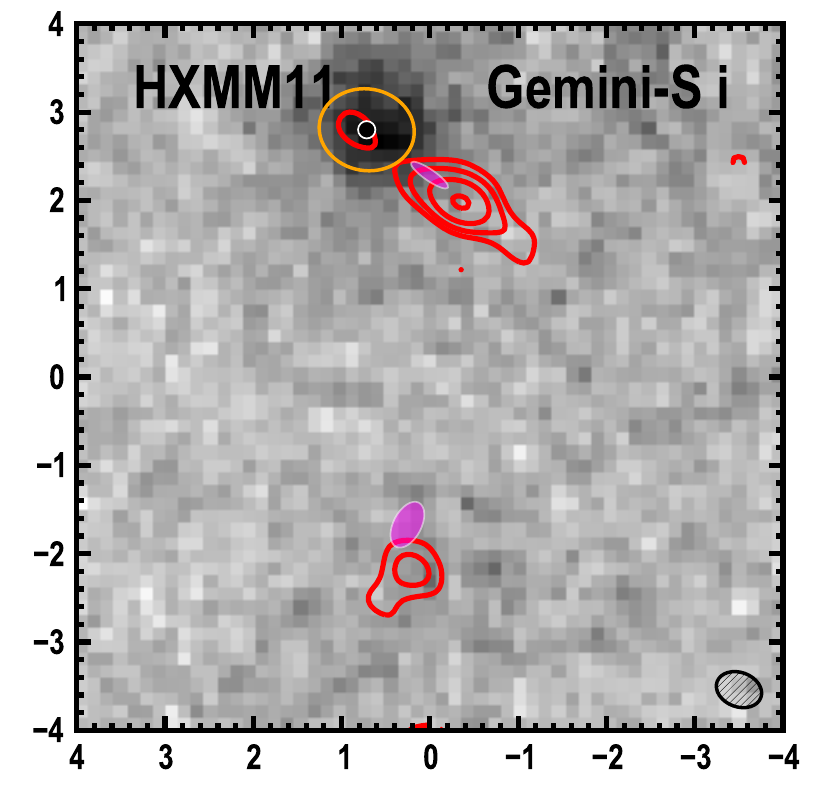}
\includegraphics[width=0.162\textwidth]{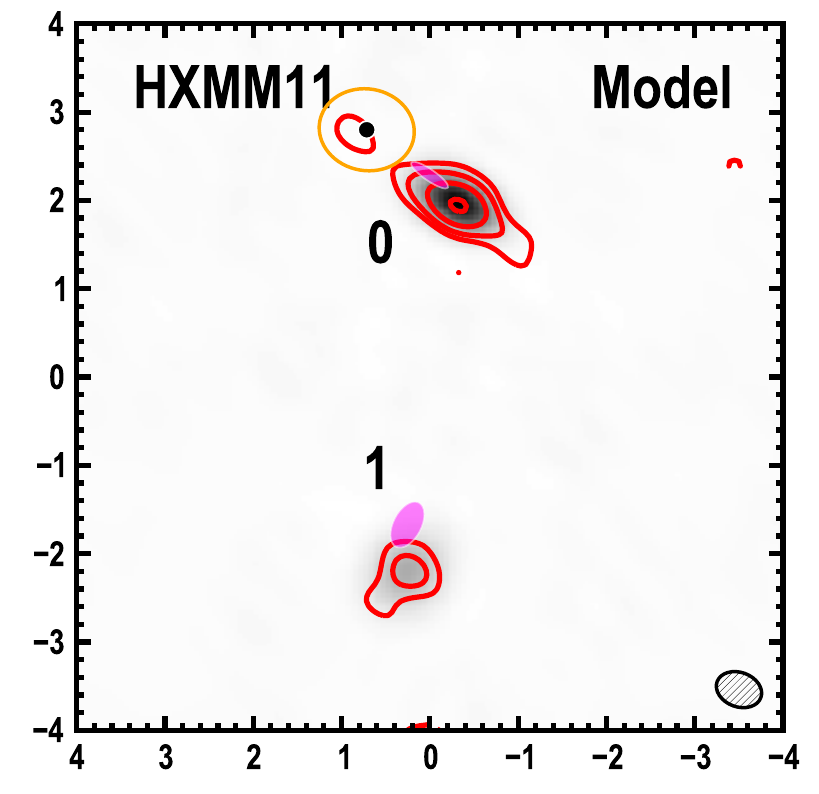}
\includegraphics[width=0.162\textwidth]{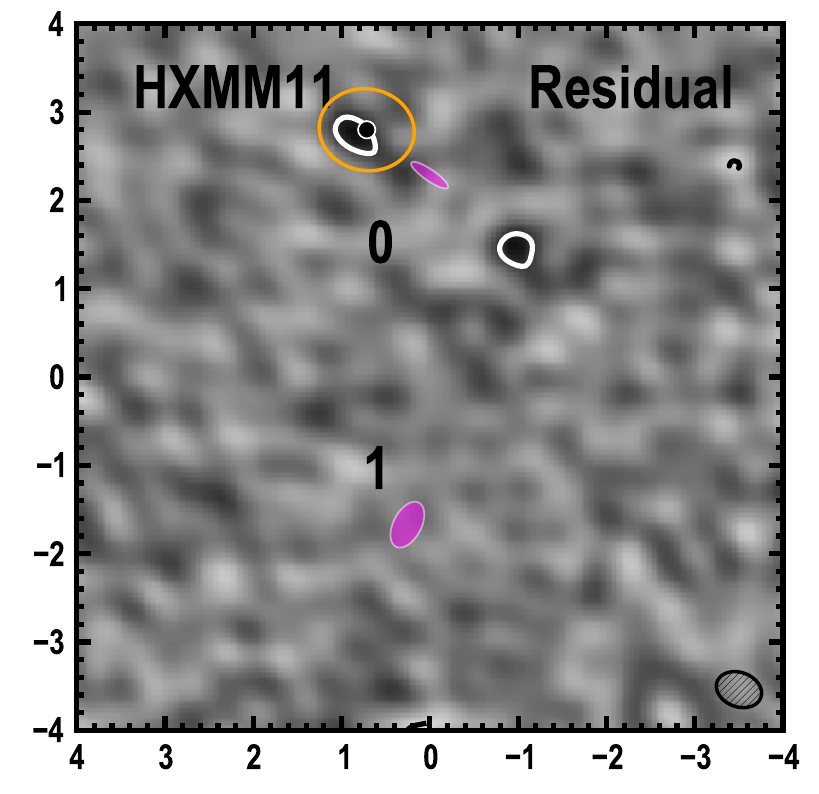}
\includegraphics[width=0.162\textwidth]{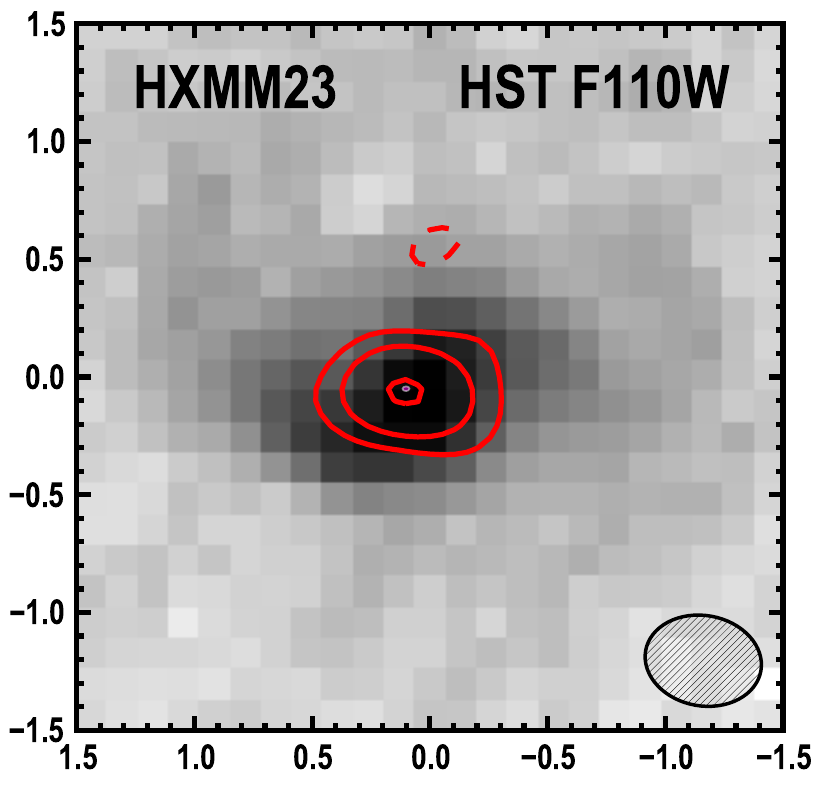}
\includegraphics[width=0.162\textwidth]{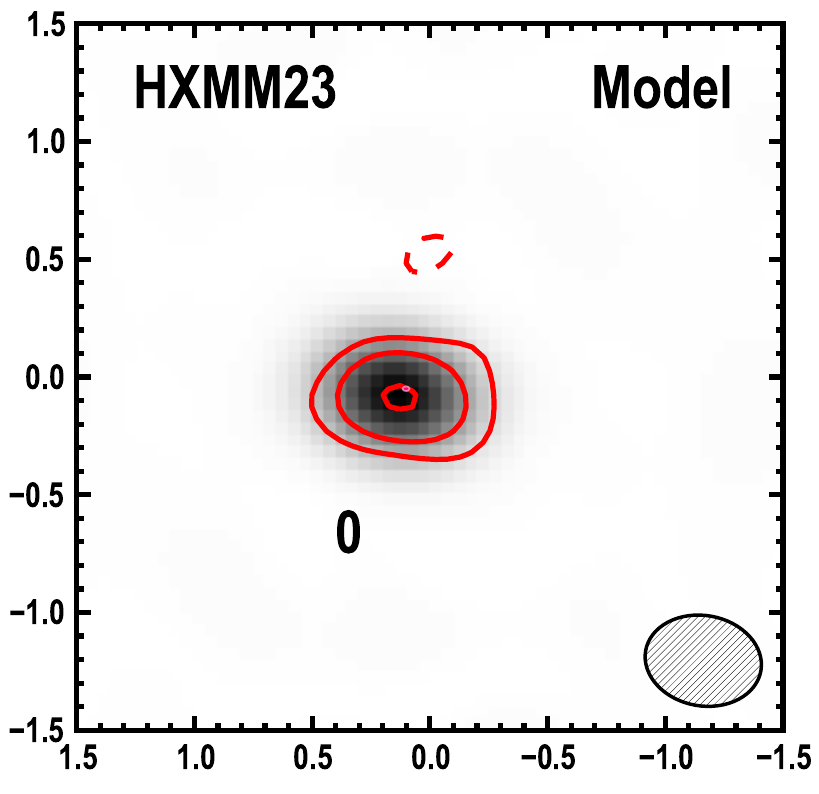}
\includegraphics[width=0.162\textwidth]{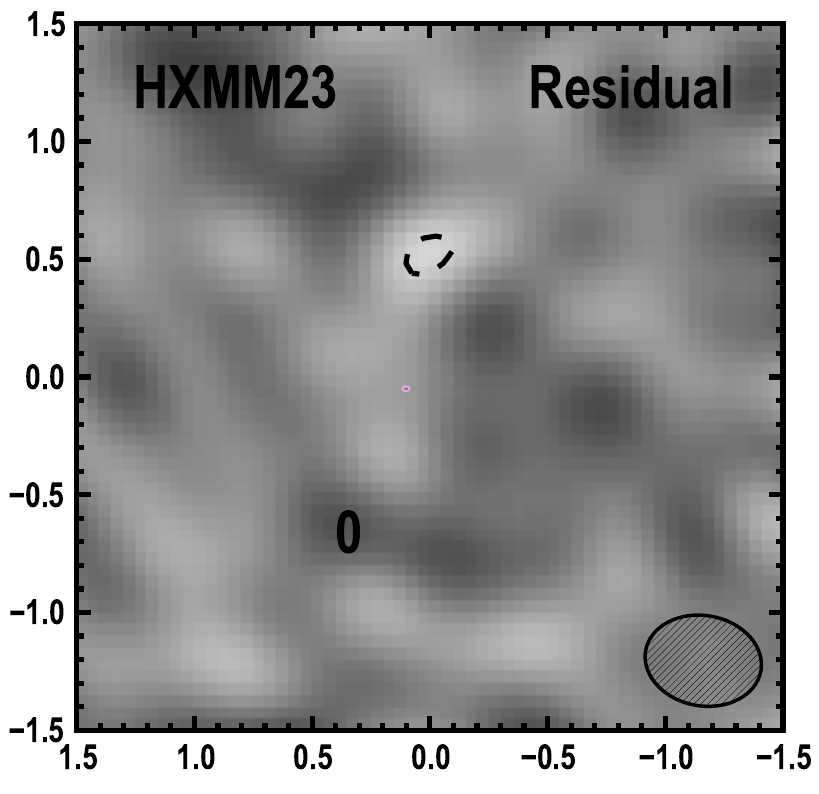}
\includegraphics[width=0.162\textwidth]{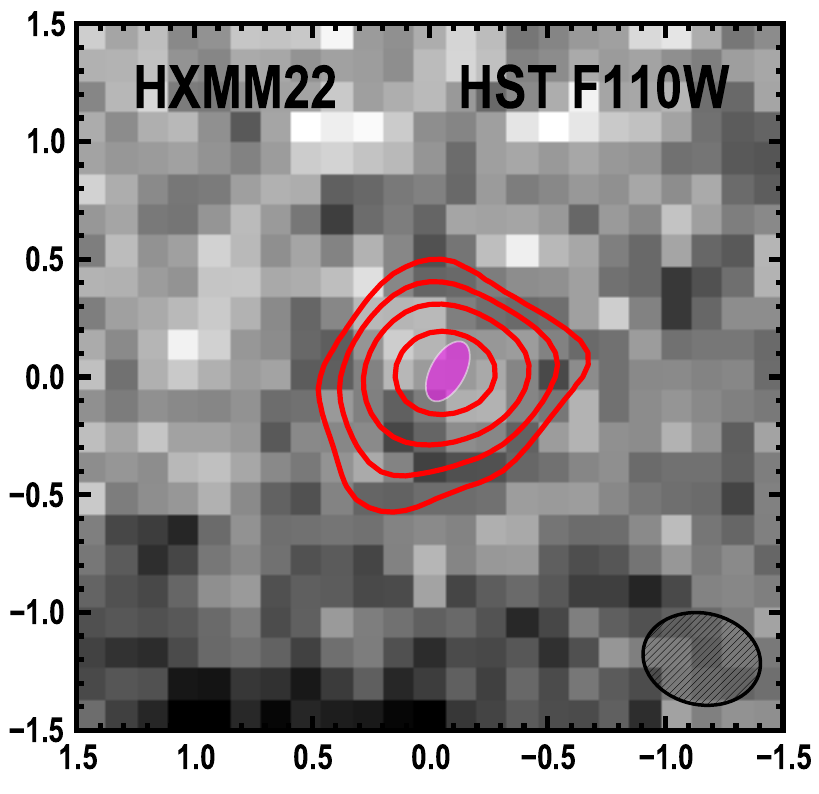}
\includegraphics[width=0.162\textwidth]{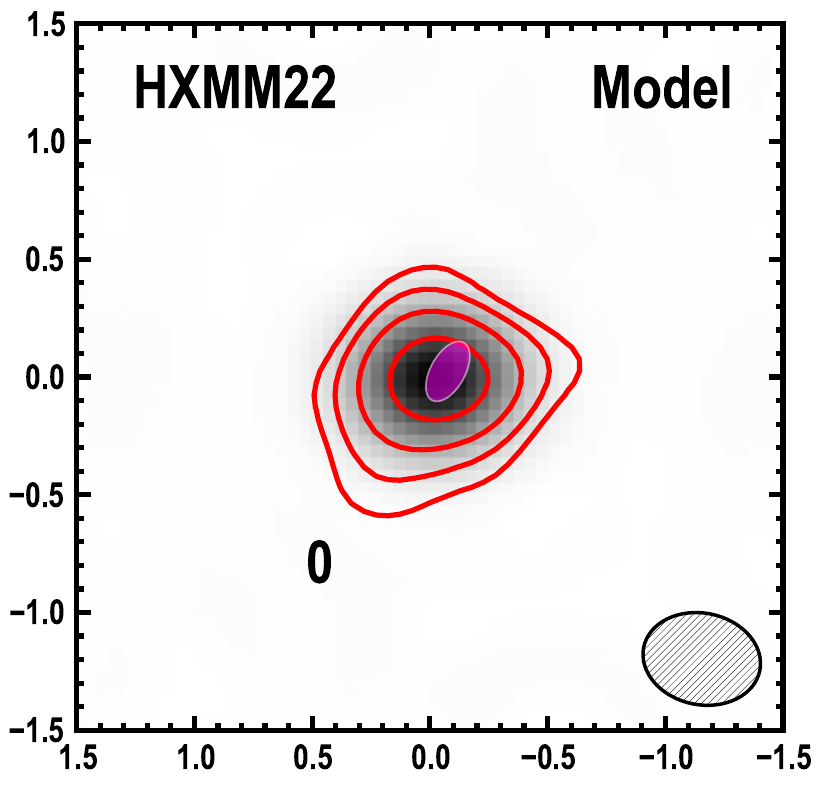}
\includegraphics[width=0.162\textwidth]{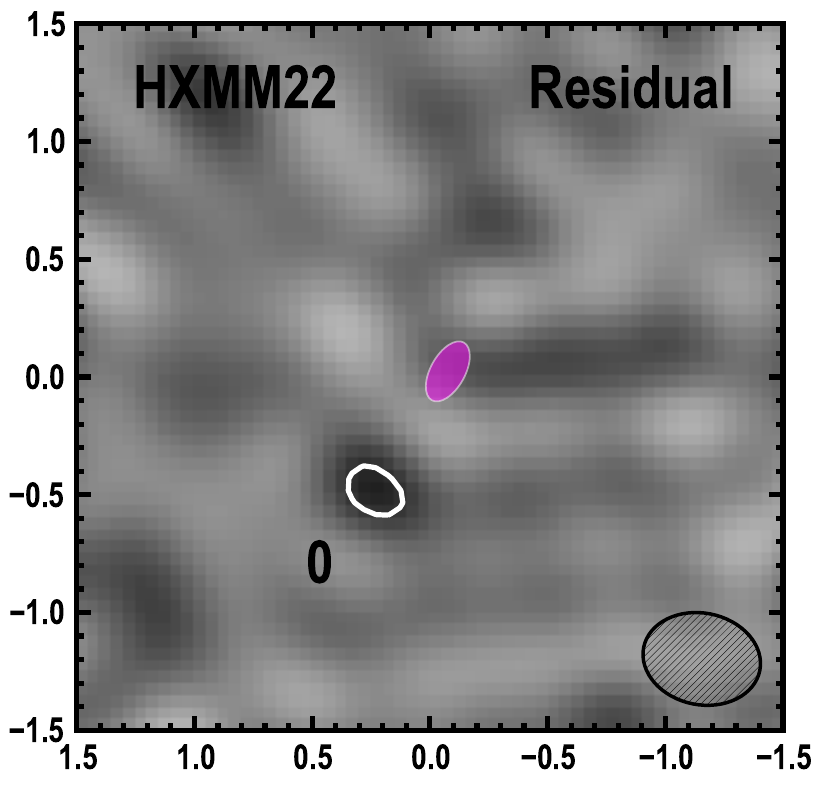}
\includegraphics[width=0.162\textwidth]{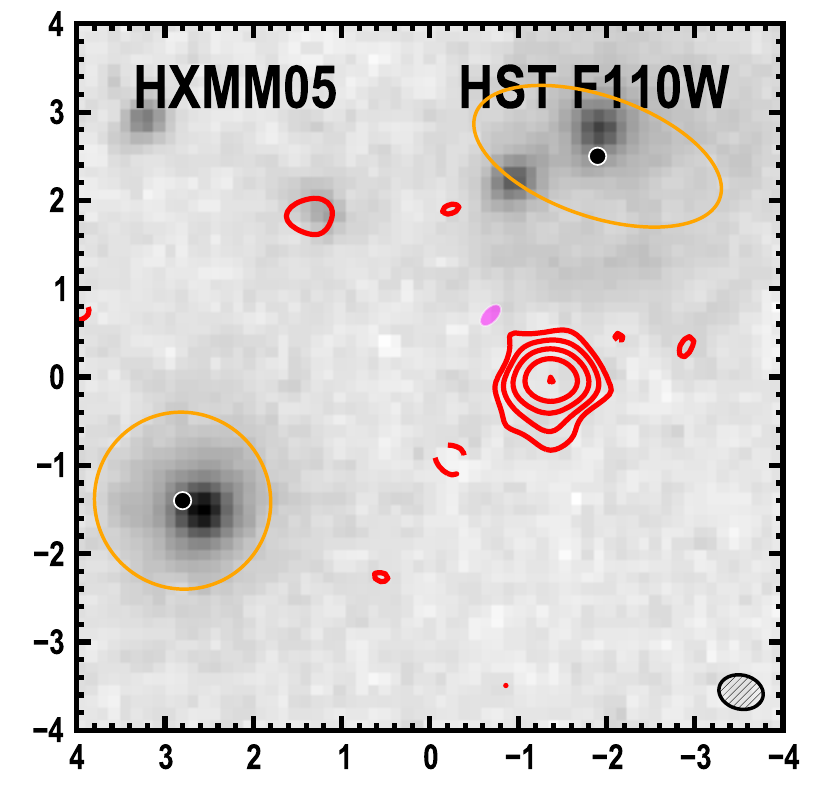}
\includegraphics[width=0.162\textwidth]{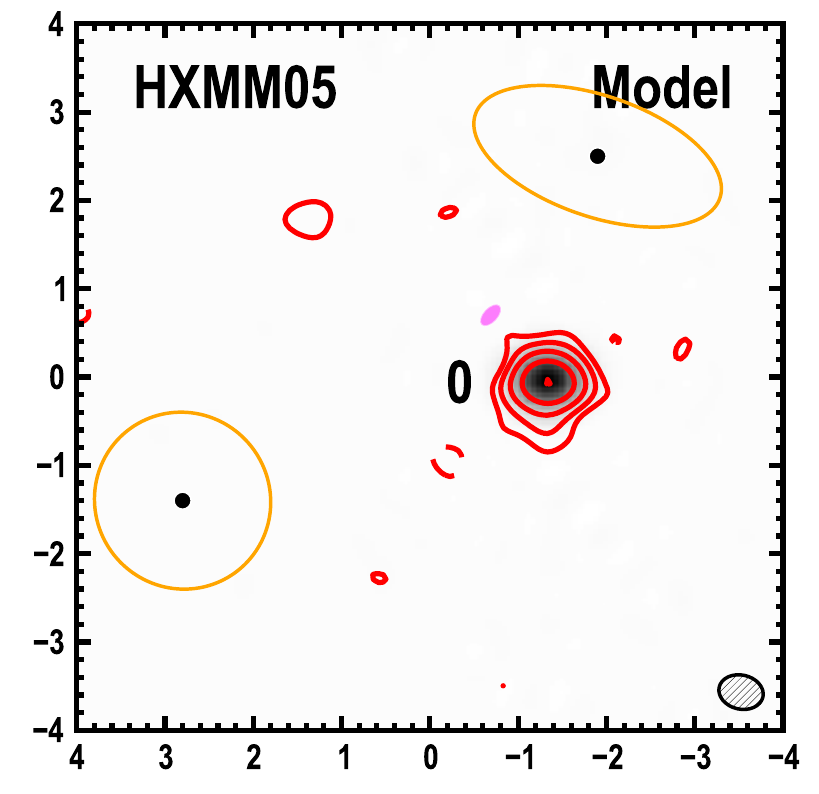}
\includegraphics[width=0.162\textwidth]{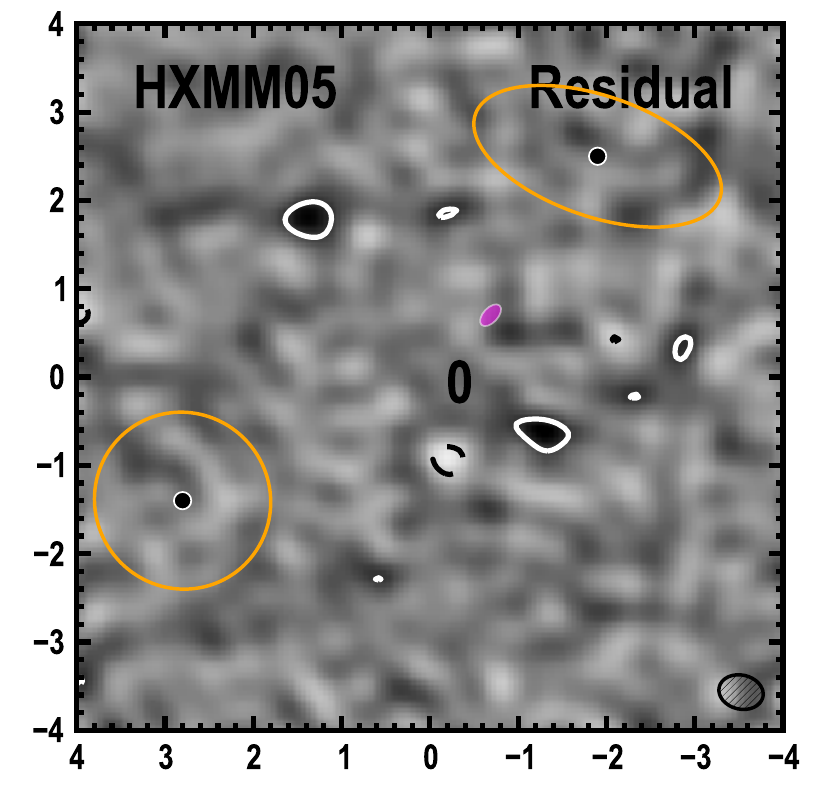}
\includegraphics[width=0.162\textwidth]{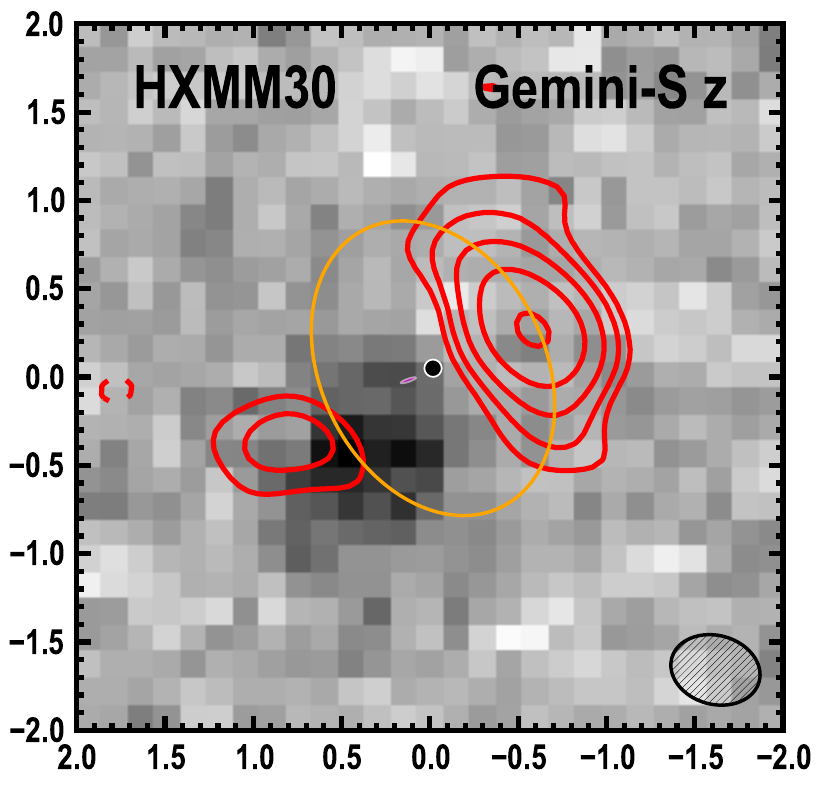}
\includegraphics[width=0.162\textwidth]{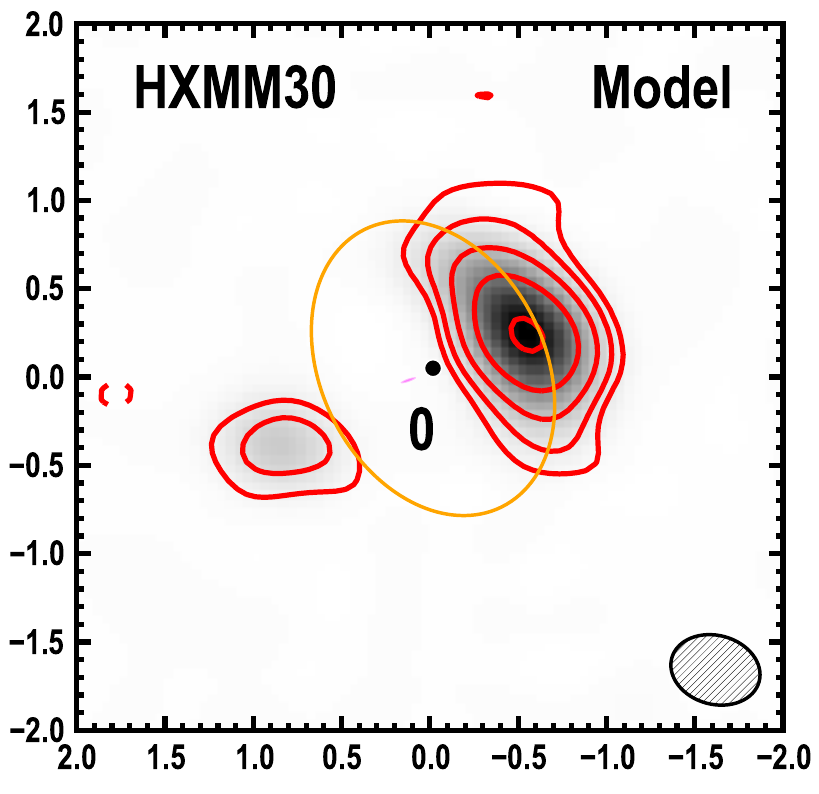}
\includegraphics[width=0.162\textwidth]{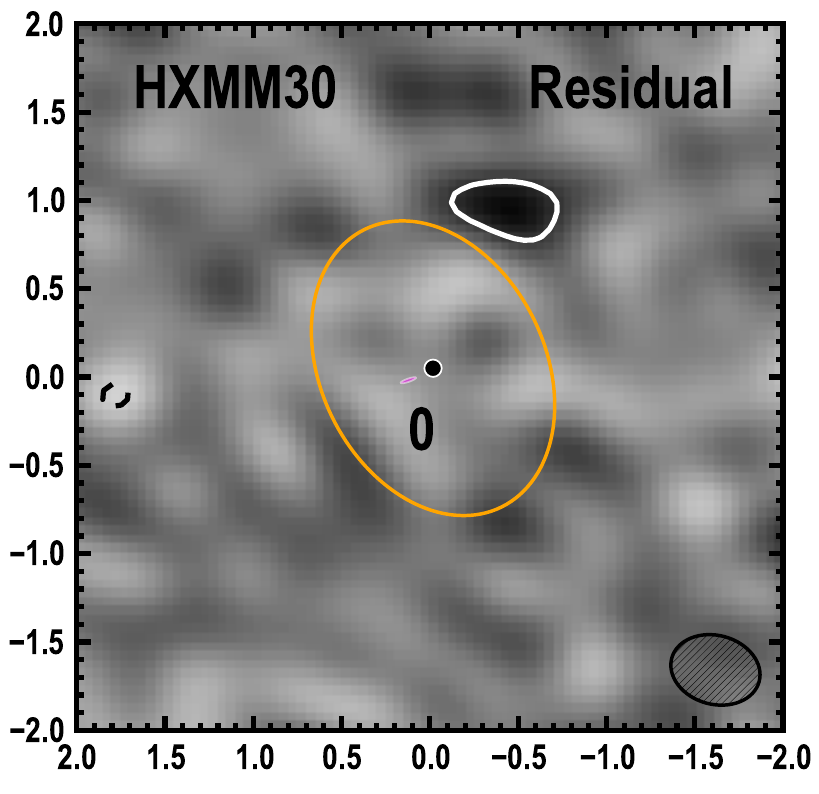}
\includegraphics[width=0.162\textwidth]{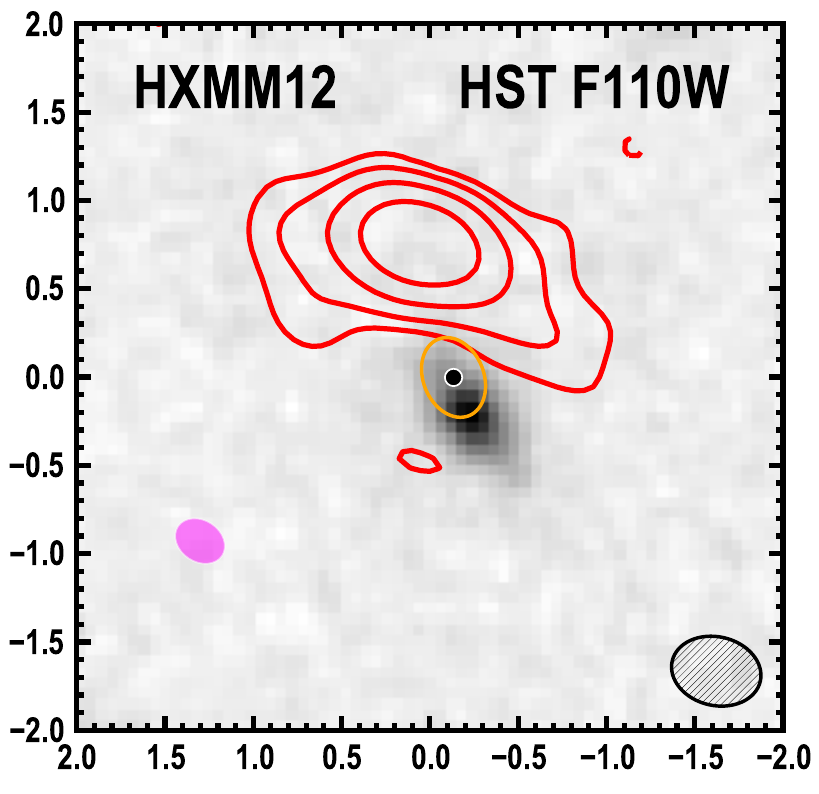}
\includegraphics[width=0.162\textwidth]{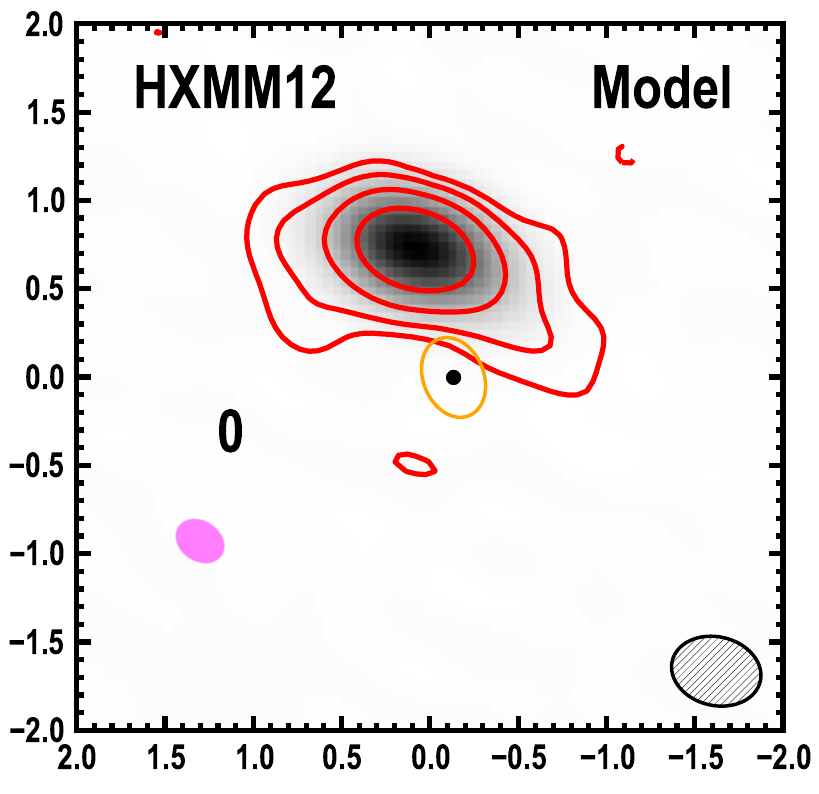}
\includegraphics[width=0.162\textwidth]{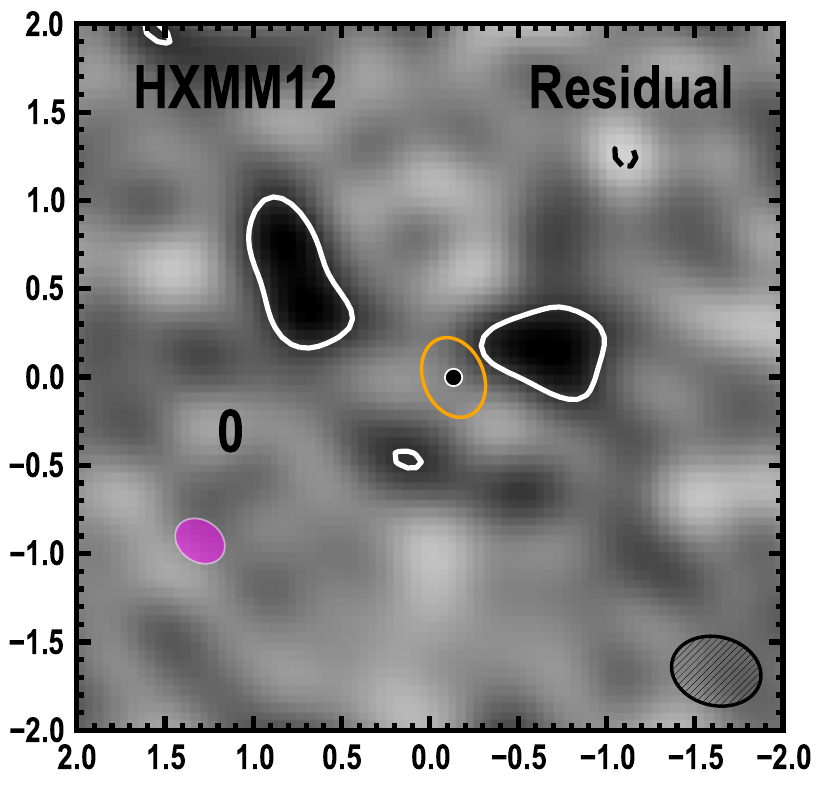}
\end{centering}

\caption{ Model fits for each target in the {\it Herschel}-ALMA sample, 3
panels per target.  North is up and east is left.  {\it Left}: ALMA 870$\mu$m
imaging (red contours, starting at $\pm 3\sigma$ and increasing by factors of
2) overlaid on best available optical or near-IR imaging (grayscale, with
telescope and filter printed in upper right corner).  The location and
morphology of all sources used in the model are represented by magenta
ellipses.  If a lens is present, its location is given by a black circle and
its critical curve is traced by an orange line.  The FWHM size of the ALMA
synthesized beam is shown in the lower left corner of each panel.  {\it
Middle}: Same as {\it left}, but showing best-fit model in grayscale.  Numbers
indicate the location of sources.  {\it Right}: Same as {\it left}, but showing
residual image obtained from subtracting best-fit model from the data.
\label{fig:uvmodels}} \addtocounter{figure}{-1}

\end{figure*}

\begin{figure*}[!tbp] 
    \begin{centering}
\includegraphics[width=0.162\textwidth]{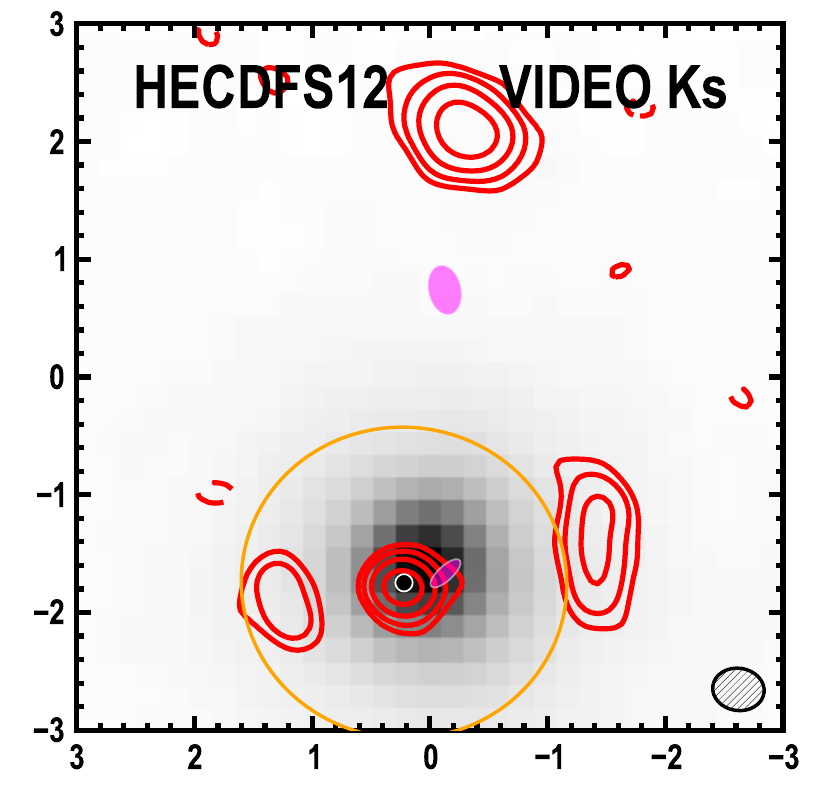}
\includegraphics[width=0.162\textwidth]{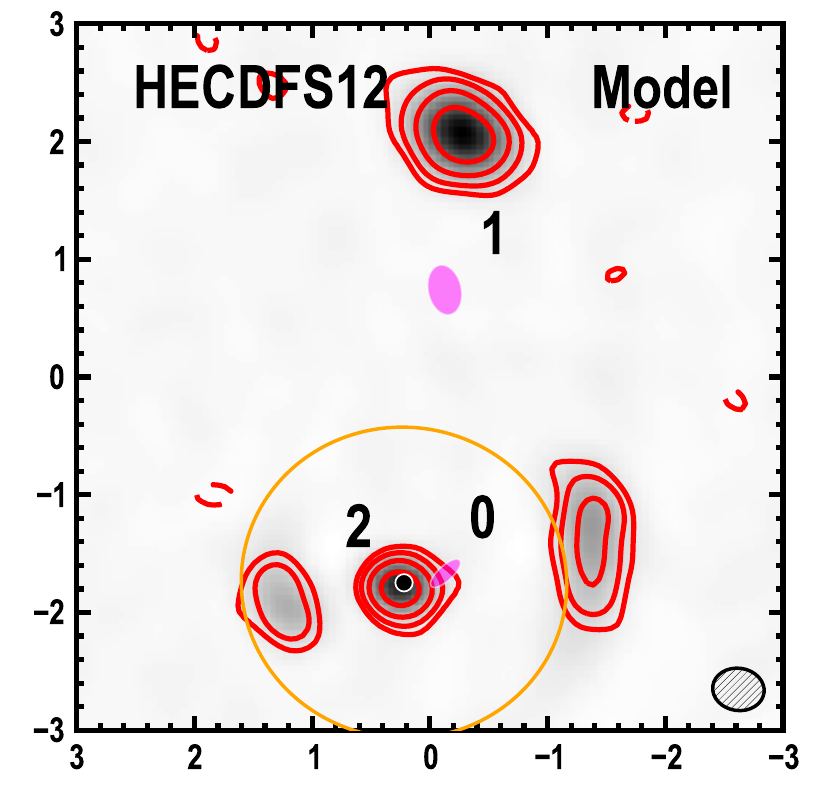}
\includegraphics[width=0.162\textwidth]{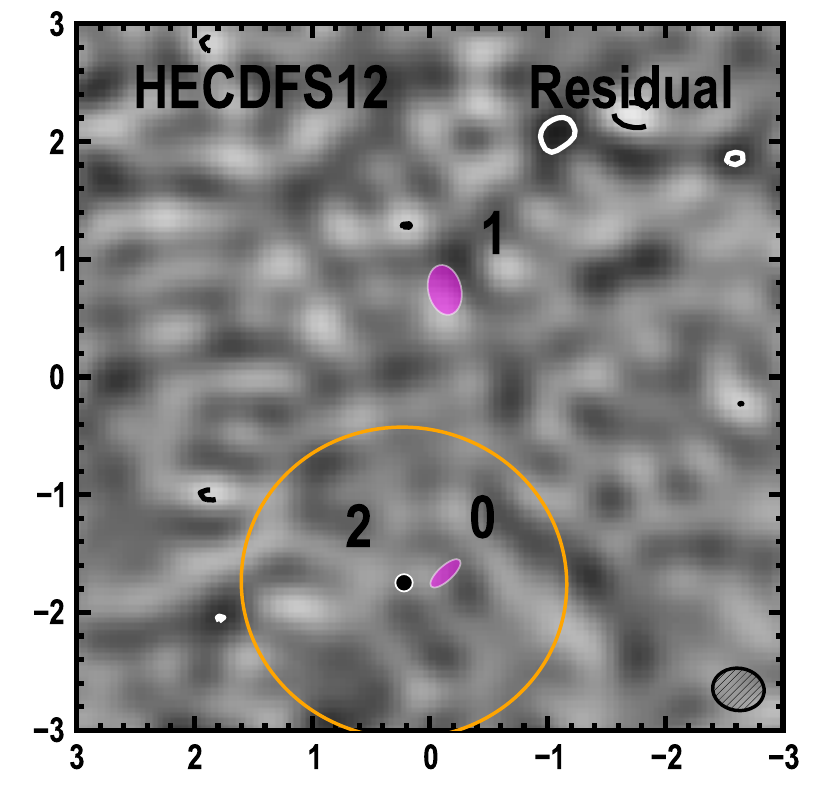}
\includegraphics[width=0.162\textwidth]{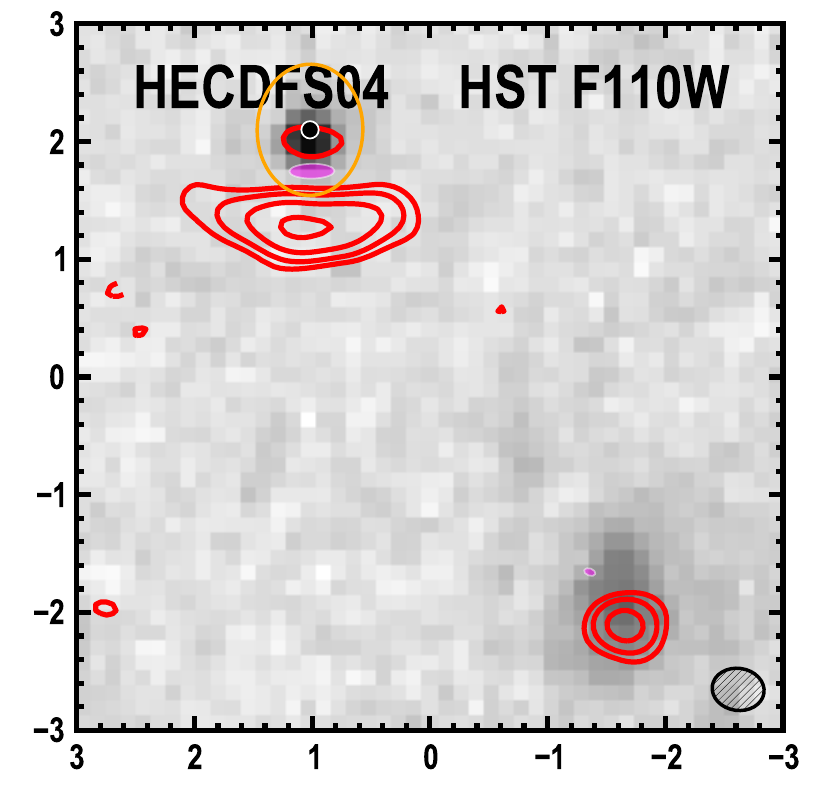}
\includegraphics[width=0.162\textwidth]{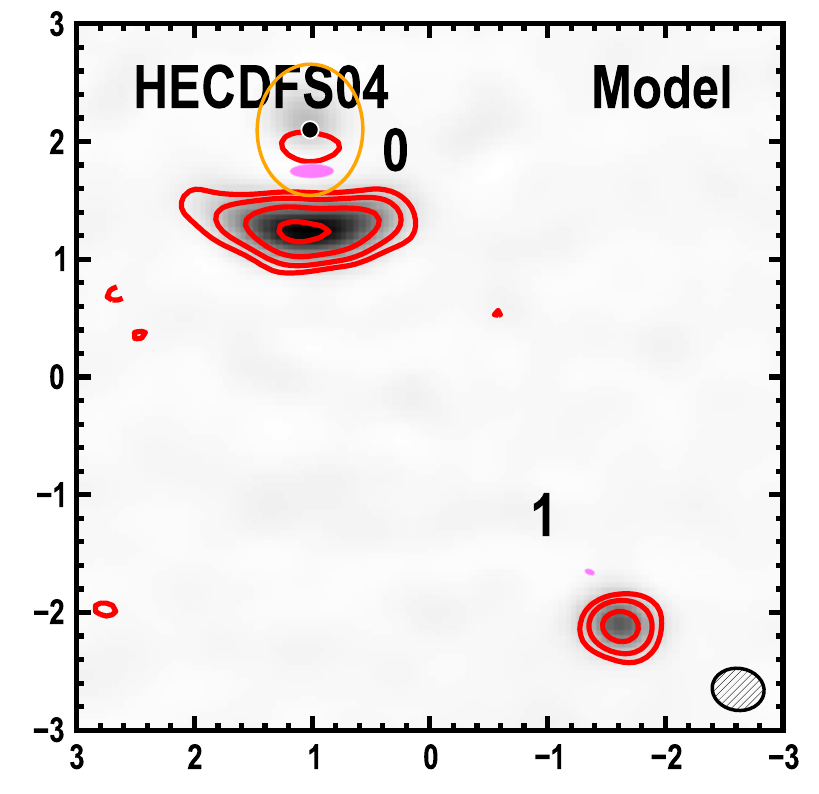}
\includegraphics[width=0.162\textwidth]{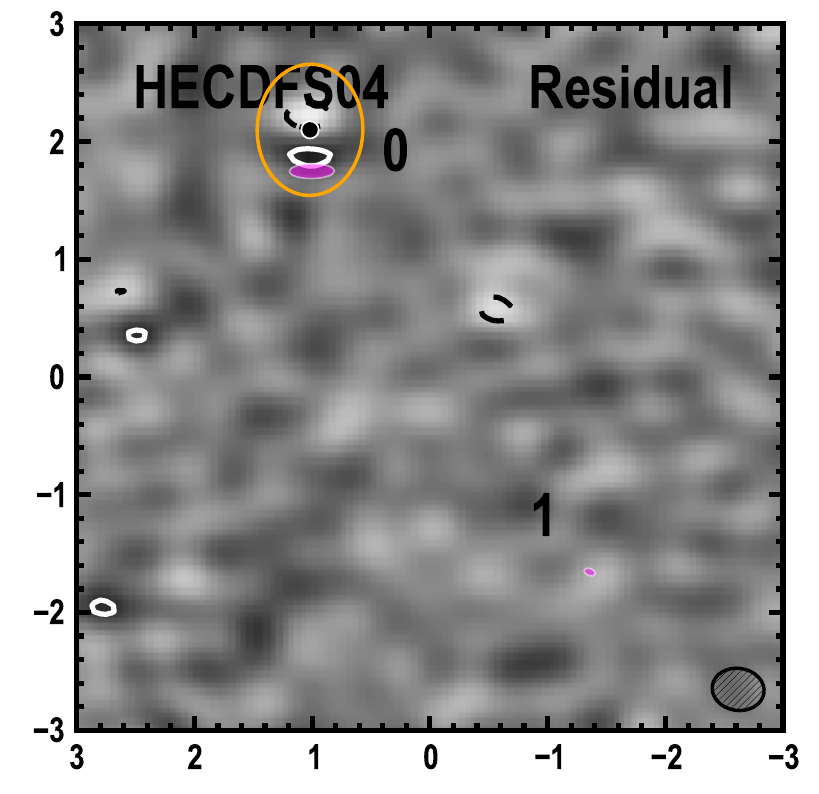}
\includegraphics[width=0.162\textwidth]{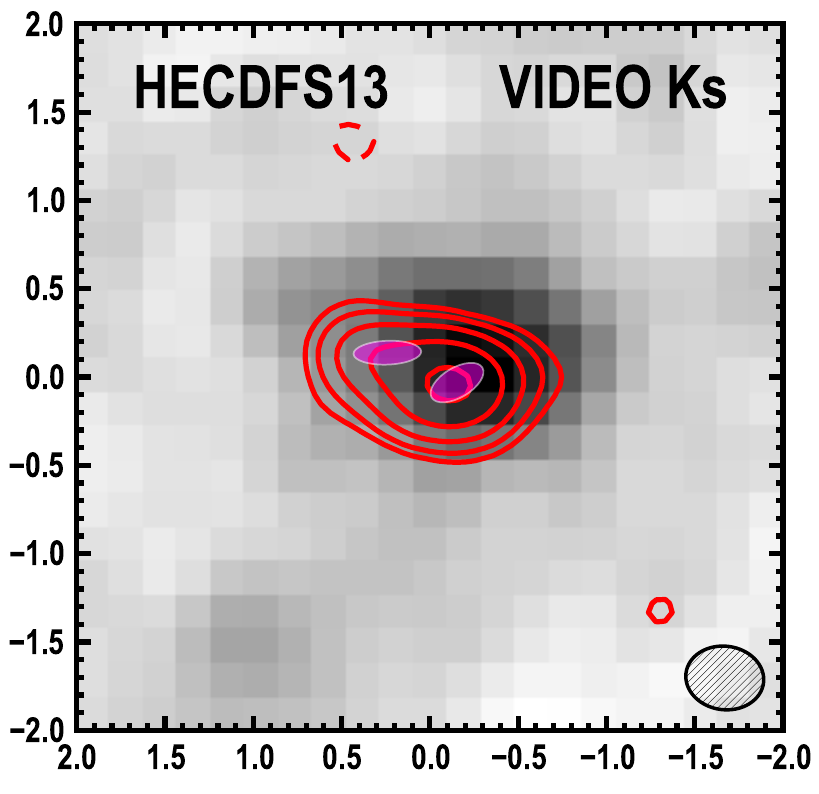}
\includegraphics[width=0.162\textwidth]{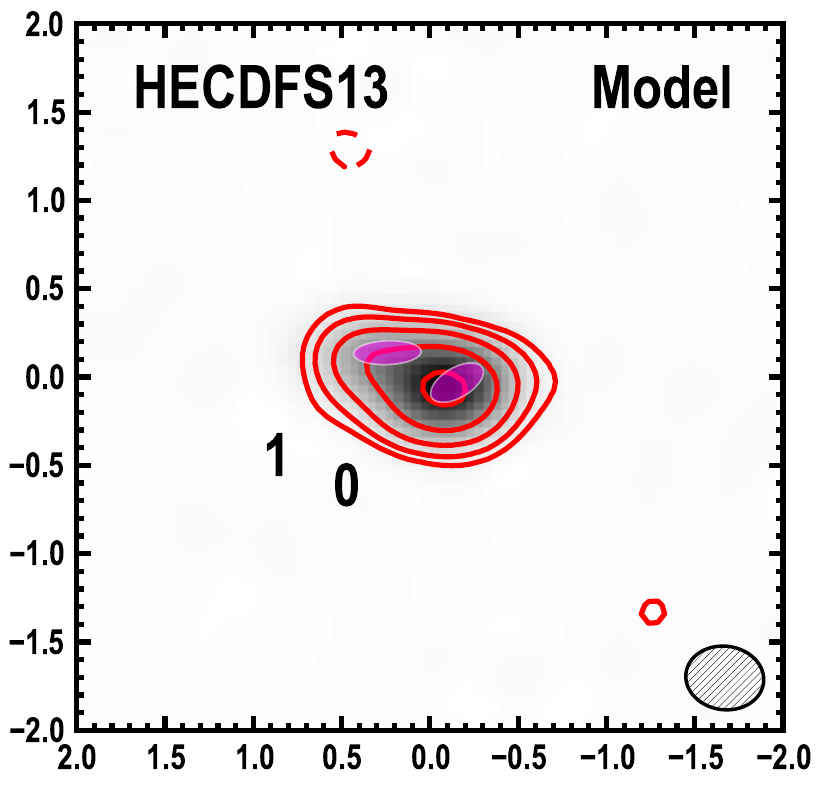}
\includegraphics[width=0.162\textwidth]{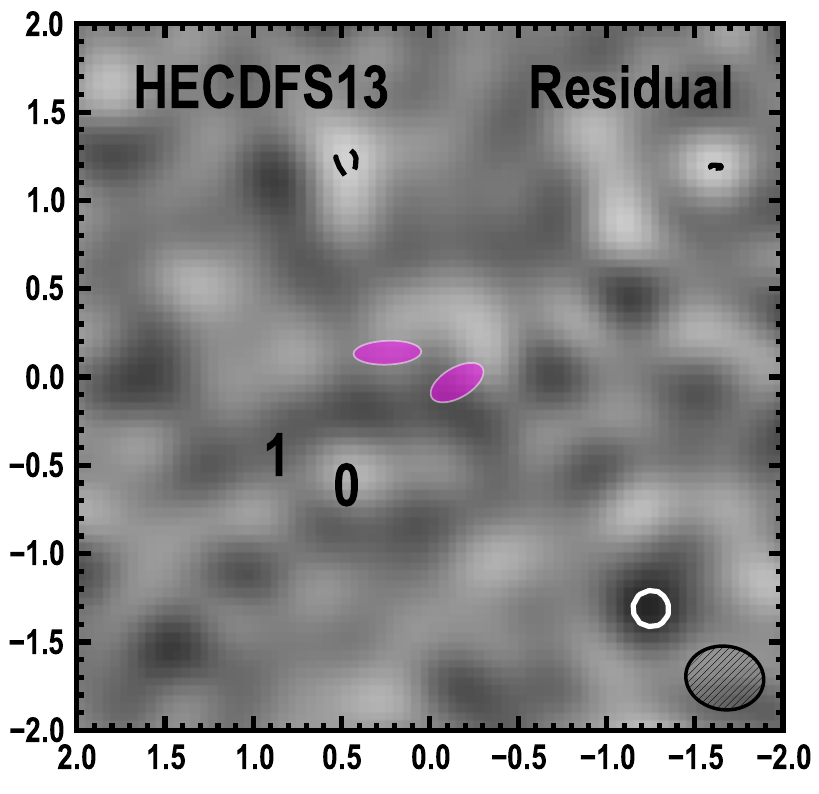}
\includegraphics[width=0.162\textwidth]{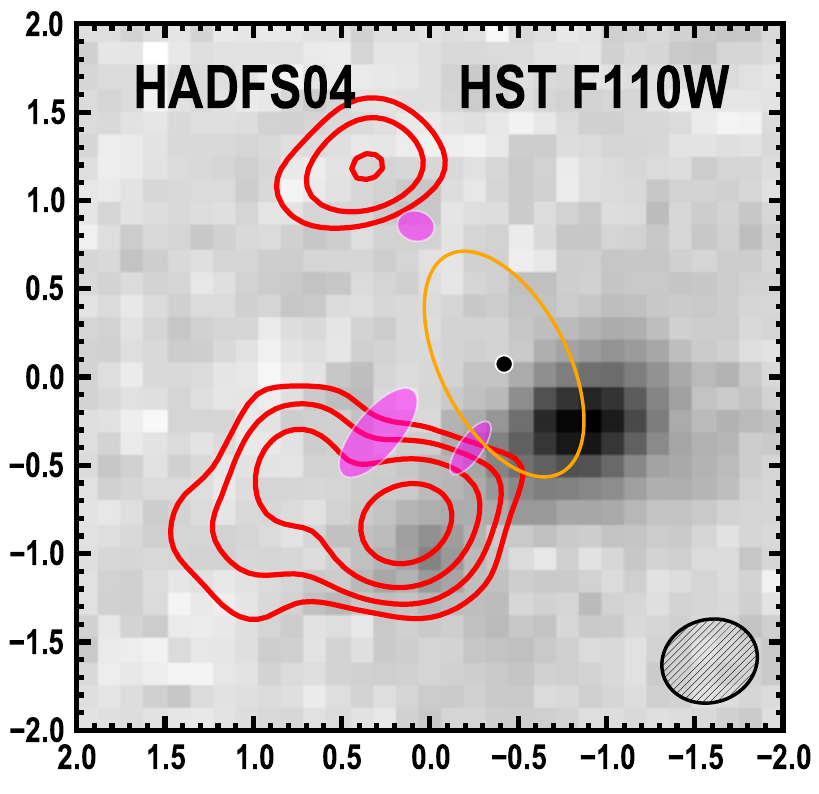}
\includegraphics[width=0.162\textwidth]{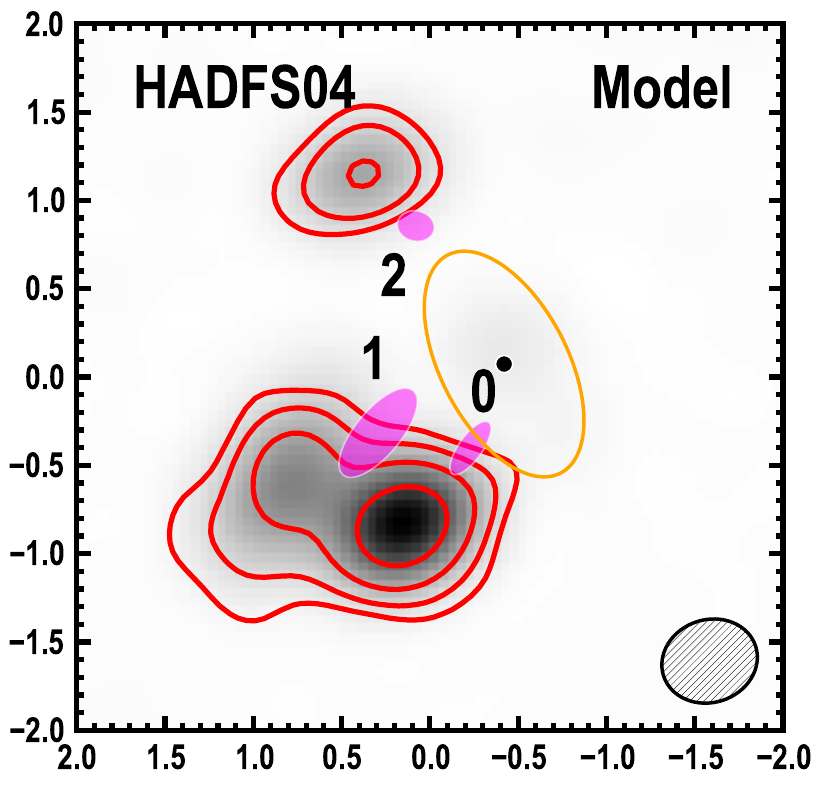}
\includegraphics[width=0.162\textwidth]{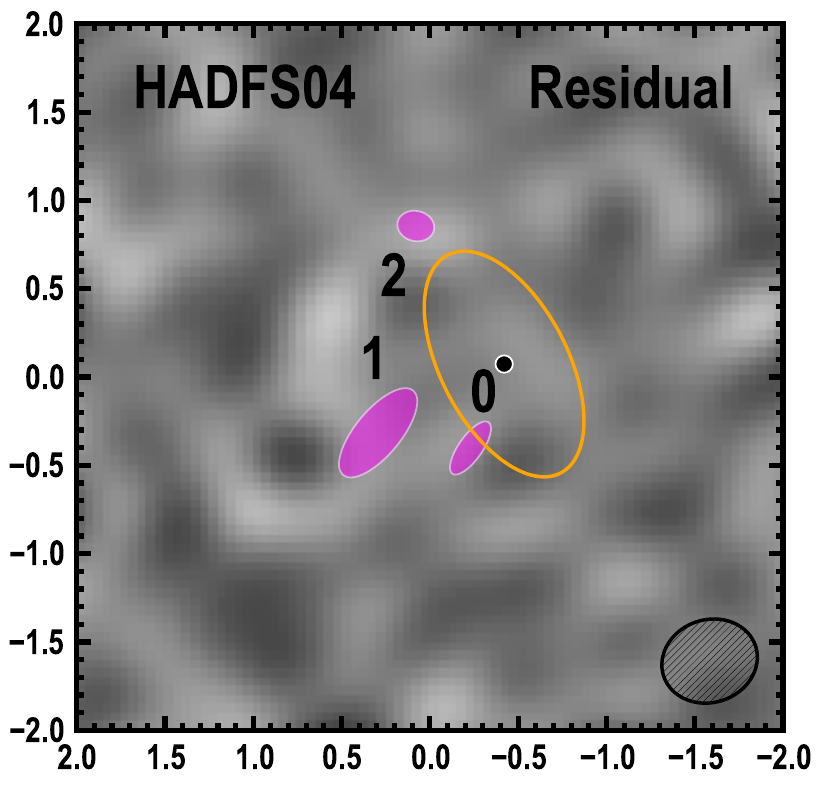}
\includegraphics[width=0.162\textwidth]{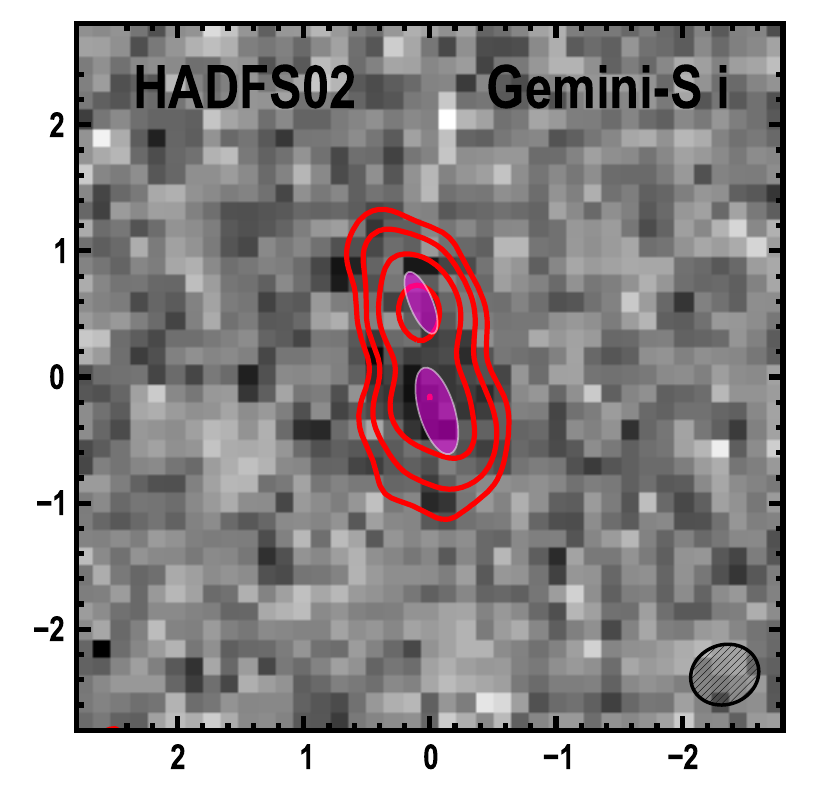}
\includegraphics[width=0.162\textwidth]{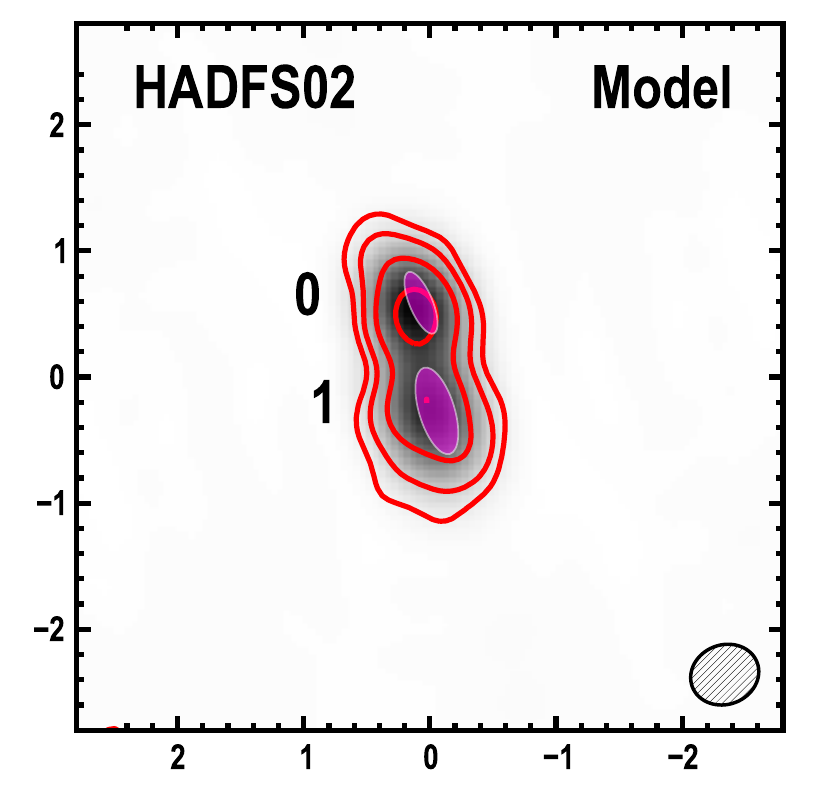}
\includegraphics[width=0.162\textwidth]{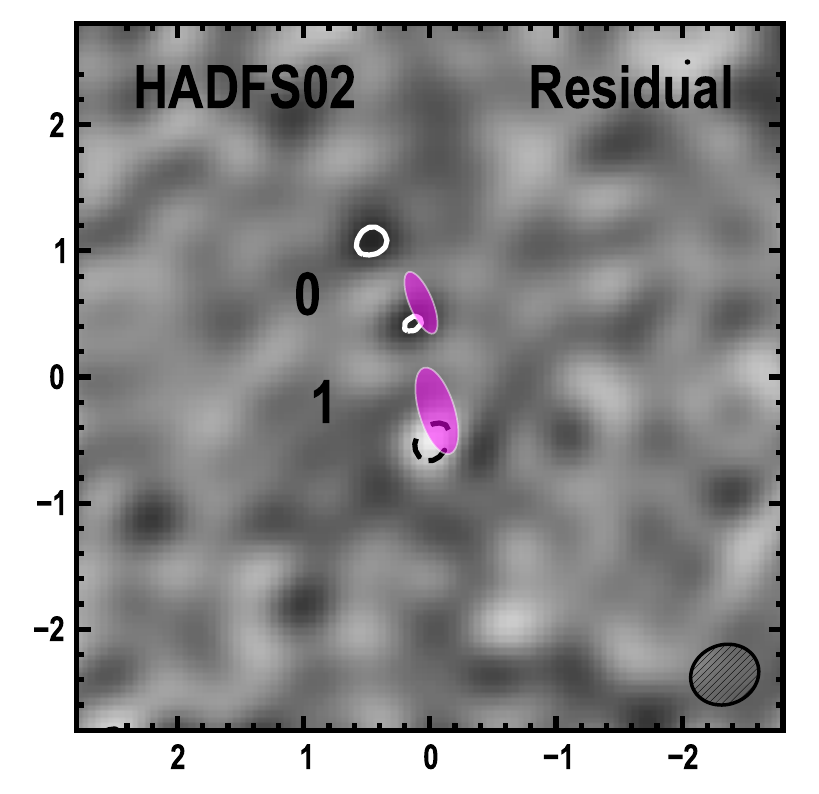}
\includegraphics[width=0.162\textwidth]{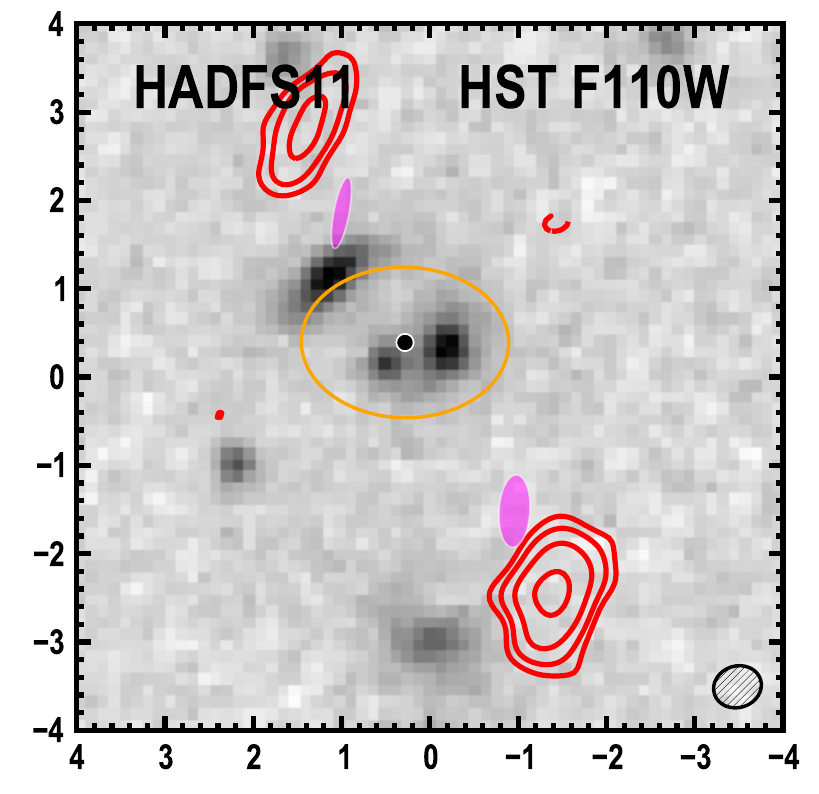}
\includegraphics[width=0.162\textwidth]{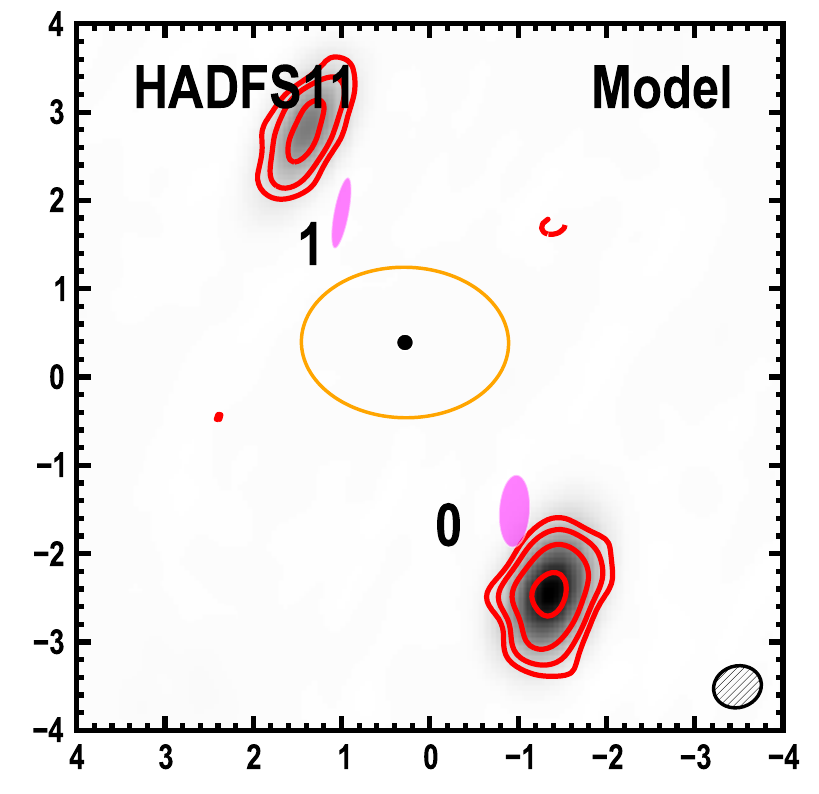}
\includegraphics[width=0.162\textwidth]{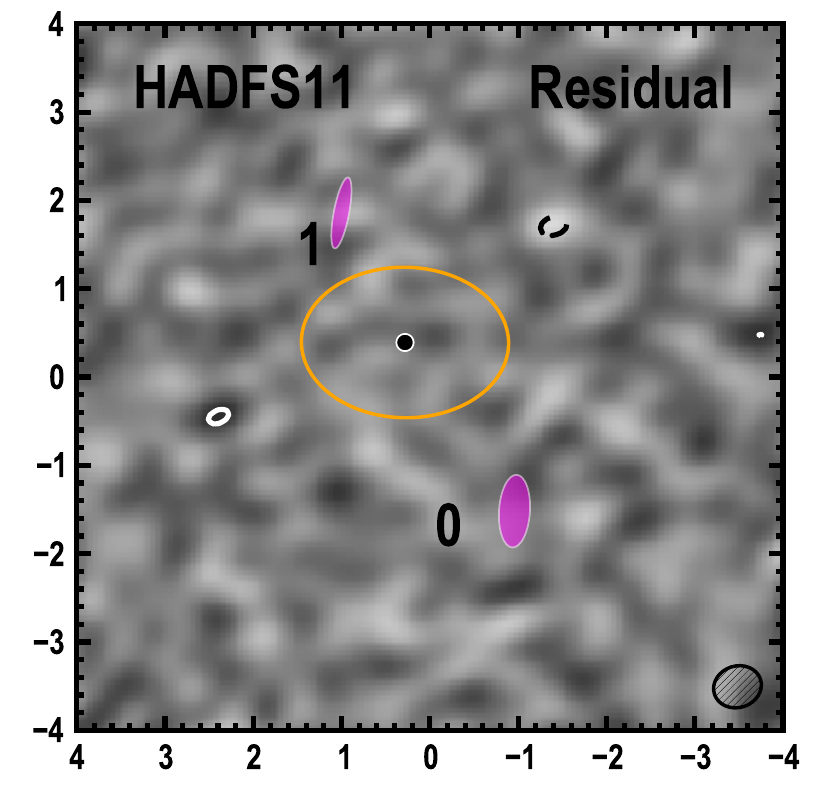}
\includegraphics[width=0.162\textwidth]{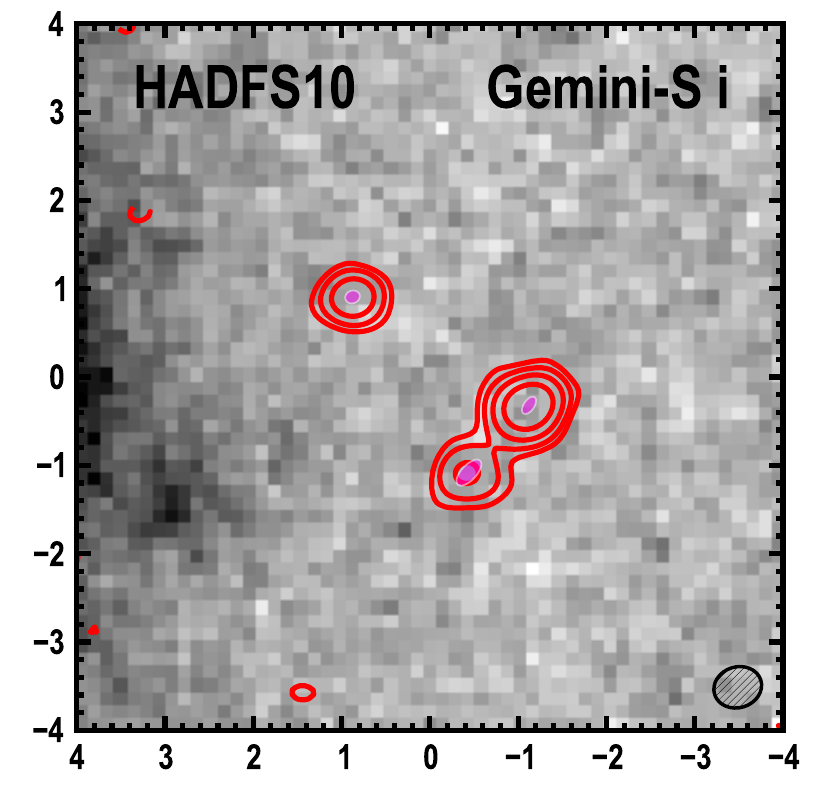}
\includegraphics[width=0.162\textwidth]{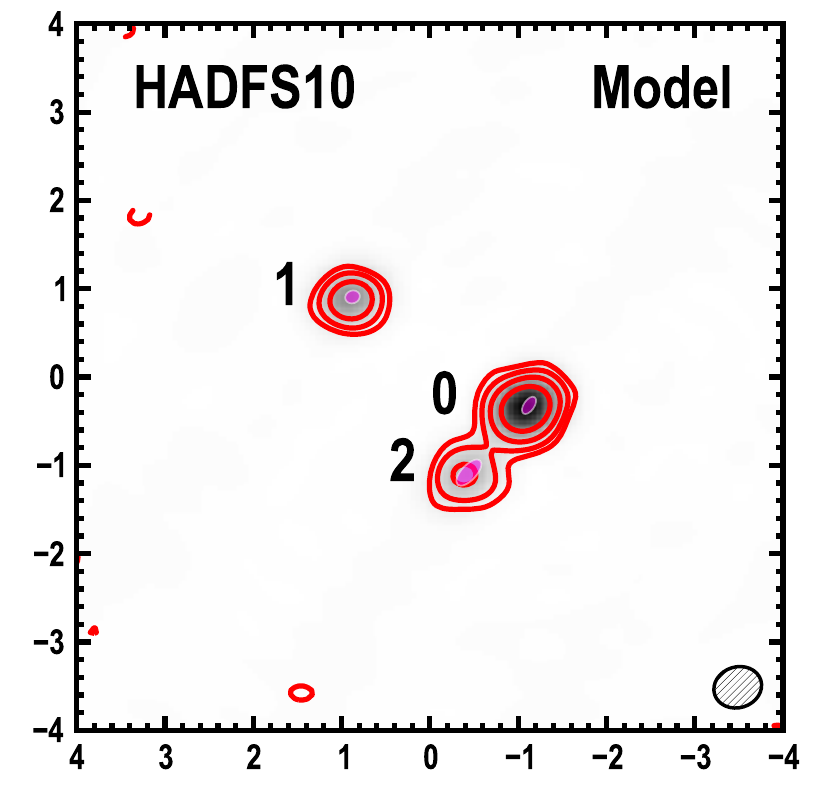}
\includegraphics[width=0.162\textwidth]{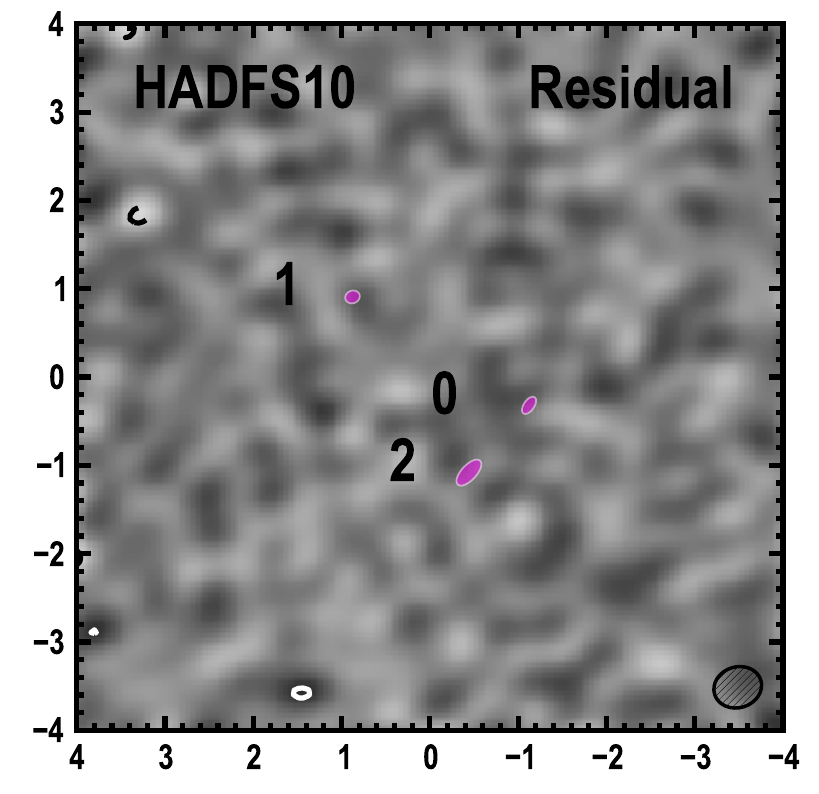}
\includegraphics[width=0.162\textwidth]{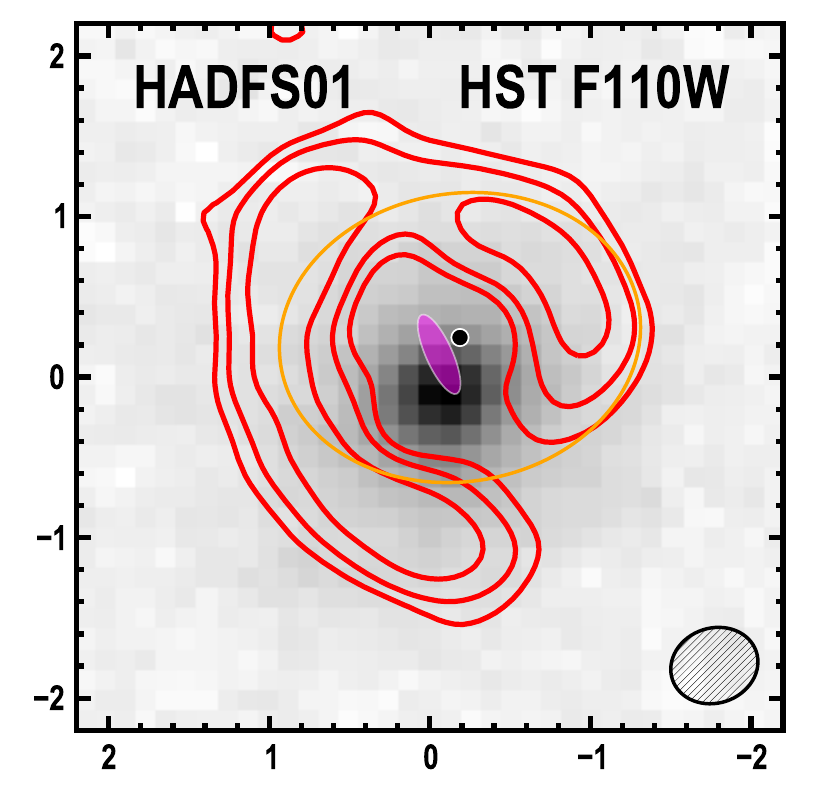}
\includegraphics[width=0.162\textwidth]{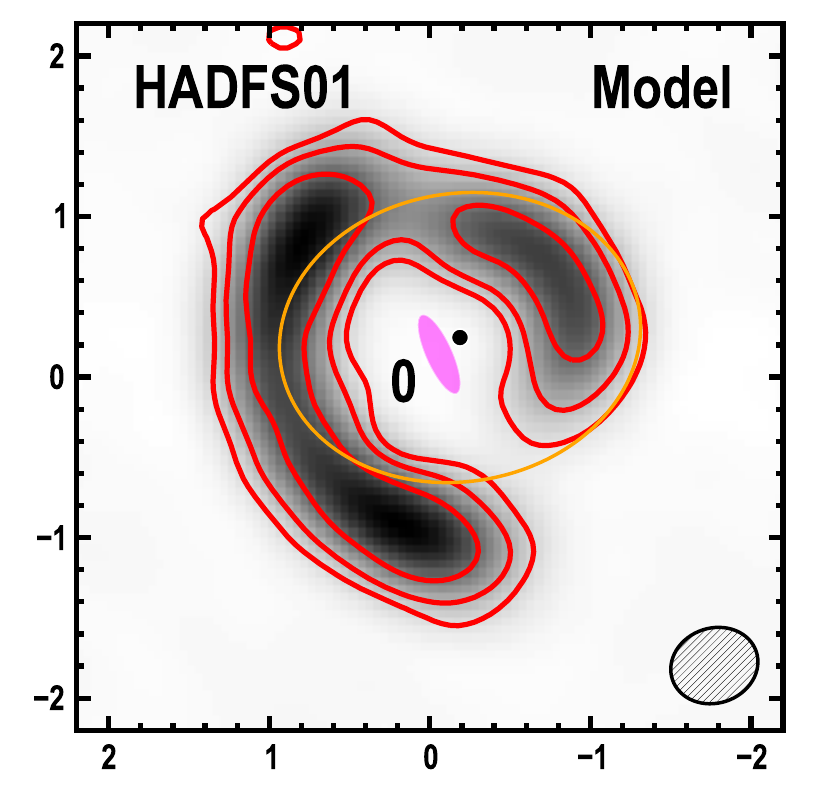}
\includegraphics[width=0.162\textwidth]{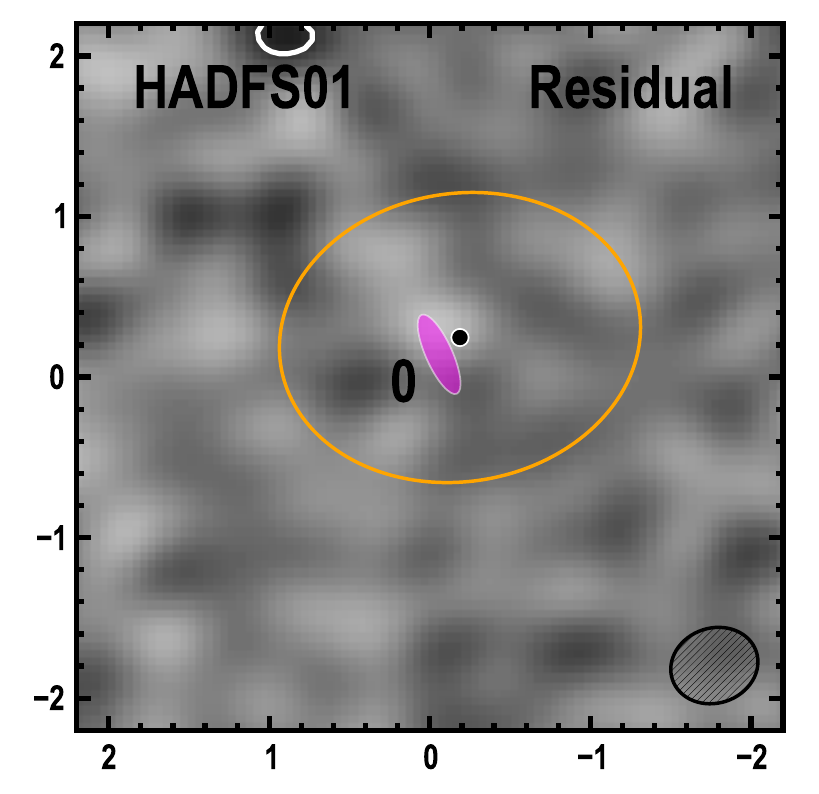}
\includegraphics[width=0.162\textwidth]{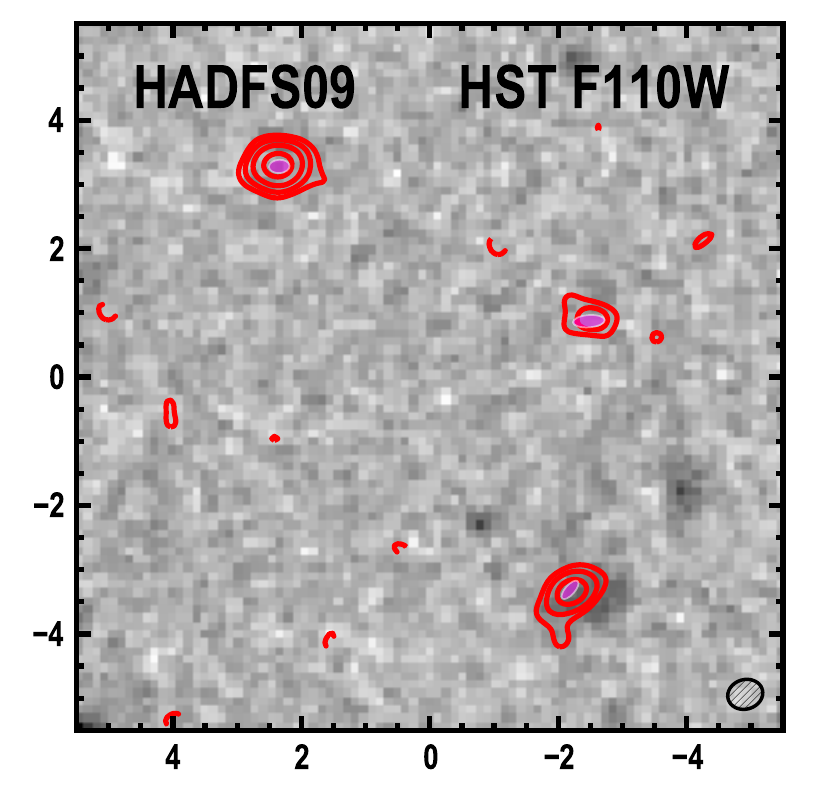}
\includegraphics[width=0.162\textwidth]{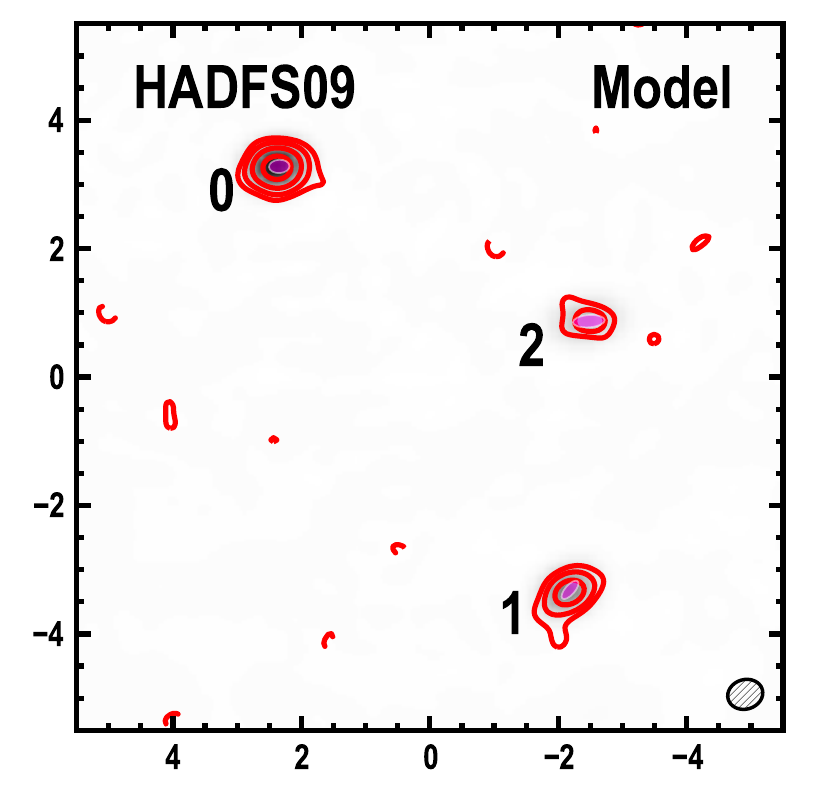}
\includegraphics[width=0.162\textwidth]{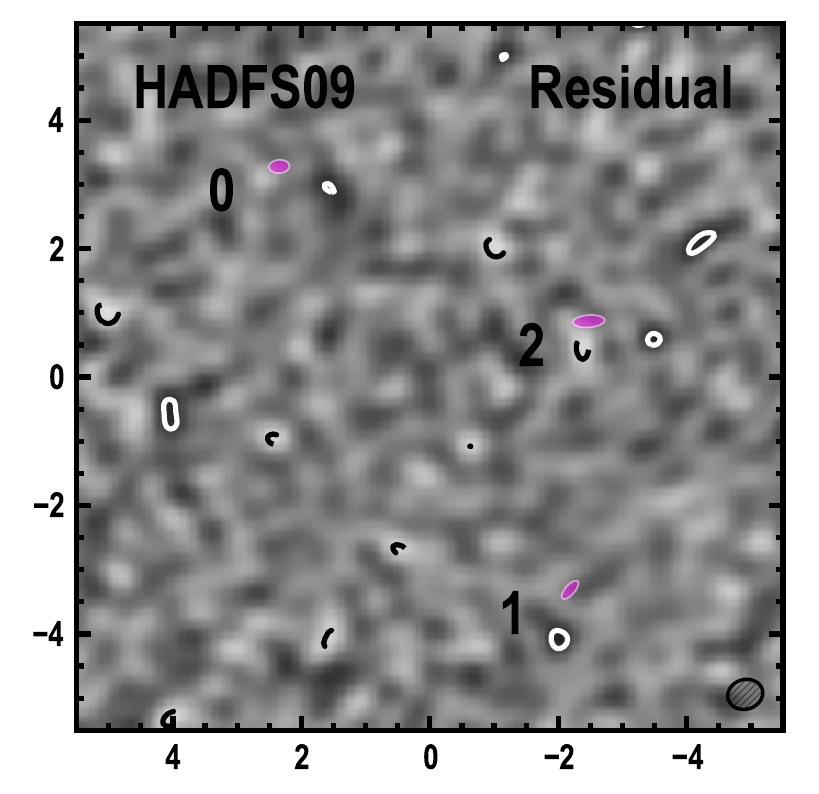}
\includegraphics[width=0.162\textwidth]{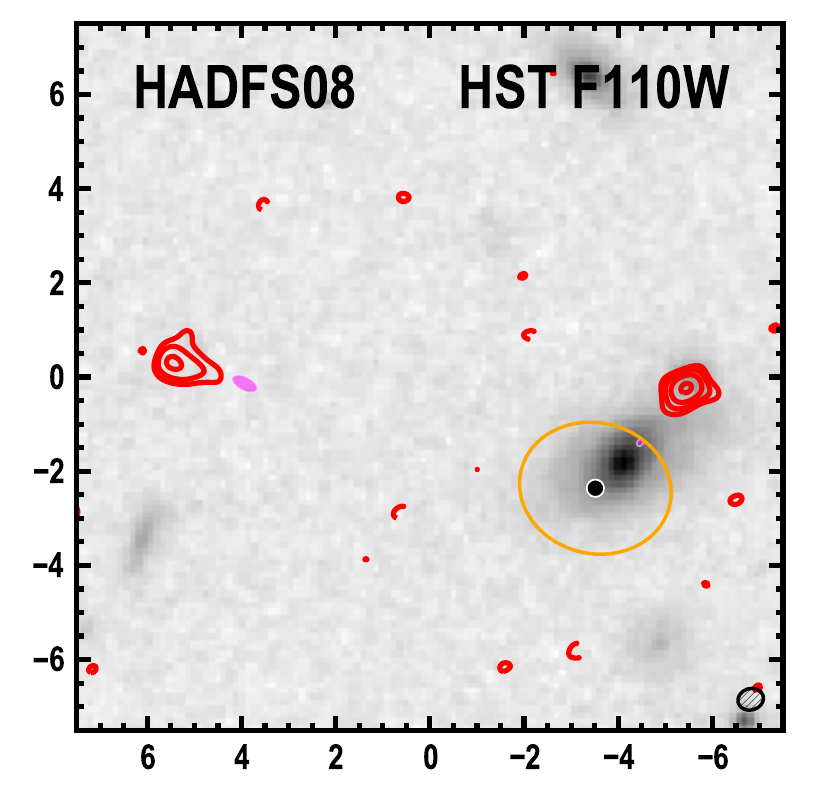}
\includegraphics[width=0.162\textwidth]{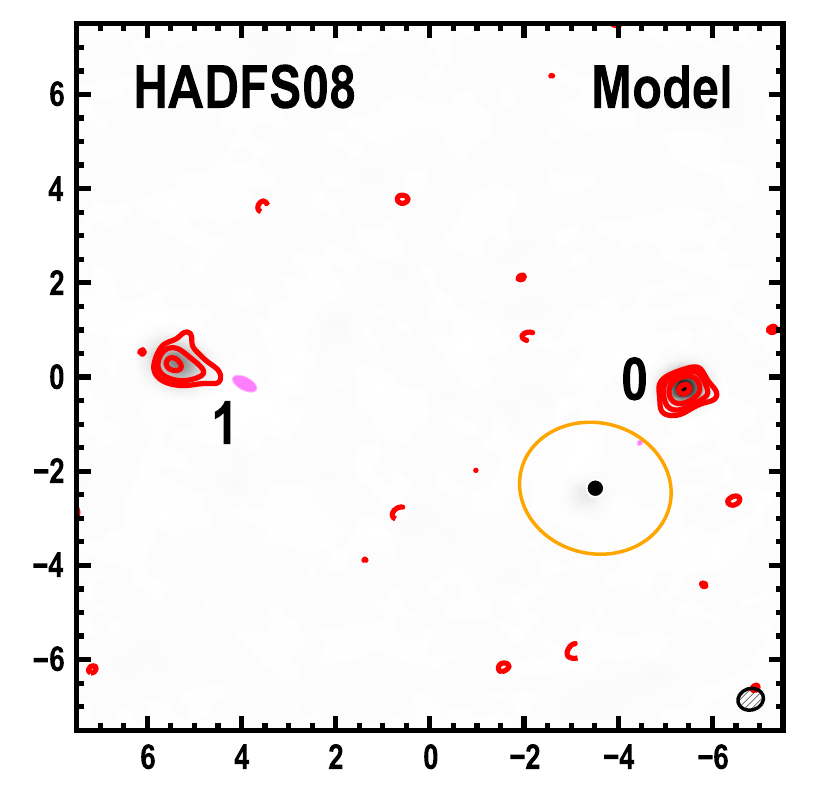}
\includegraphics[width=0.162\textwidth]{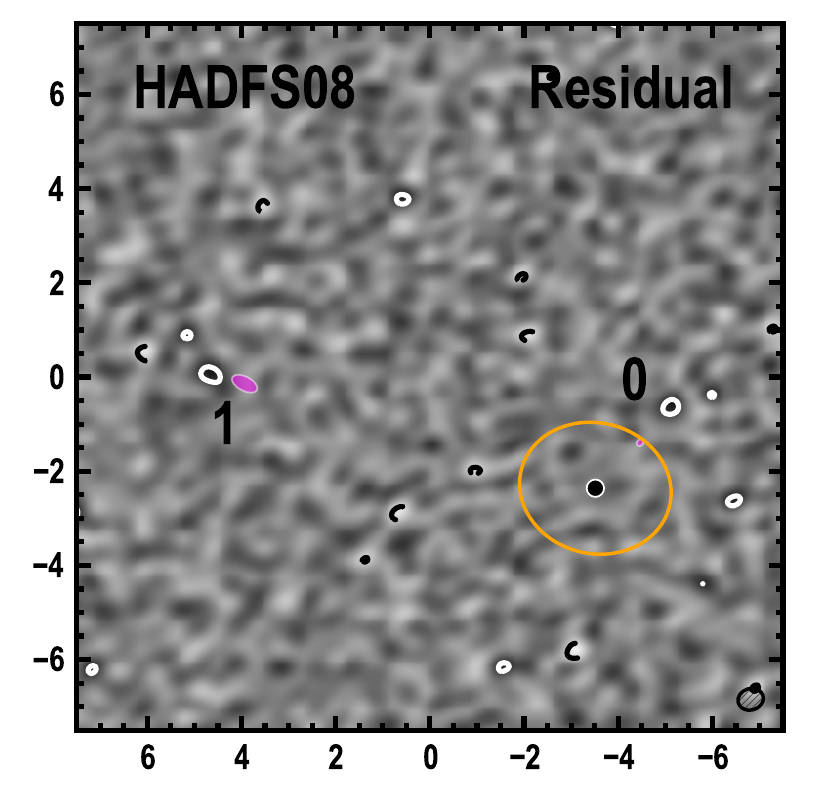}
\includegraphics[width=0.162\textwidth]{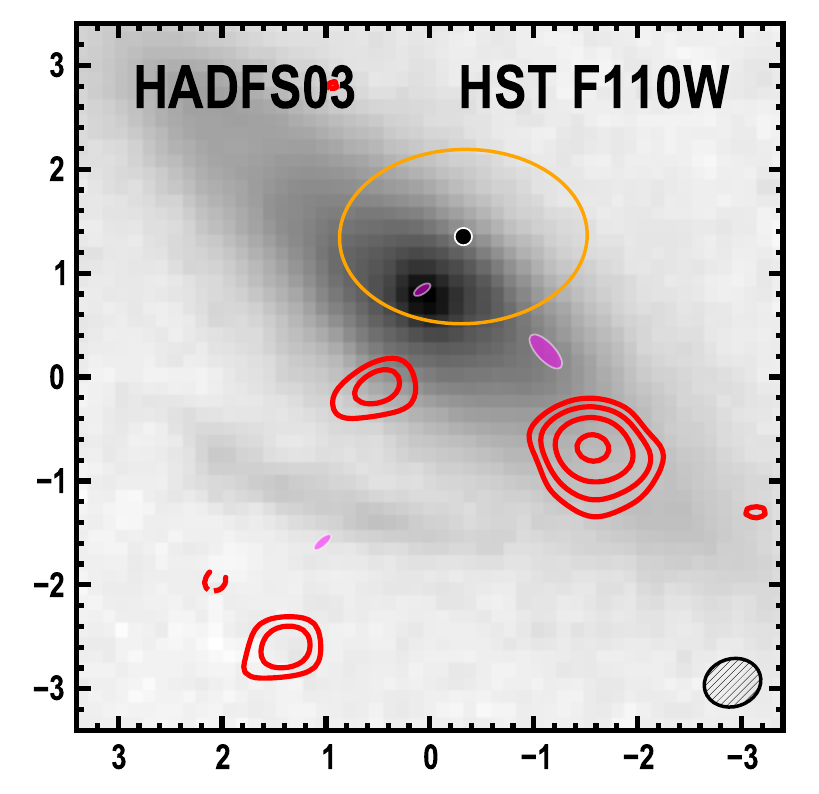}
\includegraphics[width=0.162\textwidth]{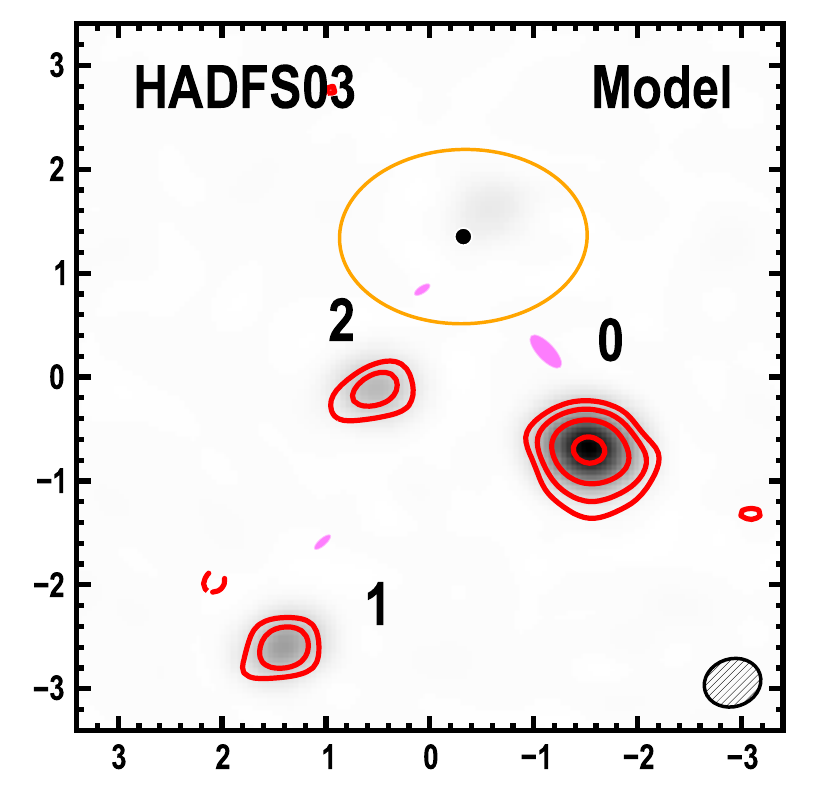}
\includegraphics[width=0.162\textwidth]{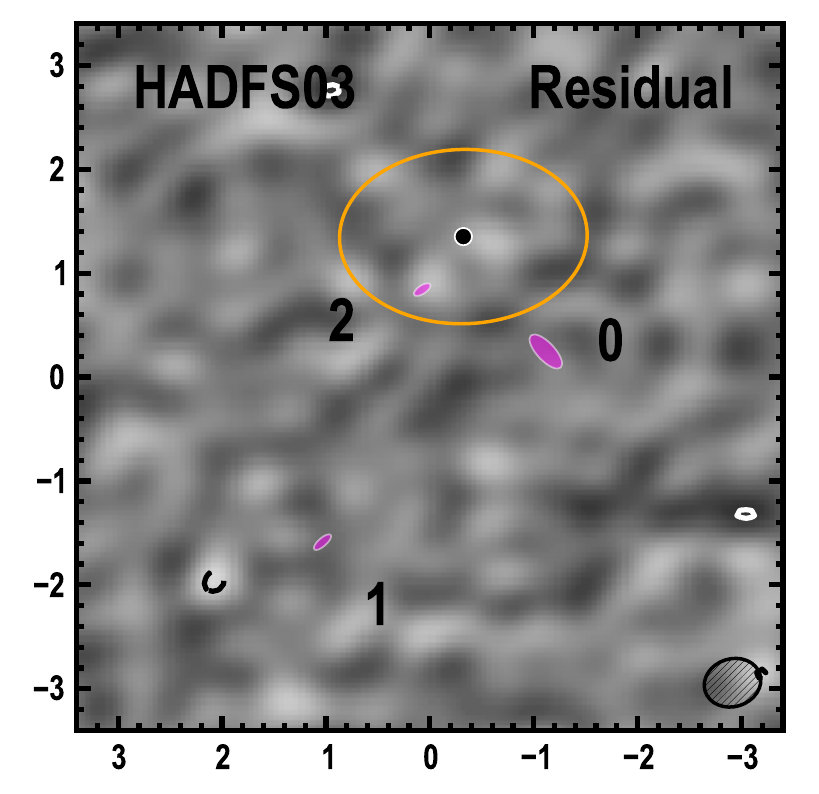}
\includegraphics[width=0.162\textwidth]{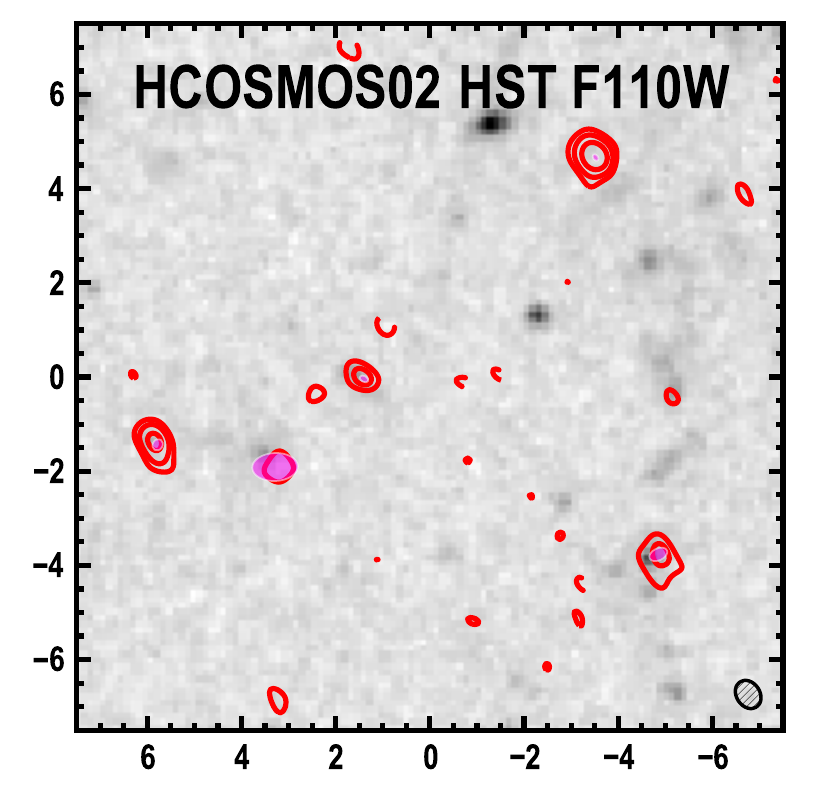}
\includegraphics[width=0.162\textwidth]{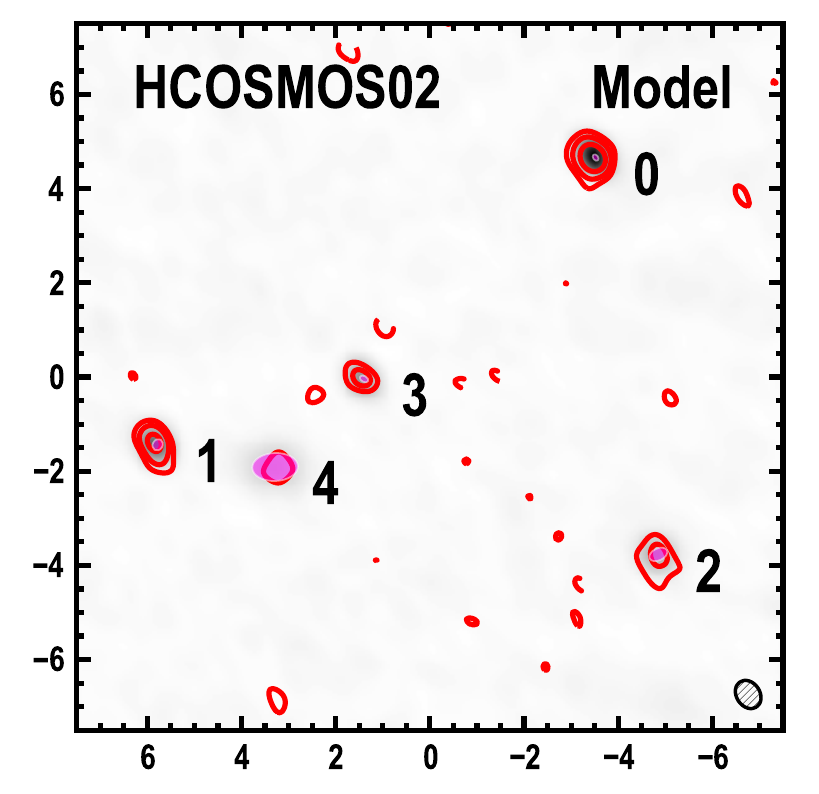}
\includegraphics[width=0.162\textwidth]{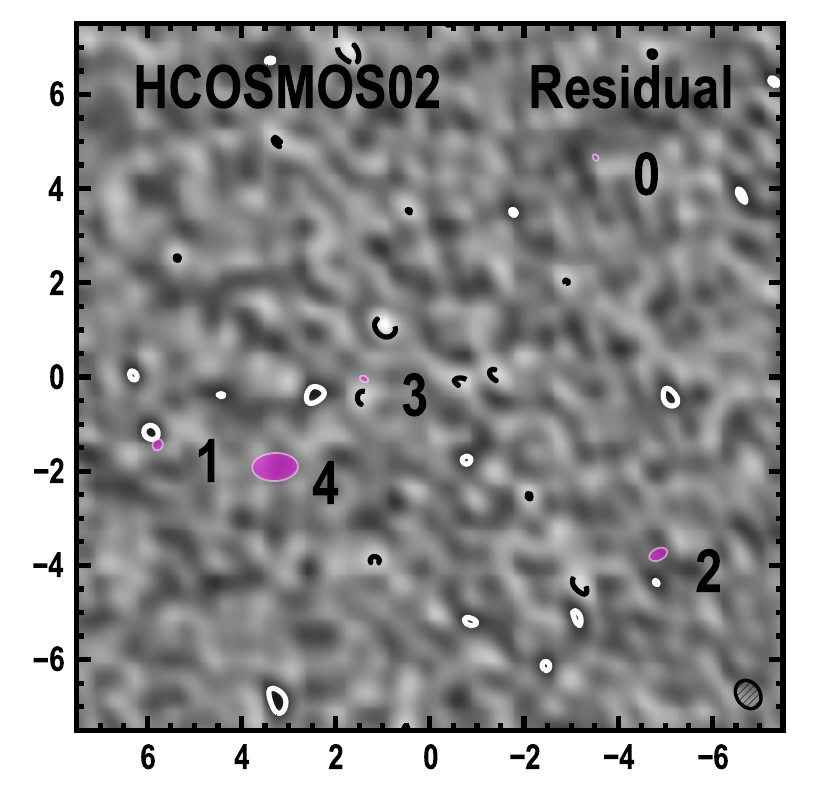}
\includegraphics[width=0.162\textwidth]{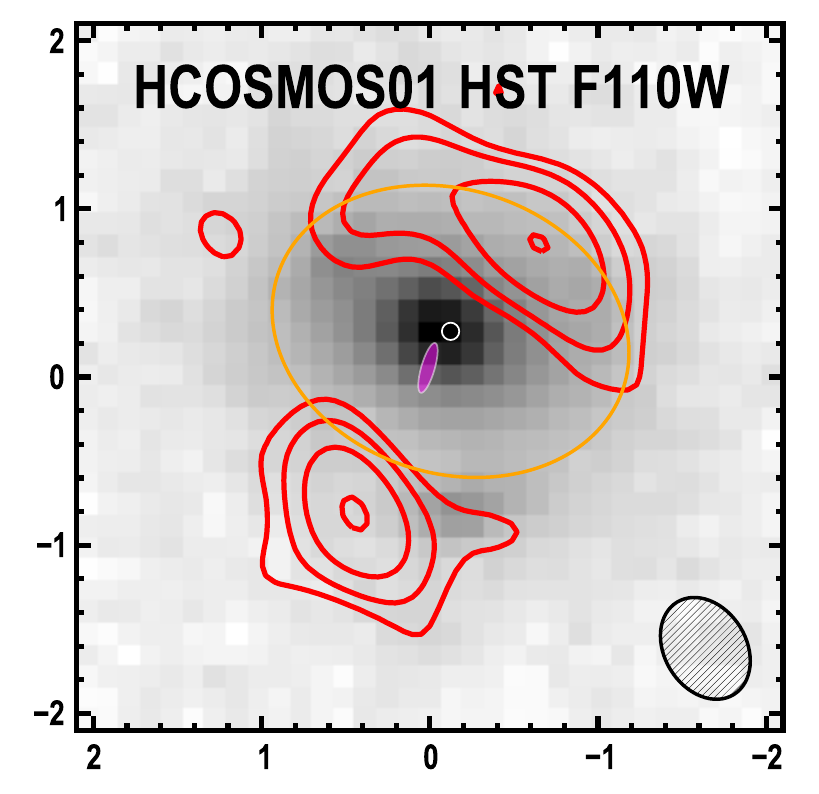}
\includegraphics[width=0.162\textwidth]{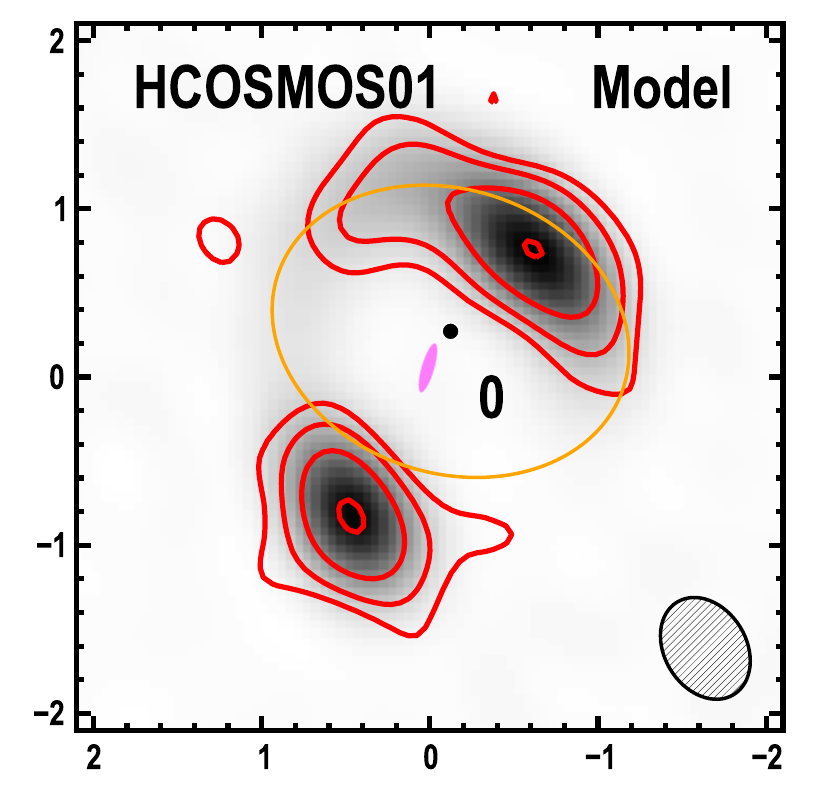}
\includegraphics[width=0.162\textwidth]{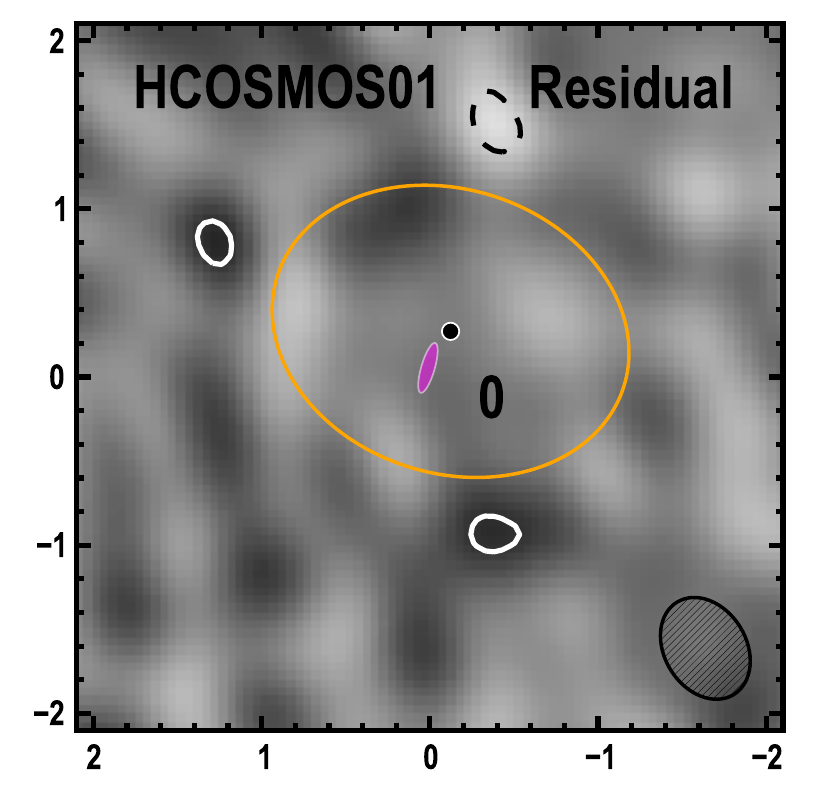}
\end{centering}

\caption{ Continued.}

\end{figure*}

\input{table_intrinsic}

\section{Results}\label{sec:results}

\subsection{De-lensing the ALMA Sample}\label{sec:lensing}

The combination of our optical or near-IR imaging and our deep, high-resolution
ALMA imaging permits us to take the first step towards mapping the foreground
structure along the line of sight to the ALMA sources.  With such maps in hand
for all of our targets, we can estimate the impact that lensing has on the
intrinsic properties of the ALMA sources.  In other words, we can ``de-lens''
the {\it Herschel}-ALMA sample.

\begin{figure}[!tbp] 
\includegraphics[width=\linewidth]{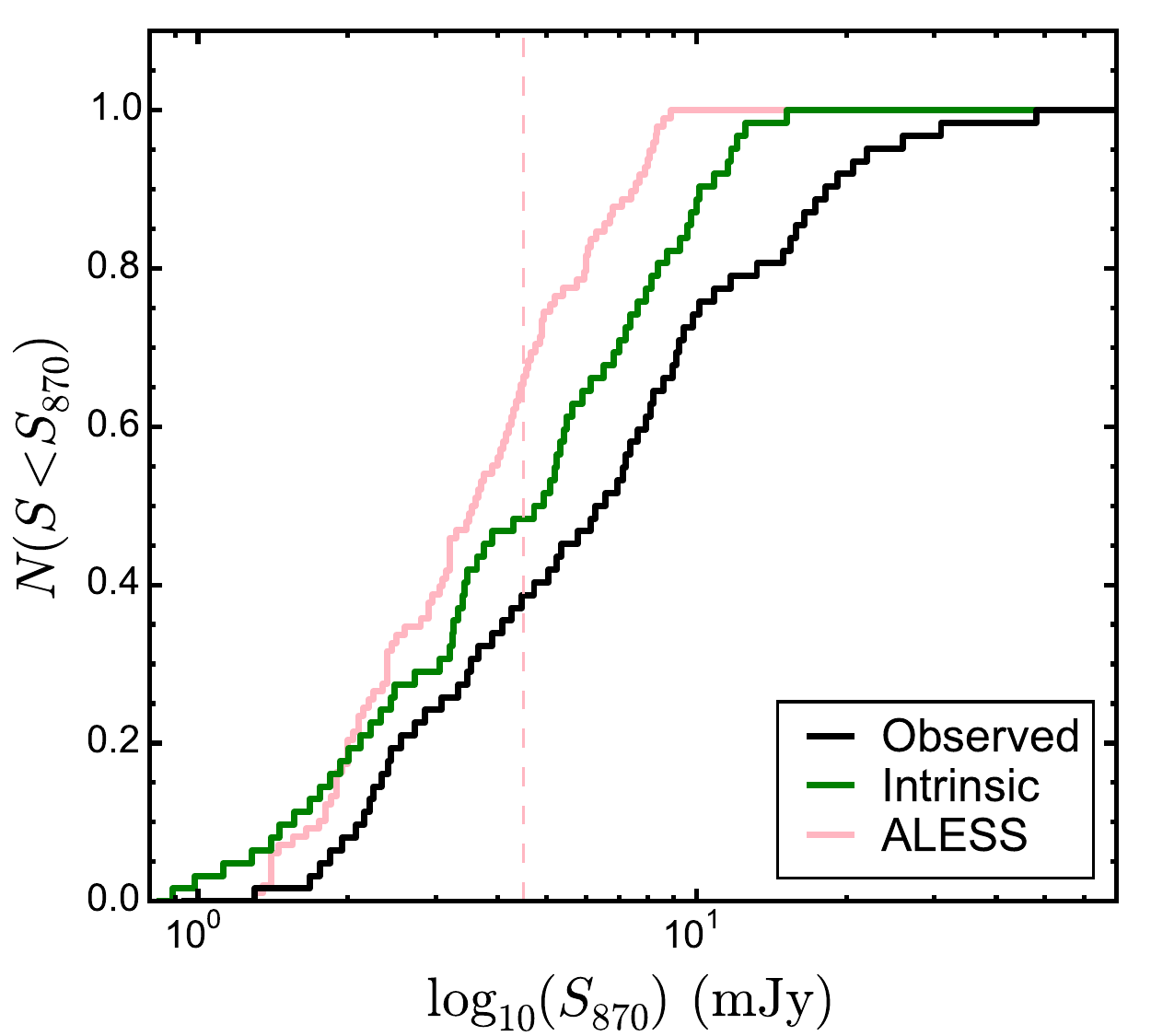}

\caption{ Cumulative distribution functions showing the effect of lensing on
the inferred flux densities of the {\it Herschel}-ALMA sample.  The median flux
density in the {\it Herschel}-ALMA sample drops by a factor of 1.3 when lensing
is taken into account.  For comparison we also show the flux density
distribution from ALESS (pink line), with the completeness limit of the LESS
survey indicated by a dashed pink line.} \label{fig:delens}

\end{figure}

Figure~\ref{fig:delens} shows the observed (i.e., apparent) and intrinsic
(i.e., de-lensed) distributions of $S_{870}$.  Lensing has the strongest effect on $S_{870}$: the median flux density
in the {\it Herschel}-ALMA sample drops by a factor of 1.6 when lensing is
taken into account. A two-sided Kolmogorov-Smirnov (KS) test yields a $p$-value
of 0.044, suggesting the apparent and intrinsic flux density distributions are
inconsistent with being drawn from the same parent population.  Even if
strongly lensed sources are removed from the sample, the median intrinsic flux
density is 1.3 times lower than the median apparent flux density.  Removing the
unlensed sources from consideration pushes this factor back to 1.6.  At these
levels, failing to correct for amplification due to gravitational lensing will
be a significant source of error, since the absolute calibration uncertainty is
typically of order 5-10\%.  When discussing the intrinsic properties of bright
sources \citep[including their number counts, e.g.][]{Wyithe:2011rm}
discovered in wide-field FIR or mm surveys, it is critical to consider the
effects of lensing.

For comparison, we also show the cumulative distribution of $S_{870}$ for the
ALESS sample (including the completeness limit of LESS of 4.5$\,$mJy).  ALESS
is the only existing sample of DSFGs with interferometric follow-up of a
sensitivity and angular resolution that is comparable to our ALMA data, so it
is the best sample with which to compare our results.  The significant overlap
in $S_{870}$ between our sample and ALESS is evidence that the DSFGs in our
sample have higher $S_{500}/S_{870}$ ratios (even when the effect of lensing in
our sample is taken into account) than the DSFGs in ALESS (recall
Figure~\ref{fig:sample}, which shows that ALESS sources have much lower
$S_{500}$ than our targets).  This difference is likely due to differences in
dust temperature and/or redshift distributions of the two samples and probably
arises from selection effects.

The effect on the other source parameters ($r_{\rm s}$, angular separation (the
angular distance between an ALMA source and the centroid of all the ALMA
sources for a given {\it Herschel} DSFG), and $q_{\rm s}$) is less pronounced.
The median source size decreases by a factor of 1.2 in the {\it Herschel}-ALMA
sample after accounting for lensing, but the two-sided KS test reveals a
$p$-value of 0.174, suggesting that we cannot rule out the null hypothesis that
both size distributions were drawn from the same parent distribution.  We find
no significant difference between the axial ratios of the apparent and
intrinsic distributions, as well as between the angular separations of apparent
and intrinsic distributions (two-sided KS test $p$-values of 0.984 and 0.920,
respectively). 


Finally, the brightest source in the {\it Herschel}-ALMA sample is HADFS11.0,
with an intrinsic flux density of $S_{870} = 17.5 \pm 0.4\,$mJy.  However,
there are also two objects with multiple sources that have separations smaller
than 1$\arcsec$, which have summed flux densities comparable to this; namely
HADFS02 (16.8$\,$mJy) and HECDFS13 (15.3$\,$mJy).  This is approaching the
values found in the most extreme systems, such as GN20
\citep[20.6$\,$mJy,][]{2006MNRAS.370.1185P} and HFLS3
\citep[15-20$\,$mJy;][]{Riechers:2013lr, Cooray:2014rm, Robson:2014xy}.  It is
a level that is extremely difficult to reproduce in simulations
\citep[e.g.,][]{Narayanan:2010lr}.  One possibility is that the objects with
multiple sources represent blends of physically unassociated systems.  We
explore this possibility via comparison to theoretical models in
Section~\ref{sec:spatialdist}, but a direct empirical test requires redshift
determinations for each source and is beyond the scope of this paper.

\subsection{Multiplicity in the ALMA Sample}\label{sec:multiplicity}

The second key result from our deep, high-resolution ALMA imaging is a firm
measurement of the rate of multiplicity in {\it Herschel} DSFGs.  We find that
20/29 {\it Herschel} DSFGs break down into multiple ALMA sources, implying a
multiplicity rate of 69\%.  
However, 5/9 of the single-component
systems are strongly lensed.  If these five are not considered, then the
multiplicity rate increases to 80\%.  Such a high rate of multiplicity is
consistent with theoretical models \citep[e.g.,][hereafter HB13]{HB13}.

In comparison, the 69 DSFGs in the MAIN ALESS catalog show a multiplicity rate
of 35 - 40\% \citep{Hodge:2013qy}.  Smoothing our ALMA images and adding
noise to match the resolution and sensitivity of ALESS results in a
multiplicity rate of 55\% (four objects with sources that are separated by
$<1\arcsec$ become single systems).  
On the other hand, the ALESS sources are much fainter overall, having a median
870$\,\mu$m flux density of $S_{870} \approx 6\,$mJy, compared to $S_{870}
=14.9\,$mJy in our {\it Herschel}-ALMA sample.  Thus, the evidence favors brighter sources
having a higher multiplicity rate.  This result is also consistent with
multiplicity studies of $S_{870}$-selected DSFGs by \citet{Ivison:2007qv},
\citet{Smolcic:2012zl}, and \citet{Barger:2012yg}, who use VLA, PdBI/1.3$\,$mm,
and SMA/870$\,\mu$m imaging to determine rates of 18\%, 22\%, and 40\%,
respectively.

\begin{figure}[!tbp] 
\includegraphics[width=\linewidth]{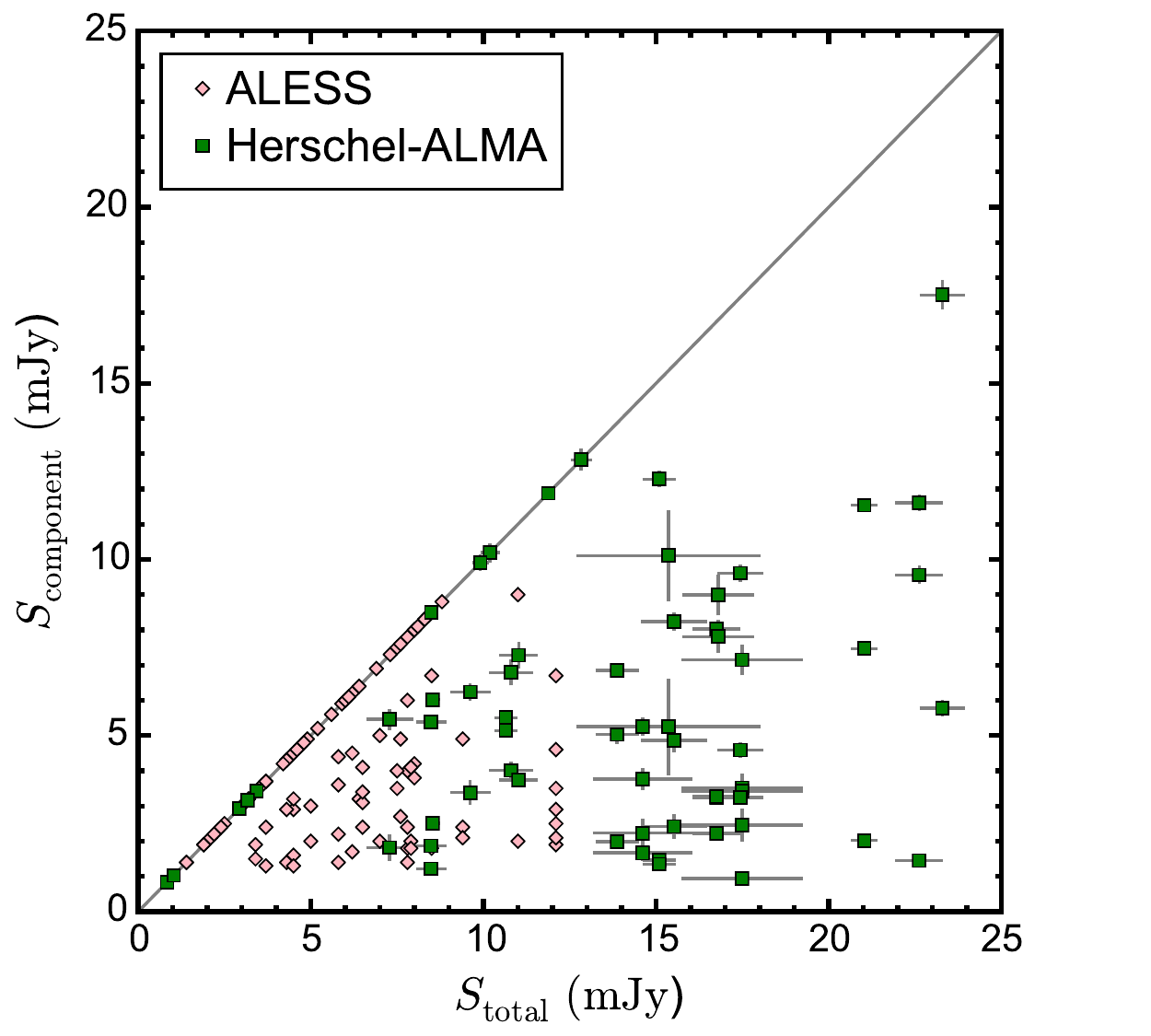}

\caption{ Comparison of the total 870$\,\mu$m flux density, $S_{\rm total}$,
with the individual component 870$\,\mu$m flux density, $S_{\rm component}$
(both of these are after accounting for lensing).  Objects falling along the
gray dashed line are single component systems (i.e., $S_{\rm total} = S_{\rm
component}$).  The solid lines trace the average ratio of component to total
flux for a given total flux.  Our sample of {\it Herschel} DSFGs ({\it
Herschel}-ALMA sample,
green squares) has a higher multiplicity and a lower average factional
contribution per component than the ALESS sample (pink diamonds), but not as
low as would be expected from a simple extrapolation of the trend in the ALESS
data alone.} \label{fig:componentflux}

\end{figure}

One useful way to characterize multiplicity is with a comparison of the total
870$\,\mu$m flux density, $S_{\rm total}$, with the individual component
870$\,\mu$m flux density, $S_{\rm component}$.  Figure~\ref{fig:componentflux}
shows these values for our {\it Herschel}-ALMA sample and compares to ALESS.
Lensing has a significant impact on the apparent flux densities of many objects
in our ALMA sample, so we are careful to show only intrinsic flux densities in
this diagram.  This diagram reflects the known result that the multipicity rate
in ALESS rises and the average fractional contribution per component decreases
with increasing $S_{\rm total}$ \citep{Hodge:2013qy}.  A simple extrapolation
of this phenomenon to the flux density regime probed by our {\it Herschel}-ALMA
sample would have suggested a very high multiplicity rate and a very low
average fractional contribution per component.  The multiplicity rate in our
sample is indeed higher, but we find that the average fractional contribution
per component hovers around 0.4 for essentially the full range in our sample.
This is a reflection of the fact that the brightest {\it Herschel} DSFGs
comprise 1-3 ALMA components, not 5-10 ALMA components as might have been
expected from a naive extrapolation of the ALESS results.  

\subsection{Spatial Distribution of Multiple Sources}\label{sec:spatialdist}

\begin{figure*}[!tbp] 
\includegraphics[height=3in]{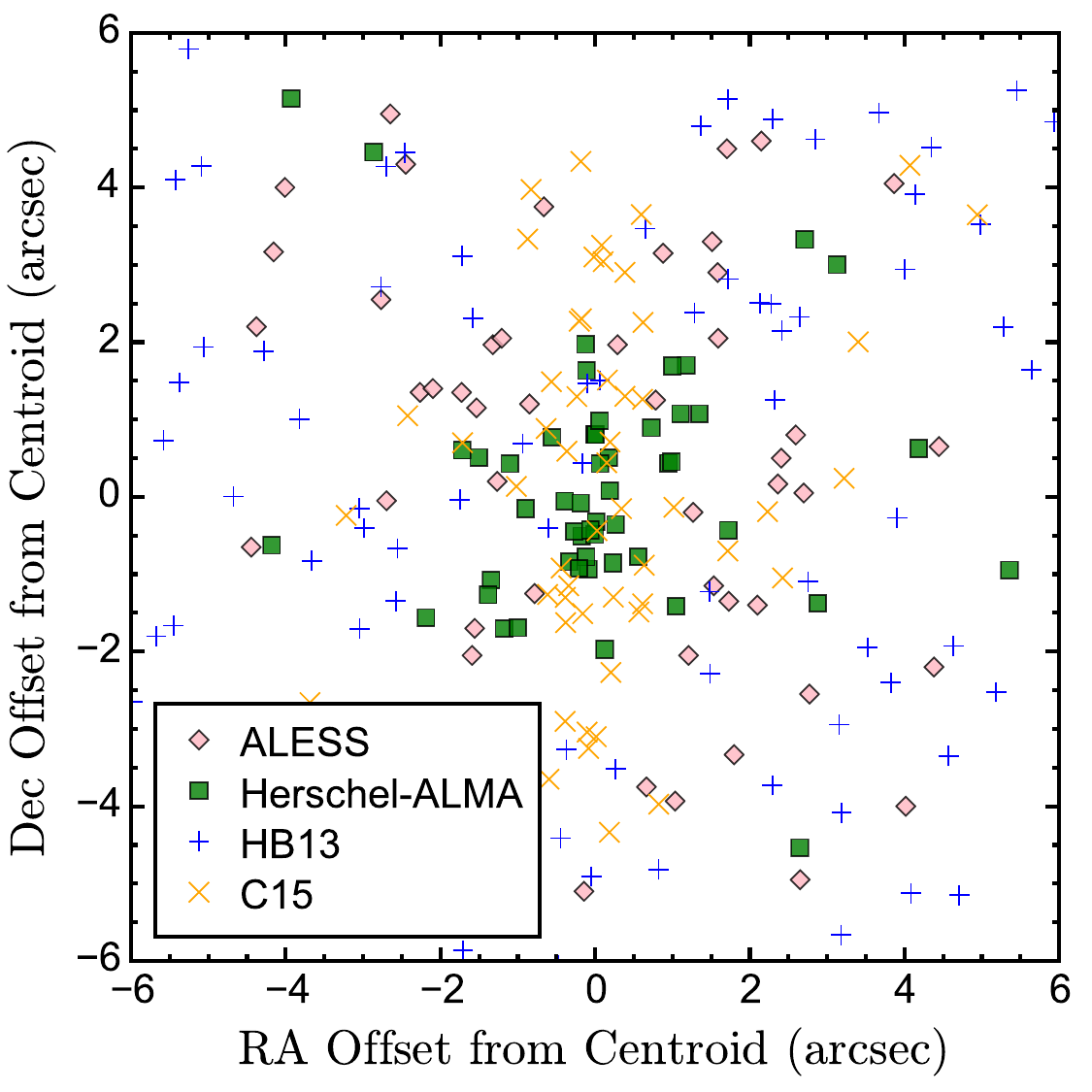}
\includegraphics[height=3in]{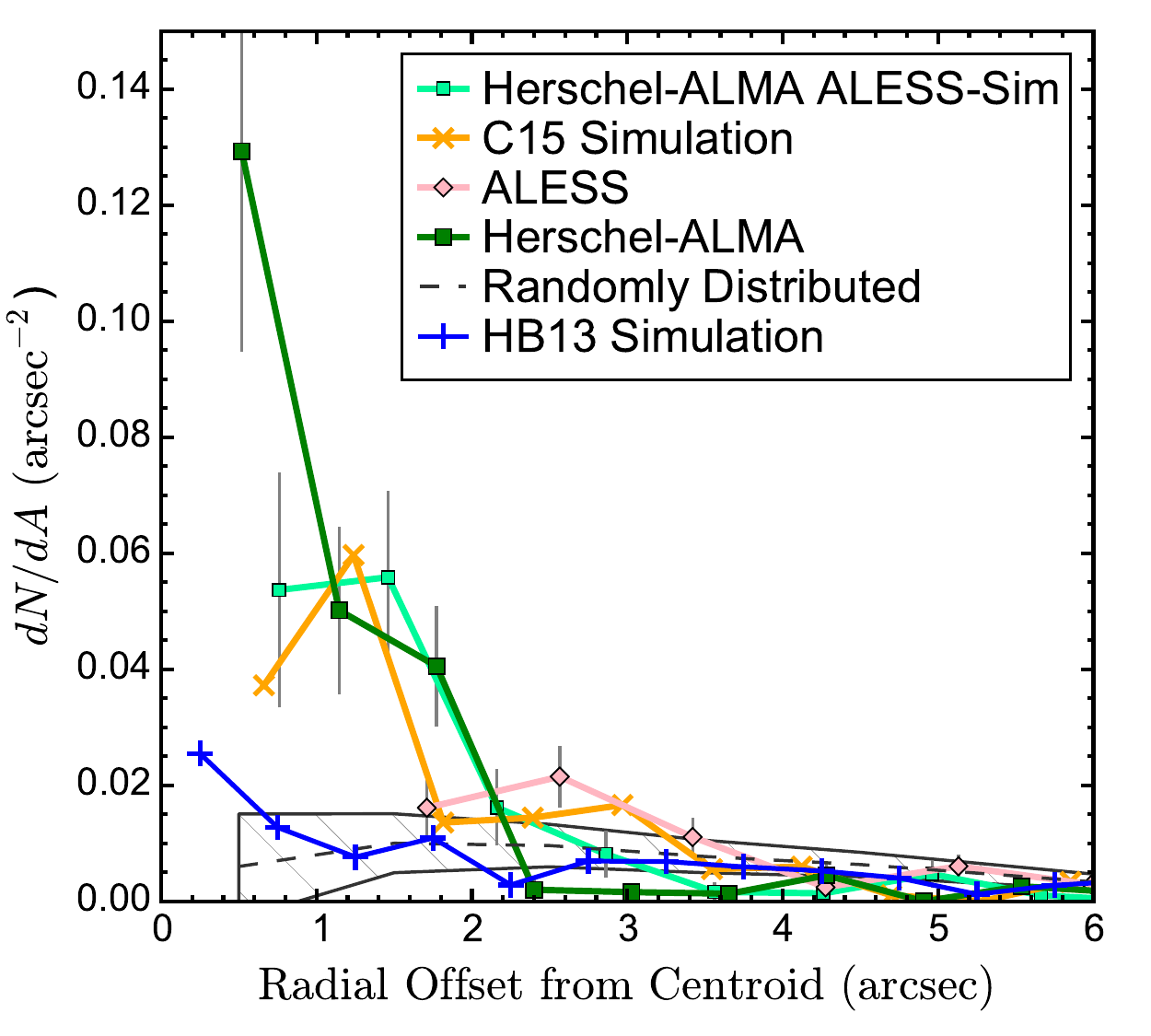}

\caption{ {\it Left}: Spatial distribution of sources with multiple
counterparts found in our {\it Herschel}-ALMA sample (green squares), in ALESS
(pink diamonds) and mock catalogs from \citetalias{HB13} and \citetalias{Cowley:2015lr}
(blue plus signs and orange crosses, respectively).  Sources identified in our
{\it Herschel}-ALMA sample lie much closer to each other than they do in either
ALESS or the \citetalias{HB13} simulations.  {\it Right}: Number of ALMA sources per
annular area as a function of angular separation from the ALMA centroid.
Symbols and colors are as in left panel.  We also show how our {\it
Herschel}-ALMA sample would appear if it had been observed with ALESS
resolution and sensitivity (light green squares).  The range of separations
that would be seen if sources were randomly distributed within the ALMA field
of view is also shown (dark dashed line and hatched region).  The {\it
Herschel}-ALMA DSFGs show a significantly stronger excess on angular separation
scales $<2\arcsec$ compared to both ALESS and the \citetalias{HB13} simulations,
even when taking into account the difference in sensitivity and spatial
resolution between our ALMA observations and those of ALESS.  The simulations
from \citetalias{Cowley:2015lr} show better agreement with the data, likely due to
the more sophisticated treatment of blending compared to \citetalias{HB13}.}
\label{fig:dNdA}

\end{figure*}

We can dig further into our ALMA data by exploring the average number of ALMA
sources per annular area ($dN/dA$) as a function of how far they are from each
other.  Figure~\ref{fig:dNdA} shows the results of this analysis for both our
{\it Herschel}-ALMA sample and ALESS.  We formulate the separation as an
angular distance between each ALMA source (using the lensing-corrected data)
and the centroid of all of the ALMA sources for that {\it Herschel} DSFG.  This
is different from the pairwise separation distance estimator used by
\citet{Hodge:2013qy} that becomes ill-defined when there are more than two ALMA
counterparts (as is often the case in our {\it Herschel}-ALMA sample).
Figure~\ref{fig:dNdA} shows $dN/dA$ values for ALESS that have been re-computed
using our method.  We also show the median and 1$\sigma$ range found from
simulated datasets for both ALESS and our {\it Herschel}-ALMA sample.  The
simulated datasets consist of 200 runs of DSFGs with the same flux density and
multiplicity as the observed datasets (both the ALESS sample and our ALMA
sample), but placed randomly within the primary beam FWHM.  We also show
predictions from simulations by \citetalias{HB13} (see below for details).

We recover the result from \citet{Hodge:2013qy} that the ALESS DSFGs are
consistent with a uniformly distributed population.  Interestingly, however,
there is a dramatic rise in $dN/dA$ for angular separations less than
$2\arcsec$ in our {\it Herschel}-ALMA sample.  Indeed, for an angular
separation of $0\farcs5$, we find an excess in $dN/dA$ by a factor of $\approx
10$ compared to a random, uniformly distributed population.  This excess
persists (although at significantly lower amplitude) even when the quality of
our ALMA observations are degraded to match the typical sensitivity, spatial
resolution, and {\it uv} coverage of ALESS (as represented by observations of
ALESS 122).  The persistence of the excess suggests that it is an intrinsic
property of the sample; i.e., that only the brightest DSFGs show an excess of
sources on small separation scales (with the caveat that we cannot rule out the
possibility of at least part of the excess being due to strong lensing from
optically dark lenses).  

An excess of sources with small separations from each other could be an
indication of interacting or merging systems.  However, it is also possible
that the sources are merely unrelated galaxies that appear blended due to
projection effects \citep{HB13, Cowley:2015lr, Munoz-Arancibia:2015qy}.
Spatially resolved spectroscopy is necessary to answer this question
definitively, but is not currently available.  Instead, to investigate these
possibilities further, we make use of mock catalogs of DSFGs that are based on
numerical simulations and presented by \citetalias{HB13} and \citetalias{Cowley:2015lr}.  

We begin with the \citetalias{HB13} simulations, summarizing the methodology used to
generate the mock catalogs here and refering the reader to \citetalias{HB13} for
full details. 

Halo catalogs are generated from the {\it Bolshoi} dark matter-only
cosmological simulation \citep{Klypin11} using the {\sc rockstar} halo finder
\citep{Behroozi13a, Behroozi13b}.  Catalogs of subhalos are created from eight
randomly chosen lightcones, each with an area of 1.4$^\circ \times 1.4^\circ$.
Galaxy properties such as stellar mass and SFR are assigned to the subhalos
using the abundance matching method of \citet{Behroozi13c}.  Dust masses are
assigned using an empirically determined redshift-dependent mass--metallicity
relation and an assumed dust-to-metal density ratio of 0.4 (see \citealt{HN13}
for details). Finally, submm flux densities are interpolated from the SFRs and
dust masses using a fitting function that is based on the results of dust
radiative transfer calculations performed on hydrodynamical simulations of
isolated and interacting galaxies \citep{H11,H12,HN13}.

A blended source is defined as any galaxy in the mock catalogs above a
threshold flux density ($S_{\rm thresh}$) that has at least one neighbor within
a projected angular distance $d_{\rm neighbor}$.  To obtain a direct comparison
with our {\it Herschel}-ALMA sample, we use $S_{\rm thresh} = 1.0\,$mJy
(corresponding to the 5$\sigma$ limit of the ALMA data) and $d_{\rm neighbor} =
40\arcsec$ (reflecting the size of the {\it Herschel} beam at 500$\,\mu$m).  We
use the known positions in the mock catalogs for all blended sources and
compute centroid and separations for every blended source using the same
methodology as we applied to our {\it Herschel}-ALMA sample and to ALESS.  

The $dN/dA$ values found in the mock catalogs are shown by the thick blue line
and plus signs in Figure \ref{fig:dNdA}.  There is a significant increase in
$dN/dA$ on separations smaller than $\approx 0\farcs5$, but the amplitude of
the increase is much lower than is apparent in our {\it Herschel}-ALMA sample. 


The \citetalias{HB13} model does not include SFR enhancements induced by starbursts
(see Section 4.5 of \citealt{HB13} for a detailed discussion of this
limitation). To explore whether interaction-induced starbursts are the origin
of the excess at small angular separations observed in our {\it Herschel}-ALMA
sample, we analyzed modified versions of the \citetalias{HB13} model that include a
crude treatment of interaction-induced SFR enhancements \citep{Miller:2015lr}.
Mock galaxies with one or more neighbors within a physical distance of 5$\,$kpc
and with a stellar mass between one-third and three times that of the galaxy
under consideration (i.e.  a `major merger') had their SFRs increased by a
factor of two. For distances smaller than 1$\,$kpc, the imposed increase was a
factor of ten.  Because these SFR enhancements are greater than suggested by
simulations \citep[e.g.][]{Cox08, H11, H14, Torrey12} or observations of local
galaxy pairs \citep[e.g.][]{Scudder12, Patton13}, we consider this test to
provide an upper limit on the possible effect of interactions on blended
sources in the \citetalias{HB13} model, although the incompleteness of the
\citet{Behroozi13b} catalogs for mergers with small separations could cause
some interacting systems to be missed.  We find an insignificant effect on the
values of $dN/dA$ when using the merger-induced model as described above.  The
main reason for this is that only two sources had their SFRs boosted by a
factor of ten, and $\approx150$ experienced a factor of two increase. In the
\citetalias{HB13} model, a factor of two increase in SFR corresponds to only a ~30\%
increase in $S_{870}$, so it is perhaps unsurprising that the weak boosts in
SFR cause little change in $dN/dA$.

Experiments with stronger interaction-induced SFR enhancements showed that very
high enhancements (e.g. a factor of 10 for separation of 5-15$\,$kpc and 100 for
separation of $<5\,$kpc) in major mergers are required to match the observed
excess in $dN/dA$ on small separations. Incorporating starbursts induced by
minor-merger could possibly reduce the required SFR enhancements. The tension
between the model prediction and observations may also indicate that a more
sophisticated treatment of blending is necessary.

To explore this possibility, we investigate mock catalogs based on the
methodology presented in \citetalias{Cowley:2015lr}.  Here, we give a brief summary
and refer the interested reader to \citetalias{Cowley:2015lr} for full details. A
new version of the {\sc galform} \citep[e.g.][Lacey et al. in
preparation]{Cole:2000fk} semi-analytic model of hierarchical galaxy formation
is used to populate halo merger trees \citep[e.g.][]{Parkinson:2008qy,
Jiang:2014lr} derived from a Millennium style $N$-body dark matter only
simulation \citep{Springel:2005lr, Guo:2013lr} with WMAP7 cosmology
\citep{Komatsu:2011fk}.  A sub-mm flux for each galaxy is calculated using a
self-consistent model based on radiative transfer and energy balance arguments.
Dust is assumed to exist in two components,  dense molecular clouds and a
diffuse ISM.  Energy absorbed from stellar radiation by each dust component is
calculated by solving the equations of radiative transfer in an assumed dust
geometry.  The dust is then assumed to emit radiation as a modified blackbody.  

Three randomly orientated 20 deg$^2$ lightcone catalogues are generated using
the method described in \citet{Merson:2013lr}.  We choose as the lower flux
limit for inclusion of simulated galaxies into our lightcone catalogue
$S_{500\mu\rm{m}}>0.1$ mJy, as this is the limit at which we recover 90\% of
the extragalactic background light (EBL) as predicted by the model (122
Jy~deg$^{-2}$). This is in excellent agreement with observations from the
\emph{COBE} satellite \citep[e.g.,][]{Puget:1996lr}, and thus ensures a
realistic 500 $\mu$m  background in the mock images.  

Mock imaging is created by binning the lightcone galaxies onto a pixelated grid
which is then convolved with a 36$\arcsec$ FWHM Gaussian (corresponding to the
\emph{Herschel}/SPIRE beam at 500$\,\mu$m).  The image is then constrained to
have a zero mean by the subtraction of a uniform background.  No instrumental
noise is added, nor are any further filtering procedures applied to the mock
image.  For the purposes of source identification, this procedure is repeated
at 250$\,\mu$m.  For this, we adjust the FWHM of the Gaussian PSF to
18$\arcsec$ and change the lower flux limit of inclusion into our lightcone to
ensure 90\% of the predicted EBL is recovered at this wavelength.    

Source positions are selected as maxima in the mock 250$\,\mu$m image, with the
position of the source being recorded as the center the maximal pixel for
simplicity.  To mimic `deblended' \emph{Herschel} photometry we record the
value of the pixel located at the position of the 250$\,\mu$m maxima in the
500$\,\mu$m images. We select all \emph{Herschel} sources satisfying
$S_{500\,\mu\rm{m}}>50\,$mJy and $z > 1$ to identify galaxies from our
lightcone catalogs within a 9$\arcsec$ radius of the source position, modelling
the ALMA primary beam profile as a Gaussian with an 18$\arcsec$ FWHM and a
maximum sensitivity of 1$\,$mJy.

The $dN/dA$ values derived from the \citetalias{Cowley:2015lr} mock catalogs are
shown by the thick orange line and crosses in Figure \ref{fig:dNdA}.  Here, the
amplitude of the increase in $dN/dA$ on separations smaller than $\approx
2\arcsec$ mimics the trend seen in the data.  However, there is a deficit of
multiple systems with separations of $0\farcs5$ or less compared to the {\it
Herschel}-ALMA sample.  This result suggests that a sophisticated treatment of
blending yields better agreement between simulations and observations but the
simulations still under-predict the number of multiple systems with small
separations.

Future work on the theoretical side should seek to determine if the application
of the \citetalias{Cowley:2015lr} blending algorithm to the \citetalias{HB13} simulations
yields similarly better agreement with the data.  On the observational side, it
is critical to establish whether {\it Herschel} sources with multiple ALMA
counterparts are physically related by measuring spectroscopic redshifts to
individual counterparts.  Fortunately, this is a viable project today with the
VLA and ALMA.

\section{Implications for the Bright End of the DSFG Luminosity
Function}\label{sec:discuss}

The distribution of magnification factors for sources found in wide-field
surveys with the brightest apparent flux densities are highly sensitive to the
shape of the intrinsic number counts at the bright end.  In this section,
we combine our ALMA and SMA measurements of magnification factors to investigate
this as it pertains to DSFGs.

\subsection{Statistical Predictions for $\mu_{870}$}\label{sec:statpredict}

Our methodology follows the procedures outlined in previous efforts to predict
magnification factors for DSFGs with a given apparent flux density
\citep[chiefly,][]{Lima:2010fk, Wardlow:2013lr, Fialkov:2015eu}.  We summarize
the essential elements here and highlight significant differences where
appropriate.  

The key components of the model are the mass density profile of the lenses,
$\rho_{\rm lens}\,(r)$, the number density of lenses as a function of mass and
redshift, $n_{\rm lens}\,(M, \, z),$ the redshift distribution of the sources,
$dN_{\rm source}/dz$, and the intrinsic number counts of the sources,
$dn_{\rm source}/dS_{870}^\prime$.  The latter component is the least certain
and also has the strongest impact on the predicted apparent luminosity
function.  For these reasons, we fix all components of the model except the
shape of the intrinsic number counts.  Our goal is to take luminosity
functions that can successfully fit observed faint DSFG number counts
\citep{Karim:2013lr} and test whether they lead to predicted magnification
factors consistent with our ALMA and SMA observations.

To describe $\rho_{\rm lens}\,(r)$, we use a superposition of a singular
isothermal sphere (SIS) and a Navarro-Frenk-White (NFW) profile
\citep{Navarro:1997ys} that is truncated at the virial radius.  The NFW profile
describes the outskirts of dark matter halos better \citep{Mandelbaum:2005lr},
while the SIS profile is preferred on smaller scales because it correctly fits
the observed flat rotational curves in galaxies \citep{Kochanek:1994fk}.  We
make sure that the resulting probability density of lensing, $P\,(\mu)$, is
normalized to unity.  To describe $n_{\rm lens}\,(M, \, z)$, we generate the
abundance of halos at each mass and redshift using the \citet{Sheth:1999kx}
formalism.  To describe $dN_{\rm source}/dz$, we adopt the following redshift
distribution which is based on photometric redshifts of optical counterparts to
ALMA sources identified in ALESS \citep{Simpson:2014lr}:

\begin{equation}
    dN/dz \propto \frac{1}{a_z\sigma_z \sqrt{2\pi}}
    \exp\left( \frac{-[\ln(a_z) - \ln(1 + z_\mu)]^2} {2\sigma_z^2 a_z} \right),
\end{equation}

\noindent where $a_z = 1 + z$, $z_\mu = 1.5$ (to reflect the relatively blue
SPIRE colors of the sample), and $\sigma_z = 0.2$.  Alternative
values for $z_\mu$ and $\sigma_z$ yield second-order perturbations which are
not significant at the level of our current analysis.

We explore two intrinsic number counts that are intended to bracket the
plausible range of values for DSFGs based on two interferometric surveys.  One
is the number counts measured in ALESS \citep{Karim:2013lr}, and the other is
from interferometric follow-up of the first AzTEC survey in COSMOS
\citep{Scott:2008qy} using the SMA \citep{Younger:2007fk, Younger:2009lr} and
PdBI \citep{Miettinen:2015lr}.  These interferometric observations have shown
that all of the sources in their surveys are either unlensed or lensed by
magnification factors $<2$ \citep[a similar result is found based on ALMA
imaging of 52 DSFGs in the Ultra Deep Survey;][]{Simpson:2015lr}.  This is why
the ALESS and COSMOS luminosity functions represent a plausible range of
intrinsic number counts for DSFGs.  These number counts are shown in the left
panel of Figure~\ref{fig:lensstats}.  Interferometric follow-up data in COSMOS
\citep{Smolcic:2012zl} and GOODS-N \citep{Barger:2012yg} are published, but
unknown completeness corrections in the single-dish surveys on which these
follow-up datasets are based precludes their use here.

In detail, we use a broken power-law of the form

\begin{equation}
\frac{dn}{dS^\prime} = N_\ast\left(\frac{S^\prime}{S_\ast}\right)^{-\beta_1},~~~~\textrm{for}~S<S_\ast,
\end{equation} 

\begin{displaymath}
\frac{dn}{dS^\prime} = N_\ast\left(\frac{S^\prime}{S_\ast}\right)^{-\beta_2},~~~~\textrm{for}~S>S_\ast.
\end{displaymath}

Table~\ref{tab:models} provides values for the parameters of the broken
power-law for the ALESS and COSMOS number counts.  The data and
corresponding number counts are shown in the left panel of
Figure~\ref{fig:lensstats}.

The product of the model is the lensing optical depth for a given lensing
galaxy and source galaxy,  $f_\mu$. The lensing probability with magnification
larger than $\mu$ is then calculated via $P(>\mu) = 1-\exp(-f_\mu)$ and the
differential probability distribution is $P(\mu) = -dP(>\mu)/d\mu$.  The sum
over the distribution of source redshifts and lens masses and redshifts yields
the total probability distribution function.

\input{table_lf}

The fundamental measurement provided by the spatially resolved ALMA and SMA
imaging and associated lens models is the magnification factor of a source with
a given apparent $S_{870}$.  We use the combined sample to compute the average
magnification as a function of $S_{870}$ from the data:
$\langle\mu_{870}\rangle$.  The same quantity can also be directly computed
from our model as 

\begin{equation}
    \langle\mu_{870}\rangle = \int_0^\infty \mu P(\mu|S_{870})  d\mu,
 \end{equation}

\noindent where the probability for lensing with magnification $\mu$ given the
apparent flux is:

\begin{equation}
    P(\mu|S_{870}) = \frac{1}{N} \frac{P(\mu)}{\mu} \frac{dn}{dS_{870}^\prime}
\left(\frac{S_{870}}{\mu}\right),
\end{equation} 

\noindent and 

\begin{equation}
    N = \int \frac{P(\mu)}{\mu} \frac{dn}{dS_{870}^\prime}
    \left(\frac{S_{870}}{\mu}\right)d\mu.
\end{equation}

\noindent Here $ dn/dS_{870}$ is the observed number counts and
$dn/dS_{870}^\prime$ is the intrinsic number counts.

As part of the lens models, the ALMA and SMA imaging also provide the
probability that a source with a given apparent $S_{870}$ experiences a
magnification above some threshold value, $\mu_{\rm min}$: $P\,(\mu > \mu_{\rm
min})$.  It is therefore of interest to make a similar prediction from our
model.  We use the following to do this:

\begin{equation}
    P(\mu>\mu_{\rm min}) = \frac{\int_\mu^\infty P(\mu|S_{obs})}{\int_0^\infty
    P(\mu|S_{obs})}.
\end{equation}

\subsection{Comparing Models with Data}

\begin{figure*}[!tbp] 
\includegraphics[width=0.335\textwidth]{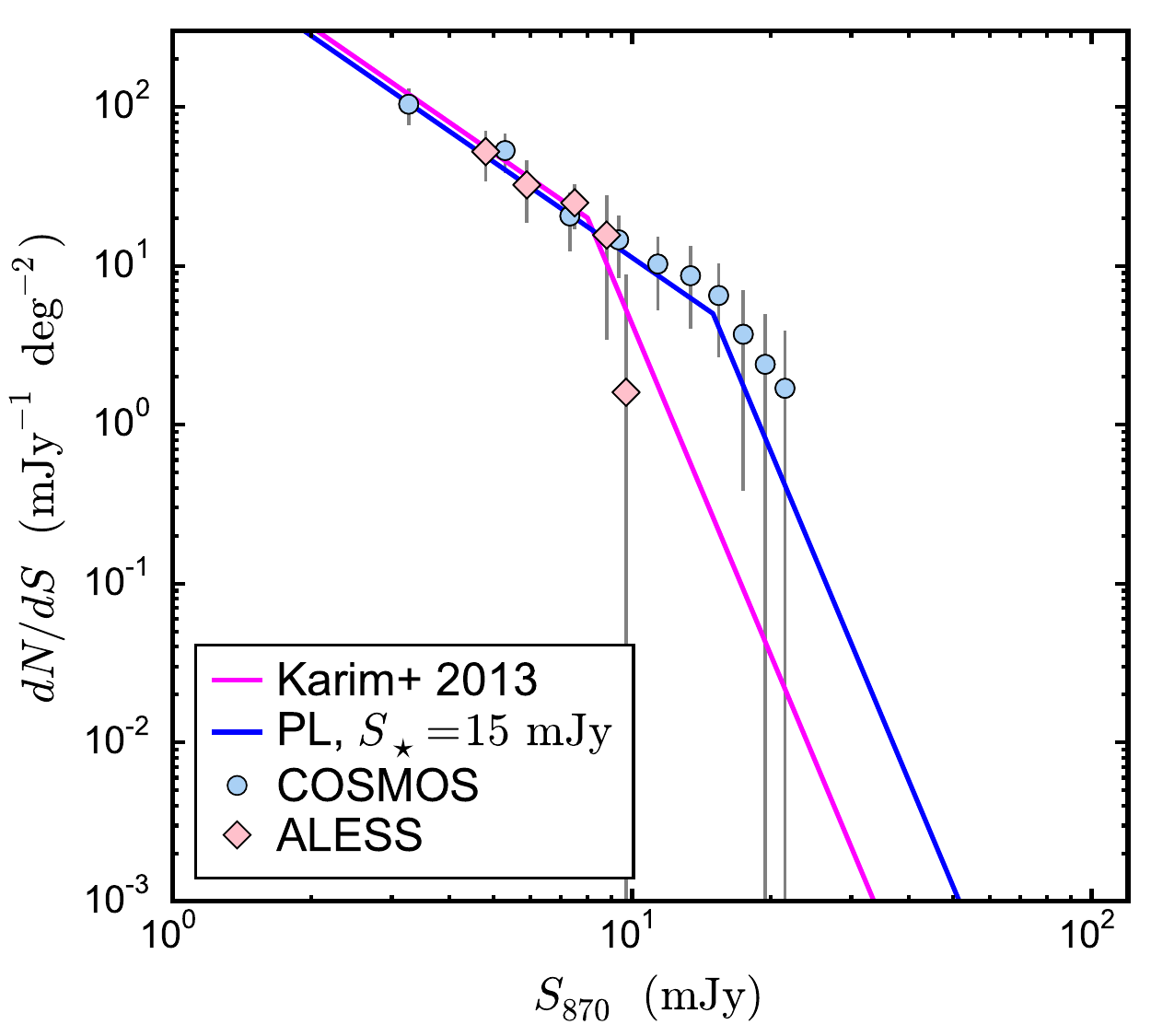}
\includegraphics[width=0.335\textwidth]{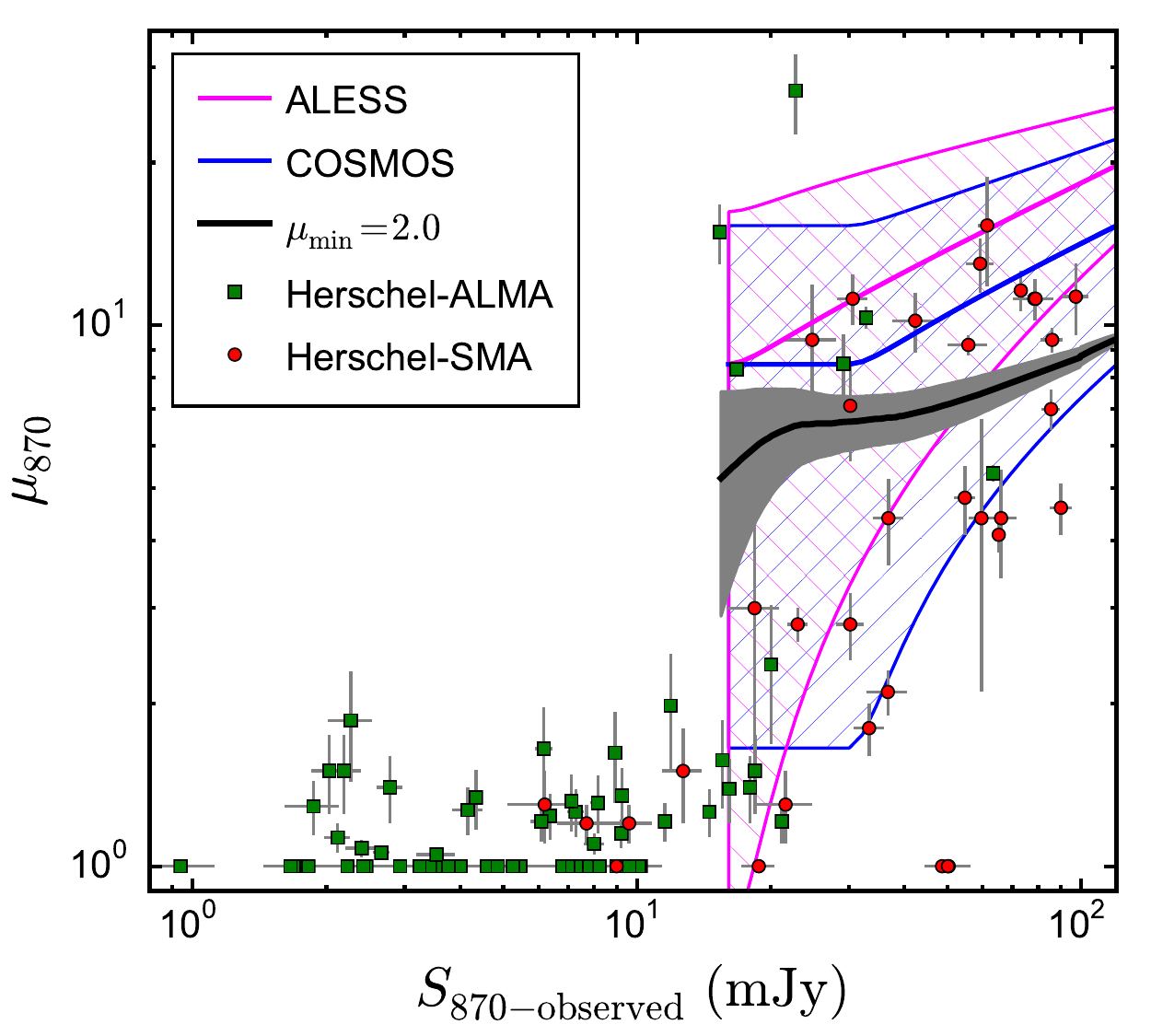}
\includegraphics[width=0.335\textwidth]{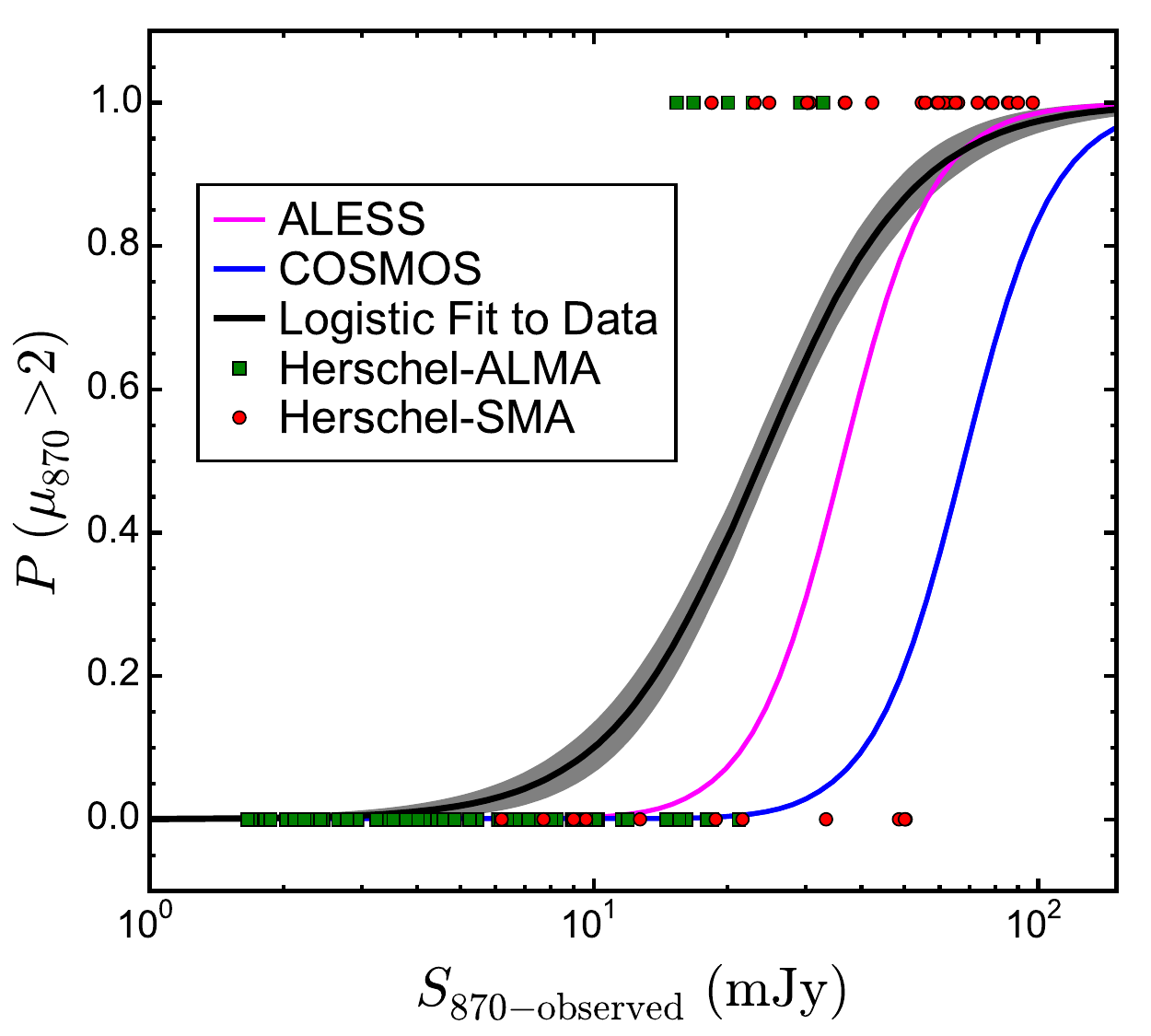}

\caption{ {\it Left}: Observed number counts from interferometer
follow-up of mm sources in COSMOS \citep[black circles;][]{Younger:2007fk,
Younger:2009lr, Miettinen:2015lr} and from ALESS \citep[pink
diamonds;][]{Karim:2013lr}.  The magneta and blue lines represent the range of
plausible intrinsic number counts for DSFGs given these two datasets.
{\it Middle}: Magnification factors at 870$\,\mu$m as a function of apparent
$S_{870}$ for every source in our ALMA (green squares) and SMA (red circles)
samples.  The black line represents a running average of the magnification
measurements when sources with $\mu_{870} > 2.0$ are considered (grey shaded
region highlights the 1$\sigma$ uncertainty), and is truncated at the minimum
flux density at which we have observed $\mu_{870} > 2.0$.  Colored lines show
the average magnification factor, $\langle\mu_{870}\rangle$, predicted by our
models for the two intrinsic number counts shown on the left.  The model
number counts predict higher $\langle\mu_{870}\rangle$ than are seen in
the data.  {\it Right}: Probability that a source with a given $S_{870}$
experiences $\mu_{870} > 2.0$.  The black line (grey shaded region) shows a
logistic regression fit to the SMA and ALMA data (grey shaded region highlights
the 1$\sigma$ uncertainty).  Colored lines are the same as in the middle and
left panels.  The intrinsic number counts that provide a good fit to the
COSMOS data predict too many unlensed or weakly lensed sources with intrinsic
flux densities of $S_{870}^\prime \sim 50\,$mJy.} \label{fig:lensstats}

\end{figure*}

The middle panel of Figure~\ref{fig:lensstats} shows a direct comparison of the
measured $\mu_{870}$ values as a function of apparent $S_{870}$ for the {\it
Herschel}-ALMA and {\it Herschel}-SMA samples.  We also show a running average
of the combined sample (considering only $\mu_{870} > 2.0$ objects) to serve as
a direct comparison to our theoretical models.  We compute this by
interpolating the observed $\mu_{870}$ and $S_{870}$ onto a fine grid using the
{\sc Scipy} {\sc griddata} package and then smoothing the resulting grid using
a Gaussian filter in the Scipy {\sc gaussian\_filter} package.
Also shown in this diagram are model predictions for the average magnification
as a function of $S_{870}$, $\langle\mu_{870}\rangle$, assuming the two
intrinsic number counts for DSFGs described in Table~\ref{tab:models}.  

Both models predict higher $\langle\mu_{870}\rangle$ than are seen in the data
by factors of 5-10.  However, the dispersion in the predicted
$\langle\mu_{870}\rangle$ values for both intrinsic number counts rises
smoothly from $\sigma_\mu \approx 2$ at $S_{870} = 15\,$mJy to $\sigma_\mu
\approx 8$ at $S_{870} = 100\,$mJy, so this difference is not statistically
significant.  Furthermore, there is reason to believe that the data may be
biased against high magnification factor measurements.  In both the {\it
Herschel}-ALMA and {\it Herschel}-SMA samples, the spatial resolution is
$\approx 0\farcs5$.  This is nearly always sufficient to resolve the images of
the lensed galaxy, but it is not always sufficient to resolve the images
themselves.  Therefore, it may be the case that the lens models over-predict
the intrinsic sizes of the lensed galaxies and hence under-predict the
magnification factors.  For example, the average half-light sizes of unlensed DSFGs
have been found recently to be $\approx 0.8 \,$kpc \citep{Ikarashi:2014nr} and
$\approx1.2\,$kpc \citep{Simpson:2015lr}.  In contrast, we reported a median
half-light radius of $1.53\,$kpc in \citet{Bussmann:2013lr}.  Lens models from
higher resolution data with ALMA suggest that magnification factors could
increase by a factor of $\approx 1.5-2$ \citep{Rybak:2015fk, Tamura:2015lq,
Dye:2015fp}.  Therefore, it is plausible that both of the intrinsic luminosity
functions tested here provide statistically consistent fits to the data.

A related but distinct test of the intrinsic number counts for DSFGs
comes from the probability of a given source experiencing a magnification above
some threshold value, $\mu_{\rm min}$.  Unlike the case with the average
magnification factor measurements, our ALMA and SMA data should provide a
reliable estimate of this quantity.  The results of this are shown in the right
panel of Figure~\ref{fig:lensstats}.  For clarity of presentation, we show one
choice of $\mu_{\rm min}$: $\mu_{\rm min} = 2.0$.  The shape of the curves
varies with  $\mu_{\rm min}$, but the overall results are qualitatively the
same.  The models we consider are the same as those used in the left panel of
Figure~\ref{fig:lensstats}.  Instead of computing a running average of the
data, we show a logistic regression fit to the data \citep[obtained with the
{\sc Scikit-learn} package;][]{scikitlearn}.


Both models tested in this paper exhibit a sharp transition from low
probability to high probability of being lensed, consistent with the data.
However, there are significant differences in where this transition flux
density, $S_{\rm trans}$, occurs --- i.e., where $P(\mu>\mu_{870}) > 2$.  In
the data, the logistic regression fit yields $S_{\rm trans} = 24\pm3\,$mJy
(error accounts only for statistical uncertainty in $S_{870}$ and $\mu_{870}$),
whereas the models based on the ALESS and COSMOS number counts yield
$S_{\rm trans} = 37\,$mJy $S_{\rm trans} = 69\,$mJy, respectively. 

This analysis highlights the difficulty encountered with the luminosity function
based on the COSMOS data: unlensed sources with $S_{870} > 50\,$mJy are
over-predicted and lensed sources with $S_{870} < 50\,$mJy are under-predicted.
If the ALESS number counts continue to be supported by the evidence as
additional data are obtained \citep[e.g.;][]{Simpson:2015ls}, the implications
are significant.  We should then expect to find $\approx 3$ sources satisfying
$S_{870} > 10\,$mJy in a 1$\,$deg$^2$ survey.  This is about a factor of 7 lower
than typical measurements from single-dish, broad-beam studies
\citep[e.g.,][]{Weis:2009ly}.  This suggests that very luminous galaxies such as
GN20 and HFLS3 may be more rare than previously thought.




\section{Conclusions} \label{sec:conclusions}

We present ALMA 870$\,\mu$m $0\farcs45$ imaging of 29 {\it Herschel} DSFGs
selected from 55$\,$deg$^2$ of HerMES.  The {\it Herschel} sources have
$S_{500} = 52 - 134\,$mJy, placing them in a unique phase space between the
brightest sources found by {\it Herschel} and those found in ground-based
surveys at sub-mm wavelengths that include more typical, fainter galaxies.  Our
ALMA observations reveal 62 sources down to the 5$\sigma$ limit ($\sigma
\approx 0.3\,$mJy, typically).  We make use of optical and near-IR imaging to
assess the distribution of intervening galaxies along the line of sight.  We
introduce a new, publicly available software called {\sc uvmcmcfit} and use it
to develop lens models for all ALMA sources with nearby foreground
galaxy.  Our results from this effort are summarized as follows:

\begin{enumerate}

    \item 36/62 ALMA sources experience significant amplification from a nearby
        foreground galaxy that is comparable to or greater than the absolute
        calibration uncertainty (i.e., $\mu_{870} > 1.1$).  The median
        amplification in the subset that experiences lensing is $\mu_{870} =
        1.6$.  Only 6 sources show morphology typical of strong gravitational
        lensing and could be identified as lenses from the ALMA imaging alone.
        A multi-wavelength approach is critical to identifying structure along
        the line of sight and determining an unbiased measurement of the flux
        densities in our sample.

    \item 20/29 {\it Herschel} DSFGs break down into multiple ALMA
        counterparts.  Of the 9 isolated systems, 5 are strongly lensed by
        factors of 5-30 (HECDFS12 is a non-isolated system with a strongly
        lensed source in it).  After correcting for amplification, the brightest
        source in the sample has $S_{870} = 17.5 \pm 0.4\,$mJy.  There
        is a weak trend towards even higher multiplicity at the highest total
        $S_{870}$ flux densities.

    \item When a {\it Herschel} source comprises multiple ALMA counterparts,
        these counterparts are typically located within 2$\arcsec$ of each
        other.  Their separations are significantly smaller than ALMA
        counterparts to ALESS sources as well as simulated sources from
        \citetalias{HB13} and \citetalias{Cowley:2015lr}, although the improved
        treatment of blending by the latter yields superior agreement with the
        data.  This conclusion remains true even when we degrade our
        ALMA observations to match the spatial resolution, sensitivity, and
        {\it uv} coverage of the ALESS observations.

    \item Intrinsic number counts for DSFGs with a form that matches
        observations in COSMOS \citep{Miettinen:2015lr} under-predict the
        number of lensed sources with apparent $S_{870} > 10\,$mJy.  Number
        counts based on ALESS observations provide a better match
        to our magnification measurements.  The interpretation of these results
        is complicated by the fact that our sample is likely biased towards
        blends of multiple sources within a {\it Herschel} beam.  Our primary
        goal is to draw attention to this analysis method as a means to test
        number counts of DSFGs using wide-field {\it Herschel} data. 

\end{enumerate}

If the ALESS number counts continue to provide the best predicted
magnification factors in larger samples with weaker biases, this suggests that
galaxies with intrinsic flux densities above $S_{870}^\prime \approx 10\,$mJy
are extremely rare.  One possible explanation for their rarity is that they are
simply the tip of the mass function among starbursts.  An alternative is that
they represent a very short phase in galaxy evolution.  It is interesting to
note that consistent with this idea is the high multiplicity rate in our ALMA
sample as well as the small projected separations between multiple ALMA
counterparts.  The inability of numerical simulations to reproduce the small
projected separations seen in the data might highlight a productive path
forward to improve our theoretical understanding of the enhancement in
star-formation by interactions and mergers of galaxies which are already
forming stars at a very high rate.

In the future, higher spatial resolution imaging is needed to investigate the
morphologies of individual ALMA sources.  Tidal tails, multiple nuclei, and
other signs of mergers and interactions should become evident at $0\farcs1$
resolution.  In addition, molecular spectroscopy will be critical to determine
distances to and dynamics of individual ALMA sources and hence characterize
what fraction of {\it Herschel} sources are actually physically associated with
each other (and not simply a result of projection effects along the line of
sight).  Finally, deeper optical or near-IR imaging is needed to search for
dark lenses that may have been missed in our Gemini-S or {\it HST}
imaging.

\begin{acknowledgments}

We thank the referee for useful comments that helped improve the clarity of the
manuscript.  This paper makes use of the following ALMA data: ADS/JAO.ALMA\#
2011.0.00539.S.  ALMA is a partnership of ESO (representing its member states),
NSF (USA) and NINS (Japan), together with NRC (Canada) and NSC and ASIAA
(Taiwan), in cooperation with the Republic of Chile. The Joint ALMA Observatory
is operated by ESO, AUI/NRAO and NAOJ.

The results described in this paper are based on observations obtained with
{\it Herschel}, an ESA space observatory with science instruments provided by
European-led Principal Investigator consortia and with important participation
from NASA.  

This research has made use of data from the HerMES project
(http://hermes.sussex.ac.uk/). HerMES is a Herschel Key Programme utilizing
Guaranteed Time from the SPIRE instrument team, ESAC scientists, and a mission
scientist. HerMES is described in \citet{Oliver:2012lr}.  The HerMES data
presented in this paper will be released through the {\em Herschel} Database in
Marseille (HeDaM\footnote{http://hedam.oamp.fr/HerMES}).

SPIRE has been developed by a consortium of institutes led by Cardiff Univ.
(UK) and including: Univ. Lethbridge (Canada); NAOC (China); CEA, LAM (France);
IFSI, Univ. Padua (Italy); IAC (Spain); Stockholm Observatory (Sweden);
Imperial College London, RAL, UCL-MSSL, UKATC, Univ. Sussex (UK); and Caltech,
JPL, NHSC, Univ. Colorado (USA). This development has been supported by
national funding agencies: CSA (Canada); NAOC (China); CEA, CNES, CNRS
(France); ASI (Italy); MCINN (Spain); SNSB (Sweden); STFC, UKSA (UK); and NASA
(USA).

The VIDEO imaging data are based on data products from observations
made with ESO Telescopes at the La Silla or Paranal Observatories under ESO
programme ID 179.A-2006.

This paper is partly based on observations obtained at the Gemini Observatory,
which is operated by the Association of Universities for Research in Astronomy,
Inc., under a cooperative agreement with the NSF on behalf of the Gemini
partnership: the National Science Foundation (United States), the National
Research Council (Canada), CONICYT (Chile), the Australian Research Council
(Australia), Minist\'erio da Ci\^encia, Tecnologia e Inova\c{c}\~ao (Brazil)
and Ministerio de Ciencia, Tecnolog\'ia e Innovaci\'on Productiva (Argentina).  

The National Radio Astronomy Observatory is a facility of the National Science
Foundation operated under cooperative agreement by Associated Universities,
Inc.  The Dark Cosmology Centre is funded by the Danish National Research
Foundation. R.J.I. acknowledges support from European Research Council Advanced
Investigator Grant, Cosmicism, 321302. C.C.H. is grateful to the Gordon and
Betty Moore Foundation for financial support.

Facilities: ALMA, Gemini-S.

\end{acknowledgments}



\appendix

\section{{\it Herschel}/SPIRE Photometry}\label{sec:photometry}

We present in Table~\ref{tab:photometry} SPIRE photometry ($S_{250}$,
$S_{350}$, and $S_{500}$) for each {\it Herschel} source in our sample using as
priors the ALMA and {\it Spitzer} 24$\,\mu$m counterpart positions.  Many of
the {\it Herschel} sources in our sample have multiple 24$\,\mu$m counterparts
close enough to the ALMA counterparts that they can make a significant
contribution to the SPIRE flux at the position of the ALMA counterparts.  For
this reason, it is critical to include the 24$\,\mu$m data when estimating {\it
Herschel}/SPIRE photometry.  In cases where the {\it Spitzer} 24$\,\mu$m and
ALMA counterparts spatially overlap (defined here as having a separation
smaller than $2\farcs5$), we exclude the {\it Spitzer} counterpart from the
calculations.  

Note that in the case of HELAISS02, we exclude a 24$\,\mu$m source associated
with the lensing galaxy, despite being more than $2\farcs5$ away from the
nearest ALMA counterpart.  This is because the ALMA counterparts are arranged
so that they surround the 24$\mu$m source completely.  The extended emission
seen in the SPIRE maps is therefore most likely attributable to the ALMA
counterparts rather than the 24$\,\mu$m source.

One of the key results from this analysis is that the use of StarFinder to
deconvolve the SPIRE beam will lead to the removal of a portion of the targets
with extended emission from a flux-limited sample.  This is by design, as the
goal outlined in \citet{Wardlow:2013lr} was to develop the purest sample of
lens candidates as possible from HerMES data.  Blends were considered by these
authors to be unlikely to be lensed.  This method is effective for lenses where
the Einstein radius is smaller than $\approx 2\arcsec$.  However, it selects
against deeper potential wells, such as those of groups or clusters
\citep[e.g., HLock01][]{Gavazzi:2011lr}, that produce images separated by
scales that comprise a significant fraction of the SPIRE PSF.

It is worth emphasizing that blends likely constitute an interesting path of
study for future work, as they potentially represent proto-groups or
proto-clusters during a particular active stage of their evolution.  They are
also prime examples of systems of sources that are poorly reproduced in
simulations, as evidenced by the investigations in
Section~\ref{sec:spatialdist} in this paper.  Table~\ref{tab:photometry} shows
that one possible means of selecting candidate blends is by comparing the
SUSSEXtractor and StarFinder flux densities, particularly at 500$\,\mu$m.  A
large difference between these two measurements likely indicates multiple
components separated by scales that are a significant fraction of the SPIRE
beam (e.g., HELAISS02, HXMM20, and HCOSMOS02).

\input{table_photometry}

\end{document}

%% file: table_positions.tex
\LongTables
\begin{deluxetable*}{llccccccccc}[!tbp]
\tabletypesize{\scriptsize}
\tablecolumns{11}

\tablecaption{Observed positions and flux densities of ALMA sources.  For each
{\it Herschel} source, we give the fiducial flux density in all SPIRE bands
(see main text), as well as the observed positions and flux densities of all
ALMA sources.  The statistical rms and synthesized beam FWHM in each ALMA map
($\sigma_{\rm ALMA}$ and $\Omega_{\rm ALMA}$, respectively) are also given.
Positional uncertainties (for unlensed sources) range from $\approx 0\farcs005$
for well-detected sources to $\approx 0\farcs15$ for the faintest soures in our
sample.  Uncertainties in flux density do not include the absolute calibration
uncertainty of $\approx 10\%$. Quoted uncertainties in {\it Herschel}
photometry are dominated by confusion noise \citep{Nguyen:2010fk}.}

\tablehead{
\colhead{} & 
\colhead{} & 
\colhead{RA$_{870}$} &
\colhead{Dec$_{870}$} &
\colhead{$S_{250}$} &
\colhead{$S_{350}$} &
\colhead{$S_{500}$} &
\colhead{$S_{870}$} &
\colhead{$\sigma_{\rm ALMA}$} &
\colhead{$\Omega_{\rm ALMA}$} &
\colhead{Lens}
\\
\colhead{IAU address\tablenotemark{a}} & 
\colhead{Short name} & 
\colhead{(J2000)} &
\colhead{(J2000)} &
\colhead{(mJy)} &
\colhead{(mJy)} &
\colhead{(mJy)} &
\colhead{(mJy)} &
\colhead{(mJy)} &
\colhead{($\arcsec \times \arcsec$)} &
\colhead{grade\tablenotemark{b}}
}
\startdata
J003823.6$-$433707            & HELAISS02  & 00:38:23.587 & $-$43:37:04.15  & $ 115 \pm  6$ & $ 124 \pm  6$ & $ 108 \pm  6$   &    $20.11\pm 0.45$ & 0.14 & $0.54\times0.44 $& --- \\
---                           & Source0    & 00:38:23.762 & $-$43:37:06.10  & --- & --- & ---                                 &    $ 9.22\pm 0.17$ & ---  & $      ---      $& C   \\
---                           & Source1    & 00:38:23.482 & $-$43:37:05.56  & --- & --- & ---                                 &    $ 4.34\pm 0.16$ & ---  & $      ---      $& C   \\
---                           & Source2    & 00:38:23.313 & $-$43:36:58.97  & --- & --- & ---                                 &    $ 4.16\pm 0.32$ & ---  & $      ---      $& C   \\
---                           & Source3    & 00:38:23.803 & $-$43:37:10.46  & --- & --- & ---                                 &    $ 2.40\pm 0.19$ & ---  & $      ---      $& C   \\
J021830.5$-$053124            & HXMM02     & 02:18:30.673 & $-$05:31:31.75  & $  78 \pm  7$ & $ 122 \pm  8$ & $  99 \pm  7$   &    $63.33\pm 0.58$ & 0.20 & $0.49\times0.37 $& A   \\
J021841.5$-$035002            & HXMM31     & 02:18:41.613 & $-$03:50:03.70  & $ 102 \pm  6$ & $  94 \pm  6$ & $  65 \pm  6$   &    $10.80\pm 0.46$ & 0.20 & $0.49\times0.37 $& --- \\
---                           & Source0    & 02:18:41.520 & $-$03:50:04.72  & --- & --- & ---                                 &    $ 6.79\pm 0.37$ & ---  & $      ---      $& C   \\
---                           & Source1    & 02:18:41.700 & $-$03:50:02.57  & --- & --- & ---                                 &    $ 4.01\pm 0.26$ & ---  & $      ---      $& C   \\
J021853.1$-$063325            & HXMM29     & 02:18:53.111 & $-$06:33:24.65  & $  97 \pm  6$ & $ 102 \pm  6$ & $  78 \pm  6$   &    $ 7.28\pm 0.45$ & 0.20 & $0.49\times0.37 $& --- \\
---                           & Source0    & 02:18:53.118 & $-$06:33:24.19  & --- & --- & ---                                 &    $ 5.46\pm 0.30$ & ---  & $      ---      $& C   \\
---                           & Source1    & 02:18:53.095 & $-$06:33:25.21  & --- & --- & ---                                 &    $ 1.82\pm 0.38$ & ---  & $      ---      $& C   \\
J021918.4$-$031051            & HXMM07     & 02:19:18.417 & $-$03:10:51.35  & $  89 \pm  7$ & $ 107 \pm  8$ & $  85 \pm  7$   &    $29.16\pm 0.58$ & 0.21 & $0.49\times0.38 $& A   \\
J021942.7$-$052436            & HXMM20     & 02:19:42.783 & $-$05:24:34.84  & $  72 \pm  6$ & $  85 \pm  6$ & $  66 \pm  6$   &    $17.49\pm 0.74$ & 0.20 & $0.49\times0.37 $& --- \\
---                           & Source0    & 02:19:42.629 & $-$05:24:37.11  & --- & --- & ---                                 &    $ 7.15\pm 0.44$ & ---  & $      ---      $& X   \\
---                           & Source1    & 02:19:42.838 & $-$05:24:35.11  & --- & --- & ---                                 &    $ 3.52\pm 0.41$ & ---  & $      ---      $& X   \\
---                           & Source2    & 02:19:42.769 & $-$05:24:36.48  & --- & --- & ---                                 &    $ 3.42\pm 0.26$ & ---  & $      ---      $& X   \\
---                           & Source3    & 02:19:42.682 & $-$05:24:36.82  & --- & --- & ---                                 &    $ 2.46\pm 0.47$ & ---  & $      ---      $& X   \\
---                           & Source4    & 02:19:42.955 & $-$05:24:32.22  & --- & --- & ---                                 &    $ 0.94\pm 0.18$ & ---  & $      ---      $& X   \\
J022016.5$-$060143            & HXMM01     & 02:20:16.609 & $-$06:01:43.18  & $ 179 \pm  7$ & $ 188 \pm  8$ & $ 134 \pm  7$   &    $29.56\pm 0.46$ & 0.20 & $0.48\times0.37 $& --- \\
---                           & Source0    & 02:20:16.648 & $-$06:01:41.93  & --- & --- & ---                                 &    $16.13\pm 0.31$ & ---  & $      ---      $& C   \\
---                           & Source1    & 02:20:16.571 & $-$06:01:44.56  & --- & --- & ---                                 &    $11.56\pm 0.32$ & ---  & $      ---      $& C   \\
---                           & Source2    & 02:20:16.609 & $-$06:01:40.72  & --- & --- & ---                                 &    $ 1.87\pm 0.26$ & ---  & $      ---      $& C   \\
J022021.7$-$015328            & HXMM04     & 02:20:21.756 & $-$01:53:30.92  & $ 162 \pm  7$ & $ 157 \pm  8$ & $ 125 \pm 11$   &    $20.03\pm 0.47$ & 0.23 & $0.53\times0.38 $& C   \\
J022029.2$-$064845            & HXMM09     & 02:20:29.140 & $-$06:48:46.49  & $ 129 \pm  7$ & $ 118 \pm  8$ & $  85 \pm  7$   &    $15.30\pm 0.36$ & 0.20 & $0.49\times0.37 $& --- \\
---                           & Source0    & 02:20:29.195 & $-$06:48:48.02  & --- & --- & ---                                 &    $ 8.93\pm 0.30$ & ---  & $      ---      $& C   \\
---                           & Source1    & 02:20:29.079 & $-$06:48:44.86  & --- & --- & ---                                 &    $ 6.37\pm 0.19$ & ---  & $      ---      $& C   \\
J022135.1$-$062617            & HXMM03     & 02:21:34.891 & $-$06:26:17.87  & $ 114 \pm  7$ & $ 134 \pm  8$ & $ 116 \pm  7$   &    $22.65\pm 0.36$ & 0.21 & $0.48\times0.38 $& --- \\
---                           & Source1    & 02:21:35.124 & $-$06:26:16.62  & --- & --- & ---                                 &    $18.42\pm 0.36$ & ---  & $      ---      $& C   \\
---                           & Source2    & 02:21:35.132 & $-$06:26:18.02  & --- & --- & ---                                 &    $ 2.19\pm 0.20$ & ---  & $      ---      $& C   \\
---                           & Source0    & 02:21:35.136 & $-$06:26:17.28  & --- & --- & ---                                 &    $ 2.03\pm 0.18$ & ---  & $      ---      $& C   \\
J022201.6$-$033340            & HXMM11     & 02:22:01.616 & $-$03:33:41.40  & $ 101 \pm  7$ & $ 104 \pm  8$ & $  73 \pm  7$   &    $11.72\pm 0.49$ & 0.20 & $0.52\times0.38 $& --- \\
---                           & Source0    & 02:22:01.592 & $-$03:33:39.42  & --- & --- & ---                                 &    $ 8.17\pm 0.32$ & ---  & $      ---      $& C   \\
---                           & Source1    & 02:22:01.629 & $-$03:33:43.58  & --- & --- & ---                                 &    $ 3.54\pm 0.36$ & ---  & $      ---      $& C   \\
J022205.4$-$070728            & HXMM23     & 02:22:05.362 & $-$07:07:28.10  & $ 128 \pm  6$ & $ 105 \pm  6$ & $  68 \pm  6$   &    $ 2.93\pm 0.15$ & 0.20 & $0.48\times0.37 $& X   \\
J022250.5$-$032410            & HXMM22     & 02:22:50.573 & $-$03:24:12.35  & $ 101 \pm  6$ & $  85 \pm  6$ & $  61 \pm  6$   &    $10.19\pm 0.28$ & 0.20 & $0.49\times0.38 $& C   \\
J022547.8$-$041750            & HXMM05     & 02:25:47.942 & $-$04:17:50.80  & $ 103 \pm  7$ & $ 118 \pm  8$ & $  97 \pm  7$   &    $17.96\pm 0.43$ & 0.20 & $0.50\times0.37 $& C   \\
J022944.7$-$034110            & HXMM30     & 02:29:44.740 & $-$03:41:09.57  & $  86 \pm  6$ & $  97 \pm  6$ & $  75 \pm  6$   &    $22.76\pm 0.28$ & 0.23 & $0.50\times0.38 $& A   \\
J023006.0$-$034152            & HXMM12     & 02:30:05.950 & $-$03:41:53.07  & $  98 \pm  7$ & $ 106 \pm  8$ & $  82 \pm  7$   &    $15.56\pm 0.37$ & 0.20 & $0.50\times0.38 $& C   \\
J032752.0$-$290908            & HECDFS12   & 03:27:52.011 & $-$29:09:10.40  & $  61 \pm  7$ & $  82 \pm  6$ & $  81 \pm  6$   &    $38.78\pm 0.56$ & 0.15 & $0.43\times0.35 $& --- \\
---                           & Source0    & 03:27:52.002 & $-$29:09:12.07  & --- & --- & ---                                 &    $16.76\pm 0.51$ & ---  & $      ---      $& A   \\
---                           & Source1    & 03:27:52.002 & $-$29:09:09.65  & --- & --- & ---                                 &    $14.55\pm 0.22$ & ---  & $      ---      $& C   \\
---                           & Source2    & 03:27:52.025 & $-$29:09:12.14  & --- & --- & ---                                 &    $ 7.47\pm 0.14$ & ---  & $      ---      $& X   \\
J033210.8$-$270535            & HECDFS04   & 03:32:10.840 & $-$27:05:34.18  & $  56 \pm  6$ & $  61 \pm  6$ & $  55 \pm  6$   &    $14.57\pm 0.26$ & 0.15 & $0.44\times0.35 $& --- \\
---                           & Source0    & 03:32:10.905 & $-$27:05:32.87  & --- & --- & ---                                 &    $11.91\pm 0.24$ & ---  & $      ---      $& C   \\
---                           & Source1    & 03:32:10.729 & $-$27:05:36.22  & --- & --- & ---                                 &    $ 2.66\pm 0.11$ & ---  & $      ---      $& C   \\
J033317.9$-$280907            & HECDFS13   & 03:33:18.017 & $-$28:09:07.52  & $  95 \pm  6$ & $  89 \pm  6$ & $  63 \pm  6$   &    $15.36\pm 0.27$ & 0.14 & $0.44\times0.35 $& --- \\
---                           & Source0    & 03:33:18.006 & $-$28:09:07.55  & --- & --- & ---                                 &    $10.11\pm 1.30$ & ---  & $      ---      $& X   \\
---                           & Source1    & 03:33:18.032 & $-$28:09:07.39  & --- & --- & ---                                 &    $ 5.25\pm 1.37$ & ---  & $      ---      $& X   \\
J043340.5$-$540337            & HADFS04    & 04:33:40.450 & $-$54:03:39.51  & $  74 \pm  6$ & $  93 \pm  6$ & $  84 \pm  6$   &    $18.12\pm 0.44$ & 0.19 & $0.54\times0.46 $& --- \\
---                           & Source0    & 04:33:40.455 & $-$54:03:40.29  & --- & --- & ---                                 &    $ 9.25\pm 0.30$ & ---  & $      ---      $& C   \\
---                           & Source1    & 04:33:40.501 & $-$54:03:40.05  & --- & --- & ---                                 &    $ 6.09\pm 0.33$ & ---  & $      ---      $& C   \\
---                           & Source2    & 04:33:40.472 & $-$54:03:38.33  & --- & --- & ---                                 &    $ 2.78\pm 0.19$ & ---  & $      ---      $& C   \\
J043619.3$-$552425            & HADFS02    & 04:36:19.702 & $-$55:24:25.01  & $ 102 \pm  6$ & $  97 \pm  6$ & $  81 \pm  5$   &    $16.79\pm 0.40$ & 0.19 & $0.54\times0.46 $& --- \\
---                           & Source0    & 04:36:19.706 & $-$55:24:24.41  & --- & --- & ---                                 &    $ 7.81\pm 0.47$ & ---  & $      ---      $& X   \\
---                           & Source1    & 04:36:19.698 & $-$55:24:25.27  & --- & --- & ---                                 &    $ 8.99\pm 0.58$ & ---  & $      ---      $& X   \\
J043829.7$-$541831            & HADFS11    & 04:38:30.883 & $-$54:18:29.38  & $  19 \pm  6$ & $  39 \pm  5$ & $  52 \pm  6$   &    $28.47\pm 0.64$ & 0.19 & $0.54\times0.46 $& --- \\
---                           & Source0    & 04:38:30.780 & $-$54:18:31.79  & --- & --- & ---                                 &    $21.19\pm 0.51$ & ---  & $      ---      $& C   \\
---                           & Source1    & 04:38:30.970 & $-$54:18:26.60  & --- & --- & ---                                 &    $ 7.28\pm 0.30$ & ---  & $      ---      $& C   \\
J044103.8$-$531240            & HADFS10    & 04:41:03.942 & $-$53:12:41.01  & $  47 \pm  6$ & $  58 \pm  6$ & $  58 \pm  6$   &    $17.44\pm 0.39$ & 0.20 & $0.55\times0.45 $& --- \\
---                           & Source0    & 04:41:03.866 & $-$53:12:41.33  & --- & --- & ---                                 &    $ 9.61\pm 0.25$ & ---  & $      ---      $& X   \\
---                           & Source1    & 04:41:04.000 & $-$53:12:40.10  & --- & --- & ---                                 &    $ 4.59\pm 0.23$ & ---  & $      ---      $& X   \\
---                           & Source2    & 04:41:03.912 & $-$53:12:42.09  & --- & --- & ---                                 &    $ 3.24\pm 0.19$ & ---  & $      ---      $& X   \\
J044153.9$-$540350            & HADFS01    & 04:41:53.880 & $-$54:03:53.48  & $  76 \pm  6$ & $ 100 \pm  6$ & $  94 \pm  6$   &    $32.79\pm 0.47$ & 0.19 & $0.54\times0.45 $& A   \\
J044946.9$-$525424            & HADFS09    & 04:49:46.448 & $-$52:54:26.95  & $  98 \pm  6$ & $ 102 \pm  6$ & $  72 \pm  6$   &    $15.52\pm 0.59$ & 0.19 & $0.54\times0.45 $& --- \\
---                           & Source0    & 04:49:46.603 & $-$52:54:23.66  & --- & --- & ---                                 &    $ 8.24\pm 0.26$ & ---  & $      ---      $& X   \\
---                           & Source1    & 04:49:46.301 & $-$52:54:30.26  & --- & --- & ---                                 &    $ 4.86\pm 0.34$ & ---  & $      ---      $& X   \\
---                           & Source2    & 04:49:46.280 & $-$52:54:26.06  & --- & --- & ---                                 &    $ 2.42\pm 0.35$ & ---  & $      ---      $& X   \\
J045026.5$-$524127            & HADFS08    & 04:50:27.453 & $-$52:41:25.41  & $ 142 \pm  6$ & $ 133 \pm  6$ & $  90 \pm  6$   &    $14.18\pm 0.50$ & 0.19 & $0.54\times0.45 $& --- \\
---                           & Source0    & 04:50:27.092 & $-$52:41:25.62  & --- & --- & ---                                 &    $ 6.17\pm 0.28$ & ---  & $      ---      $& C   \\
---                           & Source1    & 04:50:27.806 & $-$52:41:25.10  & --- & --- & ---                                 &    $ 8.01\pm 0.43$ & ---  & $      ---      $& C   \\
J045057.5$-$531654            & HADFS03    & 04:50:57.715 & $-$53:16:54.42  & $ 119 \pm  6$ & $ 102 \pm  6$ & $  63 \pm  6$   &    $11.50\pm 0.39$ & 0.19 & $0.54\times0.45 $& --- \\
---                           & Source0    & 04:50:57.610 & $-$53:16:55.09  & --- & --- & ---                                 &    $ 7.12\pm 0.22$ & ---  & $      ---      $& C   \\
---                           & Source1    & 04:50:57.805 & $-$53:16:56.96  & --- & --- & ---                                 &    $ 2.12\pm 0.14$ & ---  & $      ---      $& C   \\
---                           & Source2    & 04:50:57.741 & $-$53:16:54.54  & --- & --- & ---                                 &    $ 2.27\pm 0.26$ & ---  & $      ---      $& C   \\
J100056.6$+$022014            & HCOSMOS02  & 10:00:57.180 & $+$02:20:12.70  & $  70 \pm  6$ & $  85 \pm  6$ & $  71 \pm  6$   &    $14.61\pm 0.66$ & 0.15 & $0.63\times0.50 $& --- \\
---                           & Source0    & 10:00:56.946 & $+$02:20:17.35  & --- & --- & ---                                 &    $ 5.26\pm 0.26$ & ---  & $      ---      $& X   \\
---                           & Source1    & 10:00:57.565 & $+$02:20:11.26  & --- & --- & ---                                 &    $ 3.77\pm 0.32$ & ---  & $      ---      $& X   \\
---                           & Source2    & 10:00:56.855 & $+$02:20:08.93  & --- & --- & ---                                 &    $ 1.69\pm 0.25$ & ---  & $      ---      $& X   \\
---                           & Source3    & 10:00:57.274 & $+$02:20:12.66  & --- & --- & ---                                 &    $ 1.66\pm 0.21$ & ---  & $      ---      $& X   \\
---                           & Source4    & 10:00:57.400 & $+$02:20:10.83  & --- & --- & ---                                 &    $ 2.23\pm 0.41$ & ---  & $      ---      $& X   \\
J100144.1$+$025712            & HCOSMOS01  & 10:01:44.182 & $+$02:57:12.47  & $  86 \pm  6$ & $  96 \pm  6$ & $  71 \pm  6$   &    $15.35\pm 0.25$ & 0.14 & $0.64\times0.49 $& A   \\
\enddata
\label{tab:position}
\tablenotetext{a}{IAU name = 1HerMES S250 + IAU address}
\tablenotetext{b}{A = strongly lensed, C = weakly lensed, X = unlensed.  Discussion of lens grades are given in Section~\ref{sec:objectbyobject}.}
%
\end{deluxetable*}

%% file: table_lenses.tex
\LongTables
\begin{deluxetable*}{lccccc}[!tbp]
\tabletypesize{\scriptsize}
\tablecolumns{6}
\tablecaption{Lens properties from parameters of model fits to ALMA sources
(parameters are described in Section~\ref{sec:modelfitsmeth}). Parameters without uncertainties were fixed to the given value. }
\tablehead{
\colhead{} & 
\colhead{$\Delta$RA$_{870}$} &
\colhead{$\Delta$Dec$_{870}$} &
\colhead{$\theta_{\rm E}$} &
\colhead{} &
\colhead{$\phi_{\rm lens}$}
\\
\colhead{Short name} & 
\colhead{($\arcsec$)} &
\colhead{($\arcsec$)} &
\colhead{($\arcsec$)} &
\colhead{$q_{\rm lens}$} &
\colhead{(deg)}
}
\startdata
HELAISS02.Lens0 &  $-1.59\pm 0.20$ & $ 2.25\pm 0.19$ & $1.500$ & $0.790\pm0.067$ & $ 44\pm 16$  \\
HXMM02.Lens0    &  $ 0.01\pm 0.01$ & $-0.24\pm 0.01$ & $0.507\pm0.004$ & $0.596\pm0.009$ & $157\pm 10$  \\
HXMM07.Lens0    &  $-0.27\pm 0.03$ & $ 0.04\pm 0.13$ & $0.928\pm0.007$ & $0.902\pm0.024$ & $ 26\pm  7$  \\
HXMM01.Lens0    &  $ 2.05        $ & $ 0.60        $ & $0.500$ & $0.801\pm0.062$ & $ 48\pm 14$  \\
HXMM01.Lens1    &  $-2.80        $ & $ 1.00        $ & $0.500$ & $0.882\pm0.072$ & $ 90\pm 17$  \\
HXMM04.Lens0    &  $ 0.17\pm 0.03$ & $ 0.04\pm 0.03$ & $0.500$ & $0.547\pm0.050$ & $ 11\pm 16$  \\
HXMM09.Lens0    &  $ 1.40\pm 0.07$ & $ 0.19\pm 0.05$ & $1.000$ & $0.663\pm0.094$ & $ 64\pm 16$  \\
HXMM03.Lens0    &  $-2.50        $ & $-0.50        $ & $1.000$ & $1.000$ & $  0$  \\
HXMM11.Lens0    &  $ 0.82\pm 0.12$ & $ 2.95\pm 0.10$ & $0.500$ & $0.706\pm0.124$ & $ 67\pm 11$  \\
HXMM05.Lens0    &  $ 2.80        $ & $-1.40        $ & $1.000$ & $0.531\pm0.180$ & $ 45\pm 14$  \\
HXMM05.Lens1    &  $-1.90        $ & $ 2.50        $ & $1.000$ & $0.569\pm0.197$ & $ 67\pm 16$  \\
HXMM30.Lens0    &  $-0.03\pm 0.02$ & $ 0.05\pm 0.01$ & $0.743\pm0.008$ & $0.703\pm0.050$ & $ 26\pm 10$  \\
HXMM12.Lens0    &  $-0.22\pm 0.20$ & $-0.25\pm 0.24$ & $0.200$ & $0.672\pm0.090$ & $ 30\pm 16$  \\
HXMM12.Lens1    &  $ 4.50        $ & $-4.50        $ & $2.000$ & $1.000$ & $  0$  \\
HECDFS12.Lens0  &  $ 0.22        $ & $-1.75        $ & $1.354\pm0.006$ & $0.955\pm0.007$ & $ 80\pm 16$  \\
HECDFS04.Lens0  &  $ 1.01\pm 0.02$ & $ 2.10\pm 0.01$ & $0.500$ & $0.807\pm0.006$ & $176\pm 13$  \\
HADFS04.Lens0   &  $-0.56\pm 0.13$ & $ 0.11\pm 0.07$ & $0.500$ & $0.662\pm0.135$ & $ 37\pm 12$  \\
HADFS11.Lens0   &  $ 0.41\pm 0.04$ & $ 0.27\pm 0.12$ & $1.000$ & $0.723\pm0.068$ & $ 82\pm 19$  \\
HADFS01.Lens0   &  $-0.19\pm 0.01$ & $ 0.25\pm 0.01$ & $1.006\pm0.004$ & $0.794\pm0.008$ & $ 99\pm 10$  \\
HADFS08.Lens0   &  $-3.59\pm 0.06$ & $-2.32\pm 0.06$ & $1.500$ & $0.897\pm0.047$ & $ 74\pm 18$  \\
HADFS03.Lens0   &  $-0.40\pm 0.08$ & $ 1.32\pm 0.06$ & $1.000$ & $0.707\pm0.141$ & $ 93\pm 17$  \\
HCOSMOS01.Lens0 &  $-0.12\pm 0.01$ & $ 0.28\pm 0.02$ & $0.956\pm0.005$ & $0.775\pm0.025$ & $ 72\pm 10$  \\
\enddata
\label{tab:lenses}
\end{deluxetable*}

%% file: table_intrinsic.tex
\clearpage
\LongTables
\begin{deluxetable*}{llccccccc}[!tbp]
\tabletypesize{\scriptsize}
\tablecolumns{8}
\tablecaption{Intrinsic properties from parameters of model fits to ALMA
sources (parameters are described in Section~\ref{sec:modelfitsmeth}).
Uncertainties in flux densities do not include absolute calibration uncertainty
of $\approx 10\%$.  Note that some parameters such as source size may become
unreliable when the signal to noise ratio is below 10 \citep[e.g.,][]{Simpson:2015lr}.}
\tablehead{
\colhead{} & 
\colhead{$\Delta$RA$_{870}$} &
\colhead{$\Delta$Dec$_{870}$} &
\colhead{$S_{870}$} &
\colhead{$r_{\rm s}$} &
\colhead{} &
\colhead{$\phi_{\rm s}$} & 
\colhead{}
\\
\colhead{Short name} & 
\colhead{($\arcsec$)} &
\colhead{($\arcsec$)} &
\colhead{(mJy)} &
\colhead{($\arcsec$)} &
\colhead{$q_{\rm s}$} &
\colhead{(deg)} &
\colhead{$\mu_{870}$}
}
\startdata
HELAISS02.0     & $ 3.113\pm0.160$ & $-3.112\pm0.155$ & $ 8.02\pm 0.15$ & $0.096\pm0.005$ & $ 0.80\pm 0.05$ & $ 91\pm  6$ & $ 1.15\pm 0.07$ \\
HELAISS02.1     & $-0.111\pm0.114$ & $-2.172\pm0.183$ & $ 3.24\pm 0.12$ & $0.065\pm0.008$ & $ 0.84\pm 0.05$ & $ 87\pm  7$ & $ 1.34\pm 0.17$ \\
HELAISS02.2     & $-1.470\pm0.158$ & $ 1.774\pm0.145$ & $ 3.27\pm 0.25$ & $0.105\pm0.016$ & $ 0.86\pm 0.04$ & $120\pm  7$ & $ 1.27\pm 0.13$ \\
HELAISS02.3     & $ 4.039\pm0.165$ & $-7.216\pm0.174$ & $ 2.22\pm 0.18$ & $0.124\pm0.020$ & $ 0.79\pm 0.05$ & $ 77\pm  7$ & $ 1.08\pm 0.04$ \\
HXMM02.0        & $-0.278\pm0.008$ & $ 0.239\pm0.011$ & $11.88\pm 0.11$ & $0.122\pm0.003$ & $ 0.64\pm 0.02$ & $ 62\pm  2$ & $ 5.33\pm 0.19$ \\
HXMM31.0        & $-1.380\pm0.010$ & $-1.025\pm0.010$ & $ 6.79\pm 0.37$ & $0.141\pm0.011$ & $ 0.80\pm 0.12$ & $134\pm 36$ &      ---      \\
HXMM31.1        & $ 1.311\pm0.011$ & $ 1.124\pm0.010$ & $ 4.01\pm 0.26$ & $0.070\pm0.018$ & $ 0.59\pm 0.22$ & $ 52\pm 56$ &      ---      \\
HXMM29.0        & $ 0.114\pm0.009$ & $ 0.451\pm0.008$ & $ 5.46\pm 0.30$ & $0.088\pm0.012$ & $ 0.82\pm 0.14$ & $ 90\pm 44$ &      ---      \\
HXMM29.1        & $-0.236\pm0.034$ & $-0.562\pm0.030$ & $ 1.82\pm 0.38$ & $0.116\pm0.051$ & $ 0.70\pm 0.20$ & $ 88\pm 55$ &      ---      \\
HXMM07.0        & $ 0.016\pm0.238$ & $-0.016\pm0.283$ & $ 3.43\pm 0.07$ & $0.074\pm0.007$ & $ 0.32\pm 0.02$ & $ 66\pm  2$ & $ 8.49\pm 1.13$ \\
HXMM20.0        & $-2.308\pm0.012$ & $-2.275\pm0.011$ & $ 7.15\pm 0.44$ & $0.089\pm0.014$ & $ 0.63\pm 0.16$ & $ 58\pm 27$ &      ---      \\
HXMM20.1        & $ 0.828\pm0.025$ & $-0.278\pm0.023$ & $ 3.52\pm 0.41$ & $0.137\pm0.026$ & $ 0.84\pm 0.10$ & $ 74\pm 44$ &      ---      \\
HXMM20.2        & $-0.211\pm0.017$ & $-1.647\pm0.014$ & $ 3.42\pm 0.26$ & $0.058\pm0.020$ & $ 0.80\pm 0.13$ & $ 84\pm 45$ &      ---      \\
HXMM20.3        & $-1.505\pm0.157$ & $-1.981\pm0.064$ & $ 2.46\pm 0.47$ & $0.283\pm0.198$ & $ 0.67\pm 0.17$ & $ 81\pm 21$ &      ---      \\
HXMM20.4        & $ 2.588\pm0.155$ & $ 2.611\pm0.218$ & $ 0.94\pm 0.18$ & $0.459\pm0.246$ & $ 0.58\pm 0.15$ & $ 96\pm 51$ &      ---      \\
HXMM01.0        & $-1.503\pm0.013$ & $ 0.395\pm0.017$ & $11.61\pm 0.23$ & $0.090\pm0.005$ & $ 0.56\pm 0.06$ & $ 12\pm 19$ & $ 1.39\pm 0.19$ \\
HXMM01.1        & $-2.563\pm0.018$ & $-1.337\pm0.017$ & $ 9.56\pm 0.26$ & $0.116\pm0.006$ & $ 0.34\pm 0.03$ & $  2\pm  1$ & $ 1.21\pm 0.10$ \\
HXMM01.2        & $-2.622\pm0.025$ & $-0.552\pm0.025$ & $ 1.45\pm 0.20$ & $0.077\pm0.025$ & $ 0.66\pm 0.18$ & $134\pm 33$ & $ 1.29\pm 0.15$ \\
HXMM04.0        & $ 0.095\pm0.021$ & $ 0.442\pm0.025$ & $ 8.49\pm 0.20$ & $0.117\pm0.007$ & $ 0.52\pm 0.07$ & $ -2\pm  5$ & $ 2.36\pm 0.68$ \\
HXMM09.0        & $-0.392\pm0.039$ & $-0.740\pm0.051$ & $ 5.51\pm 0.19$ & $0.064\pm0.006$ & $ 0.42\pm 0.06$ & $ 75\pm  5$ & $ 1.62\pm 0.31$ \\
HXMM09.1        & $-1.507\pm0.073$ & $ 0.805\pm0.053$ & $ 5.14\pm 0.15$ & $0.033\pm0.010$ & $ 0.46\pm 0.18$ & $116\pm 14$ & $ 1.24\pm 0.12$ \\
HXMM03.0        & $ 5.180\pm0.003$ & $ 0.924\pm0.003$ & $12.28\pm 0.24$ & $0.130\pm0.004$ & $ 0.53\pm 0.03$ & $-25\pm  2$ & $ 1.50\pm 0.25$ \\
HXMM03.1        & $ 7.560\pm0.023$ & $ 2.051\pm0.028$ & $ 1.46\pm 0.13$ & $0.093\pm0.007$ & $ 0.73\pm 0.11$ & $ 22\pm 25$ & $ 1.50\pm 0.25$ \\
HXMM03.2        & $ 7.663\pm0.035$ & $ 0.755\pm0.033$ & $ 1.35\pm 0.12$ & $0.096\pm0.005$ & $ 0.73\pm 0.11$ & $-11\pm 37$ & $ 1.50\pm 0.25$ \\
HXMM11.0        & $-0.844\pm0.111$ & $-0.648\pm0.081$ & $ 6.24\pm 0.24$ & $0.106\pm0.007$ & $ 0.26\pm 0.03$ & $ 54\pm  2$ & $ 1.31\pm 0.16$ \\
HXMM11.1        & $-0.596\pm0.122$ & $-4.592\pm0.098$ & $ 3.38\pm 0.35$ & $0.168\pm0.023$ & $ 0.59\pm 0.16$ & $139\pm 41$ & $ 1.05\pm 0.03$ \\
HXMM23.0        & $ 0.101\pm0.011$ & $-0.050\pm0.009$ & $ 2.93\pm 0.15$ & $0.020\pm0.008$ & $ 0.68\pm 0.20$ & $ 89\pm 49$ &      ---      \\
HXMM22.0        & $-0.076\pm0.004$ & $ 0.024\pm0.004$ & $10.19\pm 0.28$ & $0.085\pm0.010$ & $ 0.52\pm 0.11$ & $152\pm  6$ &      ---      \\
HXMM05.0        & $-3.505\pm0.094$ & $ 1.937\pm0.081$ & $12.83\pm 0.31$ & $0.095\pm0.006$ & $ 0.59\pm 0.06$ & $142\pm  5$ & $ 1.40\pm 0.20$ \\
HXMM30.0        & $ 0.153\pm0.024$ & $-0.073\pm0.011$ & $ 0.84\pm 0.01$ & $0.019\pm0.003$ & $ 0.20\pm 0.00$ & $109\pm  1$ & $27.15\pm 4.61$ \\
HXMM12.0        & $ 1.520\pm0.168$ & $-0.683\pm0.243$ & $ 9.91\pm 0.24$ & $0.115\pm0.005$ & $ 0.72\pm 0.07$ & $ 69\pm  8$ & $ 1.57\pm 0.29$ \\
HECDFS12.0      & $-0.348\pm0.006$ & $ 0.077\pm0.004$ & $ 2.02\pm 0.06$ & $0.085\pm0.004$ & $ 0.38\pm 0.03$ & $134\pm  3$ & $ 8.29\pm 0.19$ \\
HECDFS12.1      & $-0.342\pm0.005$ & $ 2.489\pm0.008$ & $11.54\pm 0.18$ & $0.147\pm0.003$ & $ 0.65\pm 0.02$ & $ 14\pm  2$ & $ 1.26\pm 0.13$ \\
HECDFS12.2      &     0.000        &       0.000      & $ 7.47\pm 0.14$ & $0.026\pm0.009$ & $ 0.79\pm 0.15$ & $ 85\pm 63$ &      ---      \\
HECDFS04.0      & $-0.011\pm0.011$ & $-0.347\pm0.004$ & $ 6.02\pm 0.12$ & $0.096\pm0.005$ & $ 0.35\pm 0.03$ & $ 91\pm  2$ & $ 1.98\pm 0.49$ \\
HECDFS04.1      & $-2.366\pm0.024$ & $-3.752\pm0.007$ & $ 2.51\pm 0.10$ & $0.032\pm0.012$ & $ 0.68\pm 0.19$ & $ 93\pm 55$ & $ 1.06\pm 0.03$ \\
HECDFS13.0      & $-0.156\pm0.011$ & $-0.034\pm0.011$ & $10.11\pm 1.30$ & $0.099\pm0.012$ & $ 0.52\pm 0.12$ & $123\pm  7$ &      ---      \\
HECDFS13.1      & $ 0.221\pm0.061$ & $ 0.127\pm0.018$ & $ 5.25\pm 1.37$ & $0.109\pm0.024$ & $ 0.38\pm 0.08$ & $ 88\pm  7$ &      ---      \\
HADFS04.0       & $ 0.333\pm0.101$ & $-0.513\pm0.040$ & $ 6.85\pm 0.22$ & $0.091\pm0.006$ & $ 0.39\pm 0.05$ & $142\pm  4$ & $ 1.35\pm 0.17$ \\
HADFS04.1       & $ 0.865\pm0.123$ & $-0.420\pm0.041$ & $ 5.03\pm 0.27$ & $0.165\pm0.013$ & $ 0.43\pm 0.06$ & $141\pm  4$ & $ 1.21\pm 0.10$ \\
HADFS04.2       & $ 0.604\pm0.108$ & $ 0.739\pm0.077$ & $ 1.99\pm 0.14$ & $0.077\pm0.015$ & $ 0.75\pm 0.16$ & $101\pm 40$ & $ 1.40\pm 0.20$ \\
HADFS02.0       & $ 0.067\pm0.008$ & $ 0.588\pm0.015$ & $ 7.81\pm 0.47$ & $0.136\pm0.012$ & $ 0.38\pm 0.06$ & $ 23\pm  5$ &      ---      \\
HADFS02.1       & $-0.060\pm0.009$ & $-0.268\pm0.018$ & $ 8.99\pm 0.58$ & $0.193\pm0.015$ & $ 0.42\pm 0.06$ & $ 17\pm  4$ &      ---      \\
HADFS11.0       & $-1.340\pm0.043$ & $-1.816\pm0.119$ & $17.51\pm 0.42$ & $0.225\pm0.006$ & $ 0.46\pm 0.02$ & $178\pm  1$ & $ 1.21\pm 0.11$ \\
HADFS11.1       & $ 0.658\pm0.039$ & $ 1.569\pm0.111$ & $ 5.78\pm 0.24$ & $0.180\pm0.010$ & $ 0.25\pm 0.02$ & $167\pm  2$ & $ 1.26\pm 0.13$ \\
HADFS10.0       & $-1.126\pm0.005$ & $-0.319\pm0.004$ & $ 9.61\pm 0.25$ & $0.073\pm0.010$ & $ 0.67\pm 0.15$ & $133\pm 24$ &      ---      \\
HADFS10.1       & $ 0.876\pm0.011$ & $ 0.908\pm0.009$ & $ 4.59\pm 0.23$ & $0.048\pm0.019$ & $ 0.71\pm 0.19$ & $ 84\pm 43$ &      ---      \\
HADFS10.2       & $-0.437\pm0.017$ & $-1.088\pm0.016$ & $ 3.24\pm 0.19$ & $0.093\pm0.020$ & $ 0.58\pm 0.20$ & $131\pm 38$ &      ---      \\
HADFS01.0       & $ 0.131\pm0.005$ & $-0.105\pm0.006$ & $ 3.17\pm 0.05$ & $0.128\pm0.005$ & $ 0.30\pm 0.01$ & $ 24\pm  1$ & $10.34\pm 0.47$ \\
HADFS09.0       & $ 2.343\pm0.007$ & $ 3.284\pm0.005$ & $ 8.24\pm 0.26$ & $0.109\pm0.008$ & $ 0.70\pm 0.11$ & $ 92\pm 14$ &      ---      \\
HADFS09.1       & $-2.191\pm0.013$ & $-3.320\pm0.011$ & $ 4.86\pm 0.34$ & $0.099\pm0.019$ & $ 0.53\pm 0.17$ & $135\pm 24$ &      ---      \\
HADFS09.2       & $-2.503\pm0.035$ & $ 0.886\pm0.019$ & $ 2.42\pm 0.35$ & $0.122\pm0.040$ & $ 0.51\pm 0.17$ & $ 89\pm 20$ &      ---      \\
HADFS08.0       & $-0.868\pm0.050$ & $ 0.938\pm0.048$ & $ 3.74\pm 0.17$ & $0.055\pm0.010$ & $ 0.83\pm 0.09$ & $131\pm 20$ & $ 1.65\pm 0.32$ \\
HADFS08.1       & $ 7.496\pm0.058$ & $ 2.190\pm0.059$ & $ 7.28\pm 0.39$ & $0.179\pm0.012$ & $ 0.59\pm 0.08$ & $ 63\pm  6$ & $ 1.10\pm 0.05$ \\
HADFS03.0       & $-0.734\pm0.069$ & $-1.070\pm0.053$ & $ 5.39\pm 0.17$ & $0.112\pm0.006$ & $ 0.41\pm 0.05$ & $ 45\pm  3$ & $ 1.32\pm 0.16$ \\
HADFS03.1       & $ 1.415\pm0.056$ & $-2.912\pm0.059$ & $ 1.87\pm 0.13$ & $0.059\pm0.018$ & $ 0.54\pm 0.12$ & $ 94\pm 48$ & $ 1.13\pm 0.07$ \\
HADFS03.2       & $ 0.427\pm0.055$ & $-0.514\pm0.059$ & $ 1.22\pm 0.14$ & $0.084\pm0.020$ & $ 0.50\pm 0.13$ & $125\pm 14$ & $ 1.86\pm 0.43$ \\
HCOSMOS02.0     & $-3.507\pm0.012$ & $ 4.659\pm0.013$ & $ 5.26\pm 0.26$ & $0.073\pm0.017$ & $ 0.70\pm 0.12$ & $ 94\pm 34$ &      ---      \\
HCOSMOS02.1     & $ 5.780\pm0.019$ & $-1.434\pm0.026$ & $ 3.77\pm 0.32$ & $0.094\pm0.029$ & $ 0.76\pm 0.13$ & $106\pm 65$ &      ---      \\
HCOSMOS02.2     & $-4.869\pm0.049$ & $-3.769\pm0.050$ & $ 1.69\pm 0.25$ & $0.198\pm0.051$ & $ 0.65\pm 0.13$ & $ 72\pm 41$ &      ---      \\
HCOSMOS02.3     & $ 1.410\pm0.031$ & $-0.035\pm0.033$ & $ 1.66\pm 0.21$ & $0.101\pm0.042$ & $ 0.71\pm 0.13$ & $ 74\pm 42$ &      ---      \\
HCOSMOS02.4     & $ 3.301\pm0.083$ & $-1.864\pm0.060$ & $ 2.23\pm 0.41$ & $0.312\pm0.060$ & $ 0.67\pm 0.13$ & $ 78\pm 32$ &      ---      \\
HCOSMOS01.0     & $ 0.136\pm0.011$ & $-0.220\pm0.016$ & $ 1.03\pm 0.02$ & $0.068\pm0.006$ & $ 0.27\pm 0.04$ & $164\pm  2$ & $14.86\pm 1.90$ \\
\enddata
\label{tab:intrinsic}
%
\end{deluxetable*}

%% file: table_lf.tex
\begin{deluxetable}{lcccc}[!tp]
\tabletypesize{\scriptsize}
\tablecolumns{5}
\tablecaption{Parameters of DSFG luminosity functions tested in this paper. }
\tablehead{
\colhead{} & 
\colhead{$N_\star$} & 
\colhead{$S_\star$} &
\colhead{} &
\colhead{} \\
\colhead{Luminosity Function} & 
\colhead{(deg$^{-2}$)} & 
\colhead{(mJy)} &
\colhead{$\beta_1$} &
\colhead{$\beta_2$}
}
\startdata
ALESS broken power-law & 20 & 8 & 2 & 6.9 \\
COSMOS broken power-law & 20 & 15 & 2 & 6.9 \\
\enddata
\label{tab:models}
%
\end{deluxetable}

%% file: table_photometry.tex
\begin{deluxetable*}{lccccccccc}[!tbp]
\tabletypesize{\scriptsize}
\tablecolumns{10}
\tablecaption{Compilation of {\it Herschel}/SPIRE flux density measurements.
For each {\it Herschel} source, we give the fiducial flux densities
(denoted in table as ``Fiducial''), the initial flux densities obtained using
SUSSEXtractor that were then used to generate
the target list for the ALMA observations (``SUSSEXtractor''), and the flux
densities obtained subsequently using the more sophisticated deblending
algorithm from StarFinder that were then used to construct the list of lens
candidates presented in \citet{Wardlow:2013lr} (``StarFinder'').
In all cases, uncertainties are
comparable to the uncertainties given in Table~\ref{tab:position}.}
\tablehead{
\colhead{} & 
\multicolumn{3}{c}{Fiducial} &
\multicolumn{3}{c}{StarFinder} &
\multicolumn{3}{c}{SUSSEXtractor}
\\
\colhead{} & 
\colhead{$S_{250}$} &
\colhead{$S_{350}$} &
\colhead{$S_{500}$} &
\colhead{$S_{250}$} &
\colhead{$S_{350}$} &
\colhead{$S_{500}$} &
\colhead{$S_{250}$} &
\colhead{$S_{350}$} &
\colhead{$S_{500}$}
\\
\colhead{Short name} & 
\colhead{(mJy)} &
\colhead{(mJy)} &
\colhead{(mJy)} &
\colhead{(mJy)} &
\colhead{(mJy)} &
\colhead{(mJy)} &
\colhead{(mJy)} &
\colhead{(mJy)} &
\colhead{(mJy)}
}
\startdata
HELAISS02   &  115 & 124 & 108 & 114 & 101 &  76 & 105 & 128 & 103 \\
HXMM02      &   78 & 122 &  99 &  92 & 122 & 113 & 101 & 147 & 141 \\
HXMM31      &  102 &  94 &  65 & 128 & 112 &  73 & 129 & 116 &  80 \\
HXMM29      &   97 & 102 &  78 &  89 &  83 &  56 & 100 & 107 &  80 \\
HXMM07      &   89 & 107 &  85 &  91 & 104 &  86 &  92 & 104 &  83 \\
HXMM20      &   72 &  85 &  66 &  85 &  79 &  67 &  80 &  96 &  88 \\
HXMM01      &  179 & 188 & 134 & 180 & 192 & 132 & 178 & 195 & 137 \\
HXMM04      &  162 & 157 & 125 & 144 & 137 &  94 & 173 & 174 & 127 \\
HXMM09      &  129 & 118 &  85 & 120 & 115 &  84 & 135 & 113 &  95 \\
HXMM03      &  114 & 134 & 116 & 121 & 132 & 110 & 118 & 137 & 118 \\
HXMM11      &  101 & 104 &  73 & 107 & 108 &  81 & 105 & 121 &  94 \\
HXMM23      &  128 & 105 &  68 & 137 & 108 &  57 & 132 & 118 &  88 \\
HXMM22      &  101 &  85 &  61 &  97 &  82 &  62 & 147 & 128 &  89 \\
HXMM05      &  103 & 118 &  97 & 106 & 119 &  92 & 103 & 115 & 101 \\
HXMM30      &   86 &  97 &  75 &  90 & 100 &  75 &  93 & 105 &  80 \\
HXMM12      &   98 & 106 &  82 & 102 & 110 &  81 & 107 & 115 &  89 \\
HECDFS12    &   61 &  82 &  81 &  28 &  84 &  85 &  68 &  92 & 100 \\
HECDFS04    &   56 &  61 &  55 &  73 &  86 &  85 &  65 &  87 &  96 \\
HECDFS13    &   95 &  89 &  63 &  96 &  90 &  63 &  88 &  85 &  51 \\
HADFS04     &   74 &  93 &  84 &  76 &  90 &  72 &  71 &  95 &  87 \\
HADFS02     &  102 &  97 &  81 & 110 & 102 &  87 & 103 & 100 &  79 \\
HADFS11     &   19 &  39 &  52 &  57 &  78 &  75 &  57 &  87 &  97 \\
HADFS10     &   47 &  58 &  58 &  96 &  86 &  57 & 121 & 114 &  76 \\
HADFS01     &   76 & 100 &  94 &  80 & 103 &  93 &  72 & 108 &  87 \\
HADFS09     &   98 & 102 &  72 & 115 &  61 &  24 & 112 & 117 &  86 \\
HADFS08     &  142 & 133 &  90 &  88 &  81 &  50 & 126 & 130 & 102 \\
HADFS03     &  119 & 102 &  63 & 138 & 114 &  73 & 134 & 124 &  86 \\
HCOSMOS02   &   70 &  85 &  71 &  71 &  64 &  41 &  82 &  99 &  89 \\
HCOSMOS01   &   86 &  96 &  71 &  91 & 100 &  74 &  89 &  99 &  73 \\
\enddata
\label{tab:photometry}
\end{deluxetable*}